\documentclass[a4paper,11pt]{article}
\usepackage{jheppub} 
\usepackage{lineno}

\usepackage{comment}
\usepackage{amsmath}
\usepackage{amssymb}
\usepackage{physics}
\usepackage{graphicx}
\usepackage{adjustbox}
\usepackage{xcolor}
\usepackage{cancel}
\usepackage{mathtools}
\usepackage{tikz}
\usepackage{makecell}
\usepackage{ tabularray }
\usepackage{empheq}
\usepackage{slashed}

\usepackage{nicematrix}
\usepackage{tikz}

\usepackage[table]{xcolor}
\definecolor{myGray}{HTML}{DDDDDD}

\newcommand{\bphi}{\bar{\phi}}
\newcommand{\bPhi}{\bar{\Phi}}
\newcommand{\bu}{{\bar{u}}}
\newcommand{\bw}{{\bar{w}}}
\newcommand{\bz}{{\bar{z}}}

\newcommand{\bj}{{\bar{\jmath}}}

\makeatletter
\newcommand{\linethrough}{\mathpalette\@thickbar}
\newcommand{\@thickbar}[2]{{#1\mkern0mu\vbox{
    \sbox\z@{$#1#2\mkern-1.5mu$}%
    \dimen@=\dimexpr\ht\tw@-\ht\z@+2\p@\relax 
    \hrule\@height0.5\p@ 
    \vskip\dimen@
    \box\z@}}
}
\makeatother
\newcommand{\mathstrike}[1]{\ensuremath{\linethrough{#1}}}

\newcommand{\intL}{\int \! \frac{d^D L}{(2 \pi)^D}}
\newcommand{\intLp}{\int \! \frac{d^D L'}{(2 \pi)^D}}

\newcommand{\lr}[1]{\langle  {#1} \rangle}
\newcommand{\br}[1]{ \lr{\alpha| #1 | \alpha} }

\newcommand{\bT}{\mathbf{T}}
\newcommand{\ep}{\epsilon}
\newcommand{\vep}{\varepsilon}

\newcommand{\ppm}{$(+$$+$$-)$ }
\newcommand{\mmp}{$(-$$-$$+)$ }
\newcommand{\bigsquare}{\raisebox{-0.1\baselineskip}{\Large\ensuremath{\blacksquare}}}

\newcommand{\mI}{I}

\newcommand{\tlambda}{\tilde{\lambda}}
\newcommand{\talpha}{\tilde{\alpha}}
\newcommand{\tbeta}{\tilde{\beta}}

\newcommand{\txi}{\tilde{\xi}}

\newcommand{\phiclJ}{\varphi_{{\rm cl},J}}
\newcommand{\phiclj}{\varphi_{{\rm cl},j}}

\newcommand{\tvarphi}{\tilde{\varphi}}
\newcommand{\tJ}{\tilde{J}}

\newcommand{\mD}{\mathcal{D}}

\newcommand{\mJ}{\mathcal{J}}

\newcommand{\bigPhi}{\raisebox{-0.1\baselineskip}{\Large\ensuremath{\Phi}}}
\newcommand{\bigphi}{\raisebox{-0\baselineskip}{\Large\ensuremath{\phi}}}

\newcommand{\mystack}[2]{\begin{subarray}{l} #1 \\ #2 \vphantom{\Big\rvert} \end{subarray}}
\newcommand{\mystackthree}[3]{\begin{subarray}{l} #1 \\ #2 \vphantom{\Big\rvert} \\ #3 \end{subarray}}

\newcommand{\bh}{\bar{h}}

\newcommand{\Lw}{\text{L}\myw_{1+\infty}}
\newcommand{\Ls}{\text{L}\mathfrak{s}}

\newcommand{\bvert}{\big\rvert}

\newcommand{\gym}{g}
\definecolor{myHexColor}{HTML}{AAAAAA}

\DeclareFontFamily{U}{jkpmia}{}
\DeclareFontShape{U}{jkpmia}{m}{it}{<->s*jkpmia}{}
\DeclareFontShape{U}{jkpmia}{bx}{it}{<->s*jkpbmia}{}
\DeclareMathAlphabet{\mathfrakalt}{U}{jkpmia}{m}{it}
\SetMathAlphabet{\mathfrakalt}{bold}{U}{jkpmia}{bx}{it}

\newcommand{\myw}{\mathfrakalt{w}}

\title{The facts on quantum self-dual Einstein gravity and self-dual Yang-Mills theory}

\author{Noah Miller}

\affiliation{Princeton Gravity Initiative, Princeton University}
\affiliation{School of Natural Sciences, Institute for Advanced Study}

\emailAdd{noahmiller@ias.edu}

\abstract{

Self-dual Einstein gravity is a one-loop exact, UV and IR finite, diffeomorphism invariant, non-unitary theory of quantum gravity in four dimensions. The only generically non-zero amplitudes are the one-loop-all-plus amplitudes, and they are equal to those of pure Einstein gravity.

In these notes we review SDG and SDYM pedagogically. We walk the reader through key calculations, fill in some gaps in the literature, and point out directions for future research. Topics include: derivation of the self-dual actions from the full Einstein gravity/YM actions in lightcone gauge, decoupling of ghosts, Berends-Giele currents, vanishing of tree-level-one-minus amplitudes, celestial $\Lw$ symmetry, perturbiners, double-off-shell current calculation of one-loop-all-plus SDYM and SDG amplitudes, dimension shifting, unitarity cuts, the BDPR gravity formula, one-loop color ordering, perturbative dilatation symmetry of SDG, and the covariant formulations of the self-dual theories as well as their their gauge fixing. We also review some selected results from the celestial holography program.

}

\begin{document}
\maketitle
\flushbottom

\newpage

\section{Introduction}

Classically, the defining equations of self-dual Yang-Mills (SDYM) and self-dual Einstein gravity (SDG) are
\begin{equation}\label{eq1}
    F_{\mu \nu} = \frac{i}{2} \vep_{\mu \nu \alpha \beta} F^{\alpha \beta}, \hspace{1 cm} R_{\mu \nu \rho \sigma} = \frac{i}{2} \vep_{\mu \nu \alpha \beta} R^{\alpha \beta}_{\;\;\;\; \rho \sigma}.
\end{equation}
Gauge fields (spacetimes) satisfying the above equation can be thought of as containing only positive-helicity gluon waves (gravitational waves) and no negative-helicity waves. 

In lightcone gauge, both SDYM and SDG are described by the simple cubic actions
\begin{equation}\label{actionfirst1}
    S_{\rm SDYM}[\Phi, \bPhi] = - \int d^4 x \; \mathrm{tr} \; \bPhi \big( \Box \Phi + i \sqrt{2} \gym [\partial_u \Phi, \partial_w \Phi]  \big) \, ,
\end{equation}
and
\begin{equation}\label{actionfirst2}
    S_{\rm SDG}[\phi,\bphi] = - \int d^4 x \;  \bphi \left( \Box \, \phi - \frac{\kappa}{2} \{ \partial_u \phi, \partial_w \phi \}  \right)\, .
\end{equation}
Later we shall revisit these actions systemically, but for now it is only important to know that $\Phi$ and $\bPhi$ ($\phi$ and $\bphi$) are scalar fields corresponding to positive and negative-helicity gluons (gravitons) respectively. These actions are truncations of the full Yang-Mills/Einstein-Hilbert actions to the cubic \ppm vertex. The positive-helicity field $\Phi$ ($\phi$) defines a self-dual background which the negative-helicity field $\bPhi$ ($\bphi$) propagates linearly on top of. Technically, only when the negative-helicity field is zero, $\bPhi = 0$ ($\bphi = 0$), does the corresponding solution satisfy \eqref{eq1}, but it is customary to refer to the above actions as the ``self-dual actions'' nonetheless.

Notably, these actions compute subsets of amplitudes in pure Yang-Mills theory/Einstein gravity, as we show in table \ref{fig:mytable}.

{

\begin{figure}[h]\label{fig:tableofamps}
\begin{center}


{\;\;\;\;\;\; Amplitudes in pure Yang-Mills theory/pure Einstein gravity}
\vspace{0.5 cm}

    \begin{NiceTabular}{c|c|c|c|c|c|c}
    \CodeBefore
      \rectanglecolor{myGray}{3-3}{3-4} 
      \cellcolor{myGray}{4-3}           
    \Body
    \Block{1-2}{} & \Block{1-6}{\makecell{number of negative-helicity particles:}} &\\
    \cline{3-6}
    \Block{1-2}{} & & 0 & 1 & 2 & \dots &\\
    \cline{2-6}
    \Block{4-1}<\rotate>{number of loops:} &  0 & 0 & \makecell{0 \\ \text{(except 3-pt)}} & MHV & $\ddots$ &\\
    \cline{2-6}
    & 1 & \makecell{one-loop\\-all-plus} & $\ddots$ & $\ddots$ & $\ddots$ &\\
    \cline{2-6}
    & 2 & $\ddots$ & $\ddots$ & $\ddots$ & $\ddots$ &\\
    \cline{2-6}
    & $\vdots$ & $\ddots$ & $\ddots$ & $\ddots$ & $\ddots$ &\\ \cline{2-6}
\end{NiceTabular}

\begin{equation*}
    \;\;\; \textcolor{myGray}{\bigsquare} \; = \text{ self-dual sector}
\end{equation*}

\caption{\label{fig:mytable} Important classes of amplitudes in either pure Yang-Mills theory or pure Einstein gravity, labeled by the number of negative-helicity external gluons or gravitons and the number of loops. The amplitudes in the ``self-dual sector'' are computed by the SDYM and SDG actions.}

\end{center}

\end{figure}
}

From the self-dual actions, one can show that at tree-level, the all-plus amplitudes are zero simply because there are no tree-level-all-plus Feynman diagrams. There do exist tree-level-one-minus diagrams, but the amplitudes vanish because the diagrams sum to zero\footnote{Actually, it was recently shown that the tree-level-one-minus amplitudes are not identically zero but are instead distributional, living on a 2-dimensional subspace of the 4-dimensional constraint of momentum conservation in $(2,2)$ signature. \cite{Guevara:2026qzd, Guevara:2026qwa,Brandhuber:2026njb}} except at three-points. All tree-level amplitudes with two or more negative-helicity particles vanish identically in the self-dual theories because these diagrams cannot be built from the \ppm vertex alone, but in the full theories these amplitudes are of course not zero. In particular, at tree-level the ``maximal helicity violating'' (MHV) amplitudes, with two negative-helicity particles, are the first class of non-zero tree amplitudes in the full theories.

Interestingly, SDYM and SDG possess only one class of non-zero amplitudes: the one-loop-all-plus amplitudes, which are equal to the one-loop-all-plus amplitudes in the full theories. SDYM and SDG are therefore, on their own, one-loop-exact theories. The full theories of course contain higher loop contributions to the all-plus amplitudes.

Some basic properties of the Yang-Mills and gravity one-loop-all-plus amplitudes are that they (1) are UV and IR finite, (2) are rational, and (3) have only `holomorphic collinear' singularities and no `multi-particle' singularities. In particular, this implies that SDG is a one-loop-exact finite theory of pure quantum gravity in 4-dimensions. Notably, it does not contain the infinite number of matter fields that one would see in string theory when compactified to 4-dimensions \cite{Krasnov:2016emc}.

It should be noted however that SDYM and SDG are non-unitary theories, as the Hamiltonians corresponding to the self-dual actions are not self-adjoint. This is related to the fact that any non-trivial gauge field/non-flat metric which satisfies \eqref{eq1} will be complex. The non-unitarity of the theories is however invisible at the one-loop-level, because the one-loop amplitudes are rational and the tree-level amplitudes are zero. The non-unitarity can be seen in the fact that the higher-loop amplitudes are zero while the one-loop amplitudes are not.

This document began as a set of personal notes while the author was working on an upcoming paper with Guevara and Himwich \cite{toappear} on self-dual loop integrals. Hopefully it is a useful reference for a wider audience. It was written with a few goals in mind. The first goal was to create a pedagogical and self-contained introduction to SDYM and SDG in the lightcone gauge. The second goal was to review some of the old literature on the subject with updated presentations and filled-in proofs, some of which required a non-trivial degree of effort to reproduce. It is a bit suboptimal that this document ended up becoming extremely long, although we attempted to write it so that the reader should be able to skim most of it and comfortably zero-in on the calculations that interest them.

The reader should also see a previous note on the subject by Krasnov \cite{Krasnov:2016emc}. One topic that note discusses which we do not is the the one-loop-determinant of the theories on non-flat backgrounds. The note of Krasnov formulates SDG in the non-zero cosmological constant case using the ``pure connection'' \cite{Delfino:2012aj,Krasnov:2024qkh} formalism, where a flat limit is taken after. Our note however is only interested in the exactly zero cosmological constant case in lightcone gauge. See a later work \cite{Krasnov:2021cva} on the relationship between the two formalisms.

Interest in SDYM and SDG has grown in recent years due to their relevance to the field of celestial holography. It is a tertiary goal of this document to give an overview of the results and perspectives which have come out of this program. We note that there is actually a decades-long history of twistor theorists attempting to formulate classical SDG in terms of a 2d boundary dual theory based on $\myw_{1+\infty}$ symmetry. See for instance in the fascinating ``proto-celestial holography'' papers \cite{Park:1989vq,Park:1990vi,Garcia-Compean:1995ekz,Husain:1993dp,Husain:1994dw,Jevicki:1998zb,Jevicki:1997pm,Ward:1990ty,Yamagishi:1990ax}. However, given that the tree-level self-dual amplitudes are zero, one needs to go beyond the classical level in order to make this duality non-trivial at the S-matrix level, and there had not been work in that direction until recently.

In general, most of the interest surrounding the \textit{quantum} self-dual theories comes from the fact that the \textit{classical} self-dual theories are integrable, enjoying an infinite dimensional set of twistorial symmetries \cite{atiyah1977instantons,atiyah1994construction,mason1990h,dunajski19982d,dunajski2009solitons,mason1996integrability,hitchin2013integrable,ward1990twistor,dunajski1998nonlinear,ward1985integrable}. It is natural to wonder what this infinity of classical symmetries does at the quantum level. A breakthrough by Costello \cite{Costello:2021bah} in understanding this connection birthed the ``Celstial Chiral Algebra'' (CCA) program \cite{Costello:2022wso,Costello:2022upu} for computing amplitudes, which we briefly discuss in section \ref{sec:twisted}. Nevertheless there is quite likely more to be understood about the anomaly.

In the field of celestial holography, the infinity of symmetries, which we call $\Ls$ and $\Lw$ \cite{Ward:1977ta,penrose1976nonlinear,penrose1976nonlinear2}, was (re?)discovered by studying the holomorphic collinear limits of gluons/gravitons as they approached each other in the S-matrix \cite{Guevara:2021abz,Strominger:2021mtt}. This was an important finding because it showed a way in which the symmetries lived on beyond the self-dual sector and also connected them to the tower of soft theorems \cite{Adamo:2021lrv,Li:2018gnc}. It also gave a simple-to-understand 2d OPE description of these symmetries, yielding a clear organizing principle with which to try to construct boundary duals. In the Yang-Mills case, the CCA in principle gives a prescription for using $\Ls$ OPEs to calculate tree-level Yang-Mills amplitudes beyond the self-dual sector \cite{Costello:2022wso,Costello:2022upu,Costello:2023vyy}. In a recent paper \cite{Guevara:2025tsm} it was relatedly shown how the $\Lw$ OPEs can be used to generate the tree-level gravity MHV amplitudes.\footnote{For more celestial works about these symmetries in, and beyond, the classical self-dual sector, see \cite{Freidel:2023gue,Freidel:2021ytz,Freidel:2021dfs,Kmec:2024nmu,Cresto:2024fhd,Cresto:2024mne,Geiller:2024bgf,Donnay:2024qwq,Kmec:2026dis} and \cite{Pano:2023slc,Pranzetti:2026pdg,Cresto:2024fhd} respectively.} 

In the celestial holography program, quantum SDYM and SDG in particular are natural theories to study because one would plausibly expect that the $\Ls$/$\Lw$ symmetries could fix the full S-matrix of these theories completely.

While in a certain sense, ``everything’’ is known about SDYM and SDG because there exist well-known all-multiplicity expressions for the one-loop-all-plus amplitudes, if one actually reviews the current methods we have for computing said amplitudes, one gets a strong feeling that there should be some more insightful, more geometrical method for calculating them. This is especially true for SDG which has remained cloaked in a certain air of mystery. For instance, at the time of writing this note, the famous all-multiplicity expression for the amplitudes, due to Bern, Dixon, Perelstein, Rozowsky (BDPR) \cite{Bern:1998sv,Bern:1998xc} remains a conjecture. (There is however another, less impressive, all-$n$ formula for these amplitudes derivable from double-off-shell currents we write down in section \ref{sec:sdgdoubleoffshell}). 

Based on the discussion in this introduction, it may seem as though the $\Lw$ algebra is a uniquely gravitational object, either being a symmetry manifest to the self-dual theory or a symmetry arising from the soft graviton expansion in the full tree-level theory. However, recent work on lightray operators has shown that this symmetry is actually present in \textit{any} CFT${}_4$ \cite{Himwich:2026exq,Himwich:2025ekg}! See also \cite{Hu:2022txx,Hu:2023geb} and \cite{Gonzalez:2025ene,Strominger:2026yrh,Sheta:2025oep,Oertel:2026wsm}. This is quite strange, given that neither tree-level Einstein gravity nor self-dual Einstein gravity has a bulk conformal symmetry. Having said that, it is of course the case that a theory of free gravitons has a bulk conformal symmetry, and somehow it seems as though the soft expansion is probing this symmetry as well. What exactly this result means is a fascinating open question.

A summary of the document is given below.

\begin{itemize}
    \item In section \ref{sec:overview} we review classical SDYM and SDG in lightcone gauge.
    \item In section \ref{sec:quantumactions} we review the basic diagrammatics of the quantum self-dual actions and explain why they compute the same tree-level-one-minus and one-loop-all-plus amplitudes as the full theories.
    \item In section \ref{sec:YMactionderivation} we review the derivation of the SDYM action from the YM action in lightcone gauge.
    \item In section \ref{sec:GRactionderivation} we provide the derivation of the SDG action from the Einstein-Hilbert action in lightcone gauge, including the derivation of the decoupling of the ghosts.
    \item In section \ref{sec:YMpolarization} we calculate the polarization factors we need to include for external particles in the lightcone scalar Feynman rules.
    \item In section \ref{sec:FeynmanRules} we write down all the Feynman rules for SDYM and SDG.
    \item In section \ref{sec:BG} we introduce the all-plus off-shell Berends-Giele currents. We also provide the closed-form expressions for these currents and review some collinear-limits-based recursive formulae. We also review the proof of the vanishing of the tree-level-one-minus amplitudes.
    \item In section \ref{sec:properties} we review the basic properties of the one-loop-all-plus amplitudes, including their finiteness, rationality, and holomorphic collinear splitting functions.
    \item In section \ref{sec:loopintegrals} we discuss bubble and triangle loop integrals.
    \item In section \ref{sec:doubleoffshell} we review Mahlon's original all-multiplicity calculation of the one-loop-all-plus amplitudes in SDYM using double-off-shell currents, and show how the analogous calculation works in SDG as well.
    \item In section \ref{sec:unitaritycuts} we briefly review the unitarity-cuts method for computing one-loop-all-plus amplitudes.
    \item In section \ref{sec:BDPR} we review the formula of Bern, Dixon, Perelstein, and Rozowsky for the one-loop-all-plus gravity amplitudes.
    \item In section \ref{secLw} we review the self-dual perturbiners and how the classical celestial $\Ls$/$\Lw$ symmetries can be seen within them.
    \item In section \ref{sec:anomaly} we briefly discuss literature on the ``anomaly interpretation'' of the one-loop-all-plus amplitudes and provide a brief summary of the Celestial Chiral Algebra program.
    \item In section \ref{sec:moreliterature} we review further literature on SDYM and SDG.
    \item In appendix \ref{sec:covariant} we review the Lorentz-covariant formulations of SDYM and SDG and explain how they can be gauge-fixed to lightcone gauge. We also discuss the dilatation invariance of SDG.
    \item In appendix \ref{sec:firstheavenly} we write down Plebański's first heavenly equation, compare it to the very similar-looking second heavenly equation, and explain its connection to MHV amplitudes and the half-soft function.
    \item In appendix \ref{app:color} we review the color ordering structure of the one-loop-all-plus Yang-Mills amplitudes.
    \item In appendix \ref{sec:berendssdg} we give the proof that the gravity all-plus current tree formula satisfies its Berends-Giele recursion relation.
    \item In appendix \ref{sec:collinearsdg} we give the proof of the collinear-limits-based recursive formula for the gravity all-plus current.
    \item In appendix \ref{sec:derivationdoubleoffshell} we prove Mahlon's closed-form expression for the SDYM double-off-shell current.
    \item In appendix \ref{sec:OLAPprop} we review a few properties of the SDYM one-loop-all-plus formula, namely that it is cyclic invariant and has the correct holomorphic collinear splitting function.
    \item In appendix \ref{sec:classicalsolns} we give a primer on the relationship between off-shell Berends-Giele currents, classical perturbiner solutions, and amplitudes.
\end{itemize}

\section{Classical SDYM and SDG overview}\label{sec:overview}

In this work we use spacetime signature $(+$,$-$,$-$,$-)$. We define the totally anti-symmetric Levi-Civita pseudotensor, used in \eqref{eq1}, to be
\begin{equation}
    \vep_{\mu \nu \rho \sigma} \equiv \sqrt{-g}[\mu \nu \rho \sigma], \hspace{1 cm} \vep^{\mu \nu \rho \sigma} = - \frac{1}{\sqrt{-g}}[\mu \nu \rho \sigma],
\end{equation}
where $[\mu \nu \rho \sigma]$ is the completely anti-symmetric symbol such that $[0123] = 1$. 

Note that because the Levi-Civita pseudotensor always has the same number of indices as the number of spacetime dimensions, \eqref{eq1} only makes sense in 4-dimensions. SDYM and SDG therefore can only be defined in 4-dimensions.

Any gauge field $A_\mu$ that satisfies $F_{\mu \nu} = \frac{i}{2} \vep_{\mu \nu \alpha \beta} F^{\alpha \beta}$ automatically satisfies the Yang-Mills equation of motion. Let's remind ourselves why this is. We define the field strength tensor as
\begin{equation}
    F_{\mu \nu} = \partial_\mu A_\nu - \partial_\nu A_\mu - i \gym [A_\mu, A_\nu].
\end{equation}
Note that $g$, the YM coupling constant, should not be confused with the determinant of the metric. The covariant derivative acts on the field strength tensor as
\begin{equation}
    D_\mu F_{\nu \rho} = \partial_\mu F_{\nu \rho} - i \gym [A_\mu, F_{\nu \rho}].
\end{equation}
It can be checked that $F_{\mu \nu}$ always satisfies the Bianchi identity
\begin{equation}
    \vep^{\mu \nu \rho \sigma} D_{\mu} F_{\nu \rho} = 0.
\end{equation}
If the gauge field is self-dual then the Bianchi identity above instantly implies that the Yang-Mills equation of motion
\begin{equation}
    D_\mu F^{\mu \nu} = 0
\end{equation}
is also satisfied.

Now we turn to gravity. We will analogously show that any metric $g_{\mu \nu}$ which satisfies $R_{\mu \nu} = \frac{i}{2} \vep_{\mu \nu \rho \sigma} R^{\rho \sigma}$ automatically satisfies the vacuum Einstein equation $R_{\mu \nu} = 0$. To see this, apply the duality equation twice
\begin{equation}
    R^{\mu \nu}_{\;\;\;\; \rho \sigma} = - \frac{1}{4} \vep^{\mu \nu \alpha \beta} \vep_{\rho \sigma \gamma \delta} R_{\alpha \beta}^{\;\;\;\; \gamma \delta}
\end{equation}
and then, using the identity
\begin{equation}
    \vep^{\mu \nu \alpha \beta} \vep_{\gamma \delta \rho \sigma} = - 4! \, \delta^{[\mu}_\gamma \delta^\nu_\delta \delta^\alpha_\rho \delta^{\beta]}_\sigma
\end{equation}
one will find that the Ricci tensor is zero
\begin{equation}
    R^\mu_{\;\; \nu \mu \sigma} = - \frac{3}{2} \delta^{[\gamma}_\sigma \delta^\alpha_\nu \delta^{\beta]}_\delta R_{\alpha \beta \gamma}^{\;\;\;\;\;\;\; \delta} = 0
\end{equation}
due to the Bianchi identity $R_{[\alpha \beta \gamma]}^{\;\;\;\;\;\;\;\; \delta} = 0$.

The equations of SDYM and SDG become simpler if we work in the lightcone coordinates, which we denote $(u, \bu, w, \bw)$:

\begin{equation}\label{lightconecoordinates}
\begin{aligned}
    u = \frac{x^0 - x^3}{2}, \quad\quad
    \bu = \frac{x^0 + x^3}{2}, \quad\quad
    w = \frac{x^1 + i x^2}{2}, \quad\quad     \bw = \frac{x^1 - i x^2}{2}.
\end{aligned}
\end{equation}
Note that $u$ and $\bu$ are independent real variables with $(u)^* \neq \bu$ in Lorentzian $(1,3)$ signature. We have adopted this convention for notational symmetry. In $(2,2)$ signature however these variables really would be complex conjugates of each other.

In lightcone coordinates, the flat metric is
\begin{equation}\label{etadef}
    \eta_{\mu \nu} = \begin{pmatrix} 
     0&  2 & 0 & 0 \\
     2&  0& 0 & 0 \\
     0&  0& 0 & -2 \\
     0&  0& -2 & 0 
    \end{pmatrix}
\end{equation}
and we define
\begin{equation}
    \Box = \partial^\mu \partial_\mu = \partial_u \partial_\bu - \partial_w \partial_\bw.
\end{equation}

In these (complex) coordinates $\vep_{u\bu w\bw} = 4i$ and $F_{\mu \nu} = \frac{i}{2} \vep_{\mu \nu \alpha \beta} F^{\alpha \beta}$ reduces to
\begin{equation}\label{eq2p5}
    F_{uw} = 0, \hspace{0.5 cm} F_{u \bu} = F_{w \bw}, \hspace{0.5 cm} F_{\bu \bw} = 0.
\end{equation}
We now choose the lightcone gauge condition, which is
\begin{equation}\label{ymlcg}
    A_u = 0.
\end{equation}
In this gauge, $F_{u w} = 0$ implies $A_w = 0$. $F_{u \bu} = F_{w \bw}$ is equivalent to $\partial_u A_\bu = \partial_w A_\bw$, which implies that we can write $A_\bu$ and $A_\bw$ as the derivatives of a $\mathfrak{g}$-valued  scalar field $\Phi$:
\begin{equation}
    A_\bu = \sqrt{2} \partial_w \Phi, \hspace{1 cm} A_\bw = \sqrt{2} \partial_u \Phi.
\end{equation}
$\Phi$ is called the Chalmers-Siegel scalar. We have now shown that in lightcone gauge all self-dual gauge fields can be written as
\begin{equation}
    A_\mu = \begin{pmatrix} A_u \\ A_\bu \\ A_w \\ A_\bw \end{pmatrix} = \sqrt{2} \begin{pmatrix} 0 \\ \partial_w \Phi \\ 0 \\ \partial_u \Phi \end{pmatrix}.
\end{equation}
Note that the above $A_\mu$ also satisfies the harmonic gauge condition $\partial_\mu A^\mu = 0$.

The third equation of \eqref{eq2p5} is
\begin{equation}\label{Phieom}
    \tfrac{1}{\sqrt{2}} F_{\bu \bw} = \Box \, \Phi + i \sqrt{2} \gym [\partial_u \Phi, \partial_w \Phi] = 0.
\end{equation}
This is the equation of motion for $\Phi$.

An important exact solution to \eqref{Phieom} is $\Phi = T^a  e^{i p \cdot x}$, where $T^a \in \mathfrak{g}$,  $p^2 = 0$. This solution corresponds to a finite-sized positive-helicity gluon wave. Classical self-dual Yang-Mills theory can be thought of as a sub-sector of Yang-Mills theory containing only positive-helicity gluons. This is intuitively why SDYM is described by a single field after gauge fixing, instead of two.

Now we discuss gravity. It was proven by Plebański \cite{plebanski1975some} that for any self-dual spacetime, one can find a coordinate system $(u,\bu,w,\bw)$ in which
\begin{equation}\label{eq210}
    g_{\mu \nu} = \begin{pmatrix} 
    0 & 2 & 0 & 0 \\
    2 & 0 & 0 & 0 \\
    0 & 0 & 0 & -2 \\
    0 & 0 & -2 & 0
    \end{pmatrix} + 2 \kappa \begin{pmatrix}
	0 & 0 & 0 & 0 \\
	0 & \partial_w^2 \phi & 0 & \partial_u \partial_w \phi \\
	0 & 0 & 0 & 0 \\
	0 & \partial_u \partial_w \phi & 0 & \partial_u^2 \phi
	\end{pmatrix}
\end{equation}
where $\phi$ is a scalar field which satisfies ``Plebański's second heavenly equation''
\begin{equation}\label{pleb2eom}
    \Box \, \phi - \frac{\kappa}{2} \{ \partial_u \phi, \partial_w \phi \} = 0.
\end{equation}
Here, $\{ \cdot, \cdot\}$ is defined as the Poisson bracket of the ($u$, $w$) coordinates:
\begin{equation}
    \{ f, g\} \equiv \partial_u  f \, \partial_w g - \partial_u  g \, \partial_w f.
\end{equation}
The proof that all self-dual metrics can be written in this way is more involved than the proof of the analogous statement in SDYM we gave above. We previously wrote it out in Appendix B of \cite{mypaper}.

If we write \eqref{eq210} as $g_{\mu \nu} = \eta_{\mu \nu} + \kappa h_{\mu \nu}$, then this metric perturbation also satisfies
\begin{equation}
     \eta^{\mu \nu} h_{\mu \nu} = g^{\mu \nu} h_{\mu \nu} = 0 , \hspace{1 cm}
     \eta^{\mu \nu} \partial_\mu h_{\nu \rho} =  g^{\mu \nu} \partial_\mu h_{\nu \rho} =  0, \hspace{1 cm} \det g = \det \eta,
\end{equation}
which means \eqref{eq210} is in harmonic gauge as well.

Just like in the Yang-Mills case, an important exact solution to \eqref{pleb2eom} is $\phi = e^{ip \cdot x}$, where $p^2 = 0$. This solution corresponds to a finite-sized positive-helicity gravitational wave, and classical self-dual gravity can be thought of as a sub-sector of general relativity containing only positive-helicity gravitational waves.

Let us comment on the spacetime signature. While we work in $(1,3)$ signature in this note, consider how \eqref{eq1} changes in signature $(4-m,m)$, where $m$ is the number of $-$'s in the metric. In general signature with real coordinates, $\vep_{\mu \nu \rho \sigma} = \sqrt{|g|}[\mu \nu \rho \sigma]$ and $\vep^{\mu \nu \rho \sigma} = (-1)^m \sqrt{|g|}^{-1} [\mu \nu \rho \sigma]$. The factor of $(-1)^m$ comes from the fact that when one raises all the components of $\vep_{\mu \nu \rho \sigma}$ with the inverse metric, the permutation formula for the determinant implies that one is effectively multiplying the tensor by the inverse determinant $g^{-1}$, where in particular $ g^{-1} = (-1)^m |g|^{-1}$.

Let us introduce the notation of the Hodge star, which we define to act on 2-forms $C_{\mu \nu} = C_{[\mu \nu]}$ as 
\begin{equation}
    (\star C)_{\mu \nu} =  \frac{1}{2}\vep_{\mu \nu \rho \sigma} C^{\rho \sigma}.
\end{equation}
From the formula $\frac{1}{4} \vep_{\mu \nu \alpha \beta} \vep^{\alpha \beta \rho \sigma} = (-1)^m \delta^{[\rho}_\mu \delta^{\sigma]}_\nu$, one has that
\begin{equation}
    \star^2 = (-1)^m.
\end{equation}
So we can see that $\star^2 = 1$ in $(2,2)$ and $(4,0)$ signature, but $\star^2 = -1$ in $(1,3)$ signature. This means that in $(2,2)$ and $(4,0)$ signature, $\star$ has eigenvalues $\pm 1$ but in $(1,3)$ signature it has eigenvalues $\pm i$.

Therefore, in $(2,2)$ and $(4,0)$ signature, one defines the self-duality and anti-self-duality conditions to be $(\star C)_{\mu \nu} = C_{\mu \nu}$ and $(\star C)_{\mu \nu} = -C_{\mu \nu}$ respectively, but in $(1,3)$ signature we will define them to be $(\star C)_{\mu \nu} = -i C_{\mu \nu}$ and $(\star C)_{\mu \nu} = +i C_{\mu \nu}$ respectively. So only in the $(2,2)$ and $(4,0)$ cases can there exist real non-zero (anti-)self-dual two-forms, and only in these signatures can one find real non-pure-gauge/non-flat solutions to \eqref{eq1}. In $(1,3)$ signature, by contrast, the only real solutions to the SDYM and SDG equations \eqref{eq1} occur when $A_\mu$ is pure gauge and $g_{\mu \nu}$ is flat. All other solutions must be complex. This should however not be too surprising. For instance, the metric of a single positive-helicity gravitational wave travelling throughout space, with $g_{\mu \nu} = \eta_{\mu \nu} +  \vep^+_{\mu \nu} e^{ip \cdot x}$, must necessarily be complex in part because the positive-helicity polarization tensor $\vep^+_{\mu \nu}$ is complex in $(1,3)$ signature.

\section{Quantum SDYM and SDG diagrammatics}\label{sec:quantumactions}

The lightcone actions for quantum SDYM and SDG are given in \eqref{actionfirst1} and \eqref{actionfirst2}.
These actions are due to Chalmers and Siegel \cite{Chalmers:1996rq,Siegel:1992xp}. The SDYM action was originally found via a truncation of the $\mathcal{N}=4$ super Yang-Mills action \cite{Mandelstam:1982cb,Brink:1982pd} in lightcone gauge, and analogously the SDG action is a truncation of the $\mathcal{N}=8$ supergravity action \cite{Siegel:1992wd,Brink:1982pd,Bengtsson:1983pg,Ananth:2006fh,Ananth:2008ik}:
the physical particle states of different spins in the supermultiplet are naturally organized within the lightcone superfield, and one can truncate the supersymmetric theories to their own self-dual sectors before then simply dropping all the fields which are not highest/lowest spin.
We won't however review that formalism here. We do note that in the supersymmetric theories themselves, the one-loop-all-plus amplitudes vanish due to cancellations which come from opposite-statistics particles running in the loops, as demanded by supersymmetric Ward identities \cite{Grisaru:1976vm,Grisaru:1977px}.

In any case, in the non-supersymmetric self-dual actions, the $\Phi$ and $\phi$ fields correspond to positive-helicity gluons and gravitons in lightcone gauge. The fields $\bPhi$ and $\bphi$, likewise, correspond to negative-helicity gluons and gravitons. In the action the negative fields appear as Lagrange multipliers which enforce the classical self-dual equations of motion \eqref{Phieom} and \eqref{pleb2eom}, but they are also physical particles in their own right.

\begin{figure}[h]
    \centering
    \includegraphics{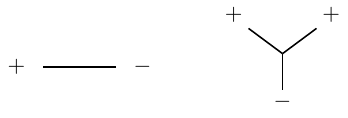}
    \caption{\label{fig:propandvertex} The $(+-)$ propagator and \ppm vertex in the self-dual theories.}
\end{figure}

\begin{figure}[h]
    \centering
    \includegraphics{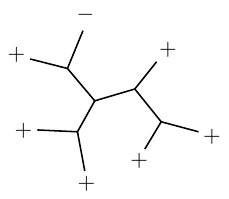}
    \caption{\label{fig:tree} An example of a tree-level Feynman diagram in SDYM or SDG.}
\end{figure}

Both actions have a $(+-)$ propagator and a single cubic \ppm vertex, shown in figure \ref{fig:propandvertex}.
With these elements, all tree-level Feynman diagrams have a single negative-helicity external leg while the rest are positive-helicity. (In our conventions we don't attach $(+-)$ propagators to external legs in the Feynman diagrams.) See figure \ref{fig:tree}.

Somewhat miraculously, after summing up these tree-level Feynman diagrams, it turns out that the tree-level-one-minus amplitudes in gauge theory and gravity vanish, except in the three point case.

Because there are no tree-level Feynman diagrams with all $+$ helicity external particles, we see the all-plus amplitudes vanish at tree-level as well.

\begin{figure}[h]
    \centering
    \includegraphics{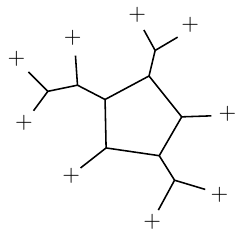}
    \caption{\label{fig:oneloopsample}An example of a one-loop Feynman diagram in SDYM or SDG.}
\end{figure}

At one-loop level, a Feynman diagram can be formed by taking the $-$ leg in figure \ref{fig:tree} and attaching it to one of the $+$ legs with the propagator. Therefore, all one-loop diagrams have only positive-helicity external legs, as in figure \ref{fig:oneloopsample}.

It is also impossible to create higher loop diagrams with the Feynman rules in figure \ref{fig:propandvertex}. Therefore, SDYM and SDG are one-loop-exact theories

As stated in the introduction, the tree-level-all-plus, tree-level-one-minus, and one-loop-all-plus amplitudes computed in SDYM and SDG are equal to the same amplitudes in pure Yang-Mills theory and pure Einstein gravity. See table \ref{fig:mytable}. But why is that?

In order to answer this question, one can use the fact that the self-dual actions are truncations of the full Yang-Mills and Einstein-Hilbert actions in lightcone gauge, keeping only the terms linear in the negative-helicity fields. In sections \ref{sec:YMactionderivation} and \ref{sec:GRactionderivation} we'll go through the derivations. Before that, however, let us simply state what kinds of vertices will appear in the full theories. 

\begin{figure}[h]
    \centering
    \includegraphics{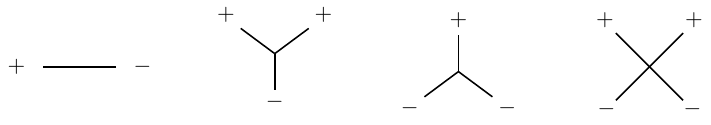}
    \caption{\label{fig:yangmillsvertices} The propagator and vertices of the \textit{full} Yang-Mills action in lightcone gauge.}
\end{figure}

\begin{figure}[h]
    \centering
    \includegraphics{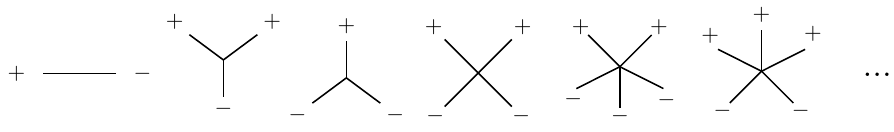}
    \caption{\label{fig:einsteinvertices} The propagator and vertices of the \textit{full} Einstein-Hilbert action in lightcone gauge. There is a $(+-)$ propagator, a \ppm and \mmp vertex, and then an infinite number of higher-pt vertices that each have at least two or more $+$'s and two or more $-$'s each. }
\end{figure}

In lightcone gauge, the full Yang-Mills action has a $(+-)$ propagator, a cubic \ppm vertex, a cubic \mmp vertex, and a quartic $(+$$+$$-$$-)$ vertex. See figure \ref{fig:yangmillsvertices}.
For gravity, once the Einstein-Hilbert action is put in lightcone gauge, there is a $(+-)$ propagator, a cubic \ppm vertex, a cubic \mmp vertex, and an infinite number of higher point vertices. Crucially, all of the higher point vertices each have two or more $+$ legs and two or more $-$ legs each. See figure \ref{fig:einsteinvertices}.

Note that with the lightcone Feynman rules, it is impossible to make a tree-level-all-plus Feynman diagram, so the tree-level-all-plus amplitudes vanish in full Yang-Mills theory and Einstein gravity just as they did in the self-dual theories.

\begin{figure}[h]
    \centering
    \includegraphics[width=0.45\linewidth]{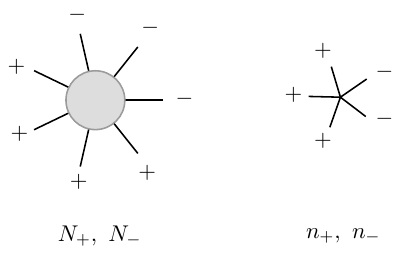}
    \caption{Consider what happens when we have a Feynman diagram with $N_+$ external pluses and $N_-$ external minuses, and attach on an extra vertex with $n_+$ pluses and $n_-$ minuses. The resulting diagram will have $N_+ + n_+ - 1$ pluses and $N_- + n_- -1$ minuses.}
    \label{fig:NNnn}
\end{figure}

Now we will explain why the tree-level-one-minus and one-loop-all-plus amplitudes in SDYM/SDG are equal to those in full YM/gravity \cite{Boels:2013bi}. Consider some tree-level diagram with, say, $N_+$ external pluses and $N_-$ external minuses. Now add on an extra vertex with $n_+$ pluses and $n_-$ minuses with the $(+-)$ propagator. See for instance figure \ref{fig:NNnn}. 


The resulting diagram will have $N_+ + n_+ - 1$ pluses and $N_- + n_- - 1$ minuses. As mentioned, SDYM/SDG contain only the \ppm vertex with $n_+ = 2$, $n_- = 1$, while all the dropped vertices have $n_- \geq 2$. This implies that the tree-level-one-minus diagrams, with $N_- = 1$, can only be built up by appending \ppm vertices to lower-point tree-level-one-minus diagrams. Any higher-point vertex would add at least one more external minus to the diagram. Therefore the tree-level-one-minus amplitudes from the truncated SDYM/SDG theories match those from the full theory.

Furthermore, because the one-loop-all-plus diagrams come from closing a $-$ and a $+$ from a tree-level-one-minus diagram, we can also clearly see that the one-loop-all-plus amplitudes in SDYM/SDG match those from the full theories as well.

As a historical note, the first paper which directly showed the scalar lightcone SDYM action \eqref{actionfirst1} computes the one-loop-all-plus amplitudes in pure YM theory was due to Cangemi \cite{Cangemi:1996rx,Cangemi:1996pf}. Before Cangemi's paper, the one-loop-all-plus amplitudes in pure YM had already been computed by Mahlon \cite{Mahlon:1993si,Mahlon:1993fe}. Actually, what Mahlon calculated were the sum of diagrams in which a massless QCD quark travels around the loop instead of the gluon one would have in pure Yang-Mills theory. However, an argument using supersymmetry shows that the one-loop-all-plus amplitudes are proportional to $(1 + \tfrac{N_s}{N_c} - \tfrac{N_f}{N_c})$, where $N_s$ is the number of complex scalars in the theory that can run around the loop, $N_f$ is the number of quarks, and $N_c$ is the number of gluon colors, so Mahlon's calculation was proportional to the pure YM calculation. Cangemi himself then used a supersymmetry argument to show that the scalar lightcone SDYM action \eqref{actionfirst1} reproduces the quark-loop (or if you like, gluon-loop) calculation of Mahlon, using the fact that the lightcone scalar of the SDYM action is just like the complex scalar in the supermultiplet. This is of course natural because the lightcone action itself was originally found via truncating the $\mathcal{N}=4$ super Yang-Mills action to the self-dual sector \cite{Chalmers:1996rq} before then truncating it further to the two bosonic gluon fields.

To conclude the section, we mention that there is another set of self-dual actions which have only one positive-helicity field. In the one-field actions, the barred fields $\bPhi$, $\bphi$, are replaced with $\Phi$, $\phi$, a $\frac{1}{2}$ is placed in front of the kinetic term and a $\frac{1}{3}$ is placed in front of the cubic interaction term. For SDYM this one-field Lagrangian was studied by Leznov, Mukhtarov \cite{leznov1988equivalence,leznov1987deformation} and Parke \cite{Parkes:1992rz}. For the gravity analog see for instance \cite{plebanski1975some,Plebanski:1996np}. However the one-field actions are ``worse'' because they contain amplitudes beyond one-loop with no physical significance. In gravity, there is also another one-field formulation one may encounter, especially in works on the N=2 string \cite{Ooguri:1990ww,Ooguri:1991fp,Ooguri:1991ie,Ooguri:1995cp}, based on Plebański's ``first heavenly equation'' of the Kähler scalar. See for instance our appendix \ref{sec:firstheavenly}.

\section{Deriving the SDYM action from the pure Yang-Mills action} \label{sec:YMactionderivation}

In this section, we will show the principled derivation of the lightcone SDYM action starting from the full Yang-Mills action, following the write-ups in \cite{Boels:2013bi,Chowdhury:2024dcy}.

Let us briefly explain how this derivation works before jumping into it. We begin by gauge fixing the path integral via the condition $A_u = 0$. The Faddeev-Popov ghosts decouple from $A_\mu$ in this gauge, meaning the ghosts can be ignored for the purposes of computing scattering amplitudes. Furthermore, the gauge-fixed action will be quadratic in $A_\bu$, so we can perform the path integral over $A_\bu$ exactly by plugging its classical solution back into the action. This leaves us with an action over two physical degrees of freedom, $A_\bw$ and $A_w$, representing positive and negative-helicity gluons respectively. Truncation of the action to the term linear in $A_w$ will give us the self-dual action.

We now start the computation. $A_\mu$ and $F_{\mu \nu}$ are $\mathfrak{g}$-valued fields $A_\mu = A^a_\mu T^a$, $F_{\mu \nu} = F^a_{\mu \nu} T^a$. In this note we use the conventions
\begin{equation}
    [T^a, T^b] = i f^{abc} T^c, \hspace{1 cm} \tr(T^a T^b) =   \delta^{ab},
\end{equation}
which imply that $f^{abc} = f^{bca}$.

The pure YM action in lightcone coordinates is
\begin{equation}
\begin{aligned}
    S_{\rm YM}[A] &= -\frac{1}{4} \int d^4 x \; \tr( F^{\mu \nu} F_{\mu \nu}) \\
    &= \frac{1}{8} \int d^4 x \; \mathrm{tr} \left( F_{u \bu}^2 + F_{w \bw}^2 + 2(F_{\bu \bw} F_{u w} + F_{u \bw} F_{\bu w} )\right).
\end{aligned}
\end{equation}
The YM action is invariant under the gauge transformations
\begin{equation}\label{deltaomega}
    \delta_{\omega} A_\mu^a = \partial_\mu \omega^a + \gym f^{abc} A^b_\mu \omega^c.
\end{equation}
We will now impose the lightcone gauge condition via the constraints
\begin{equation}
    G^a[A] \equiv A_u^a.
\end{equation}

This can be accomplished by adding in three extra fields: a bosonic Lagrange multiplier $B^a$ that enforces $G^a[A] = 0$ and two fermionic ghost fields $b^a$ and $c^a$ to make the Faddeev-Popov determinant. The full action is now
\begin{equation}
    S_{\rm YM}^{\text{(gauge  fixed)}} [A, B, b, c] = S_{\rm YM}[A] + \int d^4 x \, B^a G^a[A] + i \int d^4 x \; b^a \, \delta_c G^a[A].
\end{equation}
Here, $\delta_c G^a$ is the variation of the constraint under an infinitesimal gauge transformation \eqref{deltaomega} where $c^a$ plays the role of the gauge parameter $\omega^a$.

The ghost action is calculated to be
\begin{equation}
    b^a \delta_c G^a[A] \Big\rvert_{A_u=0} = b^a \partial_u c^a
\end{equation}
which we can see is $A_\mu$-independent. Therefore, our fermionic ghosts decouple from $A_\mu$ in the path integral, and we are free to ignore them as they only affect the path integral by an overall constant.

Setting $A_u = 0$, the action becomes
\begin{equation}
    S_{\rm YM}[A_\bu, A_w, A_\bw] = \frac{1}{8} \int d^4 x \; \mathrm{tr} \left( (\partial_u A_\bu )^2 + F_{w \bw}^2 + 2(F_{\bu \bw} \partial_u A_w + \partial_u A_\bw  F_{\bu w} )\right)
\end{equation}
and is quadratic $A_\bu$. This means we can integrate out $A_\bu$ in the path integral exactly by substituting in its value from its equation of motion, which is
\begin{equation}
    A_\bu = \frac{\partial_\bw}{\partial_u} A_w + \frac{\partial_w}{\partial_u} A_\bw - i \gym \frac{1}{\partial_u^2} ( [A_\bw, \partial_u A_w] + [A_w, \partial_u A_\bw] ).
\end{equation}
Note that symbol $\frac{1}{\partial_u}$ is an inverse derivative in the sense of integration, so $ \partial_u (\frac{1}{\partial_u}) f(u)= f(u)$. Note that fields in the path integral are understood to vanish at the boundary, and this provides boundary conditions for the inverse derivative. The inverse derivative is division by momentum when Fourier transformed.

After many integrations by parts (see \cite{Chowdhury:2024dcy} for more details) the action becomes 
\begin{equation}\label{YMlcg}
\begin{aligned}
    S_{\rm YM}[A_w, A_\bw] = \frac{1}{2}\int d^4 x \; \mathrm{tr} \Big(  & A_w \Box A_\bw - i \gym \, \partial_u A_w [ \frac{\partial_w}{\partial_u} A_\bw , A_\bw] - i \gym \, \partial_u A_\bw [ \frac{\partial_\bw}{\partial_u} A_w , A_w ]\\
    &+\gym^2 \, [A_w, \partial_u A_\bw] \frac{1}{\partial_u^2 }[\partial_u A_w, A_\bw]  \Big).
\end{aligned}
\end{equation}
As promised, this action has a $(+-)$ propagator, a \ppm vertex, a \mmp vertex, and a $(+$$+$$-$$-)$ vertex as we depicted in figure \ref{fig:yangmillsvertices}.

The self-dual action is defined by truncating the action to the linear terms in $A_w$.
\begin{equation}\label{sdymA}
    S_{\rm SDYM}[A_w, A_\bw] = \frac{1}{2} \int d^4 x \; \mathrm{tr} \Big( A_w \Box A_\bw - i \gym  (\partial_u A_w) [ \frac{\partial_w}{\partial_u} A_\bw , A_\bw] \Big)
\end{equation}
If we parameterize the two components of the gauge field by
\begin{equation}
    A_\bw = \sqrt{2} \partial_u \Phi, \hspace{1 cm} A_w = - \sqrt{2} \frac{1}{\partial_u} \bPhi,
\end{equation}
with $\Phi = \Phi^a T^a$, $\bPhi = \bPhi^a T^a$, then the action becomes
\begin{equation}
    S_{\rm SDYM}[\Phi, \bPhi] =  - \int d^4 x \, \mathrm{tr} \; \bPhi \big( \Box \Phi+ i \gym \sqrt{2}  [\partial_u \Phi, \partial_w \Phi] \big).
\end{equation}

\section{Deriving the SDG action from the Einstein-Hilbert action}\label{sec:GRactionderivation}

In this section, we will derive the self-dual action from the full Einstein-Hilbert action. We do this by enforcing lightcone gauge and expanding the action to the linear terms in the negative-helicity graviton degree of freedom we'll call $\bh$. We found the reference \cite{Chowdhury:2024dcy} to be particularly helpful in the writing of this section. See also \cite{Scherk:1974zm,Kaku:1974ja,Goroff:1983hc,Ananth:2006fh}.

We note that it is extremely tedious to put the entire Einstein-Hilbert action in lightcone gauge. While this calculation does exist in the literature (see for instance (2.28) in \cite{Chowdhury:2024dcy}) we will opt for a simpler strategy here. If all we want to do is derive the self-dual action, it is algebraically far easier to expand the action to the first order in $\bh$ at the start, rather than write out the full Einstein-Hilbert action in lightcone gauge and \textit{then} expand $\bh$. Doing the calculation this way will be sufficient to confirm that the higher point vertices all have two or more more positive and negative-helicity legs apiece, as drawn in figure \ref{fig:einsteinvertices}.

We summarize all the steps here. In section \ref{app:gravlcg} we'll define lightcone gauge. This reduces the 10 degrees of freedom of the metric to 6 functions, which will be $(\varphi, g^{uu}, g^{ui}, h, \bh)$, where the 2d index $i \in \{w, \bw\}$. Note that $g^{uu}$ and $g^{ui}$ are inverse metric components, as it is actually more convenient for us to take the inverse metric components as the fundamental degrees of freedom we are integrating over in the path integral. In section \ref{app:gravghosts} we'll show that the ghosts decouple from the metric in this gauge. In section \ref{app:gravsdgderivation}, we start by showing that, up to boundary terms, the EH action in lightcone gauge only depends on $g^{uu}$ through the term $g^{uu} R_{uu}$, meaning $g^{uu}$ is a Lagrange multiplier that enforces $R_{uu} = 0$. We then solve the $R_{uu} = 0$ constraint, which solves for $\varphi$. We will be free to set $g^{uu}$ to any convenient value we like later. Next, we'll find that, up to boundary terms, the action depends quadratically on $g^{u i}$, whose equations of motion are given by $R_{ui} = 0$. So we can perform the $g^{u i}$ path integral by solving for their classical values and plugging them back in. It will turn out that it will actually only be necessary for us to solve for $g^{ui}$ when $\bh = 0$, so we compute $R_{ui} \rvert_{\bh = 0} = 0$ and use that equation to solve for $g^{ui} \vert_{\bh = 0}$ in terms of $h$. Once this is done, we compute $R \vert_{\bh = 0}$ and find that it is not zero but actually a pure boundary term. We can however set $R \vert_{\bh = 0} = 0$ if we choose $g^{uu}$ to be a convenient value, which we now do. After these integrations, we have reduced the 6 functional degrees of freedom of the metric down to just two, $(h, \bh)$. Finally, we compute $\sqrt{-g} g^{\mu \nu} R_{\mu \nu}$ to the first order in $\bh$. Up to boundary terms, this turns out to be $\sqrt{-\eta} ( \delta_{\bh} g^{\bw \bw} ) (R_{\bw \bw} \vert_{\bh = 0})$ which readily becomes the self-dual action.

\subsection{Lightcone gauge}\label{app:gravlcg}

Defining the gravitational coupling to be
\begin{equation}
    \kappa = \sqrt{32 \pi G_N}
\end{equation}
the Einstein-Hilbert action in $(+,-,-,-)$ signature is 
\begin{equation}
    S_{\rm GR}[g] = -\frac{2}{\kappa^2} \int d^4 x \sqrt{-g} R
\end{equation}
where we use the relativist convention $R_{\alpha \beta} = \partial_\gamma \Gamma^\gamma_{\; \beta \alpha} - \partial_\beta \Gamma^\gamma_{\; \gamma \alpha} + \Gamma^\gamma_{\; \gamma \delta} \Gamma^\delta_{\beta \alpha} - \Gamma^\gamma_{\; \beta \delta} \Gamma^\delta_{\; \gamma \alpha}$.

We will write everything in the coordinates $(u, \bu, w, \bw)$, where we use the 2d index $i \in \{ w, \bw\}$. We further define the $2 \times 2$ matrix $g^{(2)}_{ij}$ to be the $(w, \bw)$ sub-block of the full 4d metric $g_{\mu \nu}$. 

Lightcone gauge in gravity is defined by the four equations
\begin{equation}
    \frac{1}{\sqrt{2}}\big(\!\det g^{(2)}\big)^{\frac{1}{4}} g^{\bu \mu} - \eta^{\bu \mu} = 0
\end{equation}
where $\eta$ is the flat metric given in \eqref{etadef}. In particular, $\eta^{\bu \mu} = \frac{1}{2} \delta^\mu_u$. Any metric that satisfies the above equation can be written in the form
\begin{equation}\label{gLCG}
    g_{\mu \nu} =  \begin{pmatrix} 0 & 2 e^{\varphi/2} & 0 & 0 \\ 2 e^{\varphi/2} & g_{\bu \bu} & g_{\bu w} & g_{\bu \bw} \\ 0 & g_{\bu w} & 2e^{\varphi} \gamma_{ww} & 2e^{\varphi} \gamma_{w\bw} \\ 0 & g_{\bu \bw} & 2e^\varphi \gamma_{w\bw} & 2e^\varphi \gamma_{\bw\bw} \end{pmatrix} , \quad
\end{equation}
with the inverse metric
\begin{equation}\label{giLCG}
    g^{\mu \nu} = \begin{pmatrix} g^{u u} & \tfrac{1}{2} e^{-\varphi/2} & g^{u w} & g^{u \bw} \\ \tfrac{1}{2}e^{-\varphi/2} & 0 & 0 & 0 \\ g^{u w} & 0 & -\tfrac{1}{2}e^{-\varphi} \gamma_{\bw\bw} & \tfrac{1}{2}e^{-\varphi} \gamma_{w\bw} \\ g^{u \bw} & 0 & \tfrac{1}{2}e^{-\varphi} \gamma_{w\bw} & -\tfrac{1}{2}e^{-\varphi} \gamma_{ww} \end{pmatrix}.
\end{equation}
Here, $\gamma_{ij}$ is a $2 \times 2$ matrix that we choose to satisfy
\begin{equation}
    \det \gamma = \gamma_{ww} \gamma_{\bw\bw} - \gamma_{w\bw}^2 = -1.
\end{equation}
(It may seem strange for us to take the determinant of $\gamma$ to be negative, but it is a natural choice because $w$ and $\bw$ are understood as complex coordinates. For the flat metric, $\gamma_{ij}dx^i dx^j = - dw d \bw$.)

Some useful equations include
\begin{equation}
    \det g =  16 \, e^{3 \varphi}, \hspace{1 cm}     \det g^{(2)}= -4 e^{2\varphi}, \hspace{1 cm}     g^{(2)}_{ij} = 2 e^\varphi \gamma_{ij}.
\end{equation}
(Even though we are in Lorentzian signature, $\det g$ is positive because $\det \gamma$ is negative due to our use of the complex coordinates $w$, $\bw$.)

Also, due to the form of the metric \eqref{gLCG}, the inverse of $g^{(2)}_{ij}$ is actually just $g^{ij}$, i.e.
\begin{equation}
    (g^{(2)})^{ij} = g^{ij}.
\end{equation}

Finally, it follows from \eqref{gLCG} and \eqref{giLCG} that the metric components $g_{\bu \bu}$ and $g_{\bu i}$ are determined by the inverse metric components $g^{uu}$ and $g^{ui}$ via
\begin{equation}
    g_{\bu \bu} = -4 \,  e^\varphi g^{uu} +  8 \, e^{2 \varphi} \gamma_{ij} g^{ui} g^{uj}\, ,
\end{equation}

\begin{equation}
    g_{\bu i} = -4 \, e^{3 \varphi/2} \gamma_{ij} g^{u j} \, .
\end{equation}

So, at this stage, we have seen that the metric in lightcone gauge is parameterized by 6 independent functions, $(\varphi, g^{uu}, g^{ui}, \gamma_{ij} \bvert_{\det \gamma = -1})$. We can further parametize $\gamma_{ij}$ by
\begin{equation}\label{gammahbh}
    \gamma_{ij} = \cosh(- \kappa \sqrt{h \bh} ) \begin{pmatrix}
        0 & -1 \\ -1 & 0
    \end{pmatrix} - \frac{\sinh(- \kappa \sqrt{h \bh})}{ \sqrt{h \bh}} \begin{pmatrix}
         \bh & 0 \\ 0 &  h
    \end{pmatrix}
\end{equation}
where $h$ and $\bh$ are two degrees of freedom corresponding to positive and negative-helicity gravitons respectively.

\subsection{Ghost decoupling} \label{app:gravghosts}

We now compute the Fadeev-Popov determinant of the gravity action in lightcone gauge. We follow the original calculation of Kaku \cite{Kaku:1974ja}, filling in some intermediate steps and fixing typographical errors.

The Einstein-Hilbert action is of course invariant under infinitesimal diffeomorphisms
\begin{equation}\label{infdiffeo}
    \delta_\xi g_{\mu \nu} = - \xi^{\rho} \partial_\rho g_{\mu \nu} - \partial_\nu \xi^\rho g_{\rho \mu} - \partial_\mu \xi^\rho g_{\rho \nu}.
\end{equation}
The full, gauge fixed, gravitational action requires adding in a Lagrange multiplier field $B_\mu$ and two fermionic ghost fields, $b_\mu$, $c^\mu$, rendering the full action
\begin{equation}
    S_{\rm GR}^{\text{(gauge fixed)}}[g, B, b, c] = S_{\rm GR}[g] + \int d^4 x B_\mu G^\mu[g] + i \int d^4 x \, b_\mu \, \delta_c G^\mu[g]
\end{equation}
where the gauge fixing constraint is
\begin{equation}\label{gravG}
    G^\mu[g] \equiv \frac{1}{\sqrt{2}}\big(\det g^{(2)}\big)^{\frac{1}{4}} g^{\bu \mu} - \eta^{\bu \mu}
\end{equation}
and $\delta_c G^\mu[g]$ is the variation of the constraint under the infinitesimal diffeomorphism \eqref{infdiffeo}, where $c^\mu$ plays the role of $\xi^\mu$.

Using the basic formulae
\begin{equation}
    \delta g^{\mu \nu} = - g^{\mu \alpha} g^{\nu \beta} \delta g_{\alpha \beta}, \hspace{1 cm} \delta \det g^{(2)} = (\det g^{(2)})g^{ij} \delta g_{ij},
\end{equation}
we have
\begin{equation}
\begin{aligned}
    \delta_\xi G^\mu[g] &= \frac{1}{\sqrt{2}} \big(\det g^{(2)} \big)^{\frac{1}{4}} \left( - g^{\mu \alpha} g^{\bu \beta} \delta_\xi g_{\alpha \beta} + g^{\mu \bu} \frac{1}{4} g^{ij} \delta_\xi g_{ij} \right) \\
    &= \left( - g^{\mu \alpha} \eta^{\bu \beta} \delta_\xi g_{\alpha \beta} + \eta^{\mu \bu} \frac{1}{4} g^{ij} \delta_\xi g_{ij} \right) \\
    &= \frac{1}{2} \left( - g^{\mu \alpha}  \delta_\xi g_{\alpha u} + \delta^\mu_u \frac{1}{4} g^{ij} \delta_\xi g_{ij} \right) .
\end{aligned}
\end{equation}
Note that we have used the constraint equation $G^\mu[g] = 0$ here, as we really want to evaluate $\delta_\xi G^\mu[g]$ on the support of the constraint itself.

We will now expand the above expression such that the dependence on the spacetime indices will be separated out explicitly into $u$, $\bu$, and $i$ independently. We will also find two of the terms cancel.  We use the detailed form of the metric and inverse metric given in \eqref{gLCG} and \eqref{giLCG}, in particular the facts $g_{\alpha u} = 2 \delta^{\bu}_\alpha e^{\varphi/2}$, $g^{\bu \alpha} = \frac{1}{2} \delta^\alpha_u e^{-\varphi/2}$, $g^{i \bu} = 0$, and $\det g^{(2)} = -4 e^{2 \varphi}$:
\begin{equation}
\begin{aligned}
    2 \, \delta_\xi G^\mu[g] \;=& \underbrace{(g^{\mu \alpha}  \partial_\rho g_{\alpha u} )\xi^\rho}_{\mystack{= \; g^{\mu \alpha} \partial_\rho( 2 \delta_\alpha^\bu e^{\varphi/2}) \xi^\rho}{=\; \cancel{ \frac{1}{2} \delta^\mu_u \xi^\rho \partial_\rho \varphi}  }} + \underbrace{ g^{\mu \alpha}  g_{\rho \alpha} (\partial_u \xi^\rho ) }_{= \; \partial_u \xi^\mu} + \underbrace{g^{\mu \alpha}  g_{\rho u} (\partial_\alpha \xi^\rho) }_{\mystackthree{= \; g^{\mu \alpha}  g_{\bu u} (\partial_\alpha \xi^\bu)}{ = \; \delta^\mu_{\bu}  \partial_u \xi^{\bu} - \delta^\mu_{u}  g^{u \alpha} g_{\bu u} \partial_\alpha \xi^\bu     }{ \;\;\;\;\;\;\;\;\;\;\;\;\;\;\;\; - \delta^\mu_{i}  g^{i \alpha} g_{\bu u} \partial_\alpha \xi^\bu } }  \\ & \underbrace{  - \frac{1}{4} \delta^\mu_u ( g^{ij}  \partial_\rho g_{ij} ) \xi^\rho}_{ \mystack{ =\; - \frac{1}{4} \delta^\mu_u \xi^\rho \partial_\rho \ln \det g^{(2)}}{ = \; \cancel{- \frac{1}{2} \delta^\mu_u \xi^\rho \partial_\rho \varphi } } } \;\; \underbrace{ - \frac{1}{2} (\delta^\mu_u g^{ij} g_{j \rho}) \partial_i \xi^\rho . }_{ \mystack{= \; - \frac{1}{2} \delta^\mu_u ( g^{\alpha i} g_{\alpha \rho} - g^{u i} g_{u \rho}) \partial_i \xi^\rho }{= \; - \frac{1}{2} \delta^\mu_u \partial_i \xi^i + \frac{1}{2} \delta^\mu_u g^{ui} g_{u \bu} \partial_i \xi^\bu  }}
\end{aligned}
\end{equation}
Writing out all the remaining terms, we have
\begin{equation}
\begin{aligned}
    2 \, \delta_\xi G^\mu[g] = \;& \partial_u \xi^\mu + \delta^\mu_{\bu}  \partial_u \xi^{\bu} - \delta^\mu_{u}  g^{u \alpha} g_{\bu u} \partial_\alpha \xi^\bu  - \delta^\mu_{i}  g^{i \alpha} g_{\bu u} \partial_\alpha \xi^\bu - \frac{1}{2} \delta^\mu_u \partial_i \xi^i + \frac{1}{2} \delta^\mu_u g^{ui} g_{u \bu} \partial_i \xi^\bu,
\end{aligned}
\end{equation}
meaning the ghost action is
\begin{equation}
\begin{aligned}
    &\int d^4 x \, b_\mu \, \delta_c G^\mu[g] \\
    &= \frac{1}{2} \int d^4 x \Big[ b_\mu \partial_u c^\mu + b_\bu \partial_u c^\bu - b_u g^{u \alpha} g_{\bu u} \partial_\alpha c^\bu - b_i g^{i \alpha} g_{\bu u} \partial_\alpha c^\bu - \frac{1}{2} b_u \partial_i c^i + \frac{1}{2} b_u g^{ui} g_{u \bu} \partial_i c^\bu \Big].
\end{aligned}
\end{equation}
Here is the final step. If we think about the above integrand as $b_\mu M^{\mu}_{\;\; \nu} c^\nu$ where $M^{\mu}_{\;\; \nu}$ is some matrix, then in $(u, \bu, w, \bw)$ coordinates, $M^{\mu}_{\;\; \nu}$ has the following form.
\begin{equation}
\begin{array}{cc}
  &  \overset{\scriptstyle \nu}{\xrightarrow{\hspace{1.8cm}}} \\
{\scriptstyle \mu} \left\downarrow \vphantom{\begin{matrix} 0\\0\\0\\0 \end{matrix}} \right. &
\begin{pmatrix}
        * & \circledast & * & * \\
        0 & * & 0 & 0 \\
        0 & \circledast & * & 0 \\
        0 & \circledast & 0 & * 
    \end{pmatrix}  
\end{array}
\hspace{1 cm} \begin{matrix*}[l]
        \hspace{0.055cm}*\hspace{0.055cm} = \text{metric independent entry} \\ \circledast = \text{metric dependent entry}
    \end{matrix*}
\end{equation}
However, the determinant of this matrix only depends on the diagonal entries, which are all metric-independent. Therefore, the fermionic ghost path integral will not change if we remove the off-diagonal terms, implying
\begin{equation}
\begin{aligned}
    \Delta_{\rm FP}^{-1}[g] & = \int \mathcal{D} b \mathcal{D} c \, \exp(i \int d^4 x \; b_\mu \delta_c G^\mu[g]) \\
    &= \int \mathcal{D} b \mathcal{D} c \, \exp( \frac{i}{2} \int d^4 x \; \big( b_\mu \partial_u c^\mu + b_\bu \partial_u c^\bu \big)).
\end{aligned}
\end{equation}
We can therefore see that the Faddeev-Popov ghosts decouple from the metric, meaning they only affect the path integral by an overall constant and can be ignored.

\subsection{Expanding the EH action to the first order in $\bh$}\label{app:gravsdgderivation}

In lightcone gauge, the Einstein-Hilbert (EH) action takes the form
\begin{equation}
    S_{\rm GR}[g] = -\frac{2}{\kappa^2} \int d^4 x \sqrt{-g} ( g^{uu} R_{uu} + 2 g^{u \bu} R_{u \bu} + 2 g^{u i} R_{ui} + g^{ij} R_{ij}).
\end{equation}
As pointed out previously, in this gauge the metric \eqref{gLCG} depends on the 6 independent functions $(\varphi, g^{uu}, g^{ui}, \gamma_{ij} \bvert_{\det \gamma = -1})$, where $\gamma_{ij}$ can further be parameterized by the two functions $(h, \bh)$ via equation \eqref{gammahbh}. In this section we will show the number of degrees of freedom can be reduced from 6 functions to just 2, namely $(h,\bh)$. We will then throw out all terms second order or higher in $\bh$ and in doing so we will arrive at the SDG action in lightcone gauge.

A well-known identity that we will use multiple times in this section is that, for any infinitesimal variation of the metric,\footnote{One can prove this identity as follows \cite{Poisson:2009pwt}. Recall $R_{\alpha \beta} = \partial_\gamma \Gamma^\gamma_{\; \beta \alpha} - \partial_\beta \Gamma^\gamma_{\; \gamma \alpha} + \Gamma^\gamma_{\; \gamma \delta} \Gamma^\delta_{\beta \alpha} - \Gamma^\gamma_{\; \beta \delta} \Gamma^\delta_{\; \gamma \alpha}$. Now let us work locally in a frame where all of the first derivatives of the metric at some point are 0, meaning that all of the Christoffel symbols (but not their derivatives) are zero. We use $\overset{*}{=}$ to denote equality in this coordinate frame at this point. So we have $\delta R_{\alpha \beta} \overset{*}{=} \partial_\gamma (\delta \Gamma^\gamma_{\; \beta \alpha}) - \partial_\beta (\delta \Gamma^\gamma_{\; \gamma \alpha}).$ But because the Christoffel symbols are zero, we also have $\delta R_{\alpha \beta} \overset{*}{=} \nabla_\gamma (\delta \Gamma^\gamma_{\; \beta \alpha}) - \nabla_\beta (\delta \Gamma^\gamma_{\; \gamma \alpha})$. If we recall that Christoffel symbols transform under coordinate transormation as $\Gamma^{\alpha'}_{\; \beta' \gamma'} = \pdv{x^{\alpha'}}{x^\alpha} \pdv{x^{\beta}}{x^{\beta'}} \pdv{x^{\gamma}}{x^{\gamma'}} \Gamma^{\alpha}_{\; \beta \gamma} + \frac{\partial^2 x^{\alpha'}}{\partial x^\beta \partial x^\gamma} \pdv{x^\beta}{x^{\beta'}} \pdv{x^\gamma}{x^{\gamma'}}$, we see that the difference of any two Christoffel symbols transforms as a tensor, so $\delta \Gamma^\gamma_{\;\; \beta \alpha}$ is a tensor and because we can promote $\overset{*}{=}$ to $=$ in any tensorial equation, we have $\delta R_{\alpha \beta} = \nabla_\gamma (\delta \Gamma^\gamma_{\; \beta \alpha}) - \nabla_\beta (\delta \Gamma^\gamma_{\; \gamma \alpha})$.
}
\begin{equation}\label{linRicci}
    \delta R_{\alpha \beta} = \nabla_\gamma (\delta \Gamma^\gamma_{\; \beta \alpha}) - \nabla_\beta (\delta \Gamma^\gamma_{\; \gamma \alpha}).
\end{equation}
If we vary $\delta g^{uu}$ but keep the other metric variables constant, i.e.
\begin{equation}
    \delta g^{uu} \neq 0, \hspace{0.5 cm} \delta \varphi = 0, \hspace{0.5 cm} \delta g^{ui} = 0, \hspace{0.5 cm} \delta \gamma_{ij} = 0,
\end{equation}
then because $\det g = 16 \, e^{3 \varphi}$ does not depend on $g^{uu}$, under this variation $\delta \sqrt{-g} = 0$ and the full EH Lagrangian varies as
\begin{equation}\label{eq624}
\begin{aligned}
    \delta ( \sqrt{-g} g^{\alpha \beta} R_{\alpha \beta} ) &= \sqrt{-g} \delta g^{uu} R_{uu} + \sqrt{-g} g^{\alpha \beta} \delta R_{\alpha \beta} \\
    &= \sqrt{-g} \delta g^{uu} R_{uu} + \sqrt{-g} \left( \nabla_{\gamma}( g^{\alpha \beta} \delta \Gamma^\gamma_{\; \beta \alpha} ) - \nabla_\beta ( g^{\alpha \beta} \delta \Gamma^\gamma_{\; \gamma \alpha} )\right) \\
    &= \sqrt{-g} \delta g^{uu} R_{uu}  +\partial_{\gamma}( \sqrt{-g} g^{\alpha \beta} \delta \Gamma^\gamma_{\; \beta \alpha} ) - \partial_\beta ( \sqrt{-g} g^{\alpha \beta} \delta \Gamma^\gamma_{\; \gamma \alpha} ) \\
    &= \sqrt{-g} \delta g^{uu} R_{uu} + \text{total derivative.}
\end{aligned}
\end{equation}
So the equation of motion for $\delta g^{uu}$ is $R_{uu} = 0$. One can then readily compute
\begin{equation}
    R_{uu} = - \partial_u^2 \varphi + \frac{1}{4} (\partial_u \gamma^{ij})(\partial_u \gamma_{ij} )
\end{equation}
using these Christoffel symbols of \eqref{gLCG}
\begin{equation}
    \Gamma^\bu_{\; u \mu} = 0, \hspace{1 cm} \Gamma^{\mu}_{uu} = \frac{1}{2}\delta^\mu_u \partial_u \varphi, \hspace{1 cm} \Gamma^i_{\; u j} = \frac{1}{2} (\partial_u \varphi ) \delta^i_j + \frac{1}{2} \gamma^{ik} \partial_u \gamma_{kj}, \hspace{1 cm} \Gamma^\mu_{\; \mu \nu} = \frac{3}{2} \partial_\nu \varphi.
\end{equation}
Notice that $R_{uu}$ is independent of $g^{uu}$. Because we showed (up to boundary terms) that the action only depends on $g^{uu}$ through the term $g^{uu} R_{uu}$, integrating over $g^{uu}$ in the path integral sets $R_{uu} = 0$. After we have integrated over $g^{uu}$, its value in the action is of course no longer meaningful. However, because we know $g^{uu}$ does not even appear in the action (up to boundary terms) once $R_{uu}$ is set to zero, we have the freedom to set it to any function we like. We will keep $g^{uu}$ undefined for now, but set it equal to a convenient value later.

Once $R_{uu}$ is set to zero, this solves for $\varphi$ as
\begin{equation}\label{varphi}
    \varphi = \frac{1}{\partial_u^2} \frac{1}{4} (\partial_u \gamma^{ij})(\partial_u \gamma_{ij})
\end{equation}
and $\varphi$ is no longer an independent degree of freedom.

Next, let's consider what happens when we vary $\delta g^{ui}$ only, keeping the other metric variables unchanged:
\begin{equation}
    \delta g^{ui} \neq 0, \hspace{0.5 cm} \delta \varphi = 0, \hspace{0.5 cm} \delta g^{uu} = 0, \hspace{0.5 cm} \delta \gamma_{ij} = 0.
\end{equation}
Because $\det g = 16 \, e^{3 \varphi}$ does not depend on $g^{ui}$, under this variation $\delta \sqrt{-g} = 0$. From the same logic as \eqref{eq624}, the EH Lagrangian varies as
\begin{equation}
    \delta (\sqrt{-g} g^{\alpha \beta} R_{\alpha \beta}) =  \sqrt{-g} \delta g^{ui} R_{ui} + \text{total derivative}.
\end{equation}
However, while $R_{uu}$ was independent of $g^{uu}$, $R_{ui}$ is not actually independent of $g^{ui}$. One can confirm that
\begin{equation}
    R_{ui} \text{ is linear in } g^{ui}
\end{equation}
which can be checked straightforwardly without actually computing the full expression for $R_{ui}$. Therefore, up to boundary terms, the full action is quadratic in $g^{ui}$. Thus we can do the path integral over $g^{ui}$ exactly by solving for $g^{ui}$ with the $R_{ui} = 0$ equation, and then plugging that value of $g^{ui}$ back into the action. Thus, the components of $\gamma_{ij}$ are the only remaining independent degrees of freedom of the metric in lightcone gauge.

It is possible to solve for $g^{ui}$ for general $\gamma_{ij}$, but it will only be necessary for us to solve for it in the case where $\bh = 0$, which is a much simpler calculation.

When $\bh = 0$, the matrix $\gamma_{ij}$ \eqref{gammahbh} is equal to
\begin{equation} \label{bh00}
    \gamma_{ij} \bvert_{\bh = 0} = \begin{pmatrix} 0 & -1 \\ -1 &  \kappa h \end{pmatrix}.
\end{equation}
Plugging this into \eqref{varphi}, we see that when $\bh = 0$ we have
\begin{equation} \label{bh01}
    \varphi \bvert_{\bh = 0} = 0.
\end{equation}
Plugging \eqref{bh00} and \eqref{bh01} into the metric, one can then compute
\begin{equation}
\begin{aligned}
    R_{uw} \bvert_{\bh = 0} &= \partial_u^2 g^{u \bw}, \\
    R_{u\bw} \bvert_{\bh = 0} &= \partial_u \left( \partial_u g^{u w} - \frac{\kappa}{2} \partial_w h -  \kappa h \partial_u g^{u \bw} \right).
\end{aligned}
\end{equation}
As promised, these expressions are linear in $g^{ui}$ and $R_{ui}\vert_{\bh = 0}=0$ allows us to solve for
\begin{equation} \label{bh02}
    g^{u \bw} \bvert_{\bh = 0} = 0, \hspace{1 cm} g^{uw}\bvert_{\bh = 0} = \frac{ \kappa}{2} \frac{\partial_w}{\partial_u} h.
\end{equation}

With \eqref{bh00} \eqref{bh01} \eqref{bh02}, one can now compute the Ricci scalar when $\bh = 0$ and find
\begin{equation}
    R\bvert_{\bh = 0} = -\frac{ \kappa}{2} \partial_w^2 h - \partial_u^2 g^{uu}.
\end{equation}
Notice that this is a pure boundary term that depends on $g^{uu}$, which we argued earlier we were free to set to whatever we like. We now choose $g^{uu}\vert_{\bh =0}$ to be
\begin{equation}
    g^{uu} \bvert_{\bh = 0} = -\frac{ \kappa}{2} \frac{\partial_w^2}{\partial_u^2} h
\end{equation}
because with this choice,
\begin{equation}
    R \bvert_{\bh = 0} = 0.
\end{equation}
After all of these steps, the metric $g_{\mu \nu} \vert_{\bh = 0}$ is now equal to
\begin{equation}\label{gsdlc}
    g_{\mu \nu} \bvert_{\bh = 0} = \begin{pmatrix} 0 & 2 & 0 & 0 \\ 2 & 0 & 0 & 0 \\
    0 & 0 & 0 & -2 \\
    0 & 0 & -2 & 0 \end{pmatrix} + 2 \kappa
    \begin{pmatrix} 
    0 & 0 & 0 & 0 \\ 
    0 & \frac{\partial_w^2}{\partial_u^2} h & 0 & \frac{\partial_w}{\partial_u} h \\
    0 & 0 & 0 & 0 \\
    0 & \frac{\partial_w}{\partial_u} h & 0 & h \end{pmatrix} .
\end{equation}
One can check that this metric satisfies the self-duality condition \eqref{eq1}.

The only non-vanishing components of $R_{\mu \nu} \bvert_{\bh = 0}$ can be computed to be

\begin{equation}\label{riccisd}
\begin{aligned}
    R_{\bu \bu} \bvert_{\bh = 0} &= -\kappa \, \partial_w^2 \left(  \Box \frac{1}{\partial_u^2} h - \frac{\kappa}{2} \{ \frac{1}{\partial_u} h,  \frac{\partial_w }{\partial_u^2} h\} \right) ,\\
    R_{\bw \bw}  \bvert_{\bh = 0} &= -\kappa \, \partial_u^2 \left(  \Box \frac{1}{\partial_u^2} h - \frac{\kappa}{2} \{ \frac{1}{\partial_u} h,  \frac{\partial_w }{\partial_u^2} h\} \right) , \\
    R_{\bu \bw}  \bvert_{\bh = 0} &= -\kappa \, \partial_u \partial_w \left(  \Box \frac{1}{\partial_u^2} h - \frac{\kappa}{2} \{ \frac{1}{\partial_u} h,  \frac{\partial_w }{\partial_u^2} h\} \right)  ,
\end{aligned}
\end{equation}
where
\begin{equation}
    \{ f, g\} \equiv \partial_u f \, \partial_w g - \partial_w f \, \partial_u f.
\end{equation}

We desire to compute the EH action to the first order in $\bh$. We denote $\delta_{\bh}$ as the linear change in the metric as $\bh$ is varied slightly.

To refresh, we have $R \bvert_{\bh = 0} = 0$, $\sqrt{-g} \bvert_{\bh = 0} = \sqrt{-\eta}$, and know $\sqrt{-g} g^{\alpha \beta} \delta R_{\alpha \beta}$ is a total derivative term due to \eqref{linRicci}. The only non-vanishing components of the Ricci tensor when $\bh = 0$ are written in \eqref{riccisd}. Using these facts and $g^{\bw \bw} = - \frac{1}{2}\kappa \bh + \mathcal{O}(\bh^2)$, we have that
\begin{equation}
\begin{aligned}
    \delta_{\bh} (\sqrt{-g} g^{\alpha \beta} R_{\alpha \beta}) \bvert_{\bh = 0} &= \sqrt{-g} ( \delta_{\bh} g^{\alpha \beta}) R_{\alpha \beta} \bvert_{\bh = 0} + \text{total derivative} \\
    &= \sqrt{-g}  ( \delta_{\bh} g^{\bw \bw}) R_{\bw \bw} \bvert_{\bh = 0} + \text{total derivative}  \\
    &= \sqrt{-\eta}  \, \frac{\kappa^2}{2} \,   \bh \, \partial_u^2 \left(  \Box \frac{1}{\partial_u^2} h - \frac{\kappa}{2} \{ \frac{1}{\partial_u} h,  \frac{\partial_w }{\partial_u^2} h\} \right) + \text{total derivative}
\end{aligned}
\end{equation}
meaning
\begin{equation}\label{grOh}
    S_{\rm GR}[h, \bh] = - \int d^4 x \, \partial_u^2 \bh \left(  \Box \frac{1}{\partial_u^2} h - \frac{\kappa}{2} \{ \frac{1}{\partial_u} h,  \frac{\partial_w }{\partial_u^2} h\} \right) + \mathcal{O}(\bh^2)
\end{equation}
where $d^4 x$ is the flat space volume element. Redefining
\begin{equation}\label{hredef}
    h = \partial_u^2 \phi, \hspace{1 cm} \bh = \frac{1}{\partial_u^2} \bphi
\end{equation}
we finally arrive at the self-dual gravity action
\begin{equation}
    S_{\rm SDG}[\phi, \bphi] = -\int d^4 x \, \bphi \left(  \Box \, \phi - \frac{\kappa}{2} \{ \partial_u \phi,  \partial_w \phi\} \right).
\end{equation}

Because this action only kept the terms linear in $\bh$, the above computation shows that all the higher point vertices must have two or more $\bh$'s per vertex. However, the full EH action is real and is invariant under complex conjugation which sends $h \leftrightarrow \bh$ and $w \leftrightarrow \bw$. This implies a few things. Because there is no quadratic $hh$ kinetic term in \eqref{grOh}, we know there is also no quadratic $\bh \bh$ kinetic term in the full theory as well. Moreover, we also know that there is a \mmp vertex analogous to the \ppm vertex. Finally, we know that if one were to Taylor expand the EH action to the first order in $h$ instead of $\bh$, all higher-point vertices would have two-or-more $+$'s per vertex just as they would also need to have two-or-more $-$'s per vertex. So we have now confirmed that the higher-point vertex structure in Einstein gravity does indeed have the structure that we drew in figure \ref{fig:einsteinvertices}.

\section{Polarization factors and spinor conventions}\label{sec:YMpolarization}

In SDYM, the fields $\Phi$ and $\bPhi$ are ($\mathfrak{g}$-valued) scalars which parameterize the ($\mathfrak{g}$-valued) vector field $A_\mu$. Positive and negative-helicity gluons are of course characterized by their color, momentum, and polarization vector $\vep^\pm_\mu$. For instance, if $A_\mu = T^a \vep^\pm_\mu e^{i p \cdot x}$ is a classical gluon wave that satisfies the free equation of motion, then what are the corresponding $\Phi$ and $\bPhi$? In particular, what overall factor must we multiply $\Phi$ and $\bPhi$ by in order to reproduce the factor of $\vep^\pm_\mu$ in $A_\mu$? In this section we answer this question, and use the opportunity to introduce spinor-helicity notation. 

First, we review spinor-helicity notation. 
We convert between flat-space spacetime indices in $(x^0, x^1, x^2, x^3)$ coordinates and spinor indices via
\begin{equation}
    p^{A \dot A} = p^\mu \sigma_\mu^{A \dot A} = \begin{pmatrix} p^0 + p^3 & p^1 - i p^2 \\ p^1 + i p^2 & p^0 - p^3 \end{pmatrix} = \begin{pmatrix} 
       2 p^{\bu} & 2p^{\bw} \\ 2p^w & 2p^u
    \end{pmatrix}.
\end{equation}
Here, $A = 1,2$, $\dot A = \dot 1, \dot 2$ are the spinor indices. If $p^\mu_i$ is a null vector then it can be decomposed as 
\begin{equation}
    p^{A \dot A}_i = \lambda_i^A \tlambda^{\dot A}_i.
\end{equation}
We raise and lower spinor indices from the left using
\begin{equation}
\begin{aligned}
    \lambda_A &= \vep_{AB} \lambda^B = \vep_{AB} (\vep^{BC} \lambda_C) \\
    \tlambda_{\dot A} &= \vep_{\dot A \dot B} \tlambda^{\dot B} = \vep_{\dot A \dot B} (\vep^{\dot B \dot C} \lambda_{\dot C})
\end{aligned}
\end{equation}
where $\vep_{AB}$, $\vep^{AB}$, $\vep_{\dot A \dot B}$, $\vep^{\dot A \dot B}$ are the 2d antisymmetric tensors defined via
\begin{equation}
    \vep^{1 2} = - \vep_{12} = \vep^{\dot 1 \dot 2} = - \vep_{\dot 1 \dot 2} = 1.
\end{equation}
The four-vectors of Pauli matrices are
\begin{equation}
\begin{aligned}
    \sigma_\mu^{A\dot A} &\equiv (1, \sigma_x, \sigma_y, \sigma_z) \, , \\
    \sigma^\mu_{A\dot A} &= (1, \sigma_x, -\sigma_y, \sigma_z) \, ,
\end{aligned}
\end{equation}
where $\sigma^\mu_{A \dot A} = \vep_{A B} \vep_{\dot A \dot B} \eta^{\mu \nu} \sigma_{\nu}^{B \dot B}$. In this instance $\eta^{\mu \nu}$ is $\text{diag}(+,-,-,-)$. Objects with naturally lowered indices, such as derivatives $\partial_\mu$, can be decomposed into spinor indices via
\begin{equation}
    \partial_{A \dot A} = \partial_\mu \sigma^\mu_{A \dot A} = \begin{pmatrix} \partial_0 + \partial_3 & \partial_1 + i \partial_2 \\ \partial_1 - i \partial_2 & \partial_0 - \partial_3 \end{pmatrix} = \begin{pmatrix} \partial_\bu & \partial_\bw \\ \partial_w & \partial_u \end{pmatrix}
\end{equation}
where we write the final equality in lightcone coordinates \eqref{lightconecoordinates}.

From the identities
\begin{align}
    \sigma^\mu_{A \dot A} \sigma_\mu^{B \dot B} = 2 \delta_A^B \delta_{\dot A}^{\dot B}\, , \hspace{1 cm} \sigma^\mu_{A \dot A} \sigma^{A \dot A}_\nu = 2 \delta^\mu_\nu\, ,
\end{align}
one has
\begin{equation}
    V^\mu W_\mu = \frac{1}{2} V^{A \dot A} W_{A \dot A}
\end{equation}
for $V^{A \dot A} = V^\mu \sigma_\mu^{A \dot A}$ and $W^{A \dot A} = W^\mu \sigma_\mu^{A \dot A}$. If we define
\begin{equation}
    \lr{ij} \equiv \vep_{ A  B} \lambda^{ A}_i \lambda_j^{ B} \, , \hspace{1 cm} [ij] \equiv \vep_{\dot A \dot B} \tlambda^{\dot A}_i \tlambda_j^{\dot B} \, ,
\end{equation}
then the inner product of two massless momenta is given by
\begin{equation}
    p_i \cdot p_j = \frac{1}{2} \lr{ij} [ij].
\end{equation}

The lightcone gauge condition for gauge fields \eqref{ymlcg} can be expressed as $\alpha^\mu A_\mu = 0$, where we can define the null reference vector
\begin{equation}
    \alpha^\mu = (\alpha^u, \alpha^\bu, \alpha^w, \alpha^\bw) = (\tfrac{1}{2},0,0,0).
\end{equation}
The reference vector can be decomposed into two spinors as
\begin{equation}
    \alpha^{A \dot A} = \alpha^A \talpha^{\dot A}, \hspace{1 cm} \alpha^A = \begin{pmatrix} 0 \\ 1 \end{pmatrix}, \hspace{1 cm} \talpha^{\dot A} = \begin{pmatrix} 0 \\ 1 \end{pmatrix}.
\end{equation}
The polarization vectors of positive and negative-helicity particles, $\vep^\pm_{A \dot A}$, always depend on two reference spinors. In lightcone gauge, these reference spinors are naturally provided by $\alpha^A$ and $\talpha^{\dot A}$. So, we define
\begin{equation}\label{epplusminus}
    \vep^+_{A \dot A} = -i \sqrt{2} \frac{\alpha_{ A} \tlambda_{\dot A} }{\lr{\alpha \lambda}}, \hspace{ 1 cm} \vep^-_{A \dot A} = -i \sqrt{2} \frac{\lambda_{A} \talpha_{\dot A}}{[\tlambda \talpha]},
\end{equation}
such that $\vep^+ \cdot \vep^- = -1$.

In lightcone gauge,
\begin{equation}
    A_\mu = \begin{pmatrix} A_u \\ A_\bu \\ A_w \\ A_\bw \end{pmatrix}, \hspace{1 cm} A_\mu \bvert_{\bPhi = 0} = \sqrt{2} \begin{pmatrix} 0 \\ \partial_w \Phi \\ 0 \\ \partial_u \Phi \end{pmatrix} , \hspace{1 cm} A_\mu \bvert_{\Phi = 0} = -\sqrt{2} \begin{pmatrix} 0 \\ \tfrac{\partial_\bw}{\partial_u^2} \bPhi \\ \tfrac{1}{\partial_u} \bPhi \\ 0 \end{pmatrix} ,
\end{equation}
and these equations can be expressed in spinor indices as
\begin{equation}
    A_{A \dot A} \bvert_{\bPhi = 0} = - \sqrt{2} \alpha_A \alpha^B \partial_{B \dot A} \Phi, \hspace{1 cm} A_{A \dot A} \bvert_{\Phi = 0} = \sqrt{2} \frac{\talpha_{\dot A} \talpha^{\dot B} \partial_{A \dot B}}{ ( \alpha^C \talpha^{\dot C} \partial_{C \dot C} )^2 }  \bPhi. 
\end{equation}
If $\Phi$, $\bPhi$ are plane waves, we can replace $\partial_{A \dot A} \to i \lambda_A \tlambda_{\dot A}$ in the equations above and see that the polarization vectors \eqref{epplusminus} can be obtained if we multiply the $\Phi$ (positive-helicity) particles by $\frac{1}{\lr{\alpha \lambda}^2}$ and the $\bPhi$ (negative-helicity) particles by $-\lr{\alpha \lambda}^2$:
\begin{equation}
\begin{aligned}
    \Phi = + \frac{1}{\lr{\alpha \lambda}^2} T^a e^{i p \cdot x} \hspace{0.83 cm} &\text{corresponds to} \hspace{0.9 cm} A_{A \dot A}\bvert_{\bPhi = 0} = \vep^+_{A \dot A} T^a e^{i p \cdot x} \, ,\\ 
    \bPhi = -\lr{\alpha \lambda}^2 T^a e^{i p \cdot x} \hspace{0.9 cm} &\text{corresponds to} \hspace{0.9 cm} A_{A \dot A}\bvert_{\Phi = 0} = \vep^-_{A \dot A} T^a e^{i p \cdot x} \, .
\end{aligned}
\end{equation}

Now let's turn to gravity. Defining the tensor $h_{\mu \nu}$ by
\begin{equation}
    g_{\mu \nu} = \eta_{\mu \nu} + \kappa \, h_{\mu \nu}
\end{equation}
in lightcone coordinates
\begin{equation}
    h_{\mu \nu} \bvert_{\bphi = 0} = 2 \begin{pmatrix} 
    0 & 0 & 0 & 0 \\
    0 & \partial_w^2 \phi & 0 & \partial_u \partial_w \phi \\
    0 & 0 & 0 & 0 \\
    0 & \partial_u \partial_w \phi & 0 & \partial_u^2 \phi 
    \end{pmatrix} , \hspace{0.5 cm} h_{\mu \nu} \bvert_{\phi = 0} = 2 \begin{pmatrix} 
    0 & 0 & 0 & 0 \\
    0 & (\frac{\partial_\bw^2}{\partial_u^4} ) \bphi &  (\frac{\partial_\bw}{\partial_u^3}) \bphi & 0 \\
    0 & (\frac{\partial_\bw}{\partial_u^3} )\bphi & (\frac{1}{\partial_u^2}) \bphi  & 0 \\
    0 & 0 & 0 & 0
    \end{pmatrix}.
\end{equation}
(Here, we used \eqref{gsdlc} and \eqref{hredef} to get the first of the above equations. We get the second equation by noting that the flip $w \leftrightarrow \bw$, which is complex conjugation, will interchange $h \leftrightarrow \bh$.)

In spinor indices, the above equations can be rewritten as
\begin{equation}
\begin{aligned}
    h_{A \dot A B \dot B} \bvert_{\bphi = 0} &= 2 (\alpha_A \alpha_B) (\alpha^C \alpha^D)  \partial_{C \dot A}  \partial_{D \dot B} \phi \, , \\
    h_{A \dot A B \dot B} \bvert_{\phi = 0} &= 2 \frac{(\talpha_{\dot A} \talpha_{\dot B} )(\talpha^{\dot C} \talpha^{\dot D}) \partial_{A \dot C} \partial_{B \dot D}  }{ ( \alpha^E \talpha^{\dot E} \partial_{E \dot E} )^4 }  \bphi \, . 
\end{aligned}
\end{equation}
If $\phi$, $\bphi$ are plane waves, we replace $\partial_{A \dot A} \to i \lambda_A \tlambda_{\dot A}$ in the equations above, and see that if we multiply the positive-helicity $\phi$ particles by $\frac{1}{\lr{\alpha \lambda}^4}$ and the negative-helicity $\bphi$ particles by $+\lr{\alpha \lambda}^4$ we reproduce the polarization vectors from \eqref{epplusminus}: 
\begin{equation}
\begin{aligned}
    \phi = +\frac{1}{\lr{\alpha \lambda}^4} e^{i p \cdot x} \hspace{0.8 cm} &\text{corresponds to} \hspace{0.9 cm} h_{A \dot A B \dot B}\bvert_{\bphi = 0} = \vep^+_{A \dot A} \vep^+_{B \dot B}  e^{i p \cdot x}, \\ 
    \bphi = +\lr{\alpha \lambda}^4 e^{i p \cdot x} \hspace{0.9 cm} &\text{corresponds to} \hspace{0.9 cm} h_{A \dot A B \dot B}\bvert_{\phi = 0} = \vep^-_{A \dot A} \vep^-_{B \dot B}  e^{i p \cdot x}.
\end{aligned}
\end{equation}

\section{The Feynman Rules of SDYM and SDG summarized}\label{sec:FeynmanRules}

We now summarize the work up until this point and write down the diagrammatic Feynman rules. The SDYM action is
\begin{equation}
\begin{aligned}
    S_{\rm SDYM}[\Phi, \bPhi] &=  -\int d^4 x \tr( \bPhi ( \Box \Phi + i \sqrt{2}  g [ \partial_u \Phi, \partial_w \Phi]) ).
\end{aligned}
\end{equation}
The Feynman rules for the propagator and vertex for this action are (with the arrow representing the direction of momentum flow)
\begin{equation}
\begin{aligned}
\raisebox{0pt 
}{ 
\includegraphics{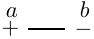}
}
        \hspace{1 cm} & \frac{i \delta^{ab}}{p^2 + i \ep} \\
\raisebox{-20pt
}{
\includegraphics{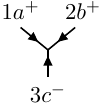}}
\label{ppmsdymvertex}
        \hspace{1 cm} & -i \sqrt{2} g f^{abc} (p_{1,u} \, p_{2,w}  - p_{1,w} \, p_{2,u}  ).
\end{aligned}
\end{equation}
As explained, ingoing external particles carry the spin-one polarization factors
\begin{equation}
\begin{aligned}
\raisebox{0pt}
{\includegraphics{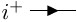} }
\hspace{1 cm} & +\!\frac{1}{\lr{\alpha i}^2}\\
\raisebox{0pt}
{\includegraphics{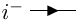} } 
\hspace{1 cm} & -\lr{\alpha i}^2 . 
\end{aligned}
\end{equation}
In this paper, we are also using a convention where we do  \textit{not} pictorially attach $(+-)$ propagators to external legs in our Feynman diagrams. If we did, then the all-plus amplitudes would be computed by summing up all-minus diagrams, which would be too confusing. 

The SDG action is
\begin{equation} 
    S_{\rm SDG}[\phi, \bphi] = - \int d^4 x \; \bphi \left( \Box \, \phi - \frac{\kappa}{2} \{ \partial_u \phi, \partial_w \phi \} \right).
\end{equation}
The Feynman rules from this action are
\begin{equation}\label{3ptsdgfeynman}
\begin{aligned}
\raisebox{0pt } 
{\includegraphics{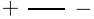}}
        \hspace{1 cm} & \frac{i }{p^2 + i \ep} \\
\raisebox{-17pt}
{\includegraphics{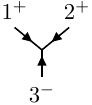}}
        \hspace{1 cm} & \frac{i \kappa}{2} (p_{1,u} \, p_{2,w} - p_{1,w} \, p_{2,u})^2
\end{aligned}
\end{equation}
and ingoing external particles carry the  spin-two polarization factors
\begin{equation}
\begin{aligned}
\raisebox{0pt}
{\includegraphics{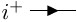}}
        \hspace{1 cm} & +\!\frac{1}{\lr{\alpha i}^4}\\
\raisebox{0pt}
{\includegraphics{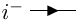}}
        \hspace{1 cm} & +\lr{\alpha i}^4 . 
\end{aligned}
\end{equation}

With spinor helicity variables, we can write the bilinear ``Poisson bracket'' combination of momenta from the 3-pt vertices as
\begin{equation}
    -(p_{1,u} \, p_{2,w}  - p_{1,w}  \, p_{2,u}  ) =  \lr{\alpha |p_1 p_2 |\alpha}.
\end{equation}
Unwrapping this notation, this bilinear spinor product is
\begin{equation}
    \lr{\alpha |p_1 p_2 |\alpha} = \alpha_A \, (p_1^\mu \, \sigma_\mu^{A \dot A}) ( p_{2,\nu} \, \sigma^\nu_{B \dot A} )\, \alpha^B,
\end{equation}
and if both of the momenta are on-shell, $p_1 = |1\rangle[1|$, $p_2 = |2\rangle[2|$, then
\begin{equation}
    \lr{\alpha |p_1 p_2 |\alpha} = \lr{\alpha 1} [12] \lr{2 \alpha}.
\end{equation}
The spinor product is antisymmetric,
\begin{equation}
    \lr{\alpha | p_1 p_2 |\alpha} = - \lr{\alpha | p_2 p_1 |\alpha}
\end{equation}
which follows from the Clifford algebra relation $ \sigma_\mu \sigma_\nu + \sigma_\nu \sigma_\mu = 2 \eta_{\mu \nu}$, which can be written out more fully as 
\begin{equation}
    \sigma_{\mu}^{A \dot A} \sigma_{\nu B \dot A} + \sigma_{\nu}^{A \dot A} \sigma_{\mu B \dot A} = 2 \eta_{\mu \nu} \delta^A_B,
\end{equation}
along with $\lr{\alpha \alpha} = 0$. This implies that any off-shell momentum $p$ pairs with itself to zero.
\begin{equation}
    \lr{\alpha | p p |\alpha} = 0
\end{equation}
Therefore, if one has three momenta that sum to zero, $p_1 + p_2 + p_3 = 0$,
\begin{equation}
    \lr{\alpha | p_1 p_2 |\alpha} = - \lr{\alpha | p_1 p_3 |\alpha}.
\end{equation}
This property implies that the three-point Feynman vertices in SDYM and SDG are permutation invariant amongst the three particles.

\subsection{Example: three-point $+$$+$$-$ amplitudes}

As an example of the Feynman rules, let us compute the $\overline{\text{MHV}}_3$ three-point $+$$+$$-$  tree-level amplitudes in gauge theory and gravity. The Feynman diagrams are shown below.
\begin{figure}[h]
    \centering
    \includegraphics{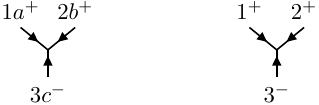}
    \caption{The three point SDYM and SDG diagrams.}
    \label{fig:three pt}
\end{figure}

In gauge theory, the amplitude is
\begin{equation}
\begin{aligned}
    \mathcal{A}^{\rm tree}(1^{a+},2^{b+},3^{c-}) &= -i \sqrt{2} g f^{a b c} \frac{\lr{\alpha 3}^2}{\lr{\alpha 1}^2 \lr{\alpha 2}^2} \lr{\alpha 1}[12]\lr{2 \alpha}.
\end{aligned}
\end{equation}
Using momentum conservation
\begin{equation}
    |1]\langle 1| + |2]\langle 2| + |3]\langle 3| = 0
\end{equation}
and sandwiching it in between $[1|$ and $|\alpha \rangle$, as well as $[2|$ and $|\alpha \rangle$, we get the two relations
\begin{equation}
    \frac{\lr{\alpha 3}}{\lr{\alpha 2}} = - \frac{[12]}{[13]}, \hspace{0.5 cm} \frac{\lr{\alpha 3}}{\lr{\alpha 1}} = - \frac{[21]}{[23]}
\end{equation}
which we can use to get rid of the reference spinor in the 3-pt amplitude:
\begin{equation}
    \mathcal{A}^{\rm tree}(1^{a+},2^{b+},3^{c-}) = i \sqrt{2} g f^{a b c} \frac{[12]^3}{[13][23]}.
\end{equation}
For gravity, we likewise can compute
\begin{equation}
\begin{aligned}
    \mathcal{M}^{\rm tree}(1^+,2^+,3^-) &= \frac{i \kappa}{2}\frac{\lr{\alpha 3}^4}{\lr{\alpha 1}^4 \lr{\alpha 2}^4} \lr{\alpha 1}^2 [12]^2 \lr{2 \alpha}^2 \\
    &= \frac{i \kappa}{2} \frac{[12]^6}{[13]^2 [23]^2}.
\end{aligned}
\end{equation}

\section{The all-plus Berends-Giele off-shell currents}\label{sec:BG}

\begin{figure}[h]
    \centering
    \includegraphics{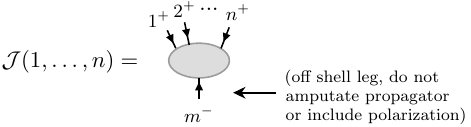}
    \caption{Definition of all-plus current.}
    \label{fig:Jdef}
\end{figure}

In this section, we introduce an important object in the self-dual theories, namely the all-plus currents denoted $\mathcal{J}(1,\ldots,n)$. These objects are defined as the sum of Feynman diagrams with $n$ positive-helicity on-shell legs and one negative-helicity off-shell leg $m$. The negative-helicity leg is defined to be un-amputated. Furthermore, leg $m$ is not multiplied by a polarization factor (i.e., $-\lr{\alpha m}^2$ for YM or $+\lr{\alpha m}^4$ for gravity) because $m$ is off-shell. See figure \ref{fig:Jdef}. We suppress the momentum conserving delta function $(2 \pi)^{4} \delta^{(4)}(p_1 + \ldots +p_n + p_m)$ from the expression, but enforce $p_m \equiv - (p_1 + \ldots + p_n)$ implicitly. Note that $p_1, \ldots, p_n$ are the independent variables of the current which then fix $p_m$.

\begin{figure}[htbp]
\centering
\includegraphics{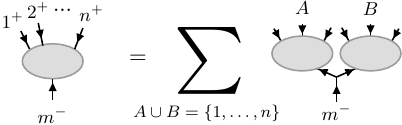}
\caption{\label{fig:ymrecursion} Schematic all-plus current recursion relation.}
\end{figure}

Because the self-dual theories only have one cubic vertex, the Berends-Giele recursion relation for $\mJ$ can be schematically represented as in figure \ref{fig:ymrecursion}.
We will present the solutions to this recursion relation for gauge theory in gravity in the next two sections.

Historically, Berends-Giele currents were introduced by Berends-Giele in \cite{Berends:1987me} to prove the Parke-Taylor formula \cite{Parke:1986gb} for MHV amplitudes in Yang-Mills theory. See also \cite{Kosower:1989xy}. In particular, they found an expression for the ``one-minus'' current, which is the current with one on-shell negative-helicity gluon, an arbitrary number of on-shell positive-helicity gluons, and one off-shell leg. Putting the off-shell leg on-shell, amputating the propagator, and contracting it with a negative-helicity polarization vector gave the Parke-Taylor formula.

Recently, a direct Berends-Giele based computation of the gravity MHV amplitudes was achieved in a nice paper by Hasuwannakit and Krasnov \cite{Hasuwannakit:2025agr}, who successfully computed the one-minus gravity current. The recursion relation for the one-minus current was originally derived using the pure connection formalism in \cite{Delfino:2014xea}.

\subsection{Yang-Mills all-plus current}

Assume our gauge group is $SU(N_c)$ where $N_c$ is the number of colors. Because the all-plus currents are tree-level objects, we can decompose the YM current $\mathcal{J}_{\rm YM}$ as a sum over color orderings via
\begin{equation}\label{JYMpartial}
    \mJ_{\rm YM}(1^{a_1},\ldots,n^{a_n})^{a_m} = \sum_{\sigma \in S_n} \tr(T^{a_{\sigma(1)}} \ldots T^{a_{\sigma(n)}} T^{a_{m}}) J_{\rm YM}(\sigma(1), \ldots, \sigma(n))
\end{equation}
where $J_{\rm YM}$ are the color-ordered ``partial currents.'' (We review the basics of color ordering in appendix \ref{app:color}.) The generator $T^{a_m}$ corresponds to the color of the minus leg. The partial currents are computed using planar Feynman diagrams where the external lines are enforced to obey clockwise color-ordering and we strip off the $i f^{abc}$ factor from the vertices, making sure to include a sign if we flip the orientation of a vertex.

If we define the ordered sets $N \equiv \{1, \ldots, n\}$ and $A = \{1, \ldots, j\}$, $B = \{j+1, \ldots, n\}$ such that $N = A \cup B$, then the recursion relation from figure \ref{fig:ymrecursion} can be written as
\begin{equation}\label{Jymrecursionrelation}
    p_N^2 J_{\rm YM}(N) = i \sqrt{2} g \sum_{\substack{j=1\\ A = \{1, \ldots, j\} \\ B=\{j+1, \ldots ,n\}  }}^{n-1} \lr{ \alpha | p_A p_B | \alpha } J_{\rm YM}(A) J_{\rm YM}(B)
\end{equation}
where we use the notation
\begin{equation}
    p_A \equiv \sum_{i \in A} p_i.
\end{equation}
The one-particle base case for the current is the polarization factor
\begin{equation}
    J_{\rm YM}(i) = \frac{1}{\lr{\alpha i}^2}.
\end{equation}
The recursion relation is solved by the Parke-Taylor-like formula

\begin{equation}\label{Jymsoln}
    \boxed{J_{\rm YM}(1, \ldots, n) = (-i \sqrt{2} g)^{n-1}\frac{1}{\lr{\alpha 1}} \frac{1}{\lr{12} \lr{23} \ldots \lr{n-1,n} } \frac{1}{\lr{\alpha n}}.}
\end{equation}

We include the proof below.

All we have to do is check that \eqref{Jymsoln} solves the recursion relation \eqref{Jymrecursionrelation}. Plugging it in and removing overall factors, we find the recursion relation is equivalent to
\begin{equation}
    -(p_1 + \ldots + p_n)^2 = \sum_{j=1}^{n-1} \lr{\alpha | (p_1 + \ldots + p_j)(p_{j+1} + \ldots + p_n) | \alpha} \frac{\lr{j,j+1}}{\lr{\alpha j} \lr{\alpha, j+1}}
\end{equation}
so all we have to do is show that the above equation holds. The RHS can be simplified as
\begin{equation}
\begin{aligned}
    \sum_{j=1}^{n-1} \sum_{a=1}^j \sum_{b=j+1}^n \frac{\lr{j,j+1}}{\lr{\alpha j} \lr{\alpha, j+1}} \lr{\alpha a}[ab] \lr{b\alpha} &= \sum_{1 \leq a < b \leq n} \lr{\alpha a}[ab] \lr{b\alpha}\sum_{j = a}^{b-1}  \frac{\lr{j,j+1}}{\lr{\alpha j} \lr{\alpha ,j+1}} \\
    &= \sum_{1 \leq a < b \leq n} \lr{\alpha a}[ab] \lr{b\alpha} \frac{\lr{a b}}{\lr{\alpha a} \lr{\alpha b}} \\
    &= -\sum_{1 \leq a < b \leq n} [ab]\lr{ab} \\
    &= -(p_1 + \ldots + p_n)^2
\end{aligned}
\end{equation}
using $2 p_a\!\cdot p_b = \lr{ab}[ab]$, and this concludes the proof. Note that in the second line we used a telescoped version of the the Schouten identity (i.e. the `eikonal' identity) which is
\begin{equation}\label{schouten3}
    \frac{\lr{12}}{\lr{\alpha 1} \lr{ \alpha 2}} + \frac{\lr{23}}{\lr{\alpha 2} \lr{ \alpha 3}} = \frac{\lr{1 3}}{\lr{\alpha 1} \lr{ \alpha 3}}.
\end{equation}

\subsubsection*{SDYM current examples}

\begin{figure}[h]
    \centering
    \includegraphics{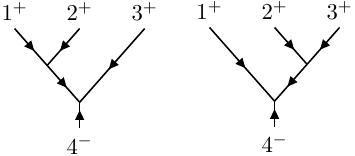}
    \caption{The two color ordered Feynman diagrams contributing to $J_{\rm YM}(1,2,3)$.}
    \label{fig:j3ym}
\end{figure}

Let us give some low-point examples for all-plus partial currents. The two-point partial current is given by a single three-point diagram, and is
\begin{equation}
    J_{\rm YM}(1,2) = \frac{\sqrt{2} g  \lr{\alpha| 12 |\alpha}}{\lr{\alpha 1}^2 \lr{\alpha 2}^2 } \frac{i}{(p_1 + p_2)^2} = - i \sqrt{2} g \frac{1}{\lr{\alpha 1}} \frac{1}{\lr{12}} \frac{1}{\lr{\alpha 2}}.
\end{equation}
The three-point partial current is given by two diagrams, drawn in figure \ref{fig:j3ym}.
The expression from the Feynman diagrams is
\begin{equation}
    J_{\rm YM}(1,2,3) = \frac{(i \sqrt{2} g)^{2}}{\lr{\alpha 1}^2 \lr{\alpha 2}^2 \lr{\alpha 3}^2} \left( \frac{\lr{\alpha | 12 |\alpha}}{(p_1 + p_2)^2} \frac{\lr{\alpha | (1+2)3 |\alpha}}{(p_1 + p_2+p_3)^2} + \frac{\lr{\alpha | 23 |\alpha}}{(p_2 + p_3)^2} \frac{\lr{\alpha | 1(2+3)|\alpha}}{(p_1 + p_2+p_3)^2}\right),
\end{equation}
and can be shown to be identically equal to the resummed Parke-Taylor-like formula
\begin{equation}
     J_{\rm YM}(1,2,3) =(-i \sqrt{2} g)^{2} \frac{1}{\lr{\alpha 1}}\frac{1}{\lr{12}\lr{23}} \frac{1}{\lr{\alpha 3}}.
\end{equation}
This equality of course holds without the use of momentum conservation.

\subsection*{Why are we calling them ``currents''?}

Let us briefly explain why we call these objects ``currents.'' In this note we work with the lightcone scalars. However, if one uses the usual vector field $A_\mu$, the all-plus current would have a free space-time index and would be equal to our expression times a factor $\lr{\alpha|\sigma_\mu (p_1 + \ldots + p_n)|\alpha}$. This factor vanishes when contracted with $(p_1 + \ldots + p_n)^\mu$, which is related to gauge invariance. After Fourier-transforming the off-shell leg, this is the equation of a conserved current, hence the name ``current.''

\subsection{Gravity all-plus current}

Now we turn to gravity. We write down the gravity all-plus current recursion relation displayed in figure \ref{fig:ymrecursion}. It is
\begin{equation}\label{gravrecursion}
    p_N^2 \mJ_{\rm G}(N) = -\frac{\kappa}{2}\sum_{A, B | A \cup B = N }   \langle \alpha | p_A p_B | \alpha \rangle^2 \mJ_{\rm G}(A) \mJ_{\rm G}(B).
\end{equation}
In gravity there is no notion of color ordering and we sum over \textit{all} subsets $A$ and $B$ such that $A \cup B = N = \{1, \ldots, n\}$, neither $A$ nor $B$ is empty and $A \cap B = \emptyset$. In this sum $A$ and $B$ are indistinguishable, so if the pair $(A,B)$ is included in the sum then we do not independently include $(B,A)$.

\begin{figure}[h]
    \centering
    \includegraphics{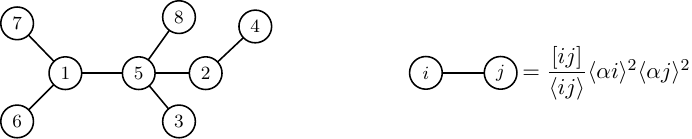}
    \caption{Example of a tree that contributes to $\mJ_{\rm G}(1,\ldots,8)$, along with the factor of a single edge.}
    \label{fig:Jtreeexample}
\end{figure}

The base case of the all-plus current is just the polarization factor
\begin{equation}
    \mJ_{\rm G}(i) = \frac{1}{\lr{\alpha i}^4 }.
\end{equation}
The recursion relation is solved by the ``tree formula''
\begin{equation}\label{JGtree}
\begin{aligned}
    \boxed{\mJ_{\rm G}(1,\ldots,n) = \left(-\frac{\kappa}{2}\right)^{n-1} \left( \prod_{a = 1}^n \frac{1}{\lr{\alpha a}^4} \right) \sum_{\text{trees}} \sum_{\text{edges } (i j)} \frac{[ij]}{\langle i j \rangle} \lr{\alpha i}^2 \lr{\alpha j}^2.}
\end{aligned}
\end{equation}
In this formula, the ``trees'' in question are not the cubic Feynman trees but rather a different kind of tree. They are trees where each \textit{node} corresponds to a positive-helicity graviton $1\ldots  n$ and every node is allowed to have arbitrarily many connections. An example is drawn in figure \ref{fig:Jtreeexample}. 
Each edge comes with a factor of $\frac{[ij]}{\lr{ij}}\lr{\alpha i}^2 \lr{\alpha j}^2$. The tree-formula can also be expressed as
\begin{equation}
\begin{aligned}
    \mJ_{\rm G}(1,\ldots,n) &= \left( - \frac{\kappa}{2}\right)^{n-1} \sum_{\text{trees}} \sum_{\text{edges } (ij)} \frac{[ij]}{\langle i j \rangle} \prod_{a=1}^n \langle \alpha a \rangle^{2(\mathrm{deg}(a) -2)}
\end{aligned}
\end{equation}
where $\mathrm{deg}(a)$ denotes the number of neighbors the node has, or its ``degree.'' The degrees of the nodes in a tree diagram satisfy the relation
\begin{equation}
    \sum_{a=1}^n (\text{deg}(a) - 2) = -2.
\end{equation}

There is another expression for the all-plus current which we'll write down. Utilizing the matrix tree theorem \cite{Feng:2012sy}, if one defines the matrix
\begin{equation}
    \Psi_{ij} = \begin{cases}
        -\frac{[ij]}{\lr{ij}}\lr{\alpha i}^2 \lr{\alpha j}^2 & \text{if } i \neq j \\
        \sum_{k=1, k \neq i}^{n} \frac{[ik]}{\lr{ik}}\lr{\alpha i}^2 \lr{\alpha k}^2 & \text{if } i = j
    \end{cases}
\end{equation}
then $\mJ_{\rm G}$ can be expressed via
\begin{equation}
    \mJ_{\rm G}(1, \ldots, n) = \left( - \frac{\kappa}{2} \right)^{n-1} \left( \prod_{a=1}^n \frac{1}{\lr{\alpha a}^4} \right)| \Psi|^i_i.
\end{equation}
Here, $|\Psi|^i_i$ is the determinant of the minor of the matrix $\Psi$ where the $i^{\rm th}$ row and column have been removed. It can be shown that the expression is independent what choice of $i = 1, \ldots, n$ one makes.

We write down the proof of the tree formula in appendix \ref{sec:berendssdg}.

\subsection*{SDG current examples}

\begin{figure}[h]
    \centering
    \includegraphics{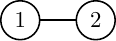}
    \caption{The one tree which contributes to $\mJ_{\rm G}(1,2).$}
    \label{fig:J2}
\end{figure}

The two-point current is given by a single cubic Feynman diagram. It is
\begin{equation}
\begin{aligned}
    \mathcal{J}_{\rm G}(1,2) = \frac{i \kappa}{2} \frac{\lr{\alpha|12|\alpha}^2}{\lr{\alpha 1}^4 \lr{\alpha 2}^4} \frac{i}{(p_1 + p_2)^2}.
\end{aligned}
\end{equation}
The expression can be simplified to
\begin{equation}\label{JG12}
\begin{aligned}
    \mathcal{J}_{\rm G}(1,2) &= - \frac{\kappa}{2} \frac{1}{\lr{\alpha 1}^2 \lr{\alpha 2}^2 } \frac{[12]}{\lr{12}}.
\end{aligned}
\end{equation}
This expression for $\mJ_{\rm G}(1,2)$ corresponds to the single tree in figure \ref{fig:J2}.

\begin{figure}[h]
    \centering
    \includegraphics{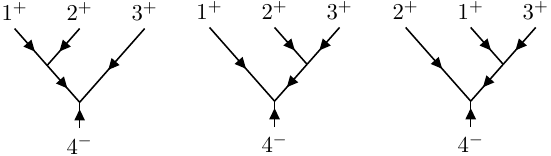}
    \caption{Three Feynman diagrams contributing to $\mJ_{\rm G}(1,2,3)$.}
    \label{fig:J3g}
\end{figure}

\begin{figure}[h]
    \centering
    \includegraphics{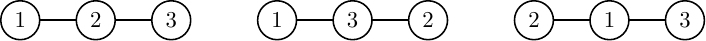}
    \caption{Three trees contributing to $\mJ_{\rm G}(1,2,3)$.}
    \label{fig:J3}
\end{figure}

Now let's look at the three-point example. Here there are three contributing Feynman diagrams, shown in figure \ref{fig:J3g}.
Summing the three Feynman diagrams, we get
\begin{equation}
\begin{aligned}
    \mJ_{\rm G}(1,2,3) = \left( - \frac{\kappa}{2} \right)^2 &\frac{1}{\lr{\alpha 1}^4 \lr{\alpha 2}^4 \lr{\alpha 3}^4} \Bigg( \frac{\lr{\alpha|12|\alpha}^2}{(p_1 + p_2)^2} \frac{\lr{\alpha|(1+2)3|\alpha}^2}{(p_1 + p_2 + p_3)^2} \\ & \\
    & + \frac{\lr{\alpha|23|\alpha}^2}{(p_2 + p_3)^2} \frac{\lr{\alpha|1(2+3)|\alpha}^2}{(p_1 + p_2 + p_3)^2} + \frac{\lr{\alpha|13|\alpha}^2}{(p_1 + p_3)^2} \frac{\lr{\alpha|2(1+3)|\alpha}^2}{(p_1 + p_2 + p_3)^2} \Bigg).
\end{aligned}
\end{equation}
This expression is identically equal to
\begin{equation}
    \mathcal{J}_{\rm G}(1,2,3) = \left( -\frac{\kappa}{2} \right)^2 \left( \frac{[12]}{\lr{12}} \frac{[23]}{\lr{23}} \frac{1}{\lr{\alpha 1}^2\lr{\alpha 3}^2} + \frac{[13]}{\lr{13}} \frac{[32]}{\lr{32}} \frac{1}{\lr{\alpha 1}^2\lr{\alpha 3}^2} + \frac{[21]}{\lr{21}} \frac{[13]}{\lr{13}} \frac{1}{\lr{\alpha 2}^2\lr{\alpha 3}^2} \right)
\end{equation}
which corresponds to the three trees drawn in figure \ref{fig:J3}. Note that this equality does not require momentum conservation.

We emphasize that the tree sum is different from the Feynman diagram sum. It is not the case that each tree corresponds to a Feynman diagram. Rather, it is the sum of trees that is equal to the sum of Feynman diagrams. The trees should be thought of as the gravitational analog of color orderings in Yang-Mills theory.

\subsection*{History of the tree formula}

The history of the tree formula \eqref{JGtree} is a bit complicated. If one likes, one could in retrospect attribute it to Bern, Dixon, Perelstein, Rozowsky \cite{Bern:1998sv}. While they did not write this formula exactly, they did give formulae for it which are ultimately equivalent, including one using Young tableaux. They were aware of the trees, or ``elk diagrams,'' at least when it came to the structure of square brackets in the numerators. Importantly, the authors also knew that their expression for the all-plus current was (essentially) equal to an expression for gravity MHV amplitudes, which they checked by comparing it to the old-fashioned KLT inspired Berends-Giele-Kuijf formula \cite{Berends:1988zp}. The trees were later independently discovered by Nguyen, Spradlin, Volovich, and Wen \cite{Nguyen:2009jk} who gave closed-form expression for MHV amplitudes as a sum over trees. Using these two papers, the tree formula for the all-plus current could then be deduced. The situation was spelled out clearly in a paper by Krasnov, Scarinci \cite{Krasnov:2013wsa} and then also Delfino \cite{Delfino:2014xea}, who showed how the tree-formula itself follows from the all-plus Berends-Giele recursion relation. See also \cite{mypaper} where it is shown how to calculate the MHV amplitudes using the all-plus current in the perturbiner framework. Various tree formulae also appear throughout the study of twistor (string) theory in many different contexts. See \cite{Adamo:2021bej,Adamo:2024hme,Skinner:2013xp,Adamo:2012xe,Adamo:2013tja} and section 6 of \cite{Adamo:2013cra}.

\subsection{Vanishing of the tree-level-one-minus amplitudes}

Now that we have computed the all-plus off-shell currents, we are in a position to see why the tree-level-one-minus amplitudes vanish (except at 3-pts).

One can extract the tree-level-one-minus amplitudes from the all-plus currents by requiring that the off-shell leg $p_m = -(p_1 + \ldots + p_n)$ be on-shell, before amputating the $p_m^2$ propagator and multiplying the expression by the minus-particle's polarization factor.

So, in gauge theory
\begin{equation}
    \mathcal{A}^{\rm tree}(1^{a_1 +}, \ldots, n^{a_n +},m^{a_m-}) =+ i  \lr{\alpha m}^2 (p_1 + \ldots + p_n)^2 \mathcal{J}_{\rm YM}(1^{a_1}, \ldots, n^{a_n})^{a_m}
\end{equation}
and in gravity
\begin{equation}\label{MgravJ}
    \mathcal{M}^{\rm tree}(1^+, \ldots, n^+,m^-) = - i  \lr{\alpha m}^4 (p_1 + \ldots + p_n)^2 \mathcal{J}_{\rm G}(1, \ldots, n).
\end{equation}
However, because, $(p_1 + \ldots + p_n)^2 = p_m^2 = 0$, this implies the tree-level-one-minus amplitudes vanish!
\begin{equation}
\begin{aligned}
    \mathcal{A}^{\rm tree}(1^{a_1+}, \ldots, n^{a_n+},m^{a_m-}) = 0\,, \hspace{0.5 cm} \mathcal{M}^{\rm tree}(1^+, \ldots, n^+,m^-) =0\,, \hspace{0.5 cm} \text{ for }n>2.
\end{aligned}
\end{equation}

The reason this happens is that the all-plus currents \eqref{Jymsoln} and \eqref{JGtree} have only \textit{holomorphic collinear singularities}, i.e. singularities of the form $\frac{1}{\lr{ij}}$ as $\lr{ij} \to 0$. In a typical theory, like the scalar $\varphi^3$ theory, tree-level amplitudes have multiparticle singularities when internal propagators $\frac{1}{P^2}$ go on shell. However, in SDYM and SDG, the multiparticle $\frac{1}{P^2}$ singularities that appear in each Feynman diagram are actually spurious, and after all the diagrams are added together the only remaining singularities are the two-particle holomorphic collinear ones. Because there is, in particular, no term in the currents like $\frac{1}{(p_1 + \ldots + p_n)^2}$, the factor of $(p_1 + \ldots + p_n)^2$ that comes from amputating the final leg does not cancel with anything in the denominator, which is why the overall expression becomes zero as $p_m^2 \to 0$. However, this argument does not render the 3-pt amplitudes zero. For instance, if we use $\mathcal{J}_{\rm G}(1,2)$ from \eqref{JG12}, we have
\begin{equation}
    \mathcal{M}^{\rm tree}(1^+,2^+,3^-) = - i \lr{\alpha 3}^4 (\lr{12}[12]) \left( - \frac{\kappa}{2} \frac{1}{\lr{\alpha 1}^2 \lr{\alpha 2}^2 } \frac{[12]}{\lr{12}}\right)
\end{equation}
which doesn't vanish due to the special nature of 3-pt kinematics.

(As an aside, another simple way to see the tree-level-one-minus amplitudes generically vanish follows from making the special choice of reference spinor $|\alpha\rangle \to |m\rangle$.)

There is however a loophole to our discussion. What if all holomorphic inner products are zero, i.e. what if $\lr{ij} = 0$ for all $i$, $j$? Then the currents would be singular, not zero. In a set of recent works \cite{Guevara:2026qzd, Guevara:2026qwa,Brandhuber:2026njb} this idea was explored and non-zero, distributional, expressions for the tree-level-one-minus amplitudes were found in the ``half collinear'' regime where all angle brackets $|i\rangle$ are proportional to each other. Note that this kinematic configuration is not possible for physical particles in $(1,3)$ signature. Actually, the non-vanishing of the three-point $+$$+$$-$ amplitudes can be seen as a special case of this story. This is because momentum conservation among three massless particles requires that either $|1\rangle \propto |2\rangle \propto |3\rangle$ or $|1] \propto |2] \propto |3]$.

\subsection{Another recursive formula for the all-plus currents from collinear limits}

The all-plus currents for gauge theory and gravity satisfy another pair of recursive formulae aside from the Berends-Giele one. These recursion relations come from the fact that the currents only have holomorphic collinear singularities, and should ultimately be understood as arising from the integrability of the classical theory.

For SDYM, this recursion relation is
\begin{equation}\label{JYMpole}
    \boxed{ \mJ_{\rm YM}(1^{a_1}, \ldots, n^{a_n})^{a_m} = \sqrt{2} g\sum_{i = 1}^{n-1}  \frac{f^{a_n a_i c} }{\lr{ni}} \frac{\lr{\alpha i}}{\lr{\alpha n}}\mJ_{\rm YM}(1^{a_1}, \ldots,   i^{c} , \ldots, (n\!-\!1)^{a_{n-1}})^{a_m}. }
\end{equation}
Note that all of the holomorphic collinear singularities of the LHS where $\lr{ni} \to 0$ for some $i$ can be read off from this formula.

At the level of the color-ordered partial currents using $[T^a, T^b] = i f^{abc} T^c$, if particle $n$ is inserted in between, say, some particles $i$ and $j$ in the color ordering, the above recursion relation is equivalent to
\begin{equation}
    J_{\rm YM}( \ldots, i, n, j, \ldots)  = i \sqrt{2} g \left( \frac{\lr{\alpha i}}{\lr{\alpha n}} \frac{1}{\lr{ni}} - \frac{\lr{\alpha j}}{\lr{\alpha n}} \frac{1}{\lr{nj}} \right) J_{\rm YM}( \ldots, i,j, \ldots ).
\end{equation}
The above formula can easily be checked to be true straight from the Schouten identity
\begin{equation}
    \left( \frac{\lr{\alpha i}}{\lr{\alpha n}} \frac{1}{\lr{ni}} - \frac{\lr{\alpha j}}{\lr{\alpha n}} \frac{1}{\lr{nj}} \right) = -\frac{\lr{ij}}{\lr{in} \lr{nj}}
\end{equation}
when applied to the Parke-Taylor-like expression for $J_{\rm YM}$ \eqref{Jymsoln}. In the special case that the particle $n$ finds itself on the edge of the color ordering, the recursion relation reads
\begin{equation}
    J_{\rm YM}(1, \ldots, n\!-\!1, n) = - i \sqrt{2} g \frac{\lr{\alpha, n\!-\!1}}{\lr{\alpha n}} \frac{1}{\lr{n\!-\!1,n}} J_{\rm YM}(1, \ldots, n\!-\!1)
\end{equation}
which is even more easily checked to be true from \eqref{Jymsoln}.

The reason that \eqref{JYMpole} holds is the meromorphicity of the all-plus current expression \eqref{Jymsoln}. In particular, both the left and right sides of the equation have all the ``correct'' holomorphic collinear poles $\lr{ni} \to 0$. Importantly, this uses the fact that $|n\rangle$ is an unconstrained variable in $\mathbb{C}P^1$. In this context, the ``correct'' pole is the one given by the 3-pt Feynman vertex, as we'll discuss more in section \ref{sec:hololimits}. Equation \eqref{JYMpole} is then simply a sum over all of the expected residues as $|n\rangle$ approaches each of the points $|i\rangle$.

In gravity, the analogous recursive formula is
\begin{empheq}[box=\fbox]{align}\label{JGpole}
  \;\;\; \vphantom{\Big\rvert}&\mJ_{\rm G}(|1\rangle, |1], \ldots, |n\rangle, |n] ) \\ \nonumber
    &= - \frac{\kappa}{2} \, \sum_{i = 1}^{n-1} \frac{\lr{\alpha i}^2}{\lr{\alpha n}^2} \frac{[ni]}{\lr{ni}} \mJ_{\rm G}(|1\rangle, |1], \ldots, |i \rangle, |i] + \frac{\lr{\alpha n}}{\lr{\alpha i}} |n], \ldots, |n\!-\!1\rangle, |n\!-\!1]). \;\;\; \vphantom{\Bigg\rvert}
\end{empheq}
Notice that this equation modifies the square bracket of particle $i$ on the RHS. The SDYM current does not depend on the square brackets, so the recursive formula \eqref{JYMpole} did not display this feature.

This formula was guessed by BDPR in \cite{Bern:1998xc}, see their (6.5), and then effectively proven to be true in their appendix C.1. It was however written down more clearly in the above form and proven by Krasnov and Scarinci in \cite{Krasnov:2013wsa,Delfino:2014xea} using natural tree/matrix identities, where it was related to Hodges' BCFW inverse-soft recursion relation \cite{Hodges:2012ym,Boucher-Veronneau:2011rwd,Dunbar:2012aj,Ma:2022qja,Arkani-Hamed:2009ljj,Nandan:2012rk} for MHV amplitudes via the NSVW formula. It was then accidentally rediscovered by the author in the context of $\Lw$ symmetry in \cite{Miller:2025wpq}. We discuss this relation more in section \ref{secLw}.

\begin{figure}[h]
    \centering
    \includegraphics{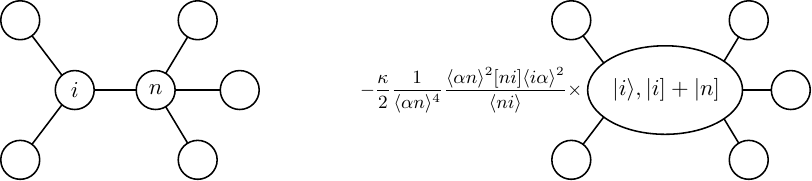}
    \caption{Left: If $|n\rangle \to |i\rangle$, the only singular trees in $\mJ_{\rm G}$ are ones where the nodes $n$ and $i$ are adjacent. Right: In this limit, we can consider $i$ and $n$ as belonging to a single node with parameters $|i\rangle$, $|i] + |n]$, multiplied by an extra factor to account for the edge between $i$ and $n$.}
    \label{fig:tree12}
\end{figure}

We've put the proof of this formula in appendix \ref{sec:collinearsdg}. Here, we'll simply explain why this formula exhibits the correct holomorphic collinear limits as $|n\rangle \to |i\rangle$. 

The key idea is drawn in figure \ref{fig:tree12}. If $|n\rangle \to |i\rangle$, then the only singular terms come from trees where the nodes $i$ and $n$ are adjacent. All the other neighboring nodes in the tree then have the option of connecting to either $i$ or $n$. We can therefore think of the $i$-$n$ edge as a single node itself, with an angle bracket of $|i\rangle$ and a square bracket of $|i] + |n]$. (If one foils out the connection between a neighboring node and the square bracket $|i] + |n]$, one reproduces the sum over trees whose edges could have connected to either $i$ or $n$.) Therefore, in the holomorphic collinear limit the tree formula \eqref{JGtree} reproduces that of the RHS of \eqref{JGpole}, which is
\begin{equation}
    \lim_{|n\rangle \to |i \rangle} \mJ_{\rm G}(1,2,3,\ldots,n) = - \frac{\kappa}{2} \frac{1}{\lr{\alpha n}^4} \frac{\lr{\alpha n}^2[ni] \lr{i \alpha}^2}{\lr{ni}} \mJ_{\rm G}(  \ldots, |i\rangle, |i] + |n], \ldots  ).
\end{equation}
This is also the correct holomorphic collinear limit one can calculate directly from the 3-pt vertex.

\section{Basic properties of the one-loop-all-plus amplitudes}\label{sec:properties}

Let us now start our discussion of the one-loop-all-plus-amplitudes of SDYM and SDG by reviewing their basic properties.

\subsection{Rationality}\label{sec:rationality}

The cutting rules state that the analytic discontinuity of a one-loop amplitude with respect to a particular kinematic invariant occurs when the momenta passing through the appropriate cut of the loop go on shell \cite{Badger:2023eqz}. So if one ``cuts'' the amplitude through the loop and demands that the two tree-level sub-amplitudes independently obey momentum conservation, and then sums over all possible internal particles that can run through the cut, one can compute the discontinuity.

\begin{figure}[h]
    \centering
    \includegraphics[width=0.75\linewidth]{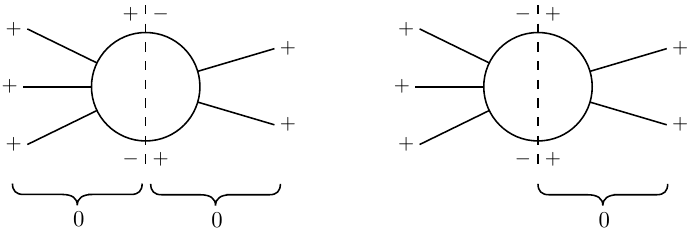}
    \caption{The possible ``cuts'' of a one-loop-all-plus amplitude. Either both sides are one-minus amplitudes, which vanish, or one side is MHV and the other is all-plus, which also vanishes.}
    \label{fig:rational}
\end{figure}

For the case of the one-loop-all-plus amplitudes in pure YM and pure gravity, the possibilities are drawn in figure \ref{fig:rational}. The discontinuity is either given by two on-shell tree-level-one-minus amplitudes, or a tree-level MHV amplitude and a tree-level-all-plus amplitude. In both cases the cuts vanish and we can plainly see that there are no analytic discontinuities.

If the amplitude has no branch cuts then its structure must be dramatically simpler than that of a generic one-loop amplitude. It cannot contain, say, polylogs or other special functions. It must be rational, i.e. be a ratio of angle and square brackets.

The one-loop-one-minus amplitudes must also be rational for the same reason.

\subsection{Finiteness}

Because the tree-level-all-plus amplitudes are zero, the one-loop-all-plus amplitudes are both IR and UV finite.
\begin{figure}[h]
    \centering
    \includegraphics{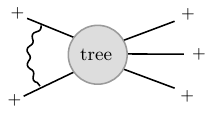}
    \caption{One-loop IR divergences in gauge theory/gravity come from internal particles connecting external legs.}
    \label{fig:IRdiv}
\end{figure}
IR divergences in YM/gravity one-loop amplitudes come from gluons/gravitons connecting two external legs. See figure \ref{fig:IRdiv}. In the case of gauge theory, the IR divergence is proportional to \cite{Bern:1998sv}
\begin{equation}
    \text{IR div.} \sim \mathcal{A}_n^{\rm tree} \sum_{i \neq j} \left[ -\frac{1}{\epsilon^2} + \frac{\ln(-s_{ij})}{\epsilon}\right]
\end{equation}
where $\epsilon$ is the parameter in dimensional regularization, and in gravity the IR divergence is proportional to 
\begin{equation}
    \text{IR div.} \sim  \mathcal{M}_n^{\rm tree} \sum_{i \neq j} \left[ s_{ij} \frac{\ln(-s_{ij})}{\epsilon}\right].
\end{equation}
For one-loop-all-plus amplitudes the IR divergence is therefore zero because the tree-level-all-plus amplitudes are zero.

Similarly, the UV divergence of the one-loop-all-plus amplitudes also vanishes because the tree-level-all-plus amplitudes are zero. Let us recall why. Think about pure Yang-Mills theory in any covariant gauge. The tree-level-all-plus amplitudes vanish because the numerators of these Feynman diagrams always contain a contraction between two positive-helicity polarization vectors, and one can always choose a residual gauge in which these vanish with $\vep^+_i \cdot \vep^+_{i'} = 0$. Now, in the action, there is of course a propagator, a 3-pt vertex, and a 4-pt vertex which all get accompanying counterterms with the same kinematic structure. Because YM is renormalizable there are no more counterterms. So, if one wants to compute a one-loop-all-plus amplitude, the contributions from the tree diagrams containing a counterterm also vanish, once again due to the contractions between the polarization vectors. Therefore, because the counterterms will not enter the computation, the original amplitude must have been finite, and the coupling is automatically the renormalized one which matches tree-level amplitudes. The story works the same for gravity as the Einstein-Hilbert action is famously renormalizable at one loop \cite{tHooft:1974toh} because the only possible counterterms that could be generated are either topological or are able to be removed by a field redefinition. At two-loops there is the Goroff-Sagnotti counterterm \cite{Goroff:1985th}.

\subsection{Holomorphic collinear limits}\label{sec:hololimits}

\begin{figure}[h]
    \centering
    \includegraphics{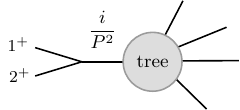}
    \caption{Tree-level amplitudes factorize as $P = p_1 + p_2$ goes on shell.}
    \label{fig:M12}
\end{figure}

Consider a tree-level YM or gravity scattering amplitude which contains two positive-helicity gluons/gravitons with momenta $p_1$ and $p_2$ with
\begin{equation}
    p_1 = |1\rangle [1|, \hspace{1 cm} p_2 = |2\rangle [2|.
\end{equation}
Now imagine we take the holomorphic collinear limit
\begin{equation}
    |1\rangle \to |2\rangle.
\end{equation}
If the two particles attach at a 3-pt vertex to an internal propagator with momentum $P$, where
\begin{equation}
    P = p_1 + p_2
\end{equation}
then in this limit
\begin{equation}
\begin{aligned}
    P  \to |2\rangle &([1| + [2|),\\ P^2 &\to 0,
\end{aligned}
\end{equation}
so $P$ goes on-shell and these diagrams develop a pole from the propagator. See figure \ref{fig:M12}.

Therefore, in this limit, the only relevant diagrams are those where particles $1^+$ and $2^+$ connect at a 3-pt vertex before attaching to the rest of the diagram. 

For gauge theory and gravity amplitudes, we express this as
\begin{equation}\label{M12eq}
\begin{aligned}
    \lim_{|1\rangle \to |2 \rangle} \mathcal{A}^{\rm tree}(1^{a+},2^{b+}, \ldots) &= \text{Split}^{1^{a+} 2^{b+}}_{{\rm YM}, P^{c-}} \, \mathcal{A}^{\rm tree}(P^{c+}, \ldots) \, ,\\
    \lim_{|1\rangle \to |2 \rangle} \mathcal{M}^{\rm tree}(1^+,2^+, \ldots) &= \text{Split}^{1^{+} 2^{+}}_{{\rm G}, P^{-}} \, \mathcal{M}^{\rm tree}(P^+, \ldots) \, .
\end{aligned}
\end{equation}
The formulae imply the tree-level amplitudes ``factorize''. Note that the helicity of the $P$ particle in the splitting function $\text{Split}^{1^+ 2^+}_{P^-}$ is negative while in the amplitude it is positive, because the propagator is of the form $(+-)$. 
\begin{figure}[h]
    \centering
    \includegraphics{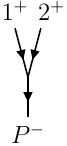}
    \caption{The diagram contributing to the $+$$+$$-$ splitting function.}
    \label{fig:split12P}
\end{figure}
We can easily compute the on-shell 3-pt splitting function using the lightcone Feynman rules from section \ref{sec:FeynmanRules}. See figure \ref{fig:split12P}. Accounting for the appropriate polarization factors, vertex, and $P$ propagator, they are simply
\begin{equation}\label{splitym}
\begin{aligned}
    \text{Split}^{1^{a+} 2^{b+}}_{{\rm YM}, P^{c-}} &= \frac{\lr{\alpha P}^2}{\lr{\alpha 1}^2 \lr{\alpha 2}^2} \times i \sqrt{2}g f^{abc} \lr{\alpha| 12|\alpha} \times \frac{i}{\lr{12}[12]} \\ & \\
    &= \sqrt{2} g f^{abc} \frac{1}{\lr{12}} \frac{\lr{\alpha P}^2}{\lr{\alpha 1} \lr{\alpha 2} }
\end{aligned}
\end{equation}
for Yang-Mills and
\begin{equation}\label{splitg}
\begin{aligned}
    \text{Split}^{1^+ 2^+}_{{\rm G}, P^-} &= \frac{\lr{\alpha P}^4}{\lr{\alpha 1}^4 \lr{\alpha 2}^4} \times \frac{i \kappa}{2} \lr{\alpha| 12|\alpha}^2 \times \frac{i}{\lr{12}[12]} \\ & \\
    &= - \frac{\kappa}{2} \frac{[12]}{\lr{12}} \frac{\lr{\alpha P}^4}{\lr{\alpha 1}^2 \lr{\alpha 2}^2 }
\end{aligned}
\end{equation}
for gravity. Similarly, one can also compute the $+$$-$$+$ splitting functions in the exact same way, and they are
\begin{equation}\label{splitpmp}
    \text{Split}^{1^{a+} 2^{b-}}_{{\rm YM}, P^{c+}} = \sqrt{2} g f^{abc} \frac{1}{\lr{12}} \frac{\lr{\alpha 2}^3}{\lr{\alpha 1} \lr{\alpha P}^2 }, \hspace{0.75cm} \text{Split}^{1^+ 2^-}_{{\rm G}, P^+} = -\frac{\kappa}{2} \frac{[12]}{\lr{12}} \frac{\lr{\alpha 2}^6}{\lr{\alpha 1}^2 \lr{\alpha P}^2 }.
\end{equation}
Also, because there is no $+$$+$$+$ vertex, the associated splitting function is zero for gauge theory and gravity:
\begin{equation}
    \text{Split}^{1^{a+} 2^{b+}}_{{\rm YM}, P^{c+}} = 0, \hspace{1cm} \text{Split}^{1^+ 2^+}_{{\rm G}, P^+} =0.
\end{equation}

\begin{figure}[h]
    \centering
    \includegraphics{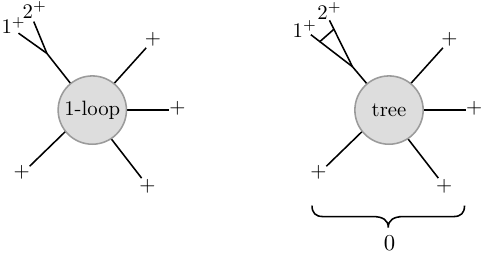}
    \caption{Two types of singular diagrams for one-loop-all-plus amplitudes as $P = p_1 + p_2$ goes on-shell. The second type of diagram vanishes because tree-level-one-minus amplitudes are zero.}
    \label{fig:oneloopsplit}
\end{figure}

Relevantly for us, the tree-level splitting functions are actually uncorrected in the one-loop-all-plus amplitudes! We can see why by looking at the structure of the self-dual lightcone diagrammatics. At one-loop level, there are two types of diagrams that become singular as $P$ goes on shell, drawn in figure \ref{fig:oneloopsplit}.

In the first diagram, particles 1 and 2 connect at tree-level and the propagator connects to a lower-point one-loop-all-plus amplitude with $P$ on-shell. In the second diagram, we see a potential loop correction to the 3-pt splitting function, but the $P$ propagator connects to a tree-level-one-minus amplitude which vanishes as $P^2 \to 0$. So only the first type of diagram contributes, and the splitting function for one-loop-all-plus amplitudes equals the tree-level splitting function \cite{Ball:2021tmb}.

Actually, even though the above conclusion is correct, our reasoning was a bit hasty. There is also another type of contribution we need to consider, the ``non-factorizing'' contribution when particles 1 and 2 both connect directly to the loop itself. One might naively think that these diagrams won't correct the splitting function because for generic values of the loop momentum $\ell$ there is no pole as $p_1 \cdot p_2 \to 0$ in the integrand. Nonetheless there do exist regimes of the loop momentum where a singularity could develop because an extra internal propagator could go on shell.

\begin{figure}[h]
    \centering
    \includegraphics{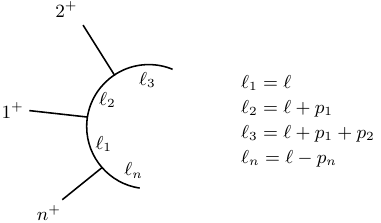}
    \caption{As $p_1 \cdot p_2 \to 0$, diagrams of this type could develop a pole in $p_1 \cdot p_2$, but don't for the one-loop-all-plus amplitudes after integration.}
    \label{fig:nonfactorize}
\end{figure}

Consider for instance the configuration drawn in figure \ref{fig:nonfactorize}, which is the potentially problematic portion of a diagram, where we integrate over $\int d^4 \ell$. Consider the integral in the $\ell \approx 0$ region (where we continue the loop momentum in the integral to have euclidean signature). In the diagram we have the four propagators $1/\ell_1^2$, $1/\ell_2^2$, $1/\ell_3^2$, $1/\ell_n^2$, and if our theory is SDYM we have the three numerators $\lr{\alpha|1 \ell|\alpha}$, $\lr{\alpha|2 (\ell + 1)|\alpha}$, $\lr{\alpha|n \ell|\alpha}$.
For small $\ell$, the four respective propagators are approximately $1/\ell_1^2 = 1/\ell^2$, $1/\ell_2^2 = 1/(2 \ell \cdot p_1)$, $1/\ell_3^2 = 1/(2 \ell \cdot (p_1 + p_2) + 2 p_1 \cdot p_2)$, and $1/\ell_n^2 = 1/(-2 \ell \cdot p_n)$. Notice that the $1/\ell_3^2$ propagator goes on shell if $p_1 \cdot p_2 \to 0$, but let's see if this becomes a divergence after integration. Counting powers of $\ell$ from the numerator and denominator, we essentially have $\int d^4 \ell \frac{\ell^2}{\ell^5}$ which vanishes over the $\ell \approx 0$ region due to the phase space factor. For SDG the integrand vanishes in this region even quicker, as the numerator is squared. So in conclusion, the one-loop-all-plus amplitudes do indeed enjoy the tree-level splitting relations
\begin{equation}\label{holosplit1loop}
\begin{aligned}
    \lim_{|1\rangle \to |2 \rangle} \mathcal{A}^\text{1-loop}(1^{a+},2^{b+}, \ldots, n^{a_n+}) &= \text{Split}^{1^{a+} 2^{b+}}_{{\rm YM}, P^{c-}} \mathcal{A}^\text{1-loop}(P^{c+}, \ldots, n^{a_n+}),\\
    \lim_{|1\rangle \to |2 \rangle} \mathcal{M}^\text{1-loop}(1^+,2^+, \ldots, n^+) &= \text{Split}^{1^{+} 2^{+}}_{{\rm G}, P^{-}} \mathcal{M}^\text{1-loop}(P^+, \ldots, n^+).
\end{aligned}
\end{equation}
(We also note that a similar argument implies that the one-loop-all-plus amplitudes have no multiparticle poles, which uses the fact that the Berends-Giele currents themselves have no such poles.) For one-loop-\textit{not}-all-plus, the story gets more complicated. We can't rely on the vanishing of the tree-level-one-minus amplitudes to render the one-loop correction to the 3-pt splitting function zero, as in figure \ref{fig:oneloopsplit}. These channels now in fact produce double poles. (In the following section \ref{sec:loopintegrals} we compute that the one-leg-off-shell-all-plus triangles for gauge theory and gravity are $J_{{\rm YM} +++}^{\text{1-loop}} \propto \frac{[12]^2}{\lr{12}}$ and $\mJ_{{\rm G} +++}^{\text{1-loop}} \propto \frac{[12]^5}{\lr{12}}$, and multiplying them by the propagator $\frac{i}{\lr{12}[12]}$ gives the double pole.) These double-poles become important if one wants to compute, say, one-loop-one-minus amplitudes using BCFW recursion \cite{Dunbar:2010xk,Bern:2005hs,Brandhuber:2007up,Alston:2012xd,Alston:2015gea,Farrow:2020voh}. See also \cite{Bittleston:2022jeq,Fernandez:2024qnu,Costello:2023vyy} for the chiral algebra perspective on the double poles.

We mention that the closely related concept of the ``full'' collinear limit is defined by taking $p_1 \to t P$, $p_2 \to (1 - t)P$, and $P = p_1 + p_2$ for some constant $t$. In the full collinear limit we have both $\lr{12} \to 0$ and $[12] \to 0$, instead of just $\lr{12} \to 0$ as in the holomorphic collinear limit. In the full collinear limit, one usually takes $|1\rangle = \sqrt{t}|P\rangle$, $|1] = \sqrt{t}|P]$, $|2\rangle = \sqrt{1-t}|P\rangle$, and $|2] = \sqrt{1-t}|P]$.

There is also one more type of limit, which we'll call the ``double-scaled'' holomorphic collinear limit. This is the limit where $[12] \to 0$, $\lr{12} \to 0$, but $[12]/\lr{12} \to \infty$. For real physical momenta, this is equivalent to the full-collinear limit where one then extracts the only terms with a ``phase singularity.'' This is defined by holding $p_1$ fixed and rotating the nearly collinear $p_2$ in a tiny circle around it. A remarkable fact about the ``double-scaled'' holomorphic collinear limit is that its splitting function is uncorrected to all loops in full Einstein gravity. This is however not the case in Yang-Mills theory, as the gravity loop integrals are ``softer'' than the Yang-Mills loop integrals because the coupling constant $\kappa$ is dimensionful. See for instance section 5.2 of \cite{Bern:1998sv} or section 4 of \cite{Bern:1998xc}. 

\section{Bubbles and triangles}\label{sec:loopintegrals}

In this section we discuss the loop integrals that arise in SDYM and SDG. This topic will be the subject of a forthcoming paper by the author, Guevara, and Himwich \cite{toappear}. Here we'll just be content with looking at the bubble and triangle.

\begin{figure}[h]
    \centering
    \includegraphics{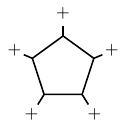}
    \caption{An all-plus $(n'\!=\!5)$-gon}
    \label{fig:mgon}
\end{figure}

\begin{figure}[h]
    \centering
    \includegraphics{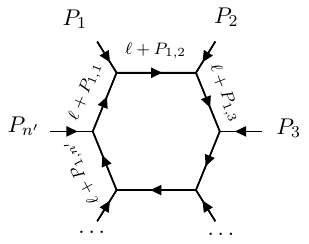}
    \caption{Momentum flowing through all-plus $n'$-gon.}
    \label{fig:mgonlabel}
\end{figure}

\noindent Consider an $n'$-gon with $n' \leq n$ where $n$ is the number of external massless particles. We denote the total off-shell momentum flowing into each leg as $P_i$ with $P_1 + \ldots + P_{n'} = 0$. We define the region momenta
\begin{equation}
    P_{1,i} \equiv \sum_{j=1}^{i-1} P_j
\end{equation}
and
\begin{equation}
    P_{i,j} \equiv P_{1,j} - P_{1,i}
\end{equation}
such that, if $j > i$,
\begin{equation}
    P_{i,j} = P_{i} + P_{i+1} + \ldots + P_{j-1}.
\end{equation}
We parametize the momenta flowing through the diagram via figure \ref{fig:mgonlabel}, where $\ell$ is the loop momentum.
We regulate the loop diagram using dimensional regularization. We denote as $L$ as the $D=4-2\epsilon$ dimensional momentum we use in the denominator of the diagram while $\ell$ is the $4$-dimensional part in the numerator. The planar SDYM $n'$-gon is then
\begin{equation}
    \mI^{n'\text{-gon}}_{\rm SDYM}(P_{1}, \ldots, P_{n'}) = 2 (-i \sqrt{2} g)^{n'}\intL \frac{1}{(L+P_{1,1})^2 \ldots (L + P_{1,n'})^2} \prod_{i=1}^{n'}\lr{\alpha | P_{i}  (\ell + P_{1,i}) | \alpha},
\end{equation}
and the SDG $n'$-gon is
\begin{equation}
    \mI^{n'\text{-gon}}_{\rm SDG}(P_{1}, \ldots, P_{n'}) = 2 \left( - \frac{\kappa}{2} \right)^{n'} \intL \frac{1}{(L+P_{1,1})^2 \ldots (L + P_{1,n'})^2} \prod_{i=1}^{n'}\lr{\alpha | P_{i} (\ell + P_{1,i})  | \alpha}^2.
\end{equation}
There is an overall factor of $2$ for the two orientations of $(+-)$ propagators running around the loop.

In \cite{toappear} we do the loop integral using standard manipulations and are left with an integral over Feynman parameters. For SDYM, the $n'$-gon is now
\begin{equation}\label{sdymngon}
    \mI^{n'\text{-gon}}_{\rm SDYM} = 2 i \frac{(-i \sqrt{2} g)^{n'}}{(4 \pi)^{\frac{D}{2}}}  \Gamma\left(n' - \tfrac{D}{2}\right) \int_{[0,1]^{n'}} d^{n'}u \frac{\delta(1-\sum_{k=1}^{n'
    }u_k)}{(\sum_{r,s} Q_{rs} u_r u_s)^{n' - \frac{D}{2}}} \prod_{i=1}^{n'} \lr{\alpha |P_i \Big( \sum_{j=1}^{n'} P_{i,j} u_j \Big) | \alpha},
\end{equation}
and for SDG
\begin{equation}\label{sdgngon}
    \mI^{n'\text{-gon}}_{\rm SDG} = 2 i \frac{(\kappa/2)^{n'}}{(4 \pi)^{\frac{D}{2}}}  \Gamma\left(n' - \tfrac{D}{2}\right) \int_{[0,1]^{n'}} d^{n'}u \frac{\delta(1-\sum_{k=1}^{n'
    }u_k)}{( \sum_{r,s} Q_{rs} u_r u_s)^{n' - \frac{D}{2}}} \prod_{i=1}^{n'} \lr{\alpha |P_i \Big( \sum_{j=1}^{n'} P_{i,j} u_j \Big) | \alpha}^2.
\end{equation}
Here we have defined the Symanzik matrix
\begin{equation}
    Q_{ij} \equiv \frac{1}{2}(P_{i,j})^2.
\end{equation}
There is much to say about these integrals for general $n'$, but here we'll just comment on the case of the bubble $n'=2$ and the triangle $n'=3$.

While the bubble has a superficial $\Gamma(\epsilon)$ divergence, the numerator
\begin{equation}
    \lr{\alpha| P_1 (P_{1,1} u_1 + P_{1,2} u_2) |\alpha} \lr{\alpha| P_2 (P_{2,1} u_1 + P_{2,2} u_2) |\alpha} = 0
\end{equation}
vanishes identically due to $\lr{\alpha| P_1 P_1 |\alpha} = 0$. Therefore, the SDYM and SDG bubbles vanish in dimensional regularization, even when $P_1$ is off-shell.\footnote{In other schemes however the bubble does not vanish identically, see for instance \cite{Brandhuber:2007vm,Thorn:2005ak,Chakrabarti:2005ny,Chakrabarti:2006mb,Chattopadhyay:2020oxe}.}
\begin{figure}[h]
    \centering
    \includegraphics{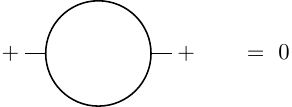}
    \caption{The bubble diagram vanishes in SDYM and SDG in dim-reg.}
    \label{fig:bubble}
\end{figure}

Now we look at the triangle with two legs on-shell, i.e. $P_1 = p_1$, $P_2 = p_2$ with $p_1^2 = p_2^2 = 0$ and $P_3 = - P_1 - P_2$.
\begin{figure}[h]
    \centering
    \includegraphics{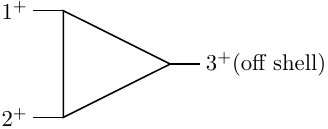}
    \caption{Triangle diagram with $p_1^2 = p_2^2 = 0$ for computing $\mJ_{+++}^{\text{1-loop}}.$}
    \label{fig:triangle}
\end{figure}
We will now define the SDYM and SDG one-loop-all-plus 3-pt currents via
\begin{equation}
\begin{aligned}
    J_{{\rm YM} +++}^{\text{1-loop}} &= \frac{1}{\lr{\alpha 1}^2} \frac{1}{\lr{\alpha 2}^2} \mI^{3-{\rm gon}}_{\rm SDYM} , \\
    \mJ_{{\rm G} +++}^{\text{1-loop}} &= \frac{1}{\lr{\alpha 1}^4} \frac{1}{\lr{\alpha 2}^4} \mI^{3-{\rm gon}}_{\rm SDG}.
\end{aligned}
\end{equation}

In order to compute the one-loop-all-plus three-point currents all we have to do is use \eqref{sdymngon} and \eqref{sdgngon}. We will also denote the measure of the integral over the simplex as $\int_{\Delta_3} \equiv \int_{[0,1]^3} d^3 u \delta(1 \!-\! u_1 \!-\! u_2 \!-\! u_3)$ for brevity. For SDYM,
\begin{equation}
\begin{aligned}
    J_{{\rm YM} +++}^{\text{1-loop}} 
    &= \frac{1}{\lr{\alpha 1}^2} \frac{1}{\lr{\alpha 2}^2} 2i \frac{(- i \sqrt{2} g)^3}{(4 \pi)^2} \int_{\Delta_3} \frac{(\lr{\alpha|1 2 |\alpha} u_3)(\lr{\alpha|2 3 |\alpha} u_1)(\lr{\alpha|3 1 |\alpha} u_1)}{(p_1 + p_2)^2 u_1 u_3}\\
    &= \frac{1}{\lr{\alpha 1}^2} \frac{1}{\lr{\alpha 2}^2} 2i \frac{(- i \sqrt{2} g)^3}{(4 \pi)^2} \frac{1}{6} \lr{\alpha|1 2 |\alpha} \lr{\alpha|2 3 |\alpha} \lr{\alpha|3 1 |\alpha} \frac{1}{\lr{12}[12]} \\
    &=  -i \frac{(- i \sqrt{2} g)^3}{3 (4 \pi)^2}  \langle \alpha 1 \rangle \langle\alpha 2 \rangle \frac{[12]^2}{\lr{12}}
\end{aligned}
\end{equation}
and for SDG
\begin{equation}
\begin{aligned}
    \mJ_{{\rm G} +++}^{\text{1-loop}} 
    &= \frac{1}{\lr{\alpha 1}^4} \frac{1}{\lr{\alpha 2}^4} 2 i \frac{(\kappa/2)^{3}}{(4 \pi)^2} \int_{\Delta_3} \frac{(\lr{\alpha|1 2 |\alpha} u_3)^2 (\lr{\alpha|2 3 |\alpha} u_1)^2(\lr{\alpha|3 1 |\alpha} u_1)^2}{(p_1 + p_2)^2 u_1 u_3} \\
    &= 2 i \frac{(\kappa/2)^{3}}{(4 \pi)^2} \frac{1}{360} \frac{[12]^5 \lr{\alpha 1}^2 \lr{\alpha 2}^2 }{\lr{12}} .
\end{aligned}
\end{equation}
Here we have used the particular simplex integrals
\begin{equation}
\begin{aligned}
    &\int_{\Delta_3} \, u_1 = \frac{1}{6}, \\ &\int_{\Delta_3}  \, u_1 u_2^2 u_3 = \frac{1}{360}.
\end{aligned}
\end{equation}

If we wish to compute the 3-pt one-loop-all-plus amplitudes, we simply put the third leg on-shell. This gives
\begin{equation}
\begin{aligned}
    A_{3;1}^{\text{1-loop}}(1^+,2^+,3^+) = \frac{1}{\lr{\alpha 3}^2} J_{{\rm YM} +++}^{{\text{1-loop}}} \Big\rvert_{p_3^2 = 0} &= 0 \\
    \mathcal{M}^{\text{1-loop}}(1^+,2^+,3^+) = \frac{1}{\lr{\alpha 3}^4} \mJ_{{\rm G} +++}^{{\text{1-loop}}} \Big\rvert_{p_3^2 = 0} &= 0
\end{aligned}
\end{equation}
where here we used $P_3^2 = p_3^2 = [12] \lr{12}$. These amplitudes are considered to vanish because there are more powers of $[12]$ in the numerator than $\lr{12}$ in the denominator. (The subscript in $A_{3;1}^{\text{1-loop}}$ denotes that this is the single-trace part of the amplitude, although at 3-pts this is also the full amplitude because $\Tr(T^a) = 0$ for $T^a \in \mathfrak{su}(N_c)$.)

Other works which study SDYM and SDG loop integrals include \cite{Boels:2013bi,Brandhuber:2006bf,Chattopadhyay:2020oxe,Monteiro:2022nqt} and the papers of Chakrabarti, Qiu, and Thorn \cite{Thorn:2005ak,Chakrabarti:2005ny,Chakrabarti:2006mb}.

\section{Double-off-shell calculation of one-loop-all-plus amplitudes}\label{sec:doubleoffshell}

\subsection{SDYM double-off-shell current calculation}

In this section we will review a historical computation and compute an all-$n$ closed form expression for the one-loop-all-plus amplitudes in Yang-Mills theory. This expression was conjectured in \cite{Bern:1993sx,Bern:1993qk} because it had the correct collinear and soft limits, and was proven by Mahlon in \cite{Mahlon:1993si,Mahlon:1993fe} using the recursive currents techniques we review here.

In particular, Mahlon computed the ``double-off-shell-current'' of tree-level diagrams with an arbitrary number of on-shell positive-helicity gluons and two off-shell quarks. The off-shell quark legs can then be sewn together and integrated over to give the contribution to one-loop-all-plus gluon QCD amplitudes where an internal massless quark is running around the loop. From supersymmetry, one can deduce that the full one-loop-all-plus amplitude is proportional to $(1 + \tfrac{N_s}{N_c} - \tfrac{N_f}{N_c})$, where $N_f$ is the number of internal quarks running around the loop and $N_s$ is the number of internal complex scalars. Therefore, Mahlon's result is $-\frac{N_f}{N_c}$ times the one-loop-all-plus amplitude in pure YM. With his double-off-shell current, Mahlon also computed many other other families of amplitudes, including the one-loop-one-minus amplitudes.

In this section, we will work in pure SDYM, translating Mahlon's quark-loop calculation into a pure gluon-loop calculation. We use the Feynman rules from the scalar SDYM lightcone action, and this slightly simplifies the calculation.

We denote the color-ordered\footnote{By color-ordered, we mean we are restricting ourselves to planar Feynman diagrams where the external momenta $\ell$, $1, \ldots, n, m$ are fixed in a clockwise pattern around the diagram, as in figure \ref{fig:JQdef}. See appendix \ref{app:color} or eq. \eqref{JYMpartial} for the analogous equation with single-off-shell currents.} all-plus double-off-shell current by $J_{\rm YM}(\ell \, ; 1, \ldots, n)$. $\ell$ is an off-shell momentum with $\ell^2 \neq 0$, while $p_1^2 = \ldots = p_n^2 = 0$. Diagrammatically, $\ell$, $1, \ldots, n$ are all positive-helicity. (We are able to assign helicity to particles which are off-shell because we are working in lightcone gauge.) There is a second off-shell negative-helicity particle called $m$, where $p_m \equiv -( \ell + p_1 + \ldots + p_n)$. The off-shell legs are of course not given polarization factors, while the on-shell legs are. We also include the propagator factors $i/\ell^2$ and $i/p_m^2$ for the off-shell legs. See figure \ref{fig:JQdef}.
\begin{figure}[h]
    \centering
    \includegraphics[width=0.61\textwidth]{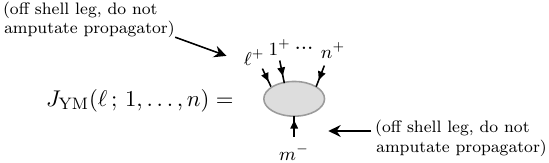}
    \caption{Color-ordered all-plus double-off-shell current definition.}
    \label{fig:JQdef}
\end{figure}

The base case for the double-off-shell current is simply
\begin{equation}
    J_{\rm YM}(\ell) =  \frac{i}{\ell^2}.
\end{equation}
Compare this to the analogous formula for the on-shell case, $J_{\rm YM}(i) = \frac{1}{\lr{\alpha i}^2}$, which includes a polarization factor.

\begin{figure}[h]
    \centering
    \includegraphics{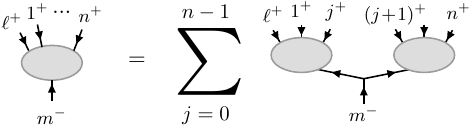}
    \caption{Double-off-shell current recursion relation \eqref{JQNYMrecursion}. }
    \label{fig:Qrecursion}
\end{figure}

We remind the reader of the formula for the single-off-shell-all-plus current that was given in eq. \eqref{Jymsoln}, which we reproduce below. 
\begin{equation}
    J_{\rm YM}(1, \ldots, n) = (-i \sqrt{2} g)^{n-1} \frac{1}{\lr{\alpha 1}} \frac{1}{\lr{12} \ldots \lr{n-1,n}} \frac{1}{\lr{\alpha n}}
\end{equation}
We can now create a Berends-Giele recursion relation for the double-off-shell current by gluing lower-point double-off-shell and single-off-shell currents together. See figure \ref{fig:Qrecursion}.
The equation for this recursion relation is
\begin{equation}\label{JQNYMrecursion}
    (\ell+p_N)^2 J_{\rm YM}(\ell \,; N) = i \sqrt{2} g \sum_{\substack{j=0\\ A = \{1, \ldots, j\} \\ B=\{j+1, \ldots ,n\}  }}^{n-1} \lr{ \alpha | (\ell+p_A) p_B | \alpha } J_{\rm YM}(\ell\,; A) J_{\rm YM}(B)
\end{equation}
where we define the ordered set $N = \{1, \ldots, n\}$. We now give the scalar lightcone gluon version of Mahlon's solution for this current.
\begin{equation}\label{JYMdoublesoln}
    \boxed{ J_{\rm YM}(\ell \, ; \, 1, \ldots, n) =(-i \sqrt{2} g)^{n} \frac{-i }{\lr{\alpha 1} \lr{12} \ldots \lr{n-1,n} \lr{\alpha n}} \left( \sum_{j=1}^{n} \frac{ \lr{\alpha |(\ell+p_{1,j} ) j|\alpha } }{(\ell + p_{1,j-1})^2(\ell + p_{1,j})^2 } \right) }
\end{equation}
We prove this formula satisfies the recursion relation in appendix \ref{sec:derivationdoubleoffshell1}. Here we use the notation\footnote{In terms of our region momenta from the previous section \ref{sec:loopintegrals}, $p_{i,j} = P_{i,j+1}$.}
\begin{equation}
    p_{i,j} \equiv p_i + p_{i+1} + \ldots + p_{j-1} + p_j
\end{equation}
where we also define
\begin{equation}
    p_{i,i-1} \equiv 0.
\end{equation}
It is easy to check that in the limit that $\ell^2 \to 0$, the double-off-shell-current becomes the single-off-shell current, modulo the dropped polarization factor and the extra propagator:
\begin{equation}
    \lim_{\ell^2 \to 0} (\tfrac{\ell^2}{i}) J_{\rm YM}(\ell\,;\, 1, \ldots, n) = \lr{\alpha \ell}^2 J_{\rm YM}(\ell ,1, \ldots, n) \Big\rvert_{\ell^2= 0}.
\end{equation}

Note that the closed-form-solution \eqref{JYMdoublesoln} only has two propagators in the denominator instead of the $n$ propagators one would naively expect from the Feynman diagrams.

In principle we should be able to sew together the two off-shell legs of this current and integrate over the loop momentum $\ell$ in order to compute the one-loop-all-plus amplitude.

However, if we want to regulate our loop integral using dimensional regularization, then our loop momentum does not live in 4 dimensions but rather in $4-2\epsilon$ dimensions, and this modifies the double-off-shell current. We denote our $4-2\epsilon$ dimensional vector as $L$, for which $\ell$ is its 4 dimensional component. The norm $L^2$ can then be expressed as
\begin{equation}
    L^2 = \ell^2 - \mu^2
\end{equation}
where $\mu^2$ is the magnitude of the $-2\epsilon$ dimensional part orthogonal to the 4 dimensional part. (The minus sign in front of $\mu^2$ is because we are working in mostly minus signature. One can also imagine $\epsilon < 0$ if one likes.) We therefore have
\begin{equation}
    (L+p)^2 = (\ell+p)^2 - \mu^2
\end{equation}
if $p$ is some other 4-dimensional vector.

In our dimensional regularization scheme, we replace all the propagators in the denominators of our Feynman diagrams $i/\ell^2$ with $i/L^2$. The numerators, however, do not get modified. They remain living in exactly $4$ dimensions, i.e. they depend on $\ell$ and not $L$. This means that the recursion relation for the double-off-shell current \eqref{JQNYMrecursion} should have $((\ell + p_N)^2 - \mu^2)$ on the LHS instead of $(\ell + p_N)^2$.

This regularization modifies the double-off shell current, and Mahlon showed it becomes
\begin{empheq}[box=\fbox]{align}
    J_{\rm YM}(L \, & ; 1, \ldots , n ) =(-i \sqrt{2} g)^{n} \frac{-i}{\lr{\alpha 1} \lr{12} \ldots \lr{n-1,n} \lr{\alpha n}} \nonumber \\
    &\times \sum_{j=1}^{n} \Bigg(    \frac{ \lr{\alpha |(\ell+p_{1,j} ) j|\alpha } }{(L + p_{1,j-1})^2(L + p_{1,j})^2 } -  \mu^2 \frac{ \lr{\alpha | p_{j,n} j  |\alpha } }{(L + p_{1,j-1})^2(L + p_{1,j})^2 (L+p_{1,n})^2 } \Bigg). \nonumber
\end{empheq}
\vspace{-0.5 cm}
\begin{equation}\label{JYMdoublesolnmu}
    \textcolor{white}{.}
\end{equation}
We prove the above formula in appendix \ref{sec:derivationdoubleoffshell2}. Let us briefly describe where the $\mu^2$ corrected term comes from. In the proof of the formula for the 4-dimensional double-off-shell current, one uses a certain Clifford algebra identity that re-expresses expressions roughly like $\lr{\alpha | (\ell+p) \ldots |\alpha} \lr{\alpha | (\ell+p) \ldots |\alpha}$ into terms roughly like $(\ell + p)^2 (\ldots)$. These terms then cancel propagators in the denominator. However, with the regulated denominators, these numerators and denominators will no longer cancel exactly, and we in particular have the relation
\begin{equation}
    \frac{(\ell + p)^2}{(L + p)^2} = 1 + \frac{\mu^2}{(L + p)^2}.
\end{equation}
The above identity is the source of the correction term in \eqref{JYMdoublesolnmu}.

Actually, \eqref{JYMdoublesolnmu} is not the fully corrected formula, but for the purposes of computing the one-loop-all-plus amplitudes it is the only part which contributes after integration. \eqref{JYMdoublesolnmu} is more precisely what one gets by going up exactly ``one-level'' in the recursive formula \eqref{JQNYMrecursion}. In other words, one obtains this formula for $J_{\rm YM}(L \, ; \, N)$ by modifying the propagator $(\ell + p_N)^2 \to (L+ p_N)^2$ on the LHS of \eqref{JQNYMrecursion}, using the naive 4 dimensional formula for $J_{\rm YM}( \ell \, ; \, A)$ \eqref{JYMdoublesoln} on the RHS while simply shifting $\ell \to L$ in its denominator. We'll explain why this is sufficient in a bit.

\begin{figure}[h]
    \centering
    \includegraphics{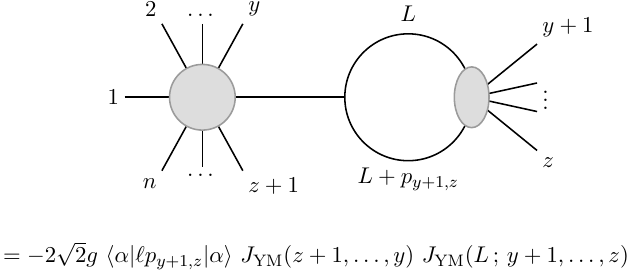}
    \caption{Configuration of currents necessary to compute one-loop-all-plus SDYM amplitudes. We need to glue a single-off-shell-all-plus current to a double-off-shell-all-plus current with a \ppm vertex in order to have all the correct tree configurations. If we just glued the double-off-shell current to itself, particle $1$ couldn't have any particles to the left of it in its tree in the color ordered diagram and $n$ couldn't have any particles to the right of it.
    }
    \label{fig:JJglue}
\end{figure}

In order to compute the one-loop-all-plus amplitudes, we want to sew the two off-shell legs together and integrate over $L$. However there is a slight issue we must address about color ordering. Note that all one-loop diagrams take the form of trees pasted on a central loop. If we were to just glue the $L$ and $m$ legs together, the color ordered trees we sum over from figure \ref{fig:JQdef} will always have particle $1$ on the leftmost side of a tree and particle $n$ on the rightmost side of a tree. You would never get $1$ or $n$ in the middle of a tree, which is a problem. We need to find a way to fix this by hand. We can accomplish this by gluing the single-off-shell current to the double off-shell current with a \ppm vertex, as shown in figure \ref{fig:JJglue}, where we take particle $1$ to always belong to the single-off-shell tree.

This means that the single-trace part of the one-loop-all-plus amplitude is given by
\begin{equation}\label{A1loopstart}
    A^{\text{1-loop}}_{n;1} = -2 \int \frac{d^D L}{(2 \pi)^D} \sum_{1 \leq y < z \leq n} \sqrt{2} g \lr{\alpha | \ell p_{y+1,z} | \alpha} J_{\rm YM}(z\!+\!1, \ldots, y) J_{\rm YM}(L\, ; y\!+\!1, \ldots, z).
\end{equation}
The factor of 2 is for the two directions the $(+-)$ propagators can flow through the loop and the overall sign is due to the direction of momentum flow through the \ppm vertex.

Let us now explain our chosen loop integration scheme. We use dimensional regularization with
\begin{equation}
    D = 4 - 2 \epsilon
\end{equation}
and split the integration measure up using
\begin{equation}
    \int \frac{d^D L}{(2 \pi)^D} = \int \frac{d^4 \ell}{(2 \pi)^4} \int \frac{d^{-2 \epsilon} \mu}{(2 \pi)^{-2 \epsilon}}.
\end{equation}
We now begin massaging \eqref{A1loopstart}. We define the Parke-Taylor factor
\begin{equation}
    \text{PT} \equiv \lr{12}\lr{23} \ldots \lr{n-1,n} \lr{n1}
\end{equation}
for economy. We then have
\begin{equation}\label{oneloopfirsteq}
\begin{aligned}
    &A^{\text{1-loop}}_{n;1} = -\frac{2 (-i\sqrt{2} g)^n}{\text{PT}} \int \frac{d^D L }{(2 \pi)^D} \sum_{1 \leq y < z \leq n} \frac{\lr{y,y+1}}{\lr{\alpha y} \lr{\alpha ,y+1}} \frac{\lr{z,z+1}}{\lr{\alpha z} \lr{\alpha, z+1}} \lr{\alpha | \ell p_{y+1,z} | \alpha} \\
    & \times \Bigg(\sum_{j=y+1}^{z}   \frac{ \lr{\alpha |(\ell+p_{y+1,j} ) j|\alpha } }{(L + p_{y+1,j-1})^2(L + p_{y+1,j})^2 } - \mu^2 \sum_{j=y+1}^{z} \frac{ \lr{\alpha |p_{j,z}  j |\alpha } }{(L + p_{y+1,j-1})^2(L + p_{y+1,j})^2 (L+p_{y+1,z})^2 } \Bigg).
\end{aligned}
\end{equation}

Let's start by integrating the first term, which is naively divergent by power counting. Shfting the integration measure, we have
\begin{equation}
    \int \frac{d^D L }{(2 \pi)^D} \frac{\lr{\alpha | \ell p_{y+1,z} | \alpha}   \lr{\alpha |(\ell+p_{y+1,j} ) j|\alpha } }{(L + p_{y+1,j-1})^2(L + p_{y+1,j})^2 } = \int \frac{d^D L }{(2 \pi)^D} \frac{\lr{\alpha | (\ell - p_{y+1,j-1}) p_{y+1,z} | \alpha}   \lr{\alpha |\ell j|\alpha } }{L^2(L + p_j )^2 } = 0.
\end{equation}
It is relatively straightforward to see that the above integral is zero due to a few standard identities. In order to evaluate this integral, one introduces a Feynman paramater to shift $\ell$ on the RHS by a vector proportional to $p_j$ in order to complete the square in the denominator. One then uses
\begin{equation}
\begin{aligned}
    \int \frac{d^4 \ell' }{(2 \pi)^4} \int \frac{d^{-2 \epsilon} \mu }{(2 \pi)^{-2 \epsilon}} \frac{{\ell'}^\nu}{({\ell'}^2 - \Delta - \mu^2)^3} &= 0 \, , \\
    \int \frac{d^4 \ell' }{(2 \pi)^4} \int \frac{d^{-2 \epsilon} \mu }{(2 \pi)^{-2 \epsilon}}  \frac{{\ell'}^\nu {\ell'}^\rho}{({\ell'}^2 - \Delta - \mu^2)^3} &\propto \eta^{\nu \rho}\, ,
\end{aligned}
\end{equation}
where $\ell'$ is the momentum after the Feynman parameter shift. Note that these formulae follow simply from the index structures of the integrands. The part quadratic in $\ell'$, which naively would have been the most divergent piece, vanishes after integration because it appears in the structure $\lr{\alpha|\ell' \ldots |\alpha } \lr{\alpha|\ell' \ldots |\alpha }$, and from the Fierz identity $\eta^{\nu \rho} \sigma_{\nu A \dot A} \sigma_{\rho B \dot B} = 2 \vep_{AB} \vep_{\dot A \dot B}$ it is zero because $\lr{\alpha \alpha} = 0$.

So, the naively divergent first ``bubble'' term in \eqref{oneloopfirsteq} is zero. Now let us turn to the $\mu^2$ corrected ``triangle'' term. We introduce Feynman parameters to put it in standard form. We use the identity
\begin{equation}
\begin{aligned}
    &\frac{1}{(L+p_{y+1,j-1})^2 (L+p_{y+1,j})^2 (L+p_{y+1,z})^2 } \\
    &= 2! \int_{[0,1]^3} d^3 u  \frac{\delta(1\!-\!u_1\!-\!u_2\!-\!u_3)}{(u_1 (L+p_{y+1,j-1})^2 + u_2(L+p_{y+1,j})^2  + u_3 (L+p_{y+1,z})^2  )^3}
\end{aligned}
\end{equation}
and foil out the denominator, finding it to be
\begin{equation}
\begin{aligned}
    (\ell^2 + 2 \ell \cdot (u_1 p_{y+1,j-1} +  u_2 p_{y+1,j} + u_3 p_{y+1,z}) + u_1 p_{y+1,j-1}^2 + u_2 p_{y+1,j}^2 + u_3 p_{y+1,z}^2 - \mu^2)^3.
\end{aligned}
\end{equation}
Now we do the shift
\begin{equation}
    \ell = \ell' - u_1 p_{y+1,j-1} - u_2 p_{y+1,j} - u_3 p_{y+1,z}
\end{equation}
to complete the square, finding that the denominator is now
\begin{equation}
    ({L'}^2 - \Delta)^3
\end{equation}
where $\Delta = (u_1 p_{y+1,j-1} + u_2 p_{y+1,j}^2 + u_3 p_{y+1,z})^2 - u_1 p_{y+1,j-1}^2 - u_2 p_{y+1,j}^2 - u_3 p_{y+1,z}^2$. We now have
\begin{equation}\label{oneloopsecondeq}
\begin{aligned}
    A^{\text{1-loop}}_{n;1} &= \frac{2 (-i\sqrt{2} g)^n}{\text{PT}} \sum_{1 \leq y < z \leq n} \frac{\lr{y,y+1}}{\lr{\alpha y} \lr{\alpha ,y+1}} \frac{\lr{z,z+1}}{\lr{\alpha z} \lr{\alpha, z+1}}   \\
    &  \times 2!\int_{[0,1]^3} d^3 u \, \delta(1\!-\!u_1\!-\!u_2\!-\!u_3)\sum_{j=y+1}^{z} \lr{\alpha |p_{j,z}  j |\alpha }   \\
    & \times \int \frac{d^D L' }{(2 \pi)^D} \frac{ \mu^2 }{({L'}^2-\Delta)^3} \lr{\alpha | (\cancel{\ell'} - u_1 p_{y+1,j-1} - u_2 p_{y+1,j} - u_3 \cancel{p_{y+1,z}}) p_{y+1,z} | \alpha}.
\end{aligned}
\end{equation}
The first cancelled-out term vanishes because the $\ell'$ integrand is odd. The second cancelled-out term vanishes from $\lr{\alpha | p_{y+1,z} p_{y+1,z}|\alpha} = 0$.

The question now becomes, how do we perform the loop integral which has an inserted factor of $\mu^2$? We will now explain how, following appendix A of \cite{Bern:1995db}.

$\mu^2$ is the $-2\epsilon$ dimensional magnitude of $L^2$ which is orthogonal to the 4-dimensional part. Note that it is in a spatial direction and not the time direction. We can consider the symbol $\mu$ to be the radial magnitude with $\mu \geq 0$. Furthermore we recall that the general expression for the angular integral of the unit $d$-sphere $S_d$ is
\begin{equation}
    \int d \Omega_d = \frac{2 \pi^{\frac{d+1}{2}}}{\Gamma(\frac{d+1}{2})}.
\end{equation}
Consider an integral over some function of the form $f(\ell^\nu, \mu^2)$ along with an insertion $(\mu^2)^r$ for some $r$. This insertion changes the radial phase space factor for the integration over $\mu$, shifting the dimension $4 - 2 \epsilon \to 4 + 2r - 2 \epsilon$. Accounting for the difference in the angular phase space factor then multiplies the expression by a factor proportional to $\epsilon$:
\begin{equation}
\begin{aligned}
    \int \frac{d^{4 - 2 \epsilon} L}{(2 \pi)^{4 - 2 \epsilon}} (\mu^2)^r f(\ell^\nu, \mu^2) &= \int \frac{d^{4} \ell}{(2 \pi)^{4}} \int \frac{d^{ - 2 \epsilon} \mu}{(2 \pi)^{ - 2 \epsilon}}  (\mu^2)^r f(\ell^\nu, \mu^2) \\
    &= \int \frac{d^{4} \ell}{(2 \pi)^{4}} \int d \Omega_{-1-2 \epsilon} \int_0^\infty d \mu \, \mu^{-1-2 \epsilon} \,  (\mu^2)^r f(\ell^\nu, \mu^2) \\
    &= (2 \pi)^{2r} \frac{\int d \Omega_{-1-2 \epsilon}}{\int d \Omega_{-1-2 \epsilon + 2r }} \int \frac{d^{4} \ell}{(2 \pi)^{4}} \int \frac{d^{ - 2 \epsilon + 2r} \mu}{(2 \pi)^{ - 2 \epsilon + 2r}} f(\ell^\nu, \mu^2) \\
    &= - \epsilon(1 - \epsilon) \ldots (r-1-\epsilon) (4 \pi)^r \int \frac{d^{4 + 2r - 2 \epsilon } L}{(2 \pi)^{4 + 2r - 2 \epsilon}} f(\ell^\nu, \mu^2).
\end{aligned}
\end{equation}
Using the well-known scalar loop integral
\begin{equation}\label{intLformula}
    \intLp \frac{1}{({L'}^2 - \Delta)^{a}} = i \frac{(-1)^{a}}{(4 \pi)^{\frac{D}{2}}} \frac{1}{\Delta^{a - \frac{D}{2}}} \frac{\Gamma(a - \frac{D}{2})}{\Gamma(a)}
\end{equation}
our desired integral in \eqref{oneloopsecondeq} becomes a triangle in $6 - 2 \epsilon$ dimensions, and the dimension shifting extracts a single finite piece after a $\epsilon/\epsilon$ cancellation:
\begin{equation}\label{triangleintegral}
     \int \frac{d^{4 - 2 \epsilon} L' }{(2 \pi)^{4 - 2 \epsilon}} \frac{ \mu^2 }{({L'}^2 - \Delta)^3} = - \epsilon \, (4 \pi) \int \frac{d^{6 - 2 \epsilon} L' }{(2 \pi)^{6 - 2 \epsilon}} \frac{ 1 }{({L'}^2- \Delta)^3} =   \frac{1}{2} \frac{i}{(4 \pi)^2} + \mathcal{O}(\epsilon).
\end{equation}

Before proceeding, let us comment on a previous point we did not justify at the time. The $4-2\epsilon$ dimensional double-off-shell current is supposed to be a solution to the full $4-2\epsilon$ dimensional recursion, i.e. the relation where one replaces $(\ell + p_N)^2 \mapsto (L + p_N)^2$ on the LHS of \eqref{JQNYMrecursion}. However our actual expression \eqref{JYMdoublesolnmu} came from going up just ``one step'' in the recursive formula, where one takes the naive 4-dimensional expression for $J_{\rm YM}(\ell;A)$ and simply replaces its denominators $(\ell + p)^2 \mapsto (L + p)^2$. We now explain why this procedure is sufficient. Note that we can calculate further corrections to the formula recursively by re-plugging in our $\mu^2$-corrected solution into the recursive formula over and over. Let us label all terms in the corrected current by $(r,a)$, where $r$ is the power of $(\mu^2)^r$ in the term and $a$ is the number of propagators. So \eqref{JYMdoublesolnmu} for instance has a $(0,2)$ term and a $(1,3)$ term. (The number of $\ell$'s in the numerator won't affect the divergence of the $L$ integrals because after shifting by Feynman parameters they all go away due to $\lr{\alpha \alpha} = 0$, so we don't have to keep track of them in the label.) As we recursively plug our formula into the Berends-Giele recursive formula, each subsequent step will create a term with an extra propagator due to the final division by $(L+p_N)^2$. The next correction, for instance, will be of the form $(1,4)$. All other subsequent corrections will also be of the form $(r,a)$ for $r < a-2$. (At any stage, it could be the case that a Clifford algebra identity could create an $(\ell + p)^2$ in the numerator, which could cancel and produce a $1 + \frac{\mu^2}{(L+p)^2}$, but note that this will not modify the previous statement.) However, from \eqref{intLformula} we can see that the corresponding loop integrals with $D = 4 + 2r - 2 \epsilon$ are finite and are killed by the overall $\epsilon$ for $r < a -2$, so these extra corrections won't contribute.

We now plug our triangle integral \eqref{triangleintegral} back into \eqref{oneloopsecondeq}. We also use the simplex integral
\begin{equation}
    \int_{[0,1]^3} d^3 u \,\delta(1\!-\!u_1\!-\!u_2\!-\!u_3) u_i = \frac{1}{6}
\end{equation}
to get a closed-form expression just in terms of spinor helicity variables,
\begin{equation}
\begin{aligned}
    A^{\text{1-loop}}_{n;1} = - \frac{i}{3 (4 \pi)^2} \frac{(-i\sqrt{2} g)^n}{\text{PT}} &\sum_{1 \leq y < j \leq z \leq n} \frac{\lr{y,y+1}}{\lr{\alpha y} \lr{\alpha ,y+1}} \frac{\lr{z,z+1}}{\lr{\alpha z} \lr{\alpha, z+1}} \\
    & \times \lr{\alpha |p_{j,z}  j |\alpha } \lr{\alpha | ( p_{y+1,j-1}+ p_{y+1,j}) p_{y+1,z} | \alpha} . 
\end{aligned}
\end{equation}
We could certainly at this point use standard tricks to further transform this expression, but we contented ourselves with simply checking numerically on a computer to high multiplicity that, with momentum conservation, it is equal to
\begin{equation}\label{YM1LAPformula}
    \boxed{A^{\text{1-loop}}_{n;1} =  -\frac{i}{3} \frac{ (-i \sqrt{2} g)^n}{(4 \pi)^2} \frac{1}{\lr{12} \ldots \lr{n1}} \sum_{1 \leq i_1 < i_2 < i_3 < i_4 \leq n} \lr{i_1 i_2}[i_2 i_3] \lr{i_3 i_4} [i_4 i_1]. }
\end{equation}
Note that the expression is rational and does not depend on the reference spinor $|\alpha\rangle$, as promised. 

The full amplitude can be written as a sum of single-trace partial amplitudes and double-trace partial amplitudes via
\begin{equation}
\begin{aligned}
    & \mathcal{A}^\text{1-loop}(1, \ldots, n) = \sum_{\sigma \in S_n / \mathbb{Z}_n} N_c \tr(T^{\sigma(1)} \ldots T^{\sigma(n)} ) A^{\text{1-loop}}_{n;1}(\sigma(1), \ldots, \sigma(n)) \\
    & + \sum_{j=2}^{\lfloor\frac{n}{2} \rfloor + 1} \sum_{\sigma \in S_n /\mathbb{Z}_{j-1} \times \mathbb{Z}_{n-j+1}}  \tr(T^{\sigma(1) } \ldots T^{\sigma(j-1)}) \tr(T^{\sigma(j)} \ldots T^{\sigma(n)}) A^{\text{1-loop}}_{n;j} (\sigma(1) \ldots \sigma(n)).
\end{aligned}
\end{equation}
We review one-loop color ordering in more detail in appendix \ref{app:color}. The point we want to make here is that the double-trace partial amplitudes are uniquely determined by the single-trace partial amplitudes via
\begin{equation}
    A_{n;j}^{\text{1-loop}}(1,2,\ldots,j\!-\!1;j, \ldots, n) = (-1)^{j-1} \sum_{\sigma \in \text{COP}\{\alpha \} \{\beta\} } A_{n;1}^{\text{1-loop}}(\sigma(1), \ldots, \sigma(n))
\end{equation}
where ``$\text{COP}\{\alpha \} \{\beta\}$'' is a certain set of color-ordered-permutations we again define in appendix \ref{app:color}. From the above formula and \eqref{YM1LAPformula}, one can check the following closed-form expression for the double-trace terms of the one-loop-all-plus amplitude holds: \cite{Dunbar:2026sle,Dunbar:2019fcq}
\begin{equation}
    A_{n;j}^{\text{1-loop}} =  -2 i \frac{(-i \sqrt{2} g)^n }{(4 \pi)^2} \frac{(p_{1,j-1}^2)^2}{(\lr{12} \ldots \lr{j-1,1})(\lr{j,j+1} \ldots \lr{n j})}.
\end{equation}

\subsection{SDG double-off-shell current calculation}\label{sec:sdgdoubleoffshell}

Let us look at the analogous calculation for gravity. The authors BDPR \cite{Bern:1998sv} presented an expression for the double-off-shell SDG current in appendix B of their paper without proof. The 4-dimensional double-off-shell current satisfies the recursion relation
\begin{equation}
    (\ell + p_N)^2 \mJ_{\rm G}(\ell \, ; \, N) = - \frac{\kappa}{2}  \sum_{\substack{ A \cup B = N \\ A \cap B = \emptyset,  B \neq \emptyset } } \lr{ \alpha | (\ell+p_A) p_B | \alpha }^2  \mJ_{\rm G}(\ell\,;\,A) \mJ_{\rm G}(B)
\end{equation}
and, with our conventions, is determined by the base case
\begin{equation}
    \mJ_{\rm G}(\ell) = \frac{i}{\ell^2}.
\end{equation}
The closed-form expression is given by
\begin{equation}
\begin{aligned}
    \mJ_{\rm G}(\ell \, ; \, N) =&\, \frac{(\kappa/2)^n (-i) }{\lr{\alpha 1} \lr{1 2} \ldots \lr{n\!-\!1,n} \lr{\alpha n}} \left( \prod_{j=1}^n \frac{\lr{\alpha | (\ell + p_{1,j}) j |\alpha } }{\lr{\alpha j}^2 } \right) \left( \sum_{m=1}^n \frac{\br{(\ell + p_{1,m})m}}{(\ell + p_{1,m-1})^2 (\ell + p_{1,m})^2} \right) \\
    & + \mathcal{P}(1,2, \ldots, n).
\end{aligned}
\end{equation}
Here $\mathcal{P}(1,2,\ldots,n)$ is a sum over all permutations of the elements $1,2,\ldots,n$. If one takes $\ell^2 \to 0$ on-shell, amputates the $\ell$ propagator and accounts for the overall polarization factors, the double-off-shell current becomes equal to the single-off-shell current:
\begin{equation}
    \lim_{\ell^2 \to 0} ( \tfrac{\ell^2}{i} ) \mJ_{\rm G}(\ell ; 1, \ldots, n) = \lr{\alpha \ell}^4 \mJ_{\rm G} (\ell, 1, \ldots, n) \Big\rvert_{\ell^2= 0}.
\end{equation}

The $4-2\epsilon$-dimensional double-off-shell current is defined by going one stage up in the recursion relation, continuing all the denominators to $4-2\epsilon$ dimensions while keeping the numerators in $4$ dimensions:
\begin{equation}
    \mJ_{\rm G}(L \, ; \, N) \equiv \frac{- \kappa/2}{(L + p_N)^2}  \sum_{\substack{ A \cup B = N \\ A \cap B = \emptyset,  B \neq \emptyset } } \lr{ \alpha | (\ell+p_A) p_B | \alpha }^2  \mJ_{\rm G}(\ell\,;\,A) \Big\rvert_{\ell \, \mapsto L} \mJ_{\rm G}(B).
\end{equation}
BDPR provide an expression for this current as well. It is
\begin{equation}\label{JGdoublemu}
\begin{aligned}
    \mJ_{\rm G}(L \, ; \, N) &= \frac{(\kappa/2)^n (-i) }{\lr{\alpha 1} \lr{1 2} \ldots \lr{n\!-\!1,n} \lr{\alpha n}} \left( \prod_{j=1}^n \frac{\lr{\alpha | (\ell + p_{1,j}) j |\alpha } }{\lr{\alpha j}^2 } \right) \\
    &\times \sum_{m=1}^n \frac{1}{(L + p_{1,m-1})^2 (L + p_{1,m})^2} \left[\br{(\ell + p_{1,m})m} - \frac{\mu^2}{(L + p_{1, n})^2} \br{p_{m,n} m } \right] \\
    &+ \mathcal{P}(1,2, \ldots, n).
\end{aligned}
\end{equation}
In their paper, they claim that they numerically checked the 4-dimensional current up to $n=8$, and checked the $\mu^2$-correction up to $n=3$. We have numerically checked the full current, including the $\mu^2$-correction, up to $n=10$ by generating random values for all the momenta and $\mu^2$.

The formula \eqref{JGdoublemu} of course does not look like the tree formula for the single-off-shell current. There is almost certainly a nicer formula for $\mathcal{J}_{\rm G}(L ; N)$ which generalizes the tree formula in some direct way. It would be interesting to find this expression.

Armed with the formula for the double-off-shell current, we can use it to compute an all-multiplicity expression for the one-loop-all-plus gravity amplitudes. This computation was sketched in words in \cite{Bern:1998sv} and we are told the authors performed the computation for $n=4,5$, but here we carry it out to all $n$. Note we cannot simply glue the two off-shell legs together due to an issue with overcounting. For instance, the configurations with $r$ trees coming off the central loop will be overcounted by a factor of $r$ because said trees can be cycled around the loop. To fix this issue, we have to glue a single-off-shell current containing one particular particle, say particle 1, to the double-off-shell vertex. The configuration is analogous to what we used in the Yang-Mills case in figure \ref{fig:JJglue}. Here, we sum over all the non-empty subsets $A$ and $B$ such that $A \cup B = N$ and $1 \in B$. We then attach the single-off-shell current $\mJ_{\rm G}(B)$ to the double-off-shell current $\mJ_{\rm G}(L;A)$:
\begin{equation}
    \mathcal{M}^{\text{1-loop}}(1^+,\ldots,n^+) = \frac{i \kappa}{2} \sum_{\substack{A \cup B = N \\ A \cap B = \emptyset \\ 1 \in B, A \neq \emptyset }} \int \frac{d^D L}{(2 \pi)^D} \br{\ell p_B}^2 \mJ_{\rm G}(L; A)  \mJ_{\rm G}(B).
\end{equation}
(An overall factor of 2 for the orientations of the $(+-)$ propagators is cancelled by a $1/2$ for graph-reflection symmetry after the currents are glued.) Using this expression, it is easy to introduce Feynman parameters and do the loop integrals exactly as we did in SDYM. The 4-dimensional ``bubble'' piece can immediately be seen to vanish due to the numerator structure. Only the $\mu^2$-corrected triangle piece remains, and any term with an $\langle \alpha | \ell' \ldots$ after the Feynman parameter shift vanishes. Only a simple scalar triangle integral in $6-2\epsilon$ dimensions remains which we already computed in \eqref{triangleintegral}. The final result is
\begin{equation}\label{1141}
\begin{aligned}
    \mathcal{M}^{\text{1-loop}}(1^+,\ldots,n^+) = \frac{-i \kappa/2}{(4 \pi)^2}  &\sum_{\substack{A \cup B = N \\ A \cap B = \emptyset \\ 1 \in B, A \neq \emptyset }} \sum_{\substack{\text{all possible}\\\text{permutations}\\ \text{of }A\text{ in }\mathcal{P}(A)}} \sum_{m=1}^{|A|}  \int_{[0,1]^3} d^3 u \, \delta(1\!-\!u_1\!-\!u_2\!-\!u_3) \\
    & \times \mJ_{\rm G}(B) \br{(u_1 p_{A} + u_2 p_{a_1,a_{m-1}} + u_3 p_{a_1,a_m})p_B}^2 \\
    & \times \frac{(\kappa/2)^{|A|}}{\lr{\alpha a_1}\lr{a_1 a_2} \ldots \lr{a_{|A|-1} a_{|A|}}  \lr{\alpha a_{|A|}}  } \br{ p_{a_m a_{|A|}} a_m } 
     \\
    & \times \prod_{j=1}^{|A|} \frac{\br{ (-u_1 p_{A} - u_2 p_{a_1,a_{m-1}} - u_3 p_{a_1,a_m}+p_{a_1,a_j})a_j}}{\lr{\alpha a_j}^2}
\end{aligned}
\end{equation}
where above we denote $p_{a_i,a_j} = p_{a_i} + p_{a_{i+1}} + \ldots p_{a_j}$. Note that we have not performed the trivial integrals over Feynman parameters. A closed form expression for these integrals is given by $\int_{[0,1]^3} d^3 u \delta(1\!-\!u_1\!-\!u_2\!-\!u_3)u_1^{n_1} u_2^{n_2} u_3^{n_3} = n_1! n_2! n_3!/(2+n_1+n_2+n_3)!$, but we haven't executed this above because it would make the formula more complicated.

Admittedly, the above formula is a bit less than inspiring, but it is at least a closed form rational expression for the one-loop-all-plus amplitudes that can be derived using Feynman diagrammatics.


\section{Unitarity cuts in $D$-dimensions}\label{sec:unitaritycuts}

In this section we will give a cursory overview of the unitarity method for computing one-loop-all-plus amplitudes \cite{Bern:1996je,Bern:1995db,Nandan:2018ody}. In particular, we'll compute the 4-pt one-loop-all-plus SDYM amplitude with this method.

It is well known that unitarity (the cutting rules) implies that the discontinuities of a one-loop amplitude are given by the on-shell cuts of the amplitude. Furthermore, as discussed in section \ref{sec:rationality}, all the cuts of the one-loop-all-plus amplitudes are zero, meaning the amplitudes are rational. So, given that the cuts are zero, how could we possibly use cuts to compute these amplitudes? The answer is that, while all of the 4-dimensional cuts are zero, the $D=4-2\epsilon$-dimensional cuts are not zero! Strictly speaking, however, it does not make sense to talk about $4-2\epsilon$-dimensional cuts using the standard unitarity theorem. This method should maybe be more precisely thought of as a way to construct simplified ansatze for Feynman integrands. In its modern incarnation, one typically performs four cuts to solve for the four components of the loop momentum, turning any loop into a sum of boxes, triangles, and bubbles \cite{Britto:2004nc}. This is of course even more remote from actual unitarity and is therefore referred to as ``generalized unitarity.'' For a description of the modern method, see for instance the work of Badger \cite{Badger:2008cm,Badger:2023eqz}. See also for instance the discussion in \cite{Boels:2013bi}. Here however we will stick to the old-fashioned method based on two-particle cuts.

The method essentially boils down to constructing Feynman integrals which have all the correct cuts in all possible channels by gluing simplified on-shell tree amplitudes together. This is especially simple when all particles are massless, although see for instance \cite{Bern:1995db} with respect to the massive case.
\begin{figure}[h]
    \centering
    \includegraphics[width=0.75\linewidth]{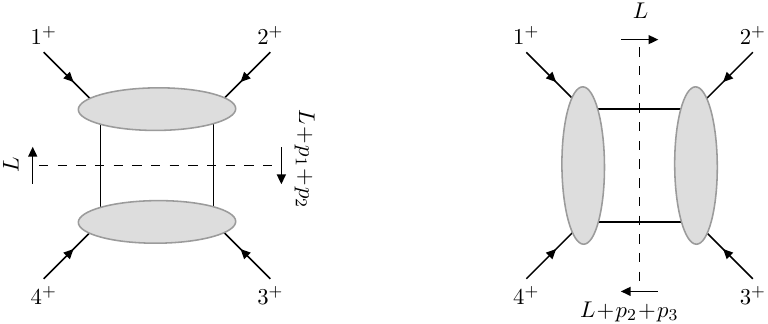}
    \caption{The two cuts of the planar 4-pt SDYM amplitude.}
    \label{fig:twocuts}
\end{figure}

For a 4-pt massless planar color-ordered amplitude, the only non-zero cuts occur in the $s_{12}$ and $s_{23}$ channels. Let's focus first on the $s_{12}$ channel, where we cut through the loop momenta $L$ and $L+p_1+p_2$. We denote the 4-dimensional part of these $4-2\epsilon$-dimensional momenta as $\ell$ and $\ell+p_1+p_2$. On this cut, we have the on-shell condition
\begin{equation}
    L^2 = 0, \hspace{1 cm} (L+p_1+p_2)^2 = 0.
\end{equation}
This is equivalent to saying the 4-dimensional parts of the loop momenta have a mass $\mu^2$:
\begin{equation}\label{ell2mu}
    \ell^2 = \mu^2, \hspace{1 cm} (\ell + p_1 + p_2)^2 = \mu^2.
\end{equation}

\begin{figure}[h]
    \centering
    \includegraphics[width=0.75\linewidth]{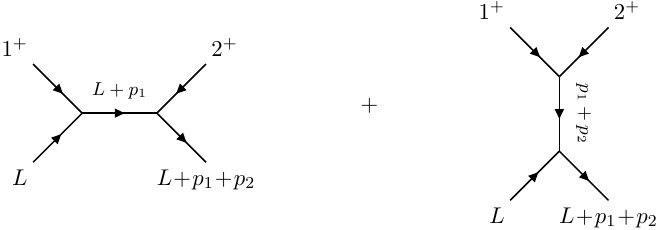}
    \caption{The two diagrams that contribute to $A^{\rm tree}(L,1^+,2^+,-L\!-\!p_1\!-\!p_2)$}
    \label{fig:twoA}
\end{figure}

Now let us define the tree-amplitude $A^{\rm tree}(L,1^+,2^+,-L-p_1-p_2)$ which satisfies the above on-shell condition. For simplicity, we imagine the $L$ and $-L-p_1-p_2$ legs to be ``scalars'' with no associated polarization factors. The two diagrams that contribute to this amplitude are drawn in figure \ref{fig:twoA}.

\noindent Summing up the two diagrams, we get the following expression:
\begin{equation}
\begin{aligned}
    & A^{\rm tree}(L, 1^+,2^+, -L\!-\!p_1\!-\!p_1) \\
    &= \frac{i (\sqrt{2} g)^2 }{\lr{\alpha 1}^2 \lr{\alpha 2}^2} \left( \frac{\br{\ell 1} \br{(\ell + 1)2 }}{(L + p_1)^2} + \frac{\br{12} \br{\ell (1 + 2)}}{ \lr{12}[12]} \right).
\end{aligned}
\end{equation}
This expression can be simplified using the following identity (proven in appendix \ref{sec:lemma})
\begin{equation}
\begin{aligned}
  &\br{\ell 1}  \br{(\ell+1)2} \\ 
&= \frac{\lr{\alpha 1} \lr{\alpha 2}}{\lr{12}} \Big(  (\ell+p_1)^2 \br{\ell(1+2)} - (\ell+p_1+p_2)^2 \br{\ell 1} - \ell^2 \br{(\ell+1) 2 } \Big).
\end{aligned}
\end{equation}
Using the cut condition \eqref{ell2mu} and the above identity, the tree amplitude simplifies to \cite{Forde:2005ue}
\begin{equation}
    A^{\rm tree}(L, 1^+,2^+, -L\!-\!p_1\!-\!p_1) =  i (\sqrt{2} g)^2 \frac{\mu^2}{(L + p_1)^2} \frac{[12]}{\lr{12}}.
\end{equation}
Note that this amplitude is proportional to $\mu^2$, which implies that it is zero in exactly 4-dimensions, as expected. Crucially, the above formula is also exact in general dimension $D$, i.e. for all possible values of $\epsilon$ not necessarily small. The use of the exact tree amplitude in general $D$ is required in order for this method to produce the correct result.

For the 4-particle amplitude, the expression is simply a sum of the two cuts.
\begin{equation}
\begin{aligned}
& A_{4;1}^{\text{1-loop}}(1^+,2^+,3^+,4^+) \\
    &= \int \frac{d^D L}{(2 \pi)^D} \frac{i}{L^2} \frac{i}{(L + p_1 + p_2)^2} A^{\rm tree}(L, 1^+,2^+, -L\!-\!p_1\!-\!p_2) A^{\rm tree}(L\!-\!p_3\!-\!p_4, 3^+,4^+ , -L)
    \\&+ \int \frac{d^D L}{(2 \pi)^D} \frac{i}{L^2} \frac{i}{(L + p_2 + p_3)^2} A^{\rm tree}(L, 2^+,3^+, -L\!-\!p_2\!-\!p_3) A^{\rm tree}(L\!-\!p_4\!-\!p_1, 4^+,1^+ , -L)
\end{aligned}
\end{equation}
(A factor of 2 for the two polarizations of particles which can run across the cut is cancelled by an identical particle phase space factor of 1/2.)

For more than 4 particles, however, the amplitude is not simply a sum over cuts as shown above. This is, of course, because the tree-amplitudes themselves generically contain multi-particle propagators which will affect the values of other cuts. In general one must engineer an expression that agree on \textit{all} cuts by creating an appropriate linear combination of basis integrals. This is now often done using software packages such as BlackHat \cite{Berger:2008ag}.

In any case, for 4 particles the expression becomes
\begin{equation}
\begin{aligned}
    A_{4;1}^{\text{1-loop}}(1^+,2^+,3^+,4^+) &= \frac{[12]}{\lr{12}} \frac{[34]}{\lr{34}} (\sqrt{2} g)^4 \int \frac{d^D L}{(2 \pi)^D} \frac{\mu^4 }{L^2 (L + p_1 + p_2)^2 (L + p_1)^2 (L - p_4)^2} \\
    &+ \frac{[23]}{\lr{23}} \frac{[41]}{\lr{41}} (\sqrt{2} g)^4 \int \frac{d^D L}{(2 \pi)^D} \frac{\mu^4 }{L^2 (L + p_2 + p_3)^2 (L + p_2)^2 (L - p_1)^2}. 
\end{aligned}
\end{equation}
The integrals are easy to evaluate and give
\begin{equation}
    A_{4;1}^{\text{1-loop}}(1^+,2^+,3^+,4^+) =\frac{-i (\sqrt{2} g)^4 }{6 (4 \pi)^2 } \left(  \frac{[12]}{\lr{12}} \frac{[34]}{\lr{34}}    + \frac{[23]}{\lr{23}} \frac{[41]}{\lr{41}} \right)
\end{equation}
which is the correct expression.

\newpage
\section{One-loop-all-plus gravity expression of Bern, Dixon, Perelstein, Rozowsky}\label{sec:BDPR}

In this section we write down the remarkable closed-form expression for the one-loop-all-plus SDG amplitudes, due to BDPR \cite{Bern:1998sv,Bern:1998xc}, and explain how the expression was found. We do not provide a proof of this formula because, at the time of writing this note, none exists. We expect the situation will be rectified in the near future.

Without further ado, the BDPR formula is
\begin{equation}\label{bdpr}
    \mathcal{M}^{\text{1-loop}}(1^+,...,n^+) = - \frac{i}{(4 \pi)^2 960} \left(-\frac{\kappa}{2}\right)^n \sum_{\substack{1 \leq a < b \leq n \\ A,B}} h(a,A,b) h(b,B,a) \tr^3[a A b B].
\end{equation} 
We checked numerically that the above expression matches equation \eqref{1141} up to $n=9$ on the support of momentum conservation. In this formula, $A$ and $B$ are indistinguishable sets such that $A \cup B \cup \{a, b\} = \{1 ... n\}$, $A \cap B = \emptyset$, $a,b \notin A,B$. We also have
\begin{equation}
    \tr[a A b B] \equiv \langle a | p_A | b] \langle b | p_B | a] + \langle b | p_A | a] \langle a | p_B | b]
\end{equation}
and define the ``half-soft function'' by
\begin{equation}\label{habdef}
    h(a, \{ 1, 2, ..., n\}, b) \equiv \frac{[12]}{\lr{12}} \frac{  \langle a | p_{1,2} | 3] \langle a | p_{1,3} |4 ] ... \langle a | p_{1,n-1} | n ]   }{  \lr{23} \lr{34} ... \lr{n-1, n} \lr{a 1} \lr{a 2} \lr{a 3} ... \lr{a n} \lr{1 b} \lr{n b} } + \mathcal{P}(2,\!3, ..., n)
\end{equation}
where $p_{1,m} = \sum_{i=1}^m p_i$ and $\mathcal{P}(2,3, \ldots, n)$ represents a sum over all permutations of the denoted elements.

There are also other nicer-looking presentations of the half-soft function \cite{Dunbar:2012aj}. These other presentations are equal to \eqref{habdef} identically, i.e. without the use of momentum conservation. Let's look at two such equivalent expressions.\footnote{To prove that the expression \eqref{habdef} is equal to these other expressions, all one has to do is show that all the expressions satisfy the same Berends-Giele recursion relation \eqref{hBerends}. Luckily for us, formula \eqref{habdef} was proven to satisfy it in appendix C.2 of \cite{Bern:1998sv}, and \eqref{eqtree} was in \cite{Krasnov:2013wsa}.}

If one defines the $n \times n $ matrix $\Psi$ via
\begin{equation}
    \Psi_{ij} = \begin{cases}
        -\dfrac{[ij]}{\lr{ij}} \lr{a i} \lr{b i} \lr{a j} \lr{bj} & i \neq j \\
        \displaystyle\sum_{\substack{k=1 \\ k \neq i}}^{n} \dfrac{[ik]}{\lr{ik}} \lr{a i} \lr{b i} \lr{a k} \lr{bk} & i = j 
    \end{cases}
\end{equation}
then the half-soft function turns out to be equal to \cite{Feng:2012sy}
\begin{equation}\label{eq2}
    h(a, \{1, 2, \ldots, n\}, b) = \left( \prod_{i=1}^n \frac{1}{\lr{ai}^2 \lr{bi}^2 } \right)| \Psi |^i_i 
\end{equation}
where $| \Psi |^i_i$ is the determinant of the minor of $\Psi$ where the row and column $i$ has been removed. The determinant is independent of $i$. Note that none of the $i$'s are equal to $a$ or $b$.

Using the matrix tree theorem, the half-soft function can also be written \cite{Dunbar:2012aj,Nguyen:2009jk} as the tree sum 
\begin{equation}\label{eqtree}
    h(a, \{1, 2, \ldots, n\}, b) = \left( \prod_{k=1}^n \frac{1}{\lr{ak}^2 \lr{bk}^2 } \right) \sum_{\substack{\text{trees with }\\ n \text{ nodes}}} \prod_{\text{edges } (ij) } \frac{[ij]}{\lr{ij}} \lr{a i}\lr{bi} \lr{a j} \lr{bj}.
\end{equation}
The ``trees'' in the above formula are graphically the same trees that we encountered with the all-plus current $\mathcal{J}_{\rm G}$ in figure \ref{fig:Jtreeexample}. Having said that, the rules for the edges are different, as they depend on $|a\rangle$ and $|b\rangle$ instead of $|\alpha\rangle$. However, if we set $|a\rangle = |b\rangle = |\alpha\rangle$, then the half-soft function is in fact equal to the all-plus current via
\begin{equation}\label{halfsoftJG}
    \mJ_{\rm G}(1,2,\ldots,n) = \left(- \frac{\kappa}{2} \right)^{n-1} h(\alpha, \{1, 2, \ldots, n\}, \alpha).
\end{equation}
As one might expect from the above relation, the half-soft function also satisfies a slightly more general form of the Berends-Giele recursion relation that $\mJ_{\rm G}$ satisfies, namely \cite{Krasnov:2013wsa}
\begin{equation}\label{hBerends}
    p_N^2 h(a, N, b) = \sum_{A,B | A \cup B = N} \lr{a| p_A p_B |a} \lr{b| p_A p_B |b} h(a, A, b) h(a, B, b)
\end{equation}
where here $N = \{1, \ldots, n\}$, $A$, $B$ are indistinguishable, and $a, b \notin N$. As a final relation, if we imagine that all of the momenta now sum to zero, the half-soft function is equal, up to a factor, to the NSVW formula for the tree-level gravity MHV amplitude \cite{Nguyen:2009jk}
\begin{equation}\label{halfsoftMHV}
    \mathcal{M}^{\rm tree}(1^{-}, 2^{-}, 3^{+}, \ldots, n^{+}) =  - i (-\kappa/2)^{n-2} \lr{12}^6 h(1, \{3, \ldots, n\}, 2)\Big\vert_{p_1 + \ldots + p_n = 0}.
\end{equation}
Why the all-plus currents and the MHV amplitudes are so closely related is a priori not so clear, although it is reasonable enough given that both expressions have similar soft and collinear properties. (In appendix \ref{sec:firstheavenly} we explain that this equality, along with the recursive formula \eqref{hBerends}, is ultimately is due to Plebański's first heavenly equation.)

Now that we have written down the BDPR formula \eqref{bdpr}, as well as some properties of the half-soft function, you may be wondering how the original authors actually came up with this formula. As we'll explain, the all-$n$ expression \eqref{bdpr} originated from an ansatz that came from a few different lines of reasoning. 

Firstly, it was conjectured by Bern, Dixon, Dunbar, and Kosower in \cite{Bern:1996ja}, and checked up to $n=6$, that the following relationship between one-loop-all-plus amplitudes in pure Yang-Mills in $D$-dimensions and one-loop MHV amplitudes in $\mathcal{N}=4$ SYM in $D+4$-dimensions holds:
\begin{equation}\label{Adimshift}
\begin{aligned}
    A_{n;1}^\text{1-loop}(1^+, \ldots, n^+) &= -2 \epsilon (1 - \epsilon) (4 \pi)^2 \left[ \frac{A_{n;1}^{\mathcal{N}=4, \text{1-loop}}(1^-, 2^-, 3^+,\ldots, n^+) }{\lr{12}^4} \Bigg\vert_{D \to D+4} \right] \\
    &= 2 \frac{A_{n;1}^{\mathcal{N}=4, \text{1-loop}}(1^-, 2^-, 3^+, \ldots, n^+) [\mu^4]}{\lr{12}^4}.
\end{aligned}
\end{equation}
The original paper \cite{Bern:1996ja} gave several pieces of evidence in favor of the all-$n$ conjecture, including the matching of the two-particle cuts of both sides to the first non-vanishing order in $\mu^2$. The full conjecture, which requires a match to all orders in $\mu^2$, was more recently revisited and proven by Britto, Jehu, and Orta in \cite{Britto:2020crg,Britto:2021tez}.

To reiterate, on the LHS of \eqref{Adimshift} we have the one-loop-all-plus amplitudes in pure Yang-Mills theory in $4 - 2\epsilon$-dimensions, while on the RHS have the one-loop $\mathcal{N}=4$ MHV amplitudes in $8-2 \epsilon$-dimensions, for all $\epsilon$. (We note that in the $8-2 \epsilon$-dimensional amplitudes, the external momenta still lie in 4-dimensions.) In particular, the MHV amplitudes in $D = 8 - 2 \epsilon$-dimensions have a $1/\epsilon$ divergence that is cancelled out by the factor of $\epsilon$, making the expression finite. Additionally, the RHS is equivalent to inserting a factor of $\mu^4$ in all of the Feynman integrals that appear in the $D =4-2\epsilon$ dimensional amplitudes.

\begin{figure}
    \centering
    \includegraphics[width=0.4\linewidth]{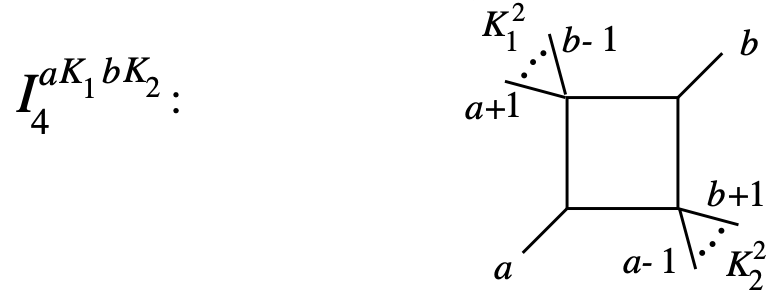}
    \caption{Box integrals with two opposite on-shell legs are the only integrals that contribute to the $\mathcal{N}=4$ SYM one-loop MHV amplitudes. Image is figure 1 of \cite{Bern:1998sv}.}
    \label{fig:Ibdpr}
\end{figure}

There are a few important facts about the one-loop $\mathcal{N} =4$ SYM amplitudes which we'll briefly review here. The first fact is that they are ``4-dimensional cut constructible,'' meaning that if one creates a linear combination of loop integrals with all the correct 4-dimensional cuts, this expression must be equal the full one-loop amplitude exactly, i.e. with all rational terms fixed uniquely. This fact was proven in \cite{Bern:1994zx}, see section 6, by checking that no linear combination of any of the the possible integrals that could appear in the supersymmetric loop integrals after Passarino-Veltman reduction are equal to a rational expression. (Actually, because there are $m-4$ powers of loop momenta in the $m$-point $\mathcal{N}=4$ integral numerators due to supersymmetric power counting, these integrals will get reduced to a sum over scalar boxes with no loop momenta in the numerators, so in fact one only needs to check that no linear combination of scalar boxes is rational.) See also \cite{Bern:1994cg}.

The second fact about the one-loop $\mathcal{N}=4$ MHV amplitudes is that they are given by a sum over ``two-mass-easy'' scalar-boxes, i.e. boxes which have massless particles in two opposite corners, which we draw in figure \ref{fig:Ibdpr}. This was originally proven in \cite{Bern:1994zx}, see section 5, through the use of two-particle cuts. Note that the non-vanishing 4-dimensional cuts of a one-loop MHV amplitude are particularly simple, and are comprised of tree-level supersymmetric Parke-Taylor MHV amplitudes on both sides. This allows one to easily find a combination of integrals with all the correct cuts, and with the use of some clever manipulations which are themselves equivalent to Passarino-Veltman reduction, one finds they are given by the scalar two-mass-easy boxes. (In retrospect, the fact that only the two-mass-easy scalar boxes contribute can more easily be shown to be true through the use of the four cuts of modern generalized unitarity, where the only non-zero cuts have two supersymmetric MHV amplitudes and two supersymmetric 3-pt amplitudes opposite each other. See for instance section 1.1 of \cite{Britto:2020crg}. See also \cite{Bullimore:2010pj,Brandhuber:2004yw}.)

Using these facts, the explicit formula for the $\mathcal{N}=4$ one-loop MHV amplitudes in $4-2\epsilon$-dimensions, as a sum over scalar box integrals to $\mathcal{O}(\epsilon)$, was found to be \cite{Bern:1994ju}


\begin{equation}
\begin{aligned}
&A_{n;1}^{\mathcal{N}=4,\text{1-loop}}(1^-, 2^-, 3^+, \ldots, n^+) = \frac{(-i \sqrt{2} g)^n}{4} \frac{\lr{12}^4 }{\lr{12} \lr{23 } \ldots \lr{n1} }\sum_{\substack{ a,b \\ \text{cyclic}}} \, \tr[a K_1 b K_2] \, I_4^{a K_1 b K_2} + \mathcal{O}(\epsilon)
\end{aligned}
\end{equation}
where we sum over all cyclic orderings where $\{a\} \cup K_1 \cup \{b\} \cup K_2 = \{1, \ldots, n\}$ and we denote $I_4^{a K_1 b K_2}$ as the scalar box integral with two on-shell legs $a$ and $b$ and off-shell legs $K_1$, $K_2$, depicted in figure \ref{fig:Ibdpr}. The box integral, and $\mu^{2r}$-shifted box integral, is
\begin{equation}
\begin{aligned}
    I_4^{a K_1 b K_2} &= \int \frac{d^D L}{(2 \pi)^D} \frac{1}{L^2 (L - p_a)^2 (L - p_a - K_1)^2 (L + K_2)^2} \\
    I_4^{a K_1 b K_2}[\mu^{2r}] &= \int \frac{d^D L}{(2 \pi)^D}\frac{(\mu^2)^{r}}{L^2 (L - p_a)^2 (L - p_a - K_1)^2 (L + K_2)^2}
    \end{aligned}
\end{equation}
where the $\mu^{2r}$ shifted integral can be computed using the dimension-shift formula
\begin{equation}
    I^{a K_1 b K_2}_4[\mu^{2r}] = - \epsilon(1 - \epsilon) \ldots (r - 1 - \epsilon)(4 \pi)^r I^{a K_1 b K_2}_{4, D \mapsto D + 2r}.
\end{equation}

Based on the Yang-Mills dimension-shift formula \eqref{Adimshift}, it was then postulated that a similar formula should hold for gravity between pure Einstein gravity one-loop-all-plus amplitudes in $D = 4 - 2 \epsilon$ dimensions and $\mathcal{N}=8$ supergravity one-loop MHV amplitudes in $D = 12 - 2 \epsilon$ dimensions, namely
\begin{equation}\label{Mdimshift}
\begin{aligned}
    &\mathcal{M}^\text{1-loop}(1^+, \ldots, n^+) \\
    &= -2 \epsilon (1 - \epsilon) (2 - \epsilon) (3 - \epsilon) (4 \pi)^4 \left[ \frac{\mathcal{M}^{\mathcal{N}=8, \text{1-loop}}(1^-, 2^-, 3^+, \ldots, n^+) }{\lr{12}^8} \Bigg\vert_{D \to D+8} \right] \\
    &= 2 \frac{\mathcal{M}^{\mathcal{N}=8, \text{1-loop}}(1^-,2^-, 3^+, \ldots, n^+) [\mu^8]}{\lr{12}^8}.
\end{aligned}
\end{equation}
From the above conjectured formula, and the known fact that supersymmetric $\mathcal{N}=4$ gauge theory one-loop MHV amplitudes are given by a sum over two-mass-easy boxes where the massive legs could reasonably be expected to behave like off-shell tree-level MHV amplitudes, we can begin to see how BDPR came up with \eqref{bdpr}, because the structure of the box integrals looks like the sum over $A$, $B$, $a$, $b$ in that formula, and the half-soft-functions themselves are like off-shell gravitational tree-level MHV amplitudes. But certainly a bit of black magic has occurred.

BDPR also showed that \eqref{bdpr} satisfied a number of consistency conditions. For instance, \eqref{bdpr} was checked to have the correct soft limits and collinear limits. To be more precise, it was showed to have the correct ``double-scaled holomorphic collinear limits'' where $[ij] \to 0$, $\lr{ij} \to 0$, $[ij]/\lr{ij} \to \infty$. (See our discussion in section \ref{sec:hololimits}). A later work by Ball, Narayan, Salzer, and Strominger \cite{Ball:2021tmb} confirmed that it also has the correct pure holomorphic collinear limits $\lr{ij} \to 0$ as well, as written in \eqref{holosplit1loop}. This check required the use of the quite non-trivial identity
\begin{equation}\label{halfsoftid}
    \sum_{b \neq a} \lr{ba} \sum_{A,B} h(a,A,b) h(a,B,b) \langle b | p_A | a] \langle a | p_B |b]^3 = 0
\end{equation}
where the sum is over $b$, $A$, $B$, such that $A \cup B \cup \{a, b\} = \{1, \ldots, n\}$, $A \cap B = \emptyset$, $a,b \notin A,B$ and $(A,B)$ is in this equation distinct from $(B,A)$. Note that this identity only holds on the support of momentum conservation. This identity was checked to hold for $n \leq 13$ in \cite{He:2014bga} and proven for all $n$ by Rao and Feng \cite{Rao:2016tgx}.

Via direct computation, BDPR also showed that their formula \eqref{bdpr} matched the correct amplitudes at $n=4,5,6$ on the support of momentum conservation. The `correct amplitudes' were computed using the $D=4-2 \epsilon$-dimensional unitarity cuts technique we reviewed in the last section. In order to obtain the $D$-dimensional gravitational tree amplitudes used in the cuts, the authors employed a dimensionally-analytically-continued form of the the KLT relations and double-copied the analogous gauge theory amplitudes. They also used supersymmetry to replace the graviton running around the loop with a complex scalar, which simplified the numerator structure of the problem. Note that this use of the complex scalar provides the same computational advantage as working with the lightcone scalar, as we do in this note. (In their appendix B, they also used the double-off-shell current method to match the amplitudes at $n=4,5$, which we straightforwardly extended to all-$n$ in \eqref{1141}.)

Let us now highlight another result from \cite{Bern:1998sv}. Now that they had computed the $n=4,5,6$ gravity amplitudes on the LHS of \eqref{Adimshift}, they in principle knew the $n=4,5,6$ $\mathcal{N}=8$ MHV amplitudes on the RHS as well. Using these expressions, they then conjectured the following formula for the $\mathcal{N}=8$ amplitudes in $D = 4 - 2 \epsilon$ (see their equation (6.18)):
\begin{equation}
\begin{aligned}
    & \mathcal{M}^{\mathcal{N}=8, \text{1-loop}}(1^-,2^-,3^+, \ldots, n^+) \\
    &= \frac{(-\kappa/2)^n}{8} \lr{12}^8 \sum_{1 \leq a < b \leq n, A, B} h(a,A,b) h(b,B,a) \tr^2[a A b B] \, I^{a A b B}_4 + \mathcal{O}(\epsilon).
\end{aligned}
\end{equation}
Quite remarkably, this expression for one-loop $\mathcal{N}=8$ MHV supergravity amplitudes is also a sum over scalar two-mass-easy boxes, just like we saw for $\mathcal{N}=4$ SYM. This is surprising because the power-counting argument that allows one to reduce all one-loop $\mathcal{N}=4$ amplitudes to scalar boxes does not apply for $\mathcal{N}=8$ supergravity. See \cite{Arkani-Hamed:2008owk}.

The above equation closely resembles the equation for the one-loop-all-plus amplitude \eqref{bdpr}, and it seems like it should be a straightforward matter to compare them via the dimension shift conjecture \eqref{Mdimshift}, but strangely the connection is not as simple as it seems. This is because dimension shifting the box integrals alone in the above equation by $\mu^8$ does not give the extra trace term $\tr[a M b N]$ as one might expect, so the true relationship is much more complicated. 

We also mention that in section 4.2 of \cite{Bern:1998sv} the authors provided a KLT ``proof'' of \eqref{Mdimshift} using \eqref{Adimshift}. Essentially, if one cuts both sides of \eqref{Adimshift}, the gauge theory cuts on the LHS must be equal to the $\mathcal{N}=4$ SYM cuts multiplied by $\mu^4$ on the RHS. One can then use the KLT double copy relations on both sides and get \eqref{Mdimshift}.

\section{Classical self-dual solutions and $\Lw$/$\Ls$ symmetry}\label{secLw}

\subsection{SDG and $\Lw$}

Classical SDYM and SDG are integrable theories, implying their solution spaces are acted on by infinite-dimensional Lie-algebras. These Lie algebras are $\Ls$ for SDYM and $\Lw$ for SDG. In this section we will discuss how these symmetries act perturbatively on the space of solutions. Note this section is a summary of the paper \cite{Miller:2025wpq}, although here we work in momentum space instead of the ``Mellin space'' more common in celestial holography papers. (The full paper also finds that a subset of $\Lw$ transformations are pure spacetime diffeomorphisms, and discusses the action of the `recursion operator,' but we won't be reviewing those stories here.)

Let us begin our discussion by defining the notion of a ``perturbiner expansion.'' A perturbiner expansion is a solution to an equation of motion which is built perturbatively out of plane waves. For economy of notation, we denote these plane waves as $\phi_i$ via
\begin{equation}
    \phi_i \equiv \epsilon_i e^{i p_i \cdot x}.
\end{equation}
These are positive-helicity gravitons. Here, $p_i$ is an on-shell momentum with $p_i^2 = 0$, and $\epsilon_i$ is an infinitesimal parameter. Each plane wave has its own infinitesimal parameter. These parameters are defined such that each one squares to zero
\begin{equation}
    (\epsilon_i)^2 = 0
\end{equation}
but the product of distinct $\epsilon_i$'s is non-zero,
\begin{equation}
    \epsilon_i \, \epsilon_{i'} \neq 0, \hspace{1 cm} \text{ if } i \neq i'. 
\end{equation}
We are now interested in perturbatively generating solutions to the SDG equation of motion
\begin{equation}\label{peom}
    \Box \, \phi - \frac{\kappa}{2} \{ \partial_u \phi, \partial_w \phi\} = 0.
\end{equation}
We notate the perturbiner expansions as
\begin{equation}
    \bigphi(\phi_1, \ldots, \phi_n) = \sum_{i=1}^n \phi_i + \ldots 
\end{equation}
The perturbiner expansion $\bigphi(\phi_1, \ldots, \phi_n)$ is defined as the classical solution generated by the following procedure. Start with a sum of free plane waves multiplied by infinitesimal parameters and solve for all the higher order terms in the $\epsilon_i$'s by recursively plugging the solution into the equation of motion \eqref{peom} repeatedly. Note that a term of order $k$ in the $\epsilon_i$'s will have $\kappa^{k-1}$ powers of the coupling. Note also that the perturbiner expansion will terminate at a term with all $n$ infinitesimal parameters, $\epsilon_1 \epsilon_2 \ldots \epsilon_n$.

Based on this definition, it is straightforward to see that a perturbiner expansion is essentially the exact same thing as a Berends-Giele current with one off-shell leg. We have provided a review of this connection in appendix \ref{sec:classicalsolns} for the uninitiated. But the upshot is that, because we already have a closed-form expression for the all-plus Berends-Giele current in SDG \eqref{JGtree}, we also have a closed-form expression for the full perturbiner expansion. If we notationally separate out the perturbiner expansion into the individual terms which only depend certain seed functions via
\begin{equation}
    \bigphi\,(\phi_1, \ldots, \phi_n) = \sum_{k = 1}^n \sum_{ \substack{\{i_1, \ldots, i_k\}   \subset \{1, \ldots, n\} }  } \bigphi^{(k)}( \phi_{i_1}, \ldots, \phi_{i_k}),
\end{equation}
where here we are summing over all subsets $\{i_1, \ldots, i_k\}$ of seed functions, then the expression for the perturbiner expansion is given by
\begin{equation}\label{phintree}
    \bigphi^{(k)}(\phi_1, \ldots, \phi_k) = \left( \frac{\kappa}{2} \right)^{k-1}  \sum_{\rm{trees}}\left( \prod_{\text{edges }(ij)} \frac{D^{ij}}{z_{ij}}\right) \left( \prod_{ a=1 }^k \phi_a \right).
\end{equation}
In the above expression for $\bigphi^{(k)}$, we sum over trees (like in figure \ref{fig:Jtreeexample}) with $k$ nodes. 

The symbol $D^{ij}$ is defined as
\begin{equation}
    D^{ij} \equiv \partial_u^{(i)} \partial^{(j)}_w - \partial_w^{(i)} \partial^{(j)}_u
\end{equation}
where $\partial_\mu^{(i)}$ is a differential symbol that only acts on the seed function $\phi_i$, but not $\phi_{i'}$ for $i \neq i'$:
\begin{equation}
    \partial_\mu^{(i)} \phi_i \equiv \partial_\mu \phi_i, \hspace{1 cm} \partial_\mu^{(i)} \phi_{i'} \equiv 0 \hspace{0.5 cm} \text{ for } i \neq i'.
\end{equation}
For example
\begin{equation}
    \partial_\mu^{(1)} ( \phi_1 \phi_2 \phi_3 ) = (\partial_\mu \phi_1) \phi_2 \phi_3,
\end{equation}
and
\begin{equation}
    D^{12}(\phi_1 \phi_2 \phi_3) = \{ \phi_1, \phi_2 \} \phi_3 .
\end{equation}
Recall the Poisson bracket is defined using the two lightcone coordinates $u$ and $w$ via
\begin{equation}\label{poisson22222}
    \{ f, g \} = \pdv{f}{u}\pdv{g}{w} -\pdv{f}{w}\pdv{g}{u}.
\end{equation}
Equation \eqref{phintree} could have also been written as a sum over cubic Feynman diagrams, but the sum over ``trees'' makes the holomorphic collinear limits manifest.

As a consequence of the integrability of SDG, the above perturbiner expansion satisfies an extra recursive formula based on collinear limits. This is the same recursive formula we noted that the Berends-Giele current satisfied in \eqref{JGpole}. It is
\begin{equation}\label{phinrecursive}
    \bigphi^{(n)} ( \phi_1, \ldots, \phi_n ) = \frac{\kappa}{2} \sum_{i = 1}^{n-1} \bigphi^{(n-1)}  ( \phi_1, \ldots, \frac{1}{z_{in}} \{ \phi_i, \phi_n\} , \ldots,  \phi_{n-1} ) .
\end{equation}

The above formula is what we call the $\Lw$ Ward identity. It is a manifestly 2-dimensional formula that generates classical solutions to the 4-dimensional theory. In fact, from the above formula, we can almost instantaneously read-off the fact that the theory is acted on by the $\Lw$ Lie algebra. In order to explain this, let us first define what the $\Lw$ Lie algebra is. It is the ``loop algebra'' of the $\myw_{1+\infty}$ algebra, where $\myw_{1+\infty}$ is simply the Lie algebra of functions on the 2d plane under the Poisson bracket. Here, we use $(u,w)$ as the coordinates of the plane, and $\{\cdot, \cdot \}$ is given by \eqref{poisson22222}.

Let us parameterize the generators of the $\myw_{1+\infty}$ Lie algebra in a suggestive way, via 2-dimensional plane-waves. Let us treat the two components of the anti-holomorphic spinor $|\tlambda]$, denoted $\tlambda_{\dot A}$ with $\dot A = \dot 1, \dot 2$, to be the two components of momentum in the 2d $(u,w)$ plane. (Actually, it turns out that these two components really will be two components of the physical momenta of a particle in lightcone coordinates, with $\tlambda_{i, \dot A} = (-\omega_i \bz_i, \omega_i) = (p_{i,w}, p_{i,u})$, but we'll get to that momentarily.)

So, we define the plane-wave generators of the $\myw_{1+\infty}$ algebra as
\begin{equation}\label{Tplanewave}
    \bT_{|\tlambda]} \equiv e^{i (\tlambda_{\dot 2} u + \tlambda_{\dot 1} w )} \in \myw_{1 + \infty}.
\end{equation}
The $\myw_{1+\infty}$ commutator of two such generators within the Poisson bracket can easily be calculated to be
\begin{equation}\label{Twcomm}
    \{ \bT_{|1]}, \bT_{|2]} \} = -[12]\,\bT_{|1] + |2]}
\end{equation}
where we recall the notation
\begin{equation}
    [12] = -\vep^{\dot A \dot B} \tlambda_{1,\dot A} \tlambda_{2,\dot B} \equiv -\tlambda_{1,\dot 1} \tlambda_{2,\dot 2} + \tlambda_{1,\dot 2} \tlambda_{2,\dot 1}.
\end{equation}

Let us now ``loopify'' the $\myw_{1+\infty}$ algebra to get the $\Lw$ algebra. To loopify a Lie algebra, we simply append an integer index to the generators that add under the commutator. (We also add 1 to the index in our conventions). So, the $\Lw$ generators are denoted
\begin{equation}
    \bT_{|\tlambda], n'} \in \Lw \hspace{1 cm} \text{for } n' \in \mathbb{Z} 
\end{equation}
and the $\Lw$ commutation relation is defined to be
\begin{equation}\label{Lwcommdef}
    [\bT_{|1], n_1} ,\bT_{|2], n_2} ] \equiv -[12] \, \bT_{|1] + |2], n_1 + n_2 + 1}.
\end{equation}

Notice that the $\Lw$ generators are labelled by three degrees of freedom: two components of $|\tlambda]$ and one integer $n'$. This is the same as the three-degrees of freedom for the on-shell momentum of a positive-helicity graviton. In fact, every $\Lw$ generator should be thought of as being equivalent to a positive-helicity graviton, as we'll explain shortly.

First, let us introduce a convenient set of conventions to discuss on-shell momenta $p = |\lambda\rangle [\tlambda|$ in lightcone coordinates. We shall now set the spinor variables $|\lambda\rangle$, $|\tlambda]$ and reference spinor $|\alpha\rangle$ to be
\begin{equation}\label{choicespinors}
    \lambda_i^A = \begin{pmatrix}
        1 \\ z_i
    \end{pmatrix}, \hspace{0.5 cm} \tlambda_i^{\dot A} = \begin{pmatrix} \omega_i \\ \omega_i \bz_i \end{pmatrix}, \hspace{0.5 cm} \alpha^A = \begin{pmatrix}
        0 \\ 1
    \end{pmatrix}.
\end{equation}
With these definitions, we have the relations
\begin{equation}
    \lr{ij} = z_{ij}, \hspace{1 cm} \lr{\alpha i} = 1, \hspace{1 cm} [12] = \omega_1 \omega_2 \bz_{12},
\end{equation}
where $z_{ij} = z_i - z_j$. The massless momenta are parameterized by $(\omega, z, \bz)$ via
\begin{equation}\label{momentum}
    p^\mu_i = (p^0_i, p^1_i, p^2_i, p^3_i) = \frac{\omega_i}{2}(1 + z \bz_i, z_i + \bz_i, -i(z_i - \bz_i), 1 - z_i \bz_i).
\end{equation}

\begin{figure}[h]
    \centering
    \includegraphics{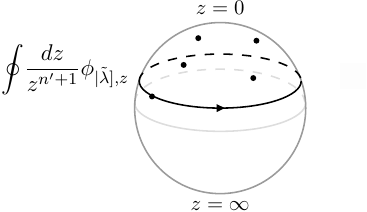}
    \caption{The action of $\bT_{|\tlambda], n'} \in \Lw$ on the spacetime. Add in one extra graviton seed function $\phi_{|\tlambda], z}$ to the perturbiner expansion and integrate it in a contour around all pre-existing seed functions, weighted by $1/z^{n'+1}$.}
    \label{fig:Lwsphere}
\end{figure}

Notice that $|\tlambda]$ depends only on $(\omega, \bz)$. Therefore, we can actually think of our massless momenta $p$ as being parameterized by $(|\tlambda], z)$, which we can denote as $p_{|\tlambda], z}$. Note that we are not treating our holomorphic and anti-holomorphic spinors on equal footing.

With these parameterizations, we shall now explain how we can use the recursive formula \eqref{phinrecursive} to read off the $\Lw$ action on classical solutions. We define the linear change that comes from ``adding in'' one more graviton seed function into the spacetime using a vertical line:
\begin{equation}
    \bigphi(\phi_1, \ldots, \phi_n | \phi_{n+1}) \equiv \bigphi(\phi_1, \ldots, \phi_n, \phi_{n+1} ) - \bigphi(\phi_1, \ldots, \phi_n ).
\end{equation}
We now write the action of $\bT_{|\tlambda], n'}$ on the spacetime, which we'll denote as $\delta_{|\tlambda], n'}$, as
\begin{equation}\label{deltaLw}
    \delta_{|\tlambda], n'} \cdot \bigphi(\phi_1, \ldots, \phi_n) \equiv \oint_C \frac{dz}{2 \pi i} \frac{1}{z^{n'+1}} \bigphi(\phi_1, \ldots, \phi_n | \phi_{|\tlambda], z}),
\end{equation}
where $C$ is a contour that surrounds $z_1, \ldots, z_n$. See figure \ref{fig:Lwsphere}.

To reiterate, an $\Lw$ transformation simply ``adds in'' one extra positive-helicity graviton into the spacetime. The momentum of this graviton is $p_{|\tlambda], z}$, where the position $z$ is integrated in a loop around all the pre-existing gravitons, and the integral is weighted by $1/z^{n'+1}$.  

Now, why does \eqref{deltaLw} satisfy the $\Lw$ commutation relation \eqref{Lwcommdef}? The proof follows the natural ``contour deformation'' argument that one encounters in introductory discussions of 2d CFTs.

If one wants to compute the action of two sequential $\Lw$ transformations, i.e. $\delta_{|\tlambda_2], n_2}  \cdot \delta_{|\tlambda_1], n_1}$, on a spacetime, one needs to choose a bigger contour for the second transformation that encloses the first contour as well as the previous seed functions and the origin. Let us say that the contour $C'$ is the bigger contour enclosing $C$. Then one has 
\begin{equation}\label{contourdiff}
\begin{aligned}
    & [\delta_{|\tlambda_2], n_2}  , \delta_{|\tlambda_1], n_1}] \cdot \bigphi(\phi_1, \ldots, \phi_n) \\
    &= \frac{1}{(2 \pi i)^2}\left( \oint_C \frac{dz_1}{z_1^{n_1+1}} \oint_{C'} \frac{dz_2}{z_2^{n_2+1}} - \oint_{C'} \frac{dz_1}{z_1^{n_1+1}} \oint_{C} \frac{dz_2}{z_2^{n_2+1}} \right) \bigphi(\phi_1, \ldots, \phi_n, \phi_{|\tlambda_1], z_1}, \phi_{|\tlambda_2], z_2} ).
\end{aligned}
\end{equation}
(In second line of the above equation, only terms in the perturbiner expansion which contain both $\phi_{|\tlambda_1], z_1}$ and $\phi_{|\tlambda_2], z_2}$ won't be projected out by the contour integrals. Terms which have neither are trivially killed by the difference of contour integrals because they don't depend on $z_1$ or $z_2$, and terms with only one of $\phi_{|\tlambda_1], z_1}$ or $\phi_{|\tlambda_2], z_2}$ are also killed by the difference of integrals, because if a term only depends on either $z_1$ or $z_2$ then the contours $C$ and $C'$ can be moved past each other and are interchangeable.)

\begin{figure}[h]
    \centering
    \includegraphics[width=0.5\linewidth]{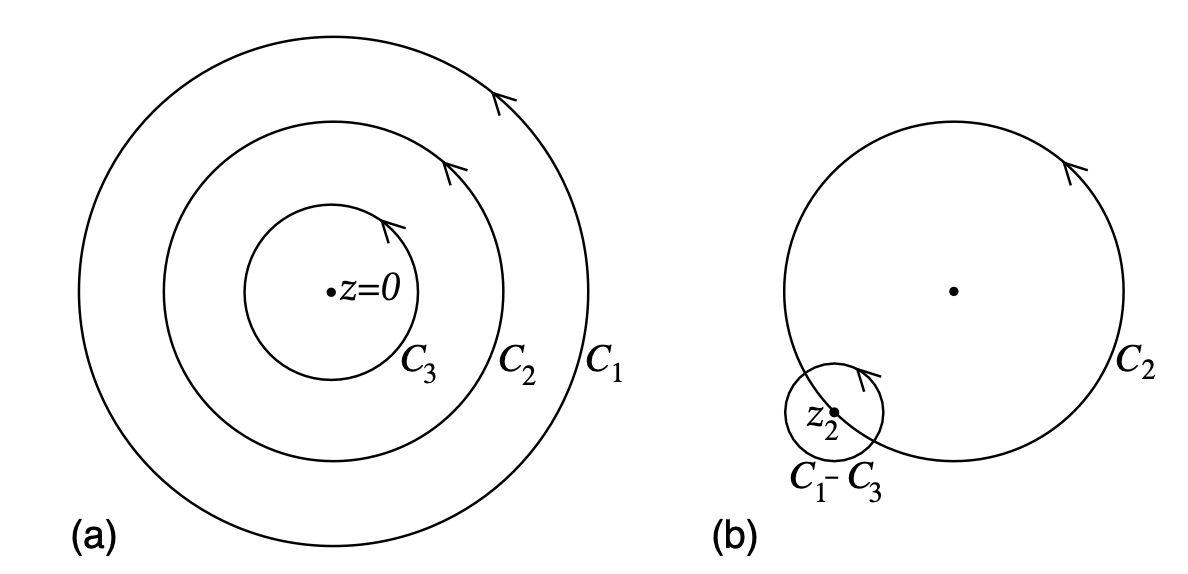}
    \caption{Figure 2.4 from Polchinski Vol. 1 \cite{Polchinski:1998rq} where he presents the 2d CFT contour deformation argument to derive the Lie algebra commutation relation from OPEs. The proof also works in the SDG context here!}
    \label{fig:polchinski}
\end{figure}
Deforming the contours in \eqref{contourdiff} in exactly the same way that they are in chapter 2 of Polchinksi Vol. 1 \cite{Polchinski:1998rq}, copied in figure \ref{fig:polchinski}, we can read off that the commutator is defined by the $z_{12} \to 0$ residue
\begin{equation}\label{deltadeltaLw}
    [\delta_{|\tlambda_1], n_1}, \delta_{|\tlambda_2], n_2} ] \cdot \bigphi(\phi_1, \ldots, \phi_n) = \oint \frac{dz}{2 \pi i} \mathop{\mathrm{Res}}_{z_2 \to z_1}  \frac{1}{z^{n_1 + 1}} \frac{1}{z^{n_2 + 1}} \bigphi(\phi_1, \ldots, \phi_n, \phi_{|\tlambda_1], z_1}, \phi_{|\tlambda_2], z_2}).
\end{equation}
The pole in $z_{12}$ can then be instantly read off from \eqref{phinrecursive} and is
\begin{equation}
    \mathop{\mathrm{Res}}_{z_2 \to z_1} \bigphi(\phi_1, \ldots, \phi_n, \phi_{|\tlambda_1], z_1}, \phi_{|\tlambda_2], z_2}) = \frac{1}{z_{12}} \frac{\kappa}{2} \bigphi(\phi_1, \ldots, \phi_n |  \{ \phi_{|\tlambda_1], z}, \phi_{|\tlambda_2], z} \} )
\end{equation}
but plugging the plane wave seed functions with the same $z$ into the $u$, $w$ commutator,\footnote{A helpful formula to check this equation is $p_i \cdot x = \omega_i ( u + z_i \bz_i \bu - \bz_i w - z_i \bw)$.}
\begin{equation}
    \{ \phi_{|\tlambda_1], z}, \phi_{|\tlambda_2], z} \} = - [12] \phi_{|\tlambda_1] + |\tlambda_2], z}
\end{equation}
which follows from $p_{|\tlambda_1],z} + p_{|\tlambda_2],z} = p_{|\tlambda_1] + |\tlambda_2],z}$,
we see that they satisfy the $\myw_{1+\infty}$ Poisson bracket relation! Plugging the above formula into \eqref{deltadeltaLw}, we now have
\begin{equation}
    [\delta_{|\tlambda_1], n_1}  , \delta_{|\tlambda_2], n_2}] = -\frac{\kappa}{2}  [12]  \delta_{|\tlambda_1]+ |\tlambda_2], n_1 + n_2 + 1}  
\end{equation}
which is now the loopified $\Lw$ \eqref{Lwcommdef} commutation relation, up to the coupling constant $\kappa/2$. So we see that our action of adding positive-helicity gravitons into self-dual spacetimes, defined via \eqref{deltaLw}, satisfies the $\Lw$ commutation relation.

It may seem strange that a non-abelian Lie algebra can arise from our operation defined by adding positive-helicity gravitons into the spacetime, given that perturbiner expansions are permutation invariant among the seed functions. The reason that the order of operations matters is that the loop must get sequentially larger with each subsequent action of the Lie algebra.

We conclude this subsection by saying that there is another basis of generators one can use for the $\myw_{1+\infty}$ algebra. One can define the basis of polynomial functions
\begin{equation}\label{Tpm}
    \bT^{p}_{m} \equiv \frac{1}{2} u^{p+m-1} w^{p-m-1}, \hspace{1 cm} \{ \bT^p_m , \bT^q_n \} = \left(m(q-1)-n(p-1)\right) \bT^{p+q-2}_{m+n},
\end{equation}
and these are the more natural generators from a celestial perspective. They correspond directly to the Mellin-transformed soft modes \cite{Miller:2025wpq}. Furthermore, the generators which correspond to positive-power monomials, i.e. $p+m-1 \geq 0$ and $p-m-1 \geq 0$, are said to live in the wedge subalgebra $\myw_{\wedge} \subset \myw_{1+\infty}$.

We also mention that the $\Lw$ algebra is deformed in the presence of a cosmological constant \cite{Taylor:2023ajd,Bittleston:2024rqe}, and can also be studied on various other non-trivial backgrounds as well. See for instance \cite{Heuveline:2025nmb,Bittleston:2023bzp,Bogna:2024gnt,Adamo:2026obu,Dunajski:2026dfe,Gomez:2026yno,Neiman:2023bkq}.

For more on gravitational perturbiners outside of lightcone gauge, see \cite{Gomez:2021shh,Gomez:2024xec,LipinskiJusinskas:2026ctz}.

\subsection{SDYM and $\Ls$}

Let us see how the analogous story works in SDYM. The seed functions are now defined as
\begin{equation}
    \Phi_i \equiv \epsilon_i T^{a_i} e^{i p_i \cdot x}
\end{equation}
where $T^a \in \mathfrak{g}$ are color generators. The SDYM perturbiner expansion
\begin{equation}
    \bigPhi(\Phi_1, \ldots, \Phi_n) = \sum_{i=1}^n \Phi_i + \ldots
\end{equation}
is defined to solve the equation of motion
\begin{equation}
    \Box \, \Phi + i \sqrt{2} \gym [\partial_u \Phi, \partial_w \Phi] = 0.
\end{equation}
Let us separate out the terms in the perturbiner expansion which depend only on subsets of seed functions via
\begin{equation}
    \bigPhi\,(\Phi_1, \ldots, \Phi_n) = \sum_{k = 1}^n \sum_{ \substack{\{i_1, \ldots, i_k\}   \subset \{1, \ldots, n\} }  } \bigPhi^{(k)}( \Phi_{i_1}, \ldots, \Phi_{i_k}).
\end{equation}
Recycling the expression for the SDYM Berends-Giele current \eqref{Jymsoln}, we can instantly write down the formula
\begin{equation}\label{psi_k}
    \bigPhi^{(k)} ( \Phi_1, \ldots, \Phi_k ) = (-i \sqrt{2} g)^{k-1} \sum_{\sigma \in S_k } \left( \prod_{i = 1}^{k-1} \frac{1}{z_{\sigma(i) \sigma(i+1) }} \right) \Phi_{\sigma(1)} \Phi_{\sigma(2)}  \ldots \Phi_{\sigma(k)}.
\end{equation}
Here $S_k$ is the permutation group of $k$ elements and we sum over all possible color orderings.

This perturbiner expansion satisfies a version of the collinear-limits based recursive formula \eqref{JYMpole}, namely
\begin{equation}\label{Phinrecursive}
     \bigPhi^{(n)}(\Phi_1, \ldots, \Phi_n) = - i \sqrt{2} g \sum_{i=1}^{n-1} \bigPhi^{(n-1)}(\Phi_1, \ldots, \frac{1}{z_{in}} [\Phi_i, \Phi_n], \ldots , \Phi_{n-1} ).
\end{equation}
This is an $\Ls$ Ward identity. Namely, if we define the following action $\delta^a_{|\tlambda], n'}$ of adding a positive-helicity gluon to the gauge field via
\begin{equation}\label{deltaLs}
    \delta^a_{|\tlambda], n'} \cdot \bigPhi(\Phi_1, \ldots, \Phi_n) \equiv \oint_C \frac{dz}{2 \pi i} \frac{1}{z^{n'+1}} \bigPhi(\Phi_1, \ldots, \Phi_n | \Phi^a_{|\tlambda], z})
\end{equation}
where we once again denote the linear change which comes from adding in an extra gluon into the list of seed functions with a bar as
\begin{equation}
    \bigPhi(\Phi_1, \ldots, \Phi_n | \Phi_{n+1}) \equiv \bigPhi(\Phi_1, \ldots, \Phi_n, \Phi_{n+1} ) - \bigPhi(\Phi_1, \ldots, \Phi_n ),
\end{equation}
then, using the same logic as the last section, we can derive the commutation relation
\begin{equation}\label{deltadeltaLs}
    [\delta^a_{|1], n_1}, \delta^b_{|2], n_2} ] = \sqrt{2} g f^{abc} \,  \delta^c_{|1]+|2], n_1 + n_2 + 1}.
\end{equation}
This is in fact nothing more than the commutation relation of the Lie algebra $\Ls$, which is the loop algebra of a Lie algebra we'll call $\mathfrak{s}$. (Many works refer to $\Ls$ as ``$S$'', but we prefer this naming convention as it is parallel with $\Lw$.) We will now define $\mathfrak{s}$ and $\Ls$ in turn.

Elements of $\mathfrak{s}$ are functions from $(u,w)$ to $\mathfrak{g}$, where the commutator is the usual $\mathfrak{g}$ commutator. It can be spanned by a basis of plane wave generators
\begin{equation}
    \bT^a_{|\tlambda]} \equiv T^a e^{i (\tlambda_{\dot 2} u + \tlambda_{\dot 1} w )} \in \mathfrak{s}.
\end{equation}
The $\mathfrak{s}$ commutation relation is then
\begin{equation}
    [\bT^a_{|1]}, \bT^b_{|2]}] = i f^{abc} \bT^c_{|1]+|2]}.
\end{equation}
To loopify the Lie algebra, we append an integer onto the generators
\begin{equation}
    \bT^a_{|\tlambda], n'} \in \Ls \hspace{1 cm} \text{for } n' \in \mathbb{Z} 
\end{equation}
and define the $\Ls$ commutation relation to be
\begin{equation}
    [\bT^a_{|1], n_1}, \bT^b_{|2], n_2}] \equiv i f^{abc} \bT^c_{|1]+|2], n_1 + n_2 + 1}.
\end{equation}
This is exactly \eqref{deltadeltaLs} up to a factor of the coupling $-i \sqrt{2} g$.

\subsection{$\Lw$ and $\Ls$ beyond tree-level-self-dual}

The integrability of classical SDG and SDYM implies that these theories have infinite dimensional $\Lw$ and $\Ls$ symmetries, but it also means that the tree-level amplitudes all vanish. So while there is a beautiful story for generating classical solutions using the $\Lw$/$\Ls$ algebras, the amplitudes associated with these solutions are zero! The sad fact is that the knife of integrability cuts both ways.

One could then ask, is there a way to see how these symmetries manifest beyond the self-dual sector? The answer, which originates from the celestial holography program, is yes. Simply put, the symmetry lives on in the holomorphic collinear limits of non-self-dual amplitudes.

\begin{figure}
    \centering
    \includegraphics{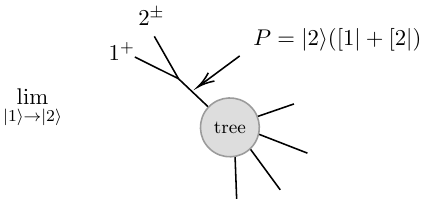}
    \caption{Holomorphic collinear limit outside of self-dual sector.}
    \label{fig:holoagain}
\end{figure}

Say, for instance, we have a tree-level amplitude containing two external massless gravitons with momenta
\begin{equation}
    p_1 = |1\rangle [1|, \hspace{1 cm} p_2 = |2\rangle [2|.
\end{equation}
Say $p_1$ is positive-helicity while $p_2$ could either have either helicity. If we take the holomorphic collinear limit
\begin{equation}
    |1\rangle \to |2\rangle
\end{equation}
then the sum of momenta $P = p_1 + p_2$ becomes
\begin{equation}
    P  \to |2\rangle ([1| + [2|).
\end{equation}
We can see that the kinematic addition of the square brackets is exactly the same structure we saw in the $\Lw$ commutation relation \eqref{Lwcommdef}. In this limit the amplitude factorizes as
\begin{equation}\label{M12pp}
    \lim_{|1\rangle \to |2 \rangle} \mathcal{M}^{\rm tree}(1^+,2^\pm, \ldots) = \text{Split}^{1^+ 2^\pm}_{{\rm G}, P^\mp} \, \mathcal{M}^{\rm tree}(P^\pm, \ldots).
\end{equation}
In section \ref{sec:hololimits} we computed the $+$$+$$-$ and $+$$-$$+$ gravitational splitting functions from the lightcone 3-pt vertex. (The $+$$+$$+$ splitting function is zero, and one can also compute that the $+$$-$$-$ splitting function vanishes when $\lr{12} \to 0$, see appendix B of \cite{Guevara:2025tsm}.) They are
\begin{equation}\label{splitgppredux}
\begin{aligned}
    \text{Split}^{1^+ 2^+}_{{\rm G}, P^-} &= -\frac{\kappa}{2} \frac{[12]}{\lr{12}} \frac{\lr{\alpha P}^4}{\lr{\alpha 1}^2 \lr{\alpha 2}^2 } = -\frac{\kappa}{2} \frac{[12]}{z_{12}} \, ,\\
    \text{Split}^{1^+ 2^-}_{{\rm G}, P^+} &= -\frac{\kappa}{2} \frac{[12]}{\lr{12}} \frac{\lr{\alpha 2}^6}{\lr{\alpha 1}^2 \lr{\alpha P}^2 } = - \frac{\kappa}{2} \frac{[12]}{z_{12}} \, .
\end{aligned}
\end{equation}
For the second equalities we plugged in the spinor parameterizations \eqref{choicespinors}.

Equation \eqref{M12pp} is interesting because it is a statement which holds beyond the self-dual sector, as all the other particles in the amplitude need not be gravitons of a particular helicity. What's more, they need not be gravitons at all. One is for instance allowed to couple Einstein gravity to any fields with minimal coupling \cite{Pate:2019lpp} and the tree-level splitting function will remain uncorrected. (The analogous statement holds for gauge theory, where in this context minimal coupling means one simply promotes derivatives to covariant derivatives.) Therefore, equation \eqref{M12pp} is exciting because it is quite general. See also \cite{Mago:2021wje} where the set of EFTs which preserve the Jacobi identity are explored, which put certain constraints on the couplings.

The relationship between \eqref{splitgppredux} to the $\Lw$ algebra is immediate. It clearly takes the form of a level-0 OPE of a 2d current with a $\myw_{1+\infty}$ color Lie-algebra, hence the putative 2d theory has an $\Lw$ symmetry. If we formally define the ``operators'' $G^\pm_{|i]}(z_i)$ to signify inserting positive-helicity gravitons in an S-matrix, then we have the OPE
\begin{equation}\label{GGope0}
    \lim_{z_{12} \to 0} G^+_{|1]}(z_1) G^\pm_{|2]}(z_2) = -\frac{\kappa}{2} \frac{[12]}{z_{12}}  G^\pm_{|1] + |2]}(z_2).
\end{equation}
In terms of operator notation, it is as if $G^+$ is a 2d conserved current of $\myw_{1+\infty}$ symmetry and $G^-$ is a charged matter operator that transforms in the adjoint representation.

These equations also of course have gauge theory analogs. If we take the holomorphic collinear limit of two gluons
\begin{equation}
    \lim_{|1\rangle \to |2 \rangle} \mathcal{A}^{\rm tree}(1^{a+},2^{b\pm}, \ldots) = \text{Split}^{1^{a+} 2^{b\pm}}_{{\rm YM}, P^{c\mp}} \, \mathcal{A}^{\rm tree}(P^{c\pm}, \ldots),
\end{equation}
from \eqref{splitym} and \eqref{splitpmp} we have
\begin{equation}
\begin{aligned}
    \text{Split}^{1^{a+} 2^{b+}}_{{\rm YM}, P^{c-}} &= \sqrt{2} g f^{abc} \frac{1}{\lr{12}} \frac{\lr{\alpha P}^2}{\lr{\alpha 1} \lr{\alpha 2} } = \sqrt{2} g \frac{f^{abc}}{z_{12}}, \\
    \text{Split}^{1^{a+} 2^{b-}}_{{\rm YM}, P^{c+}} &=\sqrt{2} g f^{abc} \frac{1}{\lr{12}} \frac{\lr{\alpha P}^2}{\lr{\alpha 1} \lr{\alpha 2} } = \sqrt{2} g \frac{f^{abc}}{z_{12}}.
\end{aligned}
\end{equation}
If we define $O^{a\pm}_{|\tlambda]}(z)$ to denote the operator which inserts a gluon into the S-matrix, then
\begin{equation}
    \lim_{z_{12} \to 0} O^{a+}_{|1]}(z_1) O^{b\pm}_{|2]}(z_2) = \sqrt{2} g \frac{f^{abc}}{z_{12}}  O^{c\pm}_{|1] + |2]}(z_2)
\end{equation}
which is the $\Ls$ equivalent of our gravitational $\Lw$ formula \eqref{GGope0}.

In this note, we have seen that the $\Lw$ and $\Ls$ OPEs generate all-plus off-shell currents and classical solutions in the self-dual sector. 
The amplitudes corresponding to these objects are of course zero. But what about beyond the self-dual sector?

The first class of amplitudes outside of the self-dual sector are the tree-level MHV amplitudes. Let us now denote the full amplitude expressions, with the momentum conserving delta function, as $\mathcal{A}$, while we denote a ``stripped'' amplitude, without the delta function, as $\overline{\mathcal{A}}$.
\begin{equation}
\begin{aligned}
    &\mathcal{A}^{\rm tree}(1^{a_1-}, 2^{a_2 -}, 3^{a_3 +}, \ldots, n^{a_n +} ) \\
    &= (2 \pi)^4 \delta^{(4)}(p_1 + \ldots + p_n) \overline{\mathcal{A}}^{\rm tree}(1^{a_1-}, 2^{a_2 -}, 3^{a_3 +}, \ldots, n^{a_n +} )
\end{aligned}
\end{equation}
\vspace{0.25 cm}

(Note that we are utilizing the ``barred = stripped amplitude'' notation for this section alone. In the majority of this note we by default dropped the delta function $(2 \pi)^4 \delta(\ldots)$ from our amplitude expressions, as is often conventional, only for brevity.)

\vspace{0.25 cm}

The stripped amplitude is of course not a unique expression, as it is only defined up to the addition of terms that vanish on the support of momentum conservation. Nevertheless, there are often natural representatives one often uses, as in the case of MHV amplitudes.

Our choice of the stripped amplitude representative will of course be the famous Parke-Taylor formula, for which the color-ordered partial amplitude is
\begin{equation}
    \overline{A}^{\rm tree}(1^+, \ldots i^- \ldots j^- \ldots n^+) = -i (-i \sqrt{2}g)^{n-2} \frac{ \lr{ij}^4}{\lr{12} \ldots \lr{n1}}.
\end{equation}
Note that this formula has no multiparticle singularities like $\frac{1}{P^2}$ associated to internal propagators going on-shell. This is because one side of such a channel would always be either a tree-level-all-plus or one-minus amplitude, which vanishes.

This stripped amplitude then satisfies a recursive ``$\Ls$ Ward identity'' which is
\begin{equation}\label{PTrecursive}
\begin{aligned}
    &\overline{\mathcal{A}}^{\rm tree}(1^{-a_1}, 2^{-a_1},\ldots, n^{+ a_n}) \\
    &= \sqrt{2} g\sum_{i = 1}^{n-1}  \frac{f^{a_n a_i c} }{\lr{ni}} \frac{\lr{\alpha i}}{\lr{\alpha n}}\overline{\mathcal{A}}^{\rm tree}(1^{-a_1}, \ldots,   i^{c} , \ldots, (n\!-\!1)^{+ a_{n-1}}).
\end{aligned}
\end{equation}
Note that we sum over all particles $1$ through $n\!-\!1$, including the two negative-helicity particles $1$ and $2$. The equation \eqref{PTrecursive} is not coming from a momentum-conserving BCFW shift, but rather is based on the meromorphicity of the Parke-Taylor formula.

This recursive formula is related to the soft theorem, as well as the WZW interpretation of the Parke-Taylor formula by Nair \cite{Nair:1988bq} and Witten's twistor string \cite{Witten:2003nn}. See also \cite{Bu:2022dis,Seet:2025mes,Melton:2024akx,Stieberger:2023fju}. More recently, it appeared in the work of Costello and Paquette \cite{Costello:2022wso} to compute amplitudes from self-dual chiral algebras, which we will mention again later in section \ref{sec:twisted}. In that work, the ``stripping'' of the delta function can be seen as an artifact of their procedure for computing amplitudes, where one first computes a form factor for an operator placed in the spacetime and then integrates over the position of the operator, giving the delta function at the end.

\begin{figure}[h]
    \centering
    \includegraphics[width=0.8\textwidth]{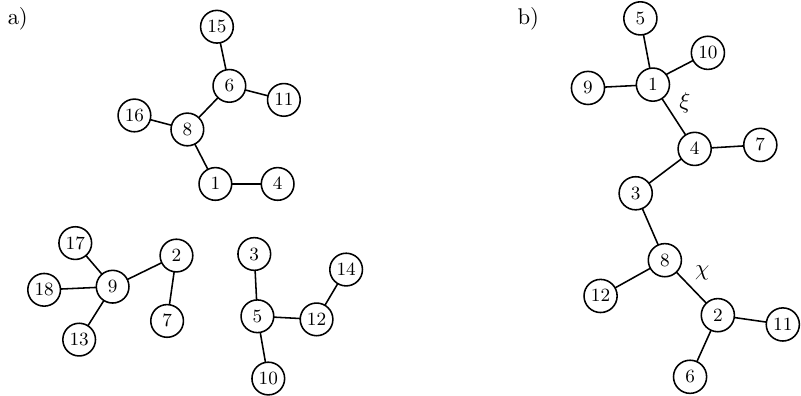}
    \caption{(a): A forest of three trees that contributes to Hodges' MHV formula $\overline{\mathcal{M}}_{n}^{\rm tree}$ for $n=18$. (b): Another graviton MHV formula ${\overline{\mathcal{M}}_{n}^{\rm tree}}'$, here for $n=12$, is a sum of $\xi,\chi$-trees. The special $\xi$ and $\chi$ edges lie on the unique path which connects node 1 to node 2.}
    \label{fig:hodge}
\end{figure}

In a recent paper by Guevara, Himwich, and the author \cite{Guevara:2025tsm}, an analogous recursive formula was found for gravitational MHV stripped amplitudes. Defining the stripped amplitude as
\begin{equation}
    \mathcal{M}^{\rm tree}(1^-, 2^-, 3^+, \ldots, n^+) = (2 \pi)^4 \delta^{(4)}(p_1 + \ldots + p_n) \overline{\mathcal{M}}^{\rm tree}(1^-, 2^-, 3^+, \ldots, n^+)
\end{equation}
we will take our representative for $\overline{\mathcal{M}}$ to be one due to Hodges \cite{Hodges:2012ym}. Defining the $n \times n$ matrix $\Psi$ by  
\begin{equation} \label{eq:Psi}
    \Psi_{ij} = \begin{cases}   - \frac{[ij]}{\langle i j \rangle} \lr{\alpha i}^2 \lr{\alpha j}^2 & i \neq j \\ \sum^{k=n}_{k=1,k \neq i}   \frac{[ik]}{\langle i k \rangle} \lr{\alpha i}^2 \lr{\alpha j}^2 & i = j\end{cases},
\end{equation} 
it turns out the graviton MHV amplitude can be written as
\begin{equation} \label{eq:HodgesDet}
    \overline{\mathcal{M}}^{\rm tree}(1^-, 2^-, 3^+, \ldots, n^+) =- i(-\kappa/2)^{n-2} \frac{\lr{12}^8}{\lr{12}^2 \lr{23}^2 \lr{31}^2} \left( \prod_{a = 4}^n \frac{1}{\lr{\alpha a}^4} \right) | \Psi|^{123}_{123},
\end{equation}
where $| \Psi|^{123}_{123}$ denotes the determinant of the $(n-3)\times(n-3)$ minor of $\Psi$ with rows and columns $1,2,3$ removed. Using the matrix tree-theorem \cite{Feng:2012sy} this formula can also be expressed as a sum over ``rooted forests,'' where nodes $1$, $2$, and $3$ are the roots for three disconnected trees. An example is drawn in figure \ref{fig:hodge}a.

Denoting a rooted forest in the set of all such rooted forests $F \in \mathcal{F}^n_{123}$, we may write
\begin{equation} \label{eq:HFF}
\begin{aligned}
    \overline{\mathcal{M}}^{\rm tree}_n &= -i (-\kappa/2)^{n-2} \frac{\lr{12}^8}{\lr{12}^2 \lr{23}^2 \lr{31}^2 } \left(\prod_{a=4}^n \frac{1}{\lr{\alpha a}^4} \right) \sum_{F \in \mathcal{F}_{123}^n} \left( \prod_{\text{edges } (ij)} \frac{[ij]}{\langle i j \rangle} \lr{\alpha i}^2 \lr{\alpha j}^2 \right) .
\end{aligned}
\end{equation}
In \cite{Guevara:2025tsm}, it was shown that this formula satisfies the recursive $\Lw$ Ward identity, which reads
\begin{equation}\label{eq:result}
    \overline{\mathcal{M}}^{\rm tree}_n = -\frac{\kappa}{2} \sum_{i = 1}^{n-1} \frac{[n i]}{\langle n i \rangle} \frac{\langle \alpha i \rangle^2}{\langle \alpha n \rangle^2}\overline{\mathcal{M}}^{\rm tree}( \, \ket{1}, |1], \, \cdots, | i \rangle, |i]{+}\frac{\langle \alpha n\rangle}{ \langle \alpha i \rangle}|n], \, \cdots, \ket{n{-}1}, |n{-}1] \, )
\end{equation}
for $n > 3$. Once again, we sum over all particles $1$ through $n\!-\!1$, including the two negative-helicity particles. Note that this recursive formula is the same type of pole-summing formula that generates the self-dual Berends-Giele currents/classical solutions \eqref{JGpole}, which is how it was found. It is superficially similar to, although not equal to, a BCFW ``inverse-soft'' recursion formula \cite{Hodges:2012ym,Boucher-Veronneau:2011rwd,Dunbar:2012aj,Ma:2022qja,Krasnov:2013wsa,Delfino:2014xea}. They differ because the BCFW-type formula uses the NSVW form of the amplitude where only the positive-helicity gravitons have associated nodes, and the negative helicity graviton spinors $|1\rangle$ and $|2 \rangle$ define the rules of the edges. As a result only the positive gravitons are summed over in that formula, but not the two negative gravitons.


There is also however \textit{another} representative for the stripped MHV amplitudes that also satisfies the exact same recursive formula. Let us denote it ${\overline{\mathcal{M}}^{\rm tree}_n}'$, which is unequal to $\overline{\mathcal{M}}^{\rm tree}_n$ off the support of momentum conservation, but equal to it on the support of momentum conservation. The formula, which originates from \cite{mypaper}, is
\begin{equation}\label{mymhvformula}
    {\overline{\mathcal{M}}^{\rm tree}_n}' = i (-\kappa/2)^{n-2} \frac{\lr{12}^4 \lr{1 \alpha}^2 \lr{2 \alpha}^2}{[\xi 1][\chi 2] } \left( \prod_{a=1}^n \frac{1}{\lr{\alpha a}^4} \right) \sum_{\xi,\chi \text{-trees} } \prod_{\text{edges }(ij)} \frac{[ij]'}{\lr{ij}}\lr{\alpha i}^2\lr{\alpha j}^2.
\end{equation}
Here, we do not sum over rooted forests but rather a different kind of tree we'll call a $\xi,\chi$-tree. We've drawn one such tree in figure \ref{fig:hodge}b. 

If $1$ and $2$ are the two negative-helicity particles, then a $\xi,\chi$ tree is a tree with $n$ nodes for which the nodes $1$ and $2$ obey a special rule. Node 1 and node 2 each have one special edge we'll call the $\xi$ edge and $\chi$ edge, respectively. The $\xi$ edge is the unique edge connected to 1 that lies on the unique path from 1 to 2. Likewise, the $\chi$ edge lies on the path from 2 to 1.

Here, $|\xi]$ and $|\chi]$ are arbitrary anti-holomorphic reference spinors. The notation $[ij]'$ in \eqref{mymhvformula} means the following: if neither $i$ nor $j$ is 1 or 2, then $[ij]' = [ij]$. Otherwise, if $i$ is $1$ (or $2$), then one is to replace $|1] \to |\xi]$ (or $|2] \to |\chi]$) if $j$ lies on the path to 2 (or 1). In other words,
\begin{equation}
\begin{aligned}
    [1j]' &= \begin{cases}
        [\xi j] & \text{ if }j\text{ is on path from $1$ to $2$} \\
        [1j] & \text{ otherwise}
    \end{cases} \\
    [2j]' &= \begin{cases}
        [\chi j] & \text{ if }i\text{ is on path from $1$ to $2$} \\
        [2j] & \text{ otherwise}
    \end{cases}.
\end{aligned}
\end{equation}
In the extra-special case that nodes 1 and 2 are neighbors, one would have $[12]' = [\xi \chi]$, although we mention that the sum over all such $\xi,\chi$-trees where 1 and 2 are neighbors is actually zero on the support of momentum conservation due to a matrix identity.

In any case, ${\overline{\mathcal{M}}^{\rm tree}_n}'$ also satisfies the recursive $\Lw$ Ward identity \eqref{eq:result} that Hodges formula ${\overline{\mathcal{M}}^{\rm tree}_n}$ does, again only for $n>3$.\footnote{Strangely, if we try to use the recursion to go from $n=2$ to $n=3$, we run into a problem. The only $n=2$ diagram has nodes 1 and 2 as neighbors, which has the $[\xi\chi]$ edge. Under the recursive Ward identity, this base case  will only generate all the diagrams where 1 and 2 remain next to each other, but as mentioned in the text these diagrams sum to zero with momentum conservation. So we generate only exactly the wrong diagrams.} It seems as though this formula could also have some chiral algebra interpretation, given that it was computed in the perturbiner framework in \cite{mypaper} by computing the expectation value of the operator $\Sigma^{AB} \wedge \Gamma_{AC} \wedge \Gamma_{B}^{\;\; C}$ in the spacetime. The reference spinors $|\xi]$ and $|\chi]$ are used to set a gauge to write down the spin connection $\Gamma_{AB}$. The full amplitude can then be generated from the Ward identity given the 3-pt seed $\frac{\lr{12}^4}{[\xi 1][\chi 2]} \frac{[\xi 3]}{\lr{13}} \frac{[3 \chi]}{\lr{32}}$. (This seed actually generates all trees where $1$ and $2$ are not neighbors, but as we said the trees where they are neighbors sum to zero anyway.) 

One natural question is, can these algebras be used to compute amplitudes beyond the MHV sector? There is a certain sense in which this has been done, again by Costello/Paquette \cite{Costello:2022wso}. By computing the position integrals of the expectation value of the appropriate number of operators $\tr(B^2)$, which deform SDYM to YM, one could use the $\Ls$ OPEs of the self-dual sector to compute any tree-level amplitude in pure YM theory if one is able to do the integrals. While this isn't the set up one would call ``holography,'' it is certainly closely related if there is only one insertion.

In any case, in this particular note we have seen that ``$\Ls$/$\Lw$ Ward Identities'' generate all the self-dual classical solutions/Berends-Giele currents, as well as tree-level stripped MHV amplitudes in both gauge theory and gravity. A natural question we should ask is, could a similar statement be true for the one-loop-all-plus amplitudes?

On the face of it, this seems quite plausible. These amplitudes are rational, and their only singularities are the holomorphic collinear ones. Furthermore, the splitting functions of these amplitudes are not loop corrected, as we showed in section \ref{sec:hololimits}, so they remain the ``pure'' $\Ls/\Lw$ OPEs we see at tree-level. These properties are quite tantalizing. Could we describe quantum self-dual gravity as the dual to some 2-dimensional boundary theory with an exact $\Lw$ symmetry?

The most basic question one could ask is, can one-loop-all-plus stripped amplitudes in gauge theory or gravity be ``generated'' by an $\Ls$/$\Lw$ Ward Identity? It's possible, but difficult to know for sure. This would require one to identify a ``good'' stripped amplitude representative which will satisfy the recursive formula, and there's no guarantee that such a representative exists. For those who are wondering, the BDPR formula \eqref{bdpr} for the gravity one-loop-all-plus amplitudes does not satisfy the recursive Ward-identity formula. This is because it only satisfies the correct splitting relation \eqref{holosplit1loop} on the support of momentum conservation, through the use of the identity \eqref{halfsoftid}, and therefore we can't treat the particle variable $|n\rangle$ as an unconstrained variable on $\mathbb{C}P^1$ and invoke the residue theorem. It could be the case that some other formula would have nicer properties. (Likewise, the canonical formula for the one-loop-all-plus gauge theory amplitudes \eqref{YM1LAPformula} also requires the use of momentum conservation to see that it has the correct holomorphic collinear limits. See appendix \ref{holcolplanarsdym}. Maybe there is some other formula which does not need momentum conservation for this, although at this time we do not know.) This should be contrasted with, say, the Parke-Taylor formula which manifestly has all the correct holomorphic collinear limits without the use of momentum conservation. Hodges' formula is a bit of a special case, because it has all the correct holomorphic collinear limits without the use of momentum conservation \textit{except} between the three particles 1,2,3,  i.e., for the limits $\lr{12} \to 0$, $\lr{23} \to 0$, or $\lr{13} \to 0$. This is enough, however, for the recursive formula \eqref{eq:result} to hold for $n>3$. 

Even if there doesn't exist a nice ``Ward identity'' for the one-loop-all-plus amplitudes, however, there could still be other ways that $\Lw$ OPE's could uniquely fix the SDG one-loop amplitudes. An interesting idea in this direction comes from the study of celestial KZ-type null state conditions, which could potentially constrain the MHV and one-loop-all-plus amplitudes uniquely \cite{Banerjee:2023jne,Banerjee:2025grp,Banerjee:2021dlm}. In addition, one can also compute the one-loop-all-plus SDYM amplitudes using chiral algebra OPEs, but not quite a ``Ward identity'', as we'll discuss in section \ref{sec:twisted}.

\section{On the ``anomaly interpretation'' of one-loop-all-plus amplitudes}\label{sec:anomaly}

In 1995, it was suggested by Bardeen that the one-loop-all-plus amplitudes in SDYM were somehow an ``anomaly'' to the symmetries of classical integrability \cite{Bardeen:1995gk}. There have been a number of works which attempted to realize this idea explicitly in different ways, some of which we've collected in this section.

\subsection{CSW/MHV rules}

One idea in a work from Brandhuber, Spence, and Travaglini \cite{Brandhuber:2006bf} is based on the CSW/MHV rules \cite{Cachazo:2004kj,Mason:2005zm} for computing YM amplitudes. In the CSW/MHV rules, tree-level YM amplitudes can be calculated as a sum over ``MHV diagrams,'' where each vertex itself is an off-shell MHV amplitude given by the Parke-Taylor formula. While it is known that these rules fail to produce the rational terms in loop amplitudes, and hence cannot compute the one-loop-all-plus amplitudes, these authors showed that if one performs a certain holomorphic change of variables of the lightcone MHV Lagrangian \cite{Mansfield:2005yd}, the action will pick up a Jacobian that allows one to compute the one-loop-all-plus amplitudes. However it does not seem as though this scheme possesses a computational advantage over standard Feynman diagrammatrics, and one still has to calculate loop integrals in the usual way. In a later work with Zoubos \cite{Brandhuber:2007vm}, they further showed the amplitudes are also captured by adding a certain bubble counterterm to the MHV Lagrangian. In fact this bubble counterterm alone computes the one-loop-all-plus amplitudes, which was apparently first observed by Bern at 4-pts \cite{Chakrabarti:2005ny}. This relies on the planarity of SDYM amplitudes in particular. The idea was revisited by Chattopadhyay and Krasnov in \cite{Chattopadhyay:2021udc,Chattopadhyay:2020oxe} where a new all-multiplicity expression for the one-loop-all-plus amplitudes was found based on it. A satisfactory understanding of how MHV rules can be used to compute general loop amplitudes remains a topic of interest, see for instance \cite{Kakkad:2022ryl,Brandhuber:2004yw}.

A twistor space investigation of similar ideas was conducted by Boels \cite{Boels:2007gv}, and in equation (4.17) of his paper, a generating function for the one-loop-all-plus amplitudes, apparently due to Mason, is motivated and written down without proof. See also \cite{Bu:2022dis}. This generating function was revisited more recently by Bogna and Mason in \cite{Bogna:2023bbd} where it was substantiated further. See for instance equation (6.4) in that paper. It seems likely that this generating function should be derivable from first principles as a Bardeen-esque anomaly term in twistor space. Section 6 of \cite{Bogna:2023bbd} contains an interesting discussion of possible approaches one might take. It would also be interesting to concoct a similar generating function for gravity, although this would be much more difficult given that there is not, as of yet, a known MHV formalism for gravity to act as a guide. 

\subsection{U(1) electric-magnetic}

Another anomaly interpretation for the one-loop-all-plus amplitudes is due to  Doran, Monteiro, and Wikeley \cite{Doran:2023cmj}. These authors show that the one-loop-all-plus amplitudes can be thought of as arising from a chiral anomaly of a $U(1)$-electric-magnetic symmetry, similar to that of the 2d Schwinger model. They also argue the amplitudes are related to the trace anomaly which explains their conformal properties. In an earlier paper by Monteiro,  Stark-Muchão, and Wikeley \cite{Monteiro:2022nqt} the amplitudes were also related to an infinite tower of non-local classically conserved currents related to integrability that were previously considered by Bardeen. These currents were used to guide the computation of the non-local ``quantum corrected'' action where off-shell integrated one-loop-all-plus $n$-gons define the quantum-corrected tree-level vertices. Interestingly, in SDYM the even parity terms of the amplitude turned out to be more easily described by such vertices.

\subsection{The Celestial Chiral Algebra program}\label{sec:twisted}

Perhaps the work which has most directly realized the spirit of Bardeen's suggestion is due to Costello \cite{Costello:2021bah}, who showed that the one-loop-all-plus amplitudes are anomalies which prevent the theory of pure SDYM from being being \textit{lifted} from 4-dimensional spacetime to 6-dimensional twistor space at the quantum level. Let us elaborate on this. It is well known that classical SDYM can be lifted to twistor space by a Penrose-Ward transform. The original 4d theory and the lifted 6d theory are equivalent to each other, but the ability to lift to twistor space makes manifest many properties of SDYM, including its classical integrability. Quantum SDYM, defined on spacetime $\mathbb{R}^4$, is of course invariant under the usual gauge transformations in four dimensions. However, if we attempt to lift the quantum theory from $\mathbb{R}^4$ to twistor space $\mathbb{R}^4 \times \mathbb{C}P^1$ in the same way that we did for the classical theory, the six-dimensional gauge transformations on twistor space are anomalous \cite{Williams:2018ows}. The source of this anomaly turns out to be the gluon one-loop-all-plus box diagram.

However, if special matter content is added to the theory, this anomaly can be cancelled and integrability can be restored. One possible type of matter which could be added is an axion with a fourth-order kinetic term and an explicit overall factor of $\hbar$ in the action, if we take the gauge group to be $SU(2)$, $SU(3)$, $SO(8)$, or any exceptional group. This cancels the one-loop gauge theory anomaly with a tree-level internal axion exchange diagram. The introduction of this axion also renders the one-loop-all-plus amplitudes zero for all $n$, not just $n=4$. So if one computes the internal-axion-tree contribution to the one-loop-all-plus amplitudes, and negates it, one computes the pure SDYM one-loop-all-plus amplitudes. This was done by Costello and Paquette in \cite{Costello:2022wso} by introducing the so-called Celestial Chiral Algebra. If one places an operator in the self-dual integrable theory to compute a form factor, one can then integrate over the position of the operator in $\mathbb{R}^4$ to compute amplitudes which can live in the non-self-dual theory. The form factors of the integrable theory can be computed easily, by simply ``doing'' the OPEs recursively in a sense reminiscent to what we've called Ward identities. These form factors have only holomorphic collinear singularities because the base theory is integrable with no amplitudes, meaning the form factors cannot have multiparticle poles. It turns out that the correct operator to integrate over to compute the internal axion contribution to the one-loop-all-plus amplitudes is $(\Box \rho)^2$, where $\rho$ is the axion field. If one wanted to compute, say, pure tree-level YM amplitudes, one would instead insert powers of $\tr(B^2)$ which deform SDYM to full YM. To compute pure YM tree-level MHV amplitudes, one inserts one instance of  $\tr(B^2)$, and every additional negative helicity gluon then requires an extra insertion of another $\tr(B^2)$. At loop-level of course internal axion exchange will affect the amplitudes. To be completely clear, the amplitudes being computed all belong to the 4-dimensional theory, and the CCA is being used as a tool for computing 4d amplitudes. The general program of computing form-factor integrands for various theories using 2d chiral algebras has become a rich area study over the last few years. See for instance \cite{Bittleston:2024efo,Charanya:2026pnh,Garner:2024tis,Bittleston:2025jmk,Fernandez:2023abp}. As a bit of extra background, the operator placed in the spacetime defines a $\mathbb{C}P^1$ defect above it in twistor space, and the particles, which are lines in twistor space, puncture this $\mathbb{C}P^1$ at isolated points which can also be thought of as points on the celestial sphere. The duality between the physics on the defect and the bulk physics is then said to be given by Koszul duality.

For gravity, it was shown that quantum SDG is similarly anomalous when lifted to twistor space, and that the anomaly can also be cancelled by a fourth-order axion \cite{Bittleston:2022jeq,Bittleston:2022nfr}, although at the time of writing this note it has not been figured out how to use the celestial chiral algebra to compute gravity amplitudes.

The axion with the fourth-order kinetic term is however somewhat undesirable due to its unphysical nature. One can also cancel the anomaly with more realistic fermionic matter in particular representations. For instance, if the gauge group is $SU(N_c)$, then the fermions can be taken to be 8 Dirac fermions transforming under the fundamental representation and one Dirac fermion transforming under the antisymmetric tensor representation of the fundamental representation. For $SU(3)$ this is equivalent to having 9 Dirac fermions in the fundamental representation. Fascinatingly, in these QCD theories one can compute all-$n$ two-loop-all-plus amplitudes using this 2d chiral algebra which are extremely difficult to compute using standard Feynman diagrammatics and loop integration \cite{Costello:2023vyy}. Because these are specially chosen theories, these particular two-loop amplitudes turn out to be rational. This is not the case for real world two-loop-all-plus QCD amplitudes. In the work \cite{Dixon:2024mzh} by Dixon and Morales, the two-loop computation was explicitly done with Feynman diagrams for $n=4$ and was shown to match the CCA result exactly in a favorable regularization scheme. We also note that in another paper  \cite{Dixon:2024tsb}, they showed that the CCA explains certain color identities involving gluons and photons in one-loop-all-plus/one-minus amplitudes. See also \cite{Morales:2025alm}. One can also cancel the anomaly by using a theory with supersymmetry \cite{Zeng:2023qqp,Charanya:2026pnh}.

We note here that the $\Ls$ algebra itself is only the leading $\mathcal{O}(\hbar^0)$ term in the full chiral algebra. The chiral algebra will also receive \textit{loop corrections}, which deform it to a ``quantum group.'' There will for instance also be a double-pole which comes from the triangle diagram, as we discussed in section \ref{sec:hololimits}, and other terms as well. Nonetheless, the full OPE does remain \textit{associative} on general grounds, and the associativity allows one to compute higher order terms in the OPE from the leading terms \cite{Fernandez:2024qnu}. In supersymmetric gauge theory the all-orders OPE was computed in \cite{Zeng:2023qqp}.

Note that the quantum-corrected double-pole, which is naturally described by the CCA, is quite difficult to deal with in BCFW loop amplitude computations. If one wants to use the BCFW method to compute rational loop amplitudes \cite{Dunbar:2010xk,Bern:2005hs,Brandhuber:2007up,Alston:2012xd,Alston:2015gea,Farrow:2020voh} one must know how to exactly account for the extra pole (and any other coincident lower-order-pole terms) using extra information in order to enact the usual contour deformation arguments. In the CCA however, they are handled quite naturally and beautifully.

\section{More literature}\label{sec:moreliterature}

Here we review a bit more literature surrounding SDYM and SDG. This is in no way exhaustive. We have in particular ignored the entire existence of twistor theory \cite{Adamo:2017qyl} twistor strings \cite{Witten:2003nn,Berkovits:2004jj,Skinner:2013xp,Mason:2009afn,Cachazo:2005ga} ambitwistor strings \cite{Mason:2013sva,Adamo:2013tsa,Geyer:2015jch,Geyer:2017ela} nonlinear twistor sigma models \cite{Adamo:2021bej,Adamo:2022mev, Adamo:2024hme} and higher spin theories \cite{Metsaev:2005ar,Krasnov:2021nsq,Skvortsov:2020gpn}.

\subsection{Old-school MHV calculation and self-duality}

In this section we discuss the original historical connection of SDYM to the Parke-Taylor formula \cite{Parke:1986gb} for MHV tree-level amplitudes, due to Selivanov and Rosly \cite{Rosly:1996vr,Rosly:1997ap,selivanov1997selfdual}.

The connection uses the notion of ``perturbiners.'' A perturbiner is a classical solution in which one begins with a sum of free plane waves, each multiplied by its own tiny infinitesimal parameter, and then recursively solves for the rest of the solution by repeatedly plugging it back into the equations of motion and solving for higher order terms in the infinitesimal parameters. 

There are two different ways that perturbiners can be used to compute amplitudes. In the first way, one realizes that classical perturbiner solutions are exactly the same as off-shell Berends-Giele currents where the momentum of the off-shell leg is swapped for a spacetime position of the solution by a Fourier transform. We review this connection in appendix \ref{sec:classicalsolns}. See also the recent review \cite{LipinskiJusinskas:2026ctz}. In the second way, one takes the full perturbiner expansion and plugs it into the action, which calculates the amplitude at tree-level. This idea has been revisited in more modern treatments in for instance \cite{Kim:2023qbl,Isen:2026xoc}. A more terse review is provided in section 4 of \cite{mypaper}.

The Selivanov and Rosly calculation uses the second scheme to compute the MHV amplitudes, where one plugs the perturbiner solution into the action and evaluates it on-shell. The amplitude is equal to the ``maximum'' term that contains each infinitesimal parameter, belonging to each of the original particle plane waves, exactly once. 
For MHV amplitudes this can be done efficiently using the following ``trick''. Note the Yang-Mills action is essentially $\tr (F \wedge \star F)$. One can add any multiple of a topological term $\tr(F \wedge F)$ to the action without affecting the scattering amplitudes. If one defines $F^\pm = F \pm i \star F$, this means that the Yang-Mills action can be argued to be equivalent to $\tr(F^- \wedge \star F^-)$. A self-dual solution, with only positive-helicity gluons satisfies $F^- = 0$. A solution with a single linearized negative-helicity gluon ``$1^-$'' propagating on a self-dual background will have $F^-_1 \propto \epsilon_1$, with $F^-_1$ proportional to the infinitesimal parameter belonging to the negative-helicity gluon. Due to the quadratic nature of $\tr(F^- \wedge \star F^-)$, and the fact that the amplitude is given by extracting the term which contains each infinitesimal parameter exactly once, we only need to solve for $F^-_{1}$ and $F^-_{2}$, and then compute $2 \tr( F^-_{1} \wedge \star F^-_{2})$ integrated over spacetime. Crucially, this ``trick'' means that we never have to calculate a perturbiner expansion where the two negative-helicity gluon terms interact with each other. It is enough to solve for the motion of a linearized negative-helicity gluon propagating on the self-dual background created by all the positive-helicity particles. In doing so, one can derive the Parke-Taylor MHV formula using only perturbiners computed in SDYM. (We mention that the spiritual generalization of this procedure to full YM can be thought of as giving the CSW rules using the twistor action \cite{Mason:2005zm}.)

In \cite{mypaper}, it was shown how perturbiners can be used to compute the gravity MHV amplitudes as well. In the gravity case, the analog of the $\tr (F^- \wedge \star F^-)$ term is then $\Sigma^{AB} \wedge \Gamma_{A C} \wedge \Gamma_{B}^{\;\; C}$ where $\Sigma^{AB}$ is an anti-self-dual 2-form and $\Gamma_{AB}$ is the anti-self-dual part of the spin connection 1-form. A closely related twistor computation was performed in  \cite{Adamo:2021bej} based on Plebański's first heavenly equation. We review this in appendix \ref{sec:firstheavenlyhalfsoft}.

\subsection{Color-kinematics duality}

There is a close relationship between the $\mathfrak{s}$ algebra/$\myw_{1+\infty}$ algebra and the kinematic algebra underlying colour-kinematics duality in the self-dual sectors \cite{Monteiro:2022lwm}. Both simply look at the three-point vertex in different ways. In lightcone gauge, the \ppm vertex in gravity is simply the kinematic part of the \ppm vertex in gauge theory squared, so the self-dual actions satisfy a particularly manifest form of color-kinematics duality \cite{Monteiro:2011pc}.  See also \cite{Raeymaekers:2025akr} at the classical perturbiner level. (There have also been a myriad of works relating self-dual classical solutions via the double copy, a few include \cite{Chacon:2020fmr,Campiglia:2021srh,Monteiro:2020plf}.) It is also found that loop amplitudes themselves satisfy BCJ double copy relations at an integrated level, not just at an integrand level \cite{Boels:2013bi,Monteiro:2022nqt}. The intricacies implicit in the full BCJ relations could potentially be viewed as the beginning of a more general story, for which the simplicity of the $\myw_{1+\infty}$-governed self-dual sector is just the tip of a much larger iceberg relating the algebraic structure of the colour-kinematics duality to that of the celestial OPE beyond the self-dual sector \cite{Monteiro:2022lwm}. 

\subsection{Conformal symmetry}\label{sec:conformalsym}
Note that tree-level (classical) Yang-Mills theory has a bulk $SO(4,2)$ conformal symmetry \cite{Bargheer:2009qu}. Furthermore, the equation that restricts one to the self-dual sector $F_{\mu \nu} = \frac{i}{2} \vep_{\mu \nu \alpha \beta} F^{\alpha \beta}$ is also conformally invariant, and because the self-dual one-loop-all-plus amplitudes are finite and there is no scale introduced by the loop integrals, these facts together imply that quantum SDYM is perturbatively a CFT${}_4$. (Non-perturbatively, however, instanton corrections will break the conformal symmetry because there is secretly a topological $\tr(F \wedge F)$ term in the action whose theta-term-coupling does run \cite{Bittleston:2025jmk,Losev:2017qrj} under renormalization.)

The explicit conformal invariance of the SDYM one-loop-all-plus amplitudes was verified by Henn, Power, and Zoia in \cite{Henn:2019mvc}. They accomplished this by finding an expression for the amplitudes which has manifest conformal invariance, the so-called Kermit formula \cite{Chicherin:2022bov}, and proved this expression to be true using BCFW recursion.

What about gravity? Well, neither the vacuum Einstein equation $R_{\mu \nu} = 0$ nor the self-duality equation $R_{\mu \nu} = \frac{i}{2} \vep_{\mu \nu \alpha \beta} R^{\alpha \beta}$ is $SO(4,2)$-conformally invariant. Having said that, both equations are dilatation invariant. So quantum SDG is fascinatingly an example of a 4-dimensional theory which has dilatation invariance but not full conformal invariance. Note that this is allowed because SDG is a non-unitary theory. We discuss this topic further in appendix \ref{sec:dilatation}.

(Actually, the self-dual gravity one-loop-all-plus amplitude is conformally invariant in the special case that $n=4$, see \cite{Doran:2023cmj} for an interpretation of this fact.)

\subsection{N=2 string}

SDYM and SDG quite miraculously appear as the low energy effective field theory of the N=2 string \cite{Ooguri:1990ww,Ooguri:1991fp,Ooguri:1991ie,Ooguri:1995cp,Siegel:1992wd}. Here N=2 refers to worldsheet supersymmetry of the string. See for instance the old review by Marcus \cite{Marcus:1992wi} and the references therein. The N=2 string, however, is a very poorly understood object. One tension is that it can be argued that (essentially) all the tree and loop amplitudes of the N=2 string are zero \cite{Berkovits:1994vy}. However, we of course know that the SDYM/SDG loop amplitudes are non-zero. This seems to be a puzzle.

\subsection{One-loop-all-plus and  $\mathcal{N}=4$ super-Yang-Mills}

Due to a supersymmetric Ward identity, all-plus and one-minus amplitudes vanish to all loop orders in supersymmetric theories. Nonetheless, the non-zero pure YM one-loop-all-plus amplitudes have various fascinating relationships to $\mathcal{N}=4$ SYM. We already discussed the dimension-shifting relation in section \ref{sec:BDPR}. 

There is however another relationship which we now mention. It is a remarkable fact that the amplitudes in $\mathcal{N}=4$ SYM are equal to expectation values of Wilson loops in the theory with polygonal null edges \cite{Alday:2007hr}. This is related to the dual conformal invariance of $\mathcal{N}=4$ SYM amplitudes. These expectation values are of course divergent, coming from the corners of the polygon, just as the SYM amplitudes have infrared divergences. If one however takes the Lagrangian as a local operator and inserts it into the Wilson loop expectation value at infinity, this object is interestingly finite. (Performing the integration over the location of the insertion recovers the amplitude.) The incredible fact is that this Wilson loop + Lagrangian insertion at infinity object is equal to the all-plus amplitude of pure YM for all loop orders \cite{Chicherin:2022bov}! Because one point (of region momentum) has been placed at infinity, the formula breaks dual conformal invariance which is indeed not a symmetry of the all-plus amplitudes. The loop integrals associated to these objects were also recently shown to be calculable using `negative geometries' \cite{Dixon:2026ipt} and it would be interesting to see if the one-loop-all-plus amplitudes themselves have a simple interpretation in these terms.

\subsection{Burns space holography}

In \cite{Costello:2022jpg,Costello:2023hmi}, Costello, Paquette, and Sharma created a top-down topological string construction of a celestial-dual to a kind of self-dual gravity. It is based on the (type I) ``B-model'' topological string theory in six-dimensions. (It is the worldsheet theory that is topological while the target space theory is holomorphic.) This theory can be placed on a Calabi-Yau background spacetime such as $\mathbb{C}^6$. One can then place a stack of $N \gg 1$ D1 branes in the $\mathbb{C}^6$, which backreact and deform the spacetime to the manifold $SL(2,\mathbb{C})$. The topological B-model on $SL(2,\mathbb{C})$ is then dual to a 2d chiral algebra living on the D1 branes, in the usual sense of the AdS/CFT correspondence. The idea of the authors was then to replace $\mathbb{C}^6$ with a twistor space so the 6d holomorphic theory would actually be equivalent to a 4d theory, creating a 4d/2d duality instead of a 6d/2d duality. However, twistor space $\mathbb{R}^4\times \mathbb{C}P^1$ cannot be a Calabi Yau manifold as is, so one has to excise two points from the $\mathbb{C}P^1$ in order for the construction to work. When we now place the $N \gg 1$ (euclidean) D1 branes in this spacetime instead of the flat $\mathbb{C}^6$, it now deforms to a spacetime which is different from $SL(2,\mathbb{C})$. Once we reduce the 6d spacetime to 4d, the two excised points from the $\mathbb{C}P^1$ allow one to define a Kähler scalar for the 4d spacetime. This 4d manifold is a self-dual spacetime called Burns space, and the gravitational closed-string sector of the B-model describes perturbations of the Kähler scalar. More precisely it is a theory of self-dual conformal gravity called Mabuchi gravity, which has a fourth-order kinetic term and is therefore non-unitary. There is also an open-string gauge theory sector as well. The D1 branes originally sat in the middle of the spacetime, but after backreaction they in a sense disappear and reappear at the celestial sphere. Crucially, Burns space is an asymptotically flat spacetime, which makes this a top-down construction of celestial holography. Note that if we had placed this particular bulk theory in flat space instead of on the curved Burns space background, all the amplitudes of this theory would be zero. (We also note that while Burns space is asymptotically flat in euclidean signature, in Lorentzian signature it becomes singular on the null cone which does hit null infinity.) This duality is also an example of ``twisted'' holography \cite{Costello:2018zrm,Costello:2019jsy}. Here, the word ``twisted'' has nothing to do with twistor space. ``Twisted'' means that one can take a certain BPS subsector of both sides of a higher dimensional AdS/CFT duality to get the duality between the boundary theory of the D1 branes with the bulk B model. On the AdS side of the larger duality one has type I string theory with a certain collection of branes.

Also, in a later work by Bittleston, Costello, Zeng \cite{Bittleston:2024efo}, another twisted-holography string construction on a set of intersecting D5 and D5' branes in type I string theory was shown to be describe a celestial chiral algebra of 4d SDYM and additional matter content.

\acknowledgments

The author indebted to his collaborators Alfredo Guevara and Mina Himwich, as this document is a spin-off of shared work. It is also a pleasure to thank Adam Ball, Roland Bittleston, Ruth Britto, Kasia Budzik, Eduardo Casali, Jaazib Charanya, Donal O'Connell, Lance Dixon, Johannes Henn, David Kosower, Lionel Mason, Anthony Morales, Giulio Salvatori, Atul Sharma, David Skinner, and Ed Witten for many enlightening conversations. We acknowledge support from the Princeton Gravity Initiative and IAS. Our research was also supported by the Simons Foundation grant SFI-MPS-T-Institutes-00010825, and by the State Treasury funds as part of a task commissioned by the Minister of Science and Higher Education under the project Organization of the Simons Semesters at the Banach Center - New Energies in 2026- 2028 (MNiSW/2025/DAP/491).
 \textit{Statement on AI use:} The author utilized Google Gemini 3.1 Pro and ``Deep think'' for notable assistance with appendices \ref{subb3} and \ref{subb4}. They were however written by human. 

\newpage 
\appendix

\section{Covariant formulation of SDYM and SDG}\label{sec:covariant}

In this note we defined SDYM and SDG as truncations of the Yang-Mills and Einstein-Hilbert action in lightcone gauge. From this we found simple ghost-free cubic actions for SDYM and SDG, but the procedure broke manifest Lorentz covariance, which is equivalent to gauge-invariance/diffeomorphism-invariance of the amplitudes. In concrete terms, the gauge choice fixed a preferred spinor $|\alpha\rangle$ which appears in the interaction vertex and the polarization vectors. Of course, because the amplitudes in SDYM/SDG match those of the full theory, the final amplitudes ultimately do not depend on $|\alpha\rangle$ and thus SDYM/SDG must be covariant even if it is not manifest at the level of the lightcone actions.

The question we now ask is, is it possible to define SDYM and SDG as truncations of Yang-Mills theory/Einstein gravity in a manifestly covariant way? The answer is of course yes, and in this section we discuss the covariant formulations for SDYM and SDG.

As a consistency check, we will also show how, by gauge fixing the self-dual covariant actions, we can again rederive the same lightcone actions from before. So, truncating the theory first and then gauge fixing produces the same Lagrangian as gauge fixing first and then truncating.

\subsection{Covariant SDYM action}

Recall that (flat) spacetime indices can be exchanged for spinor-helicity indices via 
\begin{equation}
    V_{A \dot A} = \sigma_{A \dot A}^\mu V_\mu, \hspace{1 cm} V_\mu = \frac{1}{2} \sigma_\mu^{A \dot A} V_{A \dot A}.
\end{equation} 

It is a standard exercise to show that, in spinor variables, any antisymmetric tensor like $F_{\mu \nu}$ can be decomposed as
\begin{equation}
    F_{A \dot A B \dot B} = 2 \vep_{\dot A \dot B} F_{AB} + 2 \vep_{A B} \widetilde{F}_{\dot A \dot B}
\end{equation}
where $F_{AB} = F_{(AB)}$, $\widetilde{F}_{\dot A \dot B} = \widetilde{F}_{(\dot A \dot B)}$ are symmetric tensors with 3 independent degrees of freedom each.

In fact, $F_{AB}$ represent the anti-self-dual degrees of freedom of $F_{\mu \nu}$ while $\widetilde{F}_{\dot A \dot B}$ represent the self-dual degrees of freedom. This is easy to check if one uses the expression for the flat space $\varepsilon_{\mu \nu \rho \sigma}$ pseudotensor in spinor indices,
\begin{equation}\label{epsilonidentity}
    \varepsilon_{A \dot A B \dot B C \dot C D \dot D} = 4 i \, \vep_{AC} \vep_{BD} \vep_{\dot A \dot B} \vep_{\dot C \dot D} - 4 i \, \vep_{AB} \vep_{CD} \vep_{\dot A \dot C} \vep_{\dot B \dot D}.
\end{equation}
Using $V^\mu W_\mu = \frac{1}{2} V^{A \dot A} W_{A \dot A}$, we can compute
\begin{equation}
\begin{aligned}
    F^{\mu \nu} F_{\mu \nu} &= 2 F^{AB} F_{AB} + 2 \widetilde{F}^{\dot A \dot B} \widetilde{F}_{\dot A \dot B}\,, \\
    \vep_{\mu \nu \rho \sigma} F^{\mu \nu} F^{\rho \sigma} &= 4 i \, ( F^{AB} F_{AB} - \widetilde{F}^{\dot A \dot B} \widetilde{F}_{\dot A \dot B} )\,.
\end{aligned}
\end{equation}
The second term is the well known topological term $F \wedge F$. We can add any multiple of it to the Yang-Mills action without affecting scattering amplitudes. Doing so, we can write the action of Yang-Mills theory, plus the topological term, in terms of only the anti-self-dual component $F_{AB}$ as
\begin{equation}\label{SYMA}
    S_{\rm YM}[A_{A \dot A}] = - \int d^4 x \tr( F^{AB} F_{AB} ).
\end{equation}
There is another way, however, to write this action. Introducing a new anti-self-dual field $B_{AB}$, with $B_{AB} = B_{(AB)}$, we write
\begin{equation}\label{SymBA}
    S_{\rm YM}[A_{A \dot A}, B_{AB}] = -2 \int d^4 x \tr( B^{AB} F_{AB} - \frac{1}{2} B^{AB} B_{AB} ).
\end{equation}
In this action, the independent fields are $B_{AB}$ and the usual gauge field $A_{A \dot A}$, which enters the action via $F_{AB}$ from
\begin{equation}
    - 4 F_{AB} = \partial_{A\dot A}A_{B}^{\;\; \dot A} + \partial_{B \dot A} A_{A}^{\;\; \dot A} - i g [A_{A \dot A}, A_{B}^{\;\; \dot A}].
\end{equation}
The equation of motion for $\delta B_{AB}$ from \eqref{SymBA} is
\begin{equation}
    B_{AB} = F_{AB}
\end{equation}
and plugging this back into the action, we recover \eqref{SYMA}. The above equation implies that on-shell, $B_{AB}$ should be thought of as a negative-helicity degree of freedom.

The covariant action for SDYM is defined by removing the $\tr B^2$ term in the YM action. It is
\begin{equation}\label{SsdymBA}
    \boxed{ S_{\rm SDYM}[A_{A \dot A},B_{AB}] = - 2 \int d^4 x \tr ( B^{AB} F_{AB}). }
\end{equation}
The $B_{AB}$ field is now takes the form of a Lagrange multiplier whose equation of motion enforces the self-duality equation $F_{AB} = 0$ for the field $A_{A \dot A}$.

If one varies $A_{A \dot A}$ in the SDYM action, one finds the equation of motion for $B_{AB}$ is
\begin{equation}
    \partial_{A \dot A} B^{AB} - i g [A_{A \dot A}, B^{AB}] = 0.
\end{equation}
This equation can also be written as $D_{A \dot A}B^{AB} = 0$ where here $D_{A \dot A}$ is the covariant derivative of the self-dual field $A_{A \dot A}$. If one substitutes in $F_{AB}$ for $B_{AB}$ in the above equation, one can show that this equation is a linear combination of the usual Yang-Mills e.o.m. $D_{\mu} F^{\mu \nu} = 0$ and the Bianchi identity $\vep^{\mu \nu \rho \sigma} D_{\mu} F_{\nu \rho} = 0$, on the self-dual background, that keeps only the anti-self-dual component of the field to the first linearized order. Our point is that the equation $B_{AB} = F_{AB}$, which held for the full action but seemingly fails for the self-dual action when $A_{A \dot A}$ is self-dual and $F_{AB} = 0$, is still morally true. This is because on-shell $B_{AB}$ behaves like a linearized anti-self-dual degree of freedom propagating on a self-dual background, and is evidently equal to the $F_{AB}$ of such a perturbation.

\subsection{SDYM gauge fixing}

We now gauge fix the action \eqref{SsdymBA} in lightcone gauge. The gauge fixing condition is
\begin{equation}\label{aaA}
    \alpha^A \talpha^{\dot A} A_{A \dot A} = 0.
\end{equation}
We could introduce a Lagrange multiplier and $b$ $c$ ghost system to the action. However it is clear that the ghost system will decouple because the gauge variation of the above condition is $A$-independent. See for instance our more explicit discussion in section \ref{sec:YMactionderivation}. 

Imposing the gauge fixing condition \eqref{aaA}, we will now integrate out some components of $B_{AB}$.

Let us define a new spinor $\beta^A$ such that
\begin{equation}
    \alpha^A \beta_A = 1
\end{equation}
which implies the completeness relation
\begin{equation}\label{deltaalphabeta}
    \delta^A_B = -\alpha_B \beta^A + \beta_B \alpha^A.
\end{equation}
In components,
\begin{equation}\label{alphabeta}
    \alpha^A = (0,1), \hspace{0.5 cm} \beta^A = (1,0), \hspace{0.5 cm} \alpha_A = (-1,0), \hspace{0.5 cm} \beta_A = (0,1),
\end{equation}
and also $\talpha^{\dot A} = (0,1)$. So the gauge condition \eqref{aaA} is
\begin{equation}\label{A22}
    A_{2 \dot 2} = 0.
\end{equation}
Let now decompose the $B_{AB}$ field into three independent degrees of freedom $(B, B', B'')$ as
\begin{equation}
    B_{AB} =B \alpha_A \alpha_B + B'(\alpha_A \beta_B + \beta_A \alpha_B) +  B'' \beta_A \beta_B
\end{equation}
and let us now integrate out the component $B$. The equation of motion for $\delta B = 0$ is
\begin{equation}
    0 = -2 F_{22} =  \partial_{2 \dot A}A_{2}^{\;\; \dot A} = -  \partial_{2 \dot 2}A_{2 \dot 1}
\end{equation}
which implies $A_{2 \dot 1} = 0$. Integrating $B$ out of the action, we therefore see that in the self-dual theory \eqref{A22} is promoted to the two-component condition
\begin{equation}
    A_{2 \dot A} = 0
\end{equation}
which means we can write $A_{A \dot A}$ as
\begin{equation}
    A_{A \dot A} = \alpha_A A_{\dot A}
\end{equation}
for some $A_{\dot A}$.

Now we want to integrate out $B'$. Before we do this, however, let us introduce the symbols
\begin{equation}\label{alphabetapartial}
\begin{aligned}
    (\alpha \partial)_{\dot A} &\equiv -\alpha^A \partial_{A \dot A}\, , \hspace{1 cm} (\alpha \partial)^{\dot A} \, \equiv \alpha_A \partial^{A \dot A} \, , \\
    (\beta \partial)_{\dot A} &= -\beta^A \partial_{A \dot A} \, , \hspace{1 cm} (\beta \partial)^{\dot A} = \beta_A \partial^{A \dot A} \, .
\end{aligned}
\end{equation}
(We include the relative sign so that $(\alpha \partial)_{\dot A}$ behaves as a proper spinor under raising and lowering, due to the sign in $\alpha^A \lambda_A = -\alpha_A \lambda^A$.)

With these symbols, the decomposition of the identity \eqref{deltaalphabeta} for instance implies
\begin{equation}
    \Box = \frac{1}{2} \partial_{A \dot A} \partial^{A \dot A} = ( \beta \partial)_{\dot A} (\alpha \partial)^{\dot A},
\end{equation}
where lightcone coordinates, we recall
\begin{equation}
    \partial_{A \dot A} = \begin{pmatrix} \partial_{1 \dot 1} & \partial_{1 \dot 2}\\ \partial_{2 \dot 1} & \partial_{2 \dot 2} \end{pmatrix} = \begin{pmatrix} \partial_\bu & \partial_\bw \\ \partial_w & \partial_u \end{pmatrix}.
\end{equation}
The equation of motion for $\delta B'$ is now
\begin{equation}
\begin{aligned}
    0 &= -2 (\alpha^A \beta^B + \alpha^B \beta^A)F_{AB} \\
    &=  (\alpha \partial)_{\dot A} A^{\dot A} .
\end{aligned}
\end{equation}
Integrating out $B'$ therefore implies that $A_{\dot A}$ can be expressed as
\begin{equation}
    A_{\dot A} = \sqrt{2} (\alpha \partial)_{\dot A} \Phi
\end{equation}
for some potential $\Phi$. Now we are just left with the last component $B''$. For this component, let us substitute in
\begin{equation}
     B'' = -\frac{1}{\sqrt{2}} \bPhi.
\end{equation}
We now compute that the gauge fixed action is
\begin{equation}
\begin{aligned}
    S_{\rm SDYM}[\Phi, \bPhi] &= -2 \int d^4 x \tr( B'' \beta^A \beta^B F_{AB} )\\
    &= - \int d^4 x \tr \bPhi \left( \Box \, \Phi - \frac{i g}{\sqrt{2}} [ (\alpha \partial)_{\dot A} \Phi, (\alpha \partial)^{\dot A} \Phi ] \right)
\end{aligned}
\end{equation}
and this is exactly the lightcone action.

\subsection{Covariant SDG action}\label{subb3}

In order to write down the covariant form of the self-dual gravity action, we first need to review the ``chiral'' formulation of full Einstein gravity due to Plebański \cite{plebanski1977separation}. See also the paper of Capovilla, Dell, Jacobsen, and Mason \cite{capovilla1991self}. This construction possesses a number of subtleties which we discuss below.

This construction is expressed in terms of the tetrad 1-form, the anti-self-dual 2-forms, the spin-connection 1-form, and the Riemann curvature 2-form. We wrote a pedagogical introduction to these objects in appendix B of \cite{mypaper} and so we won't review them too deeply again. Note that we use the same conventions from that appendix here. If we define the flat-space vierbein indices $a = 0,1,2,3$, which can be raised and lowered with the flat-space metric $\eta_{ab}$, then we denote the tetrad 1-form as $\theta^a$, spin-connection 1-form as $\Gamma_{ab}$, and Riemann curvature 2-form $R_{ab}$ as
\begin{equation}
\begin{aligned}
    \theta^a &= \theta^a_{\mu} dx^\mu \, .\\
    \Gamma_{ab} &= \Gamma_{ab \mu} dx^\mu \, .\\
    R_{ab} &= \frac{1}{2} R_{ab \mu \nu} dx^\mu \wedge dx^\nu\, .
\end{aligned}
\end{equation}
$\Gamma_{ab} = \Gamma_{[ab]}$, $R_{ab} = R_{[ab]}$ are antisymmetric in the vierbein indices. $R_{ab}$ is a function of $\Gamma_{ab}$, defined by
\begin{equation}
    R_{ab} \equiv \dd \Gamma_{ab} + \Gamma_{ac} \wedge \Gamma^c_{\;\; b}.
\end{equation}
Note the metric is constructed out of the tetrads via $g_{\mu \nu} = \eta_{ab} \theta^a_\mu \theta^b_\nu$. Note also $\theta^a_\mu \theta_a^{\nu} = \delta_\mu^\nu$ and $\theta^a_\mu \theta_b^\mu = \delta^a_b$.

These objects can be used to construct the usual tetradic Palatini action for Einstein general relativity, which is
\begin{equation}
    S_{\rm GR \, (Palatini)}[\Gamma_{ab}, \theta^a] = -\frac{1}{\kappa^2} \int \vep_{abcd} R^{ab} \wedge \theta^c \wedge \theta^d .
\end{equation}
$\vep_{abcd}$ is defined by $\vep_{0123} = 1$. The equations of motion from varying $\Gamma_{ab}$ and $\theta^a$ are 
\begin{equation}\label{palatinieom}
\begin{aligned}
    \vep_{abcd} D \theta^c \wedge \theta^d &= 0 \, ,\\
    \vep_{abcd} R^{ab} \wedge \theta^c &= 0 \, ,
\end{aligned}
\end{equation}
where the ``torsion'' $D \theta^a$ is defined by
\begin{equation}
    D \theta^a \equiv \dd \theta^a + \Gamma^a_{\;\; b} \wedge \theta^b.
\end{equation}
It can be shown that first equation of \eqref{palatinieom} implies the torsion-free condition $D \theta^a = 0$.\footnote{Here is why. The equation is equivalent to $T^a \wedge \theta^b - T^b \wedge \theta^a = 0$ where $T^a = D \theta^a$ is a 2-form. Take the interior product with the vector field $\theta_b^\rho \partial_\rho$. This results in the equation $-2 T^a + 4 T^a - (\iota_{\theta_b} T^b) \wedge \theta^a - T^a = 0$, or $T^a = (\iota_{\theta_b} T^b) \wedge \theta^a$, and plugging this back into the original equation we find $T^a = 0$.} When the torsion-free condition is satisfied, one can show that the components of the spin connection can be expressed as functions of the tetrads. In particular, if we define $\Gamma^a_{\;\;bc}$ via
\begin{equation}
    \Gamma_{ab} = \Gamma^a_{\;\;bc} \theta^c
\end{equation}
and denote the components of $\dd \theta^a = \frac{1}{2} C^a_{\;\;bc} \wedge \theta^b \wedge \theta^c$ using coefficients $C^a_{\;\;bc} = \partial_{[\mu} \theta_{\nu]}^a \theta^\mu_b \theta^\nu_c$, then the torsion-free condition can be shown to imply
\begin{equation}
    \Gamma_{abc} = \frac{1}{2} (C_{abc} + C_{bca} - C_{cab}).
\end{equation}
This can also be expressed as
\begin{equation}
    \Gamma^a_{\;\; bc} = \theta^a_{\; \mu} \theta_c^{\; \nu} (\partial_\nu \theta_b^{\; \mu} + \Gamma^\mu_{\nu \rho} \theta_b^{\; \rho})
\end{equation}
where $\Gamma^\mu_{\nu \rho}$ is the usual Christoffel symbol defined by the metric.

So we see that once the torsion-free condition is satisfied, then the spin-connection 1-form is equal to the usual metric-compatible expression. This then also implies that $R^{ab} = \frac{1}{2} \theta^a_{\mu} \theta^b_{\nu} R^{\mu \nu}_{\;\;\;\; \rho \sigma} dx^\rho dx^\sigma$ where $R_{\mu \nu \rho \sigma}$ is the Riemann-tensor of the metric. The second equation of \eqref{palatinieom} can then be shown to be exactly the Einstein equation $R_{\mu \nu} - \frac{1}{2} g_{\mu \nu} R = 0$.

There is also a more general form of the tetradic action. It turns out we can add any multiple of a term called the ``Holst'' term to it, resulting in the action
\begin{equation}\label{SPH}
    S_{\rm GR \, (Palatini + Holst)}[\Gamma_{ab}, \theta^a] = - \frac{1}{\kappa^2} \int ( \vep_{abcd} R^{ab} \wedge \theta^c \wedge \theta^d  + \frac{2}{\gamma} R_{ab} \wedge \theta^a \wedge \theta^b).
\end{equation}
Here $\gamma$ is an arbitrary constant. The equations of motion have now become
\begin{equation}\label{B36}
\begin{aligned}
    \vep_{abcd} D \theta^c \wedge \theta^d + \frac{2}{\gamma} D \theta_a \wedge \theta_b &= 0 \, , \\
    \vep_{abcd} R^{ab} \wedge \theta^c + \frac{2}{\gamma} R_{cd} \wedge \theta^c &= 0 \, .
\end{aligned}
\end{equation}
For generic $\gamma$, these equations of motion are actually equivalent to those from the pure Palatini action. To see this, simply act $\vep^{ab}_{\;\;\;\;ef} - \frac{1}{\gamma} \delta^a_e \delta^b_f$ on the first equation. It turns out to be $-4(1 + \frac{1}{\gamma^2})(D \theta^a \wedge \theta^b - D \theta^b \wedge \theta^a ) = 0$ which again just gives the torsion-free condition. Once we have the torsion-free condition, $R_{cd} \wedge \theta^c$ vanishes due to the Bianchi identity of the Riemann tensor, so the second equation in \eqref{B36} is once again just the Einstein equation. Note however that if $\gamma = \pm i$ the above reasoning does not hold and we need to perform a separate analysis, a fact we'll return to momentarily.

Therefore, the ``Holst'' term is, morally speaking, somewhat like the gravitational analog to $\tr(F \wedge F)$. Because the action is quadratic in $\Gamma_{ab}$, the spin connection can be integrated out exactly, and so the classical equivalence of the actions with and without the Holst term implies the quantum equivalence as well.

Because the torsion-free condition implies $R_{ab}$ is the 2-form constructed out of the Riemann curvature of the metric, one can show that the torsion-free condition then implies
\begin{equation}
     \vep_{abcd} R^{ab} \wedge \theta^c \wedge \theta^d = 2 \, R \, \mathrm{dvol}, \hspace{1 cm} R_{ab} \wedge \theta^a \wedge \theta^b = - \frac{1}{2} \vep^{\mu \nu \rho \sigma} R_{\mu \nu \rho \sigma} \, \mathrm{dvol},
\end{equation}
where
\begin{equation}
    \mathrm{dvol} = \frac{1}{4!} \vep_{abcd} \, \theta^a \wedge \theta^b \wedge \theta^c \wedge \theta^d
\end{equation}
so the Holst term $R_{ab} \wedge \theta^a \wedge \theta^b$ vanishes when $\Gamma_{ab}$ is torsion-free by Bianchi, $\vep^{\mu \nu \rho \sigma} R_{\mu \nu \rho \sigma} = 0$.

Now let's introduce the spinor-index versions of these objects. The tetrad 1-forms are denoted
\begin{equation}
    \theta^{A \dot A} = \theta^{A \dot A}_\mu dx^\mu
\end{equation}
and we can use them to define the anti-self-dual and self-dual 2-forms
\begin{equation}
    \Sigma^{AB} \equiv \theta^{A \dot A} \wedge \theta^{B}_{\;\; \dot A}, \hspace{0.5 cm} \widetilde{\Sigma}^{\dot A \dot B} \equiv \theta^{A \dot A} \wedge \theta^{\;\; \dot B}_{A}.
\end{equation}
The full torsion-free equation with spinor-indices is
\begin{equation}
    \dd \theta^{A \dot A} + \Gamma^A_{\;\; B} \wedge \theta^{B \dot A} + \widetilde{\Gamma}^{\dot A}_{\;\; \dot B} \wedge \theta^{A \dot B} = 0 \,.
\end{equation}
We now split up the spin connection 1-form into anti-self-dual and self-dual parts
\begin{equation}
\begin{aligned}
    \Gamma_{A \dot A B \dot B} &= 2 \vep_{\dot A \dot B} \Gamma_{AB} + 2 \vep_{A B} \widetilde{\Gamma}_{\dot A \dot B}
\end{aligned}
\end{equation}
as well as the curvature 2-form 
\begin{equation}
    R_{A \dot A B \dot B} = 2 \vep_{\dot A \dot B} R_{AB} + 2 \vep_{A B} \widetilde{R}_{\dot A \dot B} ,
\end{equation}
where we have
\begin{equation}
\begin{aligned}
    R_{AB} &= \dd \Gamma_{AB} + \Gamma_{A}^{\;\; C} \wedge \Gamma_{BC} \, ,\\
    \widetilde{R}_{\dot A \dot B} &= \dd \widetilde{\Gamma}_{\dot A \dot B} + \widetilde{\Gamma}_{\dot A}^{\;\; \dot C} \wedge \widetilde{\Gamma}_{\dot B \dot C} \, . 
\end{aligned}
\end{equation}
Note that $\Gamma_{AB} = \Gamma_{(AB)}$, $\widetilde{\Gamma}_{\dot A \dot B} = \widetilde{\Gamma}_{(\dot A \dot B)}$, $R_{AB} = R_{(AB)}$, $\widetilde{R}_{\dot A \dot B} = \widetilde{R}_{(\dot A \dot B)}$ are symmetric in the spinor indices.

Using the spinor indices, the usual Palatini Lagrangian is equal to
\begin{equation}
\begin{aligned}
    \vep_{abcd} R^{ab} \wedge \theta^c \wedge \theta^d &= i  R_{AB} \wedge \Sigma^{AB} - i \widetilde{R}_{\dot A \dot B} \wedge \widetilde{\Sigma}^{\dot A \dot B} \, ,
\end{aligned}
\end{equation}
and the Holst term is
\begin{equation}
\begin{aligned}
    2  \, R_{ab} \wedge \theta^a \wedge \theta^b &=  R_{AB} \wedge \Sigma^{AB} +  \widetilde{R}_{\dot A \dot B} \wedge \widetilde{\Sigma}^{\dot A \dot B}\, .
\end{aligned}
\end{equation}
So if we set $\gamma = -i$ for the Holst term, we get a new chiral action for Einstein gravity, called the Plebański action, which only depends on the anti-self-dual component of the spin connection:
\begin{equation}\label{Sgrpleb}
\begin{aligned}
    S_{\text{GR (Plebański)}}[\theta^{A \dot A},  \Gamma_{AB}] &= - \frac{2 i}{\kappa^2} \int ( \dd \Gamma_{AB} + \Gamma_A^{\;\; C} \wedge \Gamma_{BC}) \wedge \theta^{A \dot A} \wedge \theta^{B}_{\;\; \dot A} \, .
\end{aligned}
\end{equation}
Something very peculiar has occurred here. The original Palatini action depended on both $\Gamma_{AB}$ and $\widetilde{\Gamma}_{\dot A \dot B}$, while the Plebański action only depends on $\Gamma_{AB}$. Nonetheless, we claim it is an action full Einstein GR.

If $\widetilde{\Gamma}_{\dot A \dot B}$ is unfixed by the action, how can this be? For instance, the $\delta \Gamma_{AB}$ equation of motion is
\begin{equation}\label{SigmaGamma}
    \dd \Sigma^{AB} + 2 \Gamma^{(A}_{\;\;\;\; C} \wedge \Sigma^{B)C} = 0.
\end{equation}
This is a torsion-free condition for $\Gamma_{AB}$, but not $\widetilde{\Gamma}_{\dot A \dot B}$, which seems completely unfixed.

The answer is that there is, of course, a reality condition. $\Gamma_{ab}$ has 6 real degrees of freedom which are repackaged in the 3 complex degrees of freedom of $\Gamma_{AB}$. These degrees of freedom are then conjugates of $\widetilde{\Gamma}_{\dot A \dot B}$, so if $\Gamma_{AB}$ is torsion-free then $\widetilde{\Gamma}_{\dot A \dot B}$ must be torsion-free as well. (In the self-dual theory, however, we will no longer have this reality condition.)

One could also make the argument that, because \eqref{SPH} is an action for Einstein gravity for generic $\gamma$, it would be strange if there were a discontinuity in the physical observables of the theory between the limit $\gamma \to -i$ and the strict equality $\gamma = -i$, given that none of the amplitudes depend on $\gamma$.

The SDG action is then the truncation of the Einstein gravity action where one removes the term quadratic in $\Gamma_{AB}$. It is
\begin{equation}\label{Ssdggammatheta}
\begin{aligned}
    \boxed{ S_{\rm SDG}[ \theta^{A \dot A}, \Gamma_{AB}] = - \frac{2 i}{\kappa^2} \int \dd \Gamma_{AB} \wedge \theta^{A \dot A} \wedge \theta^{B}_{\;\; \dot A} \, .}
\end{aligned}
\end{equation}
As desired, this action is manifestly covariant under global Lorentz transformations of the spinor indices.

How do we see that this action describes self-dual gravity? Well, $\Gamma_{AB}$ plays the role of a Lagrange multiplier that enforces
\begin{equation}\label{dSigma0}
    \dd \Sigma^{A B}[\theta] = 0
\end{equation}
where above we indicate the explicit dependence on $\theta$ to remind the reader $\Sigma^{A B}[\theta] = \theta^{A \dot A} \wedge \theta^B_{\;\; \dot A}$. If one then uses the torsion-free condition for the spin connection \eqref{SigmaGamma}, this implies
\begin{equation}\label{GammaAB0}
    \Gamma_{AB}[\theta] = 0
\end{equation}
where here the object $\Gamma_{AB}[\theta]$ is defined to satisfy the torsion-free equation with respect to $\theta$. This then implies
\begin{equation}
    R_{AB}[\theta] = 0
\end{equation}
and this is the condition that the spacetime constructed out of $\theta$ is self-dual.

However, we mention that the geometrically-defined $\Gamma_{AB}[\theta]$ appearing in \eqref{GammaAB0} is different from the $\Gamma_{AB}$ appearing in the self-dual action \eqref{Ssdggammatheta}. This is because the torsion-free condition for $\Gamma_{AB}[\theta]$ with respect to $\theta^{A \dot A}$ \eqref{SigmaGamma} is not actually an equation of motion for $\Gamma_{AB}$ of the self-dual action \eqref{Ssdggammatheta}. The situation is exactly analogous to what we encountered with the $B_{AB}$ field in the Yang-Mills action. After truncation to the self-dual action, this field no longer satisfied $B_{AB} = F_{AB}$ and instead acted as a Lagrange multiplier enforcing the self-duality of $A_{A \dot A}$. Likewise for gravity, after the truncation, $\Gamma_{AB}$ no longer satisfies the torsion-free condition with respect to $\theta^{A \dot A}$ but now acts as a Lagrange multiplier enforcing self-duality on the spacetime defined by the tetrads $\theta^{A \dot A}$. But afterwards, the variable $\Gamma_{AB}$ on-shell also acts as an anti-self-dual degree of freedom propagating linearly on this self-dual $\theta^{A \dot A}$ background, just as $B_{AB}$ does on $A_{A \dot A}$.

We mention that there's another way to see that the equation $\dd \Sigma^{AB} = 0$, in conjunction with the definition $\Sigma^{AB} = \theta^{A \dot A} \wedge \theta^{B}_{\;\; \dot A}$, implies that the spacetime defined by $\theta^{A \dot A}$ must be self-dual. If we define the components of the anti-self-dual-2-forms by $\Sigma^{AB} = \frac{1}{2} \Sigma^{AB}_{\mu \nu} dx^\mu \wedge dx^\nu$, and define the (Lorentzian) almost-complex-structures $I$, $J$, $K$, by
\begin{equation}
    I^{\mu}_{\;\; \nu} = \frac{1}{4} g^{\mu \rho} (\Sigma^{11} + \Sigma^{22})_{\rho \nu}, \hspace{0.5 cm} J^{\mu}_{\;\; \nu} = \frac{i}{2} g^{\mu \rho} (\Sigma^{12})_{\rho \nu}, \hspace{0.5 cm} K^{\mu}_{\;\; \nu} = \frac{i}{4} g^{\mu \rho} (\Sigma^{11} - \Sigma^{22})_{\rho \nu},
\end{equation}
then from the tetrad definition of $\Sigma^{AB}$ we have the quaternionic relations $I^2 = J^2 = K^2 = IJK = -1$. Of course, these almost-complex-structures, which are just raised versions of the 2-forms $\Sigma^{AB}_{\mu \nu}$, are actual complex-structures because the $\Sigma^{AB}$ are closed \cite{Nakahara:2003nw}. The 2-forms themselves are the associated symplectic forms and the manifold is Kähler. The quaternionic relations between $I$, $J$, $K$, further imply that the spacetime is Hyper-Kähler and therefore self-dual. We reviewed this in more detail appendix B of \cite{mypaper}.

\subsection{SDG gauge fixing}\label{subb4}

Let us now explain how to gauge fix the SDG action \eqref{Ssdggammatheta} to lightcone gauge. I am not aware of a place where this exact computation has been worked out in the literature, and it is slightly subtle.

The Plebański action for GR \eqref{Sgrpleb} has two gauge symmetries: diffeomorphisms and local Lorentz transformations. Infinitesimally, the diffeomorphism symmetry is
\begin{equation}
\begin{aligned}
    \delta_{\xi} \theta^{A \dot A}_\mu &= -(\xi^\nu \partial_\nu) \theta^{A \dot A}_\mu - (\partial_\mu \xi^\nu) \theta^{A \dot A}_\nu \, , \\
    \delta_{\xi} \Gamma_{AB \mu} &= -(\xi^\nu \partial_\nu ) \Gamma_{AB \mu} - (\partial_\mu \xi^\nu) \Gamma_{AB \nu} \, .
\end{aligned}
\end{equation}
Likewise, the infinitesimal local Lorentz symmetry is
\begin{equation}
\begin{aligned}
    \delta_{\omega} \theta^{A \dot A}  &= \omega^A_{\;\; B} \theta^{B \dot A} + \widetilde{\omega}^{\dot A}_{\;\; \dot B} \theta^{A \dot B} \, , \\
    \delta_{\omega} \Gamma_{AB} &= \omega_{A}^{\;\; C}  \Gamma_{CB} +  \omega_{B}^{\;\; C} \Gamma_{AC}  + \dd \omega_{AB}\, ,
\end{aligned}
\end{equation}
where $\omega_{AB} = \omega_{(AB)}$ and $\widetilde{\omega}_{\dot A \dot B} = \widetilde{\omega}_{(\dot A \dot B)}$ are the anti-self-dual and self-dual components of the local Lorentz transformations.

The corresponding finite local Lorentz transformations act as
\begin{equation}
\begin{aligned}
    \theta^{A \dot A}  & \, \mapsto \, \Lambda^A_{\;\; B} \theta^{B \dot A} + \widetilde{\Lambda}^{\dot A}_{\;\; \dot B} \theta^{A \dot B} \, , \\
    \Gamma_{AB} & \, \mapsto \, \Lambda_{A}^{\;\; C} \Lambda_{B}^{\;\; D} \Gamma_{CD} + \Lambda_{\;\; A}^{C} \dd \Lambda_{CB}\, .
\end{aligned}
\end{equation}
The matrices $\Lambda^A_{\;\; B}$ and $\widetilde{\Lambda}^{\dot A}_{\;\; \dot B}$ have determinant 1, which is equivalent to the inverse relation under raising and lowering $\Lambda^A_{\;\; B} \Lambda_A^{\;\; C} = \delta^C_D$ and $\widetilde{\Lambda}^{\dot A}_{\;\; \dot B} \widetilde{\Lambda}_{\dot A}^{\;\; \dot C} = \delta^{\dot C}_{\dot D}$. 

Note that the ASD spin-connection $\Gamma_{AB}$ does not transform ``covariantly'' under a local Lorentz transformation due to the inhomogeneous term. However, its curvature 2-form $R_{AB} \mapsto \Lambda_A^{\;\; C} \Lambda_B^{\;\; D} R_{CD}$ does transformation covariantly, which implies the invariance of the GR Plebański action \eqref{Sgrpleb}.

An important point, however, is that the SDG action \eqref{Ssdggammatheta} is \textit{not} invariant under the local Lorentz transformations where $\Lambda^A_{\;\; B}$ varies in spacetime. It does remain invariant under transformations where $\widetilde{\Lambda}^{\dot A}_{\;\; \dot B}$ varies, or under a global Lorentz transformation where $\Lambda^A_{\;\; B}$ is constant in spacetime. So we say that the SDG action has a $1\!/2$-local-Lorentz gauge symmetry. This is because the action is not constructed out of the covariant object $R_{AB}$, but only out of the non-covariant object $\dd \Gamma_{AB}$. The non-covariance of the action can also be seen, for instance, in its equation of motion $\dd \Sigma^{AB} = 0$, which is not invariant under local Lorentz transformations with varying $\Lambda^A_{\;\; B}$.\footnote{The SDG equation $\dd \Sigma^{AB} = 0$ only holds in frames where the tetrads are chosen such that $\Gamma_{AB} = 0$. Such a frame can always be found on self-dual spacetimes, but for a more general frame one could have $\Gamma_{AB} = \Lambda_{\;\; A}^{C} \dd \Lambda_{CB}$ for some $\Lambda_{AB}$.}

Now we discuss how to fix the diffeomorphism+$1\!/2$-local-Lorentz gauge symmetry of the SDG action to lightcone gauge. This procedure is a bit subtle. Let us start by defining the ``flat space'' tetrads $\theta^{A \dot A}_0$ via
\begin{equation}\label{theta0coords}
    \theta^{A \dot A}_0 = \begin{pmatrix}
        \theta^{1 \dot 1}_0 & \theta^{1 \dot 2}_0 \\ \theta^{2 \dot 1}_0 & \theta^{2 \dot 2}_0
    \end{pmatrix} = \begin{pmatrix} 
       2 d \bu \,  & 2d \bw \,  \\ 2dw \,  & 2du \,
    \end{pmatrix}.
\end{equation}
The gauge condition we would probably like to set is $\alpha_A \talpha_{\dot A} \theta^{A \dot A} = \alpha_A \talpha_{\dot A} \theta^{A \dot A}_0$. While this would be easy if we had the full diffeomorphism+local-Lorentz symmetry, it seems we cannot fix this gauge using the diffeomorphism+$1\!/2$ local-Lorentz gauge symmetry we have. So we will take a different approach.

Let us introduce the gauge-fixing constraints
\begin{equation}
    G_{\mu \nu} \equiv \theta^{1 \dot 1}_{[\mu} \theta^{1 \dot 2}_{\nu]} - 4\delta^{\bu}_{[\mu} \delta^{\bw}_{\nu]} \, , \hspace{1 cm} G^{\dot 1}_{\mu} \equiv \theta^{1 \dot 1}_{\mu} - 2 \delta^{\bu}_\mu \, , \hspace{1 cm} G^{\dot 2} _{\mu} \equiv \theta^{1 \dot 2}_{\mu} - 2 \delta^{\bw}_\mu \, ,
\end{equation}
where $G_{\mu \nu} = G_{[\mu \nu]}$. The first constraint vanishes when $\theta^{1 \dot 1} \wedge \theta^{1 \dot 2} = 4 d \bu \wedge d \bw$, and the last two constraints vanish when $\theta^{1 \dot 1} = 2 d \bu$, $\theta^{1 \dot 2} = 2 d \bw$. So any tetrads that make the latter two constraints vanish will also make the first constraint vanish, but we have a reason for breaking them up in this way.

Let us now introduce Lagrange multipliers $B^{\mu \nu} = B^{[\mu \nu]}$ and $B_{\dot A}^\mu$ in order to enforce these constraints. There will also be the usual $b$ $c$ fermions but we won't write them out explicitly. The gauge fixed SDG action is
\begin{equation}
\begin{aligned}
    &S_{\rm SDG}^{\text{(gauge fixed)}}[\theta^{A \dot A},\Gamma_{AB}, B^{\mu \nu}, B_{\dot A}^\mu, \text{fermions}] \\
    &= -\frac{2 i}{\kappa^2} \int \left(  \dd \Gamma_{AB} \wedge \theta^{A \dot A} \wedge \theta^{B}_{\;\; \dot A}  + B^{\mu \nu} G_{\mu \nu} + B^{\mu}_{\dot A} G^{\dot A}_\mu \right) + S_{\text{fermions}}.
\end{aligned}
\end{equation}
Note that it is straightforward to verify that the ghost fermions decouple from the rest of the action. This requires checking that the variations of our constraints under the diffeomorphims and $1\!/2$-local Lorentz symmetries
\begin{equation}
    \delta_\xi G_{\mu \nu}, \hspace{0.5 cm} \delta_\xi G_\mu^{\dot A}, \hspace{0.5cm} \delta_\omega G_{\mu \nu}, \hspace{0.5 cm} \delta_\omega G^{\dot A}_\mu \, ,
\end{equation}
decouple from the other fields on the support of the constraints, which is easy to do.

Now let us use the basis of spinors $\alpha_A$, $\beta_A$ from \eqref{alphabeta} to decompose the spin-connection 1-form $\Gamma_{AB}$ into a triple of 1-forms we call $(\Gamma, \Gamma', \Gamma'')$ via
\begin{equation}
    \Gamma_{AB} = \Gamma \alpha_A \alpha_B + \Gamma'(\alpha_A \beta_B + \beta_A \alpha_B) + \Gamma'' \beta_A \beta_B .
\end{equation}
Let's start by integrating over the component $\Gamma \alpha_A \alpha_B $. The variation of the action under $\delta \Gamma$ is
\begin{equation}
    0 = - \frac{4 i}{\kappa^2} \int \delta \Gamma  \wedge \dd( \theta^{1 \dot 1} \wedge \theta^{1 \dot 2} ),
\end{equation}
so $\Gamma$ is a Lagrange multiplier that sets $\mathrm{d} (  \theta^{1 \dot 1} \wedge \theta^{1 \dot 2} ) = 0$. Now we integrate out the Lagrange multiplier $B^{\mu \nu}$. This fixes the diffeomorphism gauge symmetry, setting $\theta^{1 \dot 1} \wedge \theta^{1 \dot 2} = 4 d \bu \wedge d \bw$. Crucially, we know this can always be done via Darboux's theorem, because $\theta^{1 \dot 1} \wedge \theta^{1 \dot 2}$ is a closed rank-two 2-form. In particular, the diffeomorphism degrees of feedom were used to set
\begin{equation}
    \theta^{1 \dot 1}_{u} = 0, \hspace{0.5 cm} \theta^{1 \dot 1}_{w} = 0, \hspace{0.5 cm} \theta^{1 \dot 2}_{u} = 0, \hspace{0.5 cm} \theta^{1 \dot 2}_{w} = 0,
\end{equation}
and we now have
\begin{equation}\label{B64}
\begin{aligned}
    \theta^{1 \dot 1} &= 2 ( f_{11} \, d \bu + f_{12} \,  d \bw) \, , \\
    \theta^{1 \dot 2} &= 2 ( f_{21} \, d \bu + f_{22} \, d \bw ) \, ,
\end{aligned}
\end{equation}
for some functions $f_{11}$, $f_{12}$, $f_{21}$, $f_{22}$ satisfying
\begin{equation}
    f_{11} f_{22} - f_{12} f_{21} = 1.
\end{equation}
(Not however that we still have the gauge freedom to modify the $u,w$ coordinates, which can be changed without affecting \eqref{B64}. This corresponds to the irrelevant choice of antiholomorphic reference spinor $|\talpha]$.)

Now that we have integrated out $\Gamma$ and $B^{\mu \nu}$ and fixed our diffeomorphism symmetry, we turn to our $1\!/2$-local Lorentz symmetry, which acts on the tetrads as $\theta^{A \dot A} \to \widetilde{\Lambda}^{\dot A}_{\;\; \dot B} \theta^{A \dot B}$ for $\widetilde{\Lambda}^{\dot A}_{\;\; \dot B}$ determinant 1.

Integrating out $B_\mu^{\dot A}$ completely fixes this remaining symmetry by setting $\widetilde{\Lambda} = ( \begin{smallmatrix} f_{11} & f_{12} \\ f_{21} & f_{22} \end{smallmatrix})^{-1}$. We are now left with
\begin{equation}\label{ourthetaconstr}
\begin{aligned}
    \theta^{1 \dot 1} &= 2 d \bu \, , \\
    \theta^{1 \dot 1} &= 2 d \bw \, ,
\end{aligned}
\end{equation}
as desired. We have now gauge fixed the action. We can now express our tetrad as a perturbation on top of the flat space tetrads \eqref{theta0coords} via
\begin{equation}
    \theta^{A \dot A} = \theta_0^{A \dot A} + \kappa \, \alpha^A \chi^{\dot A}
\end{equation}
for some 1-form $\chi^{\dot A}$. Writing $\chi^{A \dot A} = \alpha^A \chi^{\dot A}$, the SDG action is now
\begin{equation}
    S_{\rm SDG}[\chi^{\dot A}, \Gamma', \Gamma''] = - \frac{2 i}{\kappa^2} \int \dd( \Gamma' (\alpha_A \beta_B + \beta_A \alpha_B) + \Gamma'' \beta_A \beta_B) \wedge \left( (\theta_0\!+\!\kappa \chi)^{A \dot A} \wedge (\theta_0\!+\!\kappa \chi)^{B}_{\;\; \dot A} \right).
\end{equation}

Using the basis of spinors $\alpha_A$ and $\beta_A$ from \eqref{alphabeta}, let us also define the 1-forms
\begin{equation}
    (\alpha \theta_0)^{\dot A} \equiv \alpha_A \theta_0^{A \dot A}, \hspace{1 cm} (\beta \theta_0)^{\dot A} = \beta_A \theta_0^{A \dot A}.
\end{equation}
Integrating out $\Gamma'$ then implies
\begin{equation}
    \dd ( (\alpha \theta_0)^{\dot A} \wedge \chi_{\dot A} ) = 0
\end{equation}
which, assuming a topologically trivial manifold and vanishing fall-offs for $\chi$ in the path integral, implies
\begin{equation}\label{onethetaconstr}
    (\alpha \theta_0)^{\dot A} \wedge \chi_{\dot A} = 0.
\end{equation}
Because $\{ (\alpha \theta_0)^{\dot A}, (\beta \theta_0)^{\dot A} \}_{\dot A = \dot 1, \dot 2}$ are a complete basis of 1-forms, we can always express $\chi_{\dot A}$ as
\begin{equation}\label{chigeneral}
    \kappa \, \chi_{\dot A} = X_{\dot A \dot B} (\alpha \theta_0)^{\dot B} + Y_{\dot A \dot B} (\beta \theta_0)^{\dot B} \vphantom{\Bigg\rvert}
\end{equation}
for some coefficients $X_{\dot A \dot B}$, $Y_{\dot A \dot B}$. The constraint \eqref{onethetaconstr} then implies that
\begin{equation}\label{Xsym}
    X_{\dot A \dot B} = X_{(\dot A \dot B)}, \hspace{0.5 cm} Y_{\dot A \dot B} = 0.
\end{equation}

Now that we have integrated out the $\Gamma$ and $\Gamma'$ components, we are just left with $\Gamma''$, and after an integration by parts the action is
\begin{equation}\label{sdggammappchi0}
\begin{aligned}
    S_{\rm SDG}[\chi^{\dot A}, \Gamma''] = \frac{4 i}{\kappa} \int  \Gamma'' \wedge  ( (\beta \theta_0)^{\dot A} + \kappa \, \chi^{\dot A} ) \wedge \dd \chi_{\dot A} \, .
\end{aligned}
\end{equation}
Let us now expand out $2 \kappa \, ( (\beta \theta_0)^{\dot A} + \kappa \,\chi^{\dot A} ) \wedge \dd \chi_{\dot A}$. To do this, we use the decomposition of the identity \eqref{deltaalphabeta} to write the exterior derivative as
\begin{equation}
\begin{aligned}
    \dd &= \frac{1}{2} \theta_0^{C \dot C} \partial_{C \dot C} \\
    &= \frac{1}{2} (\alpha \theta_0)^{\dot C}  (\beta \partial)_{\dot C} - \frac{1}{2} (\beta \theta_0)^{\dot C}  (\alpha  \partial)_{\dot C}
\end{aligned}
\end{equation}
and using this we compute
\begin{equation}
\begin{aligned}
    2 \kappa \, \dd \chi_{\dot A} &\,=\, - (\alpha \theta_0)^{\dot B} \wedge (\alpha \theta_0)^{\dot C} (\beta \partial)_{\dot C} X_{\dot A \dot B} + (\alpha \theta_0)^{\dot B} \wedge (\beta \theta_0)^{\dot C}  (\alpha  \partial)_{\dot C} X_{\dot A \dot B}.
\end{aligned}
\end{equation}
Wedging this with $(\beta \theta_0)^{\dot A} + \kappa \, \chi^{\dot A}$ gives us
\begin{equation}
\begin{aligned}
    2 \kappa ( (\beta \theta_0)^{\dot A} + \kappa \, \chi^{\dot A}) \wedge \dd \chi_{\dot A} = & -  (\beta \theta_0)^{\dot A} \wedge (\alpha \theta_0)^{\dot B} \wedge (\alpha \theta_0)^{\dot C} (\beta \partial)_{\dot C} X_{\dot A \dot B} \\
    & +  (\beta \theta_0)^{\dot A} \wedge (\alpha \theta_0)^{\dot B} \wedge (\beta \theta_0)^{\dot C}  (\alpha  \partial)_{\dot C} X_{\dot A \dot B} \\
    &- \mathstrike{(\alpha \theta_0)^{\dot D} \wedge  (\alpha \theta_0)^{\dot B} \wedge (\alpha \theta_0)^{\dot C} } ((\beta \partial)_{\dot C} X_{\dot A \dot B} ) X^{\dot A}_{\;\; \dot D} \\
    &+ (\alpha \theta_0)^{\dot D} \wedge  (\alpha \theta_0)^{\dot B} \wedge (\beta \theta_0)^{\dot C}  ((\alpha  \partial)_{\dot C} X_{\dot A \dot B}) X^{\dot A}_{\;\; \dot D}.
\end{aligned}
\end{equation}
Note that one of the terms is zero due to the relation
\begin{equation}
    (\alpha \theta_0)^{\dot A} \wedge (\alpha \theta_0)^{\dot B} \wedge (\alpha \theta_0)^{\dot C} = 0
\end{equation}
which follows simply because the dotted indices are two dimensional. Furthermore, we note the flat-space volume form is
\begin{equation}
    \mathrm{dvol} = \frac{i}{4} \theta_0^{1 \dot 1} \wedge \theta_0^{1 \dot 2} \wedge \theta_0^{2 \dot 1} \wedge \theta_0^{2 \dot 2}  
\end{equation}
and due to the index structure we also have
\begin{equation}
    (\alpha \theta_0)^{\dot A} \wedge (\alpha \theta_0)^{\dot B} \wedge (\beta \theta_0)^{\dot C} \wedge (\beta \theta_0)^{\dot D} = -4 i \, \vep^{\dot A \dot B} \vep^{\dot C \dot D} \mathrm{dvol} \vphantom{\Bigg\rvert}.
\end{equation}
Using these relations, we can calculate
\begin{align}\label{betachiexpanded}
    (\beta \theta_0)^{\dot E} \wedge 2\kappa ( (\beta \theta_0)^{\dot A} + \kappa \, \chi^{\dot A}) \wedge \dd \chi^{\dot A} &= -4 i  \,  ( (\beta \partial)_{\dot C} X^{\dot C \dot E} - ( (\alpha \partial)^{\dot E} X_{\dot A \dot B} ) X^{\dot A \dot B}  ) \, \mathrm{dvol} \vphantom{\Bigg\rvert}
\shortintertext{and}
    (\alpha \theta_0)^{\dot E} \wedge 2 \kappa ( (\beta \theta_0)^{\dot A} + \kappa \, \chi^{\dot A}) \wedge \dd \chi^{\dot A} &= -4 i   \, (\alpha \partial)_{\dot C} X^{\dot C \dot E} \, \mathrm{dvol} \, . \vphantom{\Bigg\rvert}
\end{align}
We now use the above two equations to rewrite \eqref{sdggammappchi0}. To do this, we expand out the spacetime coefficients of the 1-form $\Gamma'' = \Gamma''_{E \dot E} \theta_0^{E \dot E}$ further into components via
\begin{equation}
    \Gamma'' = (\alpha \Gamma'')_{\dot E} (\beta \theta_0)^{\dot E} -(\beta \Gamma'')_{\dot E} (\alpha \theta_0)^{\dot E}. \vphantom{\Bigg\rvert}
\end{equation}
Plugging this in, we find
\begin{equation}\label{sdggammappchi}
\begin{aligned}
    S_{\rm SDG}[\Gamma'', \chi^{\dot A}] = 
    \frac{8}{\kappa^2} \int & \mathrm{dvol} \Big(  (\alpha \Gamma'')_{\dot E} \Big[ (\beta \partial)_{\dot C} X^{\dot C \dot E} -  \frac{1}{2}(\alpha \partial)^{\dot E} (X_{\dot A \dot B}  X^{\dot A \dot B} ) \Big]  \\
    &\;\;\;\; - (\beta \Gamma'')_{\dot E} \Big[ (\alpha \partial)_{\dot C} X^{\dot C \dot E} \Big] \Big)  \, .
\end{aligned}
\end{equation}
If we integrate out $(\beta \Gamma'')_{\dot E}$, this enforces
\begin{equation}
    (\alpha \partial)_{\dot A} X^{\dot A \dot B} = 0. \vphantom{\Bigg\rvert}
\end{equation}
In conjunction with the symmetry condition  $X_{\dot A \dot B} = X_{(\dot A \dot B)}$ \eqref{Xsym}, this implies
\begin{equation}\label{XABphi}
    X_{\dot A \dot B} = \frac{\kappa}{2} (\alpha \partial)_{\dot A} (\alpha \partial)_{\dot B} \phi \vphantom{\Bigg\rvert}
\end{equation}
for some potential $\phi$. If we also define the $\bphi$ field as
\begin{equation}\label{bphiGamma}
    \bphi = \frac{4}{\kappa} (\alpha \partial)^{\dot E} (\alpha \Gamma'')_{\dot E} \, , \vphantom{\Bigg\rvert}
\end{equation}
then with these substitutions, we now have
\begin{equation}\label{sdgactionalpha}
\begin{aligned}
    S_{\rm SDG}[\phi, \bphi] =  
    -\int & \mathrm{dvol} \;   
    \bphi \Big( \Box \, \phi -   \frac{\kappa}{4}  ( (\alpha \partial)_{\dot A} (\alpha \partial)_{\dot B} \phi )( (\alpha \partial)^{\dot A} (\alpha \partial)^{\dot B} \phi ) ) \Big) \vphantom{\Bigg\rvert}
\end{aligned}
\end{equation}
which, as desired, is the lightcone action for SDG.

In terms of lightcone coordinates, if we unwrap the tetrads as a function of $\phi$, we find they are
\begin{equation}
\begin{aligned}
    \theta^{1 \dot 1} &= 2 d \bu  \\
    \theta^{1 \dot 2} &= 2 d \bw  \\
    \theta^{2 \dot 1} &= 2 d w - \kappa (\partial_u \partial_w \phi ) d \bu - \kappa (\partial_u^2 \phi) d \bw  \\
    \theta^{2 \dot 2} &= 2 d u + \kappa  ( \partial_w^2 \phi) d \bu + \kappa (\partial_u \partial_w \phi) d \bw .
\end{aligned}
\end{equation}
The tetrads can be used to calculate the self-dual metric via $g_{\mu \nu} = \frac{1}{2} \theta^{A \dot A}_\mu \theta_{A \dot A \nu}$. Doing so reproduces \eqref{eq210}.

\subsection{Another argument for the finiteness of the self-dual theories}

In this subsection we sketch a slightly hand-wavy argument that the amplitudes of SDYM and SDG are finite, using the form of the covariant actions. Recall that the actions are
\begin{align}
    S_{\rm SDYM}[A_{A \dot A}, B_{AB}] &= \frac{1}{2} \int d^4 x \tr  B^{AB} (\partial_{A\dot A}A_{B}^{\;\; \dot A} + \partial_{B \dot A} A_{A}^{\;\; \dot A} - i g [A_{A \dot A}, A_{B}^{\;\; \dot A}]) \, \\
    \shortintertext{and}
    S_{\rm SDG}[\chi^{A \dot A}, \Gamma_{AB}] &= \frac{4 i}{\kappa} \int  \Gamma_{AB} \wedge (\theta_0^{A \dot A} + \kappa \, \chi^{A \dot A}) \wedge \dd \chi^{B}_{\;\; \dot A} \, ,
\end{align}
where for the gravitational action we expanded the tetrad around flat-space via $\theta^{A \dot A} = \theta_0^{A \dot A} + \kappa \, \chi^{A \dot A}$.

If we rescale the fields as
\begin{equation}
\begin{aligned}
    B_{AB} &\to g \, B_{AB} \\
    A_{A \dot A} &\to g^{-1} A_{A \dot A}
\end{aligned}
\end{equation}
and
\begin{equation}
\begin{aligned}
    \Gamma_{AB} &\mapsto \kappa^2 \, \Gamma_{AB} \\
    \chi^{A \dot A} &\mapsto \kappa^{-1} \chi^{A \dot A} 
\end{aligned}
\end{equation}
then the couplings $g$ and $\kappa$ drop out of the actions. As the physical observables of the theory can't depend on any dimensionful factors, they ought to be finite.

Of course, at the quantum level, when we rescale fields in a path integral we should be concerned about generating an additional term in the action coming from anomalies in the measure. However, various consistency requirements imply that there are only a short list of terms that this anomaly can be. For SDYM, it turns out the anomaly is proportional to the topological term $\tr(F \wedge F)$ and for SDG it will also be proportional to a sum of topological terms. Because topological terms do not affect scattering amplitudes, the amplitudes remain unaffected by these anomalies. The topological terms do nonetheless run under the renormalization group, as discussed in for instance \cite{Bittleston:2025jmk,Losev:2017qrj,Domurcukgul:2025xgf}. Other references on beta-functions and one-loop-determinants on non-trivial backgrounds are \cite{Krasnov:2016emc,Krasnov:2015kva}.

\subsection{Dilatation invariance of SDG}\label{sec:dilatation}

In section \ref{sec:conformalsym} we mentioned that SDG is an example of a non-unitary theory which has perturbative dilatation invariance but not full $SO(4,2)$ conformal invariance. This should be compared with quantum SDYM (and tree-level YM) whose perturbative amplitudes have both dilatation and full $SO(4,2)$ invariance.

In this subsection we will explain the exact sense in which quantum SDG is dilatation invariant, as it is a bit subtle.

First let's remind ourselves how dilatations work in Yang-Mills theory. Dilatations are the spacetime map
\begin{equation}
    x \; \mapsto \; t x 
\end{equation}
for some constant parameter $t$. The 1-form field $A_\mu(x) dx^\mu$ is then naturally sent to $A_\mu(tx) d(tx^\mu)$ under a dilatation, which we can also express as 
\begin{equation}
    A_\mu(x)  \; \mapsto \; t \, A_\mu(tx).
\end{equation}
One can then explicitly verify that the Yang-Mills action $S_{\rm YM}[A] = - \frac{1}{4} \int d^4 x \tr F^{\mu \nu} F_{\mu \nu}$ is invariant under dilatations,
\begin{equation}
    S_{\rm YM}[t A_\mu(tx)] = S_{\rm YM}[ A_\mu(x)] 
\end{equation}
which is of course a special case of the larger conformal invariance of classical Yang-Mills.

At the level of spinor helicity variables, a spacetime particle wavefunction is parameterized by $\lambda$, $\tlambda$ via
\begin{equation}
    A_{A \dot A}(x, \lambda, \tlambda) = \vep_{A \dot A} e^{\frac{i}{2} \lambda_{B} \tlambda_{\dot B} x^{B \dot B}}, \hspace{0.75 cm}
    \vep^+_{A \dot A} = -i \sqrt{2} \frac{\alpha_{ A} \tlambda_{\dot A} }{\lr{\alpha \lambda}}, \hspace{ 0.75 cm} \vep^-_{A \dot A} = -i \sqrt{2} \frac{\lambda_{A} \talpha_{\dot A}}{[\tlambda \talpha]}.
\end{equation} 
Therefore, we can see that the dilatation map can be rephrased as an action on the spinor variables via 
\begin{equation}
    A_{A \dot A}(x,\lambda, \tlambda)  \; \mapsto \; t \, A_{A \dot A}(t x, \lambda, \tlambda) = t \, A_{A \dot A}(x, t^{1/2} \lambda, t^{1/2} \tlambda).
\end{equation}
Therefore, if we have a scattering amplitude of $n$ particles and differentiate this symmetry with respect to $t$ and then set $t = 1$, the classical invariance of the action with respect to dilatation implies that all of the tree-level YM amplitudes as a function of $\{\lambda_i, \tlambda_i\}_{i=1,\ldots,n}$, with the momentum conserving delta function $\delta^4(\sum_{i=1}^n \lambda_i \tlambda_i)$ included in the amplitude, will be annihilated by the dilatation operator \cite{Witten:2003nn}
\begin{equation}\label{Ddef}
    D = \frac{1}{2} \sum_{i = 1}^n \left( \lambda^{A}_i \frac{\partial}{\partial \lambda^A_i} + \tlambda^{\dot A}_i \frac{\partial}{\partial \tlambda^{\dot A}_i}  + 2 \right).
\end{equation}
This can be checked easily for any tree-level YM amplitude, such as the Parke-Taylor MHV amplitudes, by simply counting the mass-dimension of the amplitude as a function of the spinor variables where $\lambda$ and $\tlambda$ have dimension $1/2$. In addition, the YM one-loop-all-plus amplitudes \eqref{YM1LAPformula} are also annihilated by the $D$ above \cite{Henn:2019mvc}, because those amplitudes are finite and involve no dimensionful cut-offs to break the dilatation invariance of $S_{\rm YM}[A_\mu]$.

Now what about gravity? One can check that the modified dilatation operator
\begin{equation}\label{Dprimedef}
    D' \equiv  \frac{1}{2} \sum_{i = 1}^n \left( \lambda^{A}_i \frac{\partial}{\partial \lambda^A_i} + \tlambda^{\dot A}_i \frac{\partial}{\partial \tlambda^{\dot A}_i}  \right)
\end{equation}
annihilates the one-loop-all-plus gravity amplitudes, simply by power-counting the mass-dimensions from $\lambda$, $\tlambda$ in the expression \eqref{bdpr} multiplied by the delta function. (Note that the coupling constant $\kappa$ is not affected by the action of $D'$, i.e. $D' \kappa = 0$, even though $\kappa$ is in principle dimensionful. We will return to this point momentarily.) So let's now attempt to explain why the one-loop-all-plus gravity amplitudes have this modified dilatation symmetry.

As mentioned in section \ref{sec:conformalsym} the equations of motion $R_{\mu \nu} = 0$ and $R_{\mu \nu} = \frac{i}{2} \vep_{\mu \nu \alpha \beta} R^{\alpha \beta}$ are dilatation invariant. But to get invariance of the amplitudes, we need for the action itself to be invariant, not just the equations of motion. So we now ask, is the gravitational action invariant under dilatation symmetry? 

We must mention that there are in fact two types of transformations in gravity which need to be distinguished: \textit{coordinate dilatations} and \textit{scaling}. They are defined by
\begin{equation}\label{coordscalinggmunu}
\begin{aligned}
    \text{coordinate dilatation:} \hspace{0.5 cm} & g_{\mu \nu}(x) \mapsto t^2 g_{\mu \nu}(t x)\, , \\
    \text{scaling:} \hspace{0.5 cm} & g_{\mu \nu}(x) \mapsto t^2 g_{\mu \nu}(x) \, .
\end{aligned}
\end{equation}
It is important to consider both types of transformations because a coordinate dilatation alone will also rescale the \textit{background metric} $\eta_{\mu \nu}$, which we don't want to do. However, if we do a coordinate dilatation and then a compensating inverse scaling, we will get a transformation which only affects the metric perturbation $h_{\mu \nu}(x)$ as desired:
\begin{equation}
\begin{aligned}
    g_{\mu \nu}(x) &\overset{\substack{\text{coordinate} \\ \text{dilatation} \\ \textcolor{white}{.}}}{\mapsto} t^2 \eta_{\mu \nu} + t^2 \kappa h_{\mu \nu}(tx) \overset{\substack{\text{scaling}^{-1}\\ \textcolor{white}{.}}}{\mapsto} \eta_{\mu \nu} +  \kappa h_{\mu \nu}(tx) \;=\; g_{\mu \nu}(tx).
\end{aligned}
\end{equation}
This transformation, which is the composition of the coordinate dilatation and an inverse scaling, is exactly the kind which corresponds to $D'$ in \eqref{Dprimedef}! This can be read off from the position space particle wave function
\begin{equation}
    h_{A \dot A B \dot B}(x,\lambda,\tlambda) = \vep_{A \dot A} \vep_{B \dot B} \; e^{\frac{i}{2} \lambda_C \tlambda_{\dot C} x^{C \dot C}}
\end{equation}
for which
\begin{equation}
    h_{A \dot A B \dot B}(x,\lambda, \tlambda) \mapsto h_{A \dot A B \dot B}(tx,\lambda, \tlambda) = h_{A \dot A B \dot B}(x, t^{1/2} \lambda, t^{1/2} \tlambda).
\end{equation}
So we now want to understand why the one-loop-all-plus amplitudes are invariant under $D'$ on the positive-helicity gravitons.

First off, is the Einstein Hilbert action $
S_{\rm GR}[g_{\mu \nu}(x)] = -\frac{2}{ \kappa^2} \int d^4 x \sqrt{-g} R$ invariant under such a transformation? Well, under coordinate dilatation and scaling individually, the action transforms as
\begin{equation}
\begin{aligned}
    \text{coordinate dilatation: }\; & S_{\rm GR}[t^2 g_{\mu \nu}(tx)] = S_{\rm GR}[g_{\mu \nu}(x)], \\
    \text{scaling: }\; & S_{\rm GR}[ t^2 g_{\mu \nu}(x)] = t^2 S_{\rm GR}[g_{\mu \nu}(x)] .
\end{aligned}
\end{equation}
Performing a dilatation and then a compensating inverse scaling, the above two transformation laws imply
\begin{equation}
    S_{\rm GR}[g_{\mu \nu}(tx)] = t^{-2} S_{\rm GR}[g_{\mu \nu}(x)]
\end{equation}
which is evidently \textit{not} invariant. This means for example that tree-level gravity amplitudes won't be invariant under $D'$. While it is actually possible to cook up a modified notion of invariance if one also rescales $\kappa$, this isn't the approach we want to take. So it seems like we can't understand the $D'$ invariance of the one-loop-all-plus gravity amplitudes by looking at the Einstein-Hilbert action.

However, the \textit{self-dual} gravity action $S_{\rm SDG}[ \theta, \Gamma ] = -\frac{2 i}{\kappa^2} \int \dd \Gamma_{AB} \wedge \theta^{A \dot A} \wedge \theta^{B}_{\;\; \dot A}$ from \eqref{Ssdggammatheta} is a different beast entirely. First, how do coordinate dilatations and scaling act on $\theta$, $\Gamma$? The action on the tetrads $\theta^{A \dot A}(x) = \theta_\mu^{A \dot A}(x) dx^\mu$ can be read off from \eqref{coordscalinggmunu} using the relation $g_{\mu \nu} = \frac{1}{2} \theta_{\mu}^{A \dot A} \theta_{A \dot A \nu}$, and the geometrical torsion-free condition $\dd \theta^a = - \Gamma^a_{\;\; b} \wedge \theta^b$ can then be used to find the compatible transformations on $\Gamma$ as well. All told, they are
\begin{equation}
\begin{aligned}
    \text{coordinate dilatation: }& \begin{pmatrix} \theta^{A \dot A}_\mu(x) \\ \Gamma^{AB}_\mu(x) \end{pmatrix} \mapsto \begin{pmatrix} t\, \theta^{A \dot A}_\mu(tx) \\ t\, \Gamma^{AB}_\mu(tx) \end{pmatrix} ,\\
    \text{scaling: }& \begin{pmatrix} \theta^{A \dot A}_\mu(x) \\ \Gamma^{AB}_\mu(x) \end{pmatrix} \mapsto \begin{pmatrix} t\, \theta^{A \dot A}_\mu(x) \\  \Gamma^{AB}_\mu(x) \end{pmatrix} .
\end{aligned}
\end{equation}
While the SDG action is manifestly invariant under coordinate dilatations, it is of course \textit{not} invariant under the scaling transformation, just as the Einstein-Hilbert action is not. However, recall that the torsion-free condition is \textit{not} a e.o.m.\! of $S_{\rm SDG}[\theta,\Gamma]$. So we are allowed to define a modified scaling transformation, which we'll call scaling$'$, as
\begin{equation}
    \text{scaling$'$:}\; \begin{pmatrix} \theta^{A \dot A}_\mu(x) \\ \Gamma^{AB}_\mu(x) \end{pmatrix} \mapsto \begin{pmatrix} t\, \theta^{A \dot A}_\mu(x) \\  t^{-2} \Gamma^{AB}_\mu(x) \end{pmatrix} .
\end{equation}
The SDG action is now indeed invariant under scaling$'$ even though it wasn't invariant under scaling. So, if we define a modified diltation operation, which we'll call dilatation$'$, as a coordinate dilatation composed with a compensating inverse scaling$'$, which is
\begin{equation}\label{A109}
    \text{dilatation}':  \begin{pmatrix} \theta^{A \dot A}_\mu(x) \\ \Gamma^{AB}_\mu(x) \end{pmatrix} \mapsto \begin{pmatrix}  \theta^{A \dot A}_\mu(tx) \\  t^{3} \, \Gamma^{AB}_\mu(tx) \end{pmatrix} 
\end{equation}
then the SDG action is indeed invariant under dilatation$'$. We remind the reader that $\theta^{A \dot A}$ describes the positive-helicity self-dual degrees of freedom and $\Gamma_{AB}$ describes the negative helicity anti-self-dual degrees of freedom. As a consistency check, let's now look at the action of dilatation$'$ on the tetrad $\theta$ which describes the positive-helicity particles in the one-loop-all-plus amplitudes. Splitting the tetrad $\theta = \theta_0 + \kappa \chi$ into a constant flat-space background plus a perturbation, a dilatation$'$ acts as
\begin{equation}
    \text{dilatation$'$: } \; \theta^{A \dot A}_\mu(x) \mapsto \theta^{A \dot A}_\mu(tx) = \theta^{A \dot A}_{0\; \mu} + \kappa \chi^{A \dot A}_\mu(tx)
\end{equation}
and we see that dilatation$'$ doesn't affect the constant background tetrad $\theta_0$, as desired, and only affects the spacetime dependent perturbation $\chi$.

So we have now understood why $D'$ is a symmetry of the one-loop-all-plus amplitudes using $S_{\rm SDG}[\theta,\Gamma]$. The fact that $D'$ leaves the coupling $\kappa$ unchanged, even though $\kappa$ is naively dimensionful, can be traced back to the fact that dilatation$'$ acts on $\Gamma_{A B}$ with a non-engineering dimension weight that is not compatible with the torsion-free condition.

Additionally, one can also see that the lightcone action $S_{\rm SDG}[\phi,\bar{\phi}]$ action \eqref{actionfirst2} is also dilatation$'$ invariant as well, where dilatation$'$ acts on lightcone scalars $\phi(x)$, $\bar{\phi}(x)$ as
\begin{equation}
    \text{dilatation$'$: } \phi(x) \mapsto t^{-2} \phi(t x), \hspace{0.5 cm} \bar{\phi}(x) \mapsto t^4 \bar{\phi}(t x).
\end{equation}
Note that because dilatation$'$ acts on positive-helicity spinor variables as $(\lambda, \tlambda) \mapsto (t^{1/2} \lambda, t^{1/2} \tlambda)$, the above action on $\phi(x)$ can be read-off instantly from the expression for the positive-helicity graviton wavefunction $\phi(x,\lambda,\tlambda) = \frac{1}{\langle \alpha \lambda \rangle^4} e^{\frac{i}{2} \lambda_A \tlambda_{\dot A} x^{A \dot A}}$. Likewise the action of dilatation$'$ on $\bar{\phi}$ is compatible with that of $\Gamma_{AB}$ \eqref{A109} via the relation \eqref{bphiGamma}.

\section{Plebański's first heavenly equation }\label{sec:firstheavenly}

\subsection{The equation}

In this note we have exclusively worked with SDG in lightcone gauge. In this gauge, self-dual metrics take the form
\begin{equation}
    ds^2 = 4( du d\bu - dw d \bw ) + 2 \kappa (\partial_w^2 \phi ) d \bu^2 + 4 \kappa (\partial_u \partial_w \phi) d \bu d \bw + 2 \kappa ( \partial_u^2 \phi ) d \bw^2 
\end{equation}
with coordinates $(u,\bu,w,\bw)$, where $\phi$ satisfies ``Plebański's second heavenly equation'' \cite{plebanski1975some}
\begin{equation}\label{pleb2copy}
    (\partial_u \partial_{\bu} - \partial_w \partial_{\bw}) \phi - \frac{\kappa}{2} ( (\partial_u^2 \phi)( \partial_w^2 \phi ) - (\partial_u \partial_w \phi) (  \partial_u \partial_w \phi ) ) = 0.
\end{equation}

However, there is another important way one can  parameterize self-dual metrics. This has not played a role in this note but we shall mention it anyway. It turns out that there always exists a coordinate system $(y^j, \bar{y}^{\bj})$, with $j=1,2$, $\bj = \bar{1},\bar{2}$, in which any self-dual metric can be written as
\begin{equation}\label{eq213}
    ds^2 = - 4 \, \partial_{j} \partial_{\bj} \, \Omega \, dy^j  d \bar{y}^{\bj}
\end{equation}
where $\Omega$ is the Kähler scalar and satisfies ``Plebański's first heavenly equation''
\begin{equation}\label{pleb1heaven}
    \det( \partial_{j} \partial_{\bj} \, \Omega ) = (\partial_{1} \partial_{\bar{1}} \, \Omega)(\partial_{2} \partial_{\bar{2}} \, \Omega) - (\partial_{1} \partial_{\bar{2}} \, \Omega)(\partial_{2} \partial_{\bar{1}} \, \Omega) = 1.
\end{equation}
The coordinates $(y^j, \bar{y}^{\bj})$ are related to the lightcone coordinates by 
\begin{equation}
    u = - \frac{\partial \Omega}{\partial \bar{y}^{\bar{1}}}, \hspace{0.5 cm} w = \frac{\partial \Omega}{\partial \bar{y}^{\bar{2}}}, \hspace{0.5 cm} \bu = \bar{y}^{\bar{1}}, \hspace{0.5 cm} \bw = \bar{y}^{\bar{2}}.
\end{equation}
We note that we reviewed the first and second heavenly equation in more detail in Appendix B of \cite{mypaper}.

Note that the flat metric corresponds to the potential $\Omega = y^1 \bar{y}^{\bar{1}} + y^2 \bar{y}^{\bar{2}}$. If we therefore write $\Omega$ as a perturbation on top of flat space, which we'll call $\Theta$,
\begin{equation}
    \Omega = y^1 \bar{y}^{\bar{1}} + y^2 \bar{y}^{\bar{2}} + \frac{\kappa}{2} \Theta,
\end{equation}
then \eqref{pleb1heaven} becomes
\begin{equation}\label{pleb1coords}
    (\partial_{1} \partial_{\bar{1}} + \partial_{2} \partial_{\bar{2}}) \Theta - \frac{\kappa}{2} ( (\partial_{\bar{1}} \partial_2 \Theta ) (\partial_1 \partial_{\bar{2}} \Theta) - (\partial_1 \partial_{\bar{1}} \Theta) ( \partial_2 \partial_{\bar{2}} \Theta ) ) = 0.
\end{equation}
The above equation looks extremely similar to Plebański's second heavenly equation \eqref{pleb2copy}, but it is actually Plebański's first heavenly equation! This formulation of SDG is commonly used in discussions of the N=2 string \cite{Bern:1998xc}.

In the first heavenly coordinates $(y^1, y^2, \bar{y}^{\bar{1}}, \bar{y}^{\bar{2}})$ coordinates, let us redefine the derivative notation to be (for this appendix alone)
\begin{equation}
    \partial_{A \dot A} = \begin{pmatrix} \partial_{1 \dot 1} & \partial_{1 \dot 2} \\ \partial_{2 \dot 1} & \partial_{2 \dot 2} \end{pmatrix} = \begin{pmatrix} \partial_{\bar{1}} & \partial_{\bar{2}} \\ \partial_{2} & -\partial_{1} \end{pmatrix}.
\end{equation}
Let us also use the reference spinors
\begin{equation}
    \alpha^A = (0,1), \hspace{1 cm} \beta^A = (1,0) ,   
\end{equation}
from the last section, which using the notation \eqref{alphabetapartial} give
\begin{equation}
    (\alpha \partial)_{\dot A}  = (- \partial_2, \partial_1), \hspace{1 cm} (\beta \partial)_{\dot A} = (- \partial_{\bar{1}}, - \partial_{\bar{2}}).
\end{equation}
We can now re-express Plebański's first heavenly equation \eqref{pleb1coords} as, using the flat space wave operator $\Box = - \partial_1 \partial_{\bar{1}} - \partial_2 \partial_{\bar{2}}$,
\begin{equation}\label{psiheavenly}
    -\Box \Theta + \frac{\kappa}{4} (\alpha \partial)_{\dot A} (\beta \partial)_{\dot B} \Theta \, (\alpha \partial)^{\dot A} (\beta \partial)^{\dot B} \Theta = 0.
\end{equation}
Note that if we were to take the above equation and set the second reference spinor equal to the first $|\beta\rangle \to |\alpha \rangle$, we would recover Plebański's second heavenly equation as written in \eqref{sdgactionalpha}. So the first heavenly equation can be thought of as a more general form of the second heavenly equation for SDG with two holomorphic reference spinors instead of one.

\subsection{Deriving the NSVW (i.e. half-soft) MHV formula from the first heavenly equation using perturbiners}\label{sec:firstheavenlyhalfsoft}

Now let us discuss the close relationship of Plebański's first heavenly equation to the NSVW formula for MHV amplitudes \cite{Nguyen:2009jk}. This connection was understood by Adamo, Mason, and Sharma in \cite{Adamo:2021bej}, in their construction of the ``non-linear twistor sigma model,'' which notably not only computes MHV amplitudes in Minkowski space but also on arbitrary non-flat self-dual spacetime backgrounds as well \cite{Adamo:2023fbj,Adamo:2023zeh,Adamo:2022mev,Adamo:2025fqt}.

In this appendix, we'll simply show how the first heavenly equation can be used to calculate MHV amplitudes in Minkowski space in a relatively straightforward manner using perturbiners, essentially following Adamo, Mason, and Sharma but without making use of the full power of twistor theory. This calculation will also explain why the half-soft function, which on the one hand is (essentially) equal to the all-plus Berends-Giele currents \eqref{halfsoftJG}, is simultaneously also (essentially) equal to MHV amplitudes \eqref{halfsoftMHV}, which is a fact that is a priori is a bit surprising.

Imagine that we have $n-2$ positive helicity gravitons labelled $3, \ldots, n$. The first heavenly equation \eqref{psiheavenly} is equivalent the Berends-Giele recursion relation satisfied by the half-soft function via \eqref{hBerends}. This means that, in the perturbiner language, the ``top'' piece of the Kähler scalar of the self-dual spacetime perturbiner is solved by the half-soft function,
\begin{equation}\label{OmegaHalfSoft}
    \frac{\partial^{n-2}}{\partial \epsilon_3 \ldots \partial \epsilon_n} \Omega \Big\rvert_{\epsilon_i = 0 \, \forall \, i = 3,\ldots,n} \propto \lr{\alpha \beta}^2 h(\alpha, \{3, \ldots, n\}, \beta) e^{i (p_3 + \ldots + p_n) \cdot y},
\end{equation}
where in this appendix we are not being careful with overall constants like $\kappa$ and will only compute the amplitude up to a proportionality constant. (The factor of $\lr{\alpha \beta}^2$ is placed here so the LHS has no scaling dimension in $|\alpha\rangle$, $|\beta\rangle$.)

MHV amplitudes can be computed by evaluating the GR Plebański action \eqref{Sgrpleb} on-shell for perturbiner spacetimes containing exactly two negative-helicity gravitons, see for instance the discussion in section 5.1 of \cite{mypaper} for more explanation. The equation is
\begin{equation}\label{MHVgenerating}
    \mathcal{M}^{\rm tree}_{\rm MHV}(1^-,2^-,3^+,\ldots,n^+) \propto \frac{\partial^{n}}{\partial \epsilon_1 \ldots \partial \epsilon_n} \int \Sigma^{AB} \wedge \gamma^{(1)}_{AC} \wedge \gamma^{(2) C}_{B} \Big\rvert_{\epsilon_i = 0 \, \forall\, i = 1,\ldots,n}
\end{equation}
where here $\Sigma^{AB}$ are the ASD 2-forms of the self-dual background containing the $n-2$ positive-helicity gravitons, and $\gamma^{(1)}_{AB}$ and $\gamma^{(2)}_{AB}$ are linearized perturbations of the ASD spin connection $\Gamma_{AB}$ for the negative-helicity gravitons $1$ and $2$ propagating on this self-dual background.

How can one solve for the $\gamma_{AB}$ of a negative helicity graviton propagating on a self-dual background? We can do so using the following manipulations. First, note that for any spacetime satisfying the vacuum Einstein equations, one has
\begin{equation}\label{RabPsiabcd}
    R_{AB} = \Psi_{ABCD} \Sigma^{CD}, \hspace{1 cm} \nabla^A_{\;\; \dot A} \Psi_{ABCD}= 0,
\end{equation}
where the scalar $\Psi_{ABCD} = \Psi_{(ABCD)}$ is the ASD part of the Riemann curvature tensor  and the equation $\nabla^A_{\;\; \dot A} \Psi_{ABCD}= 0$ is the differential Bianchi identity, where $\nabla$ is the spin covariant derivative constructed out of $\Gamma_{AB}$. If we then perturb \eqref{RabPsiabcd} linearly around a self-dual background $\gamma_{AB} = \delta \Gamma_{AB}$, $\psi_{(ABCD)} = \delta \Psi_{(ABCD)}$, where $\Gamma_{AB} = 0$, $\Psi_{ABCD} = 0$ on the self-dual background itself, the equations become
\begin{equation}\label{gammaABeq}
    \dd \gamma_{AB} = \psi_{ABCD} \Sigma^{AB}, \hspace{1 cm} \partial^A_{\;\; \dot A} \psi_{ABCD} = 0,
\end{equation}
where the $\Sigma^{AB}$ above are the ASD 2-forms of the self-dual background.

It is now simple to solve for $\psi_{ABCD} = \psi_{(ABCD)}$ using the Bianchi identity. It is nothing but the humble plane-wave
\begin{equation}
    \psi_{ABCD} = \epsilon \frac{i}{4} \lambda_A \lambda_B \lambda_C \lambda_D \, e^{i p \cdot y}
\end{equation}
where here $y^{A \dot A} = \left( \begin{smallmatrix} 2\bar{y}^{\bar{1}} & 2\bar{y}^{\bar{2}} \\ 2y^2 & - 2y^1 \end{smallmatrix} \right)$ and $p_{A \dot A} = \lambda_A \tlambda_{\dot A}$ is the momentum of the negative-helicity graviton. $\epsilon$ is an overall infinitesimal parameter in the perturbiner sense. Using this expression for $\psi_{ABCD}$, we can now solve for $\gamma^{(1)}_{AB}$ and $\gamma^{(2)}_{AB}$. For simplicity, we shall choose the spinors 
\begin{equation}\label{alphabeta12}
    |1\rangle = |\alpha\rangle, \hspace{1 cm} |2\rangle = |\beta\rangle,
\end{equation}
for the momenta of the negative helicity gravitons, i.e.
\begin{equation}
    p_1 = |\alpha\rangle[\talpha| \, , \hspace{1 cm} p_2 = |\beta\rangle[\tbeta| \, ,
\end{equation}
for any $|\talpha]$, $|\tbeta]$. Technically the rest of our computation relies on the specific choice \eqref{alphabeta12} for the negative-helicity spinors $|1\rangle$, $|2\rangle$, but in the full twistor computation \cite{Adamo:2021bej} one can properly justify that the answer will not depend on this choice.

With these choices, it turns out that the linearized spin connections \eqref{gammaABeq} are solved by the simple expressions
\begin{equation}
\begin{aligned}
    \gamma_{AB}^{(1)} = \epsilon_1 \frac{\alpha_A \alpha_B \alpha_C \txi_{\dot C} \theta_0^{C \dot C}}{[\txi \talpha]} e^{i p_1 \cdot y}, \hspace{1 cm} \gamma_{AB}^{(2)} = \epsilon_2 \frac{\beta_A \beta_B \beta_C \txi_{\dot C} \theta_0^{C \dot C}}{[\txi \tbeta]} e^{i p_2 \cdot y},
\end{aligned}
\end{equation}
where $\txi$ is an arbitrary reference spinor,\footnote{This reference spinors sets a choice of gauge for the spin connections $\gamma_{AB}$. One can straightforwardly show that changing $\txi$ is akin to sending $\gamma_{AB} \mapsto \gamma_{AB} + \dd f_{AB}$ for a 0-form $f_{AB}$.} and $\theta_0^{A \dot A}$ are the \textit{flat-space} tetrads  $\theta_0^{A \dot A} = \dd y^{A \dot A}$, given by
\begin{equation}
\begin{aligned}
    &\theta_0^{1 \dot 1} = 2 d \bar{y}^{\bar{1}}, \hspace{1 cm} \theta_0^{2 \dot 1} = 2 dy^2, \\
    &\theta_0^{1 \dot 2} = 2 d \bar{y}^{\bar{2}}, \hspace{1 cm} \theta_0^{2 \dot 2} = -2  d y^1 .
\end{aligned}
\end{equation}
It may seem strange that the expression for $\gamma^{(1)}_{AB}$ is built out of the flat-space tetrads, so let's check that it satisfies \eqref{gammaABeq}:
\begin{equation}
\begin{aligned}
    \dd \gamma^1_{AB} &= - \epsilon_1 \frac{i}{2} \frac{\alpha_A \alpha_B \alpha_C \alpha_D \talpha_{\dot D} \tilde{\xi}_{\dot C} }{[\tilde{\xi} \talpha]} \theta_0^{C \dot C} \wedge  \theta_0^{D \dot D} e^{i p_1 \cdot y} \\
    &= \epsilon_1 \frac{i}{4} \frac{\alpha_A \alpha_B \alpha_C \alpha_D \talpha_{\dot D} \tilde{\xi}_{\dot C} }{[\tilde{\xi} \talpha]} (\vep^{CD} \widetilde{\Sigma}_0^{\dot C \dot D} + \vep^{\dot C \dot D} \Sigma_0^{CD})e^{i p_1 \cdot y} \\
    &= \epsilon_1 \frac{i}{4} \alpha_A \alpha_B \alpha_C \alpha_D  \Sigma_0^{CD} e^{i p_1 \cdot y} \\
    &= \psi_{ABCD} \Sigma^{CD}.
\end{aligned}
\end{equation}
In the last line, we used the convenient equation
\begin{equation}\label{Sigma0alphabeta}
    \alpha_A \alpha_B \Sigma^{AB} = \alpha_A \alpha_B \Sigma^{AB}_0, \hspace{1 cm} \beta_A \beta_B \Sigma^{AB} = \beta_A \beta_B \Sigma^{AB}_0,
\end{equation}
which holds only if one uses the natural basis of tetrads arising from first heavenly equation in which
\begin{equation}\label{Sigma0Sigma}
    \Sigma^{11} = \Sigma_0^{11}, \hspace{0.5 cm} \Sigma^{22} = \Sigma_0^{22}, \hspace{0.5 cm} \Sigma^{12}_0 \neq \Sigma^{12}_0.
\end{equation}
To check this, one can simply use the equation for the tetrads for the self-dual background
\begin{equation}
\begin{aligned}
    &\theta^{1 \dot 1} = 2 d \bar{y}^{\bar{1}}, \hspace{1 cm} \theta^{2 \dot 1} = 2 \partial_1 \partial_{\bar{2}} \Omega \, d y^1 + 2 \partial_2 \partial_{\bar{2}} \Omega \, dy^2, \\
    &\theta^{1 \dot 2} = 2 d \bar{y}^{\bar{2}}, \hspace{1 cm} \theta^{2 \dot 2} = -2 \partial_1 \partial_{\bar{1}} \Omega \, d y^1 - 2 \partial_2 \partial_{\bar{1}} \Omega \, dy^2,
\end{aligned}
\end{equation}
and then calculate the ASD 2-forms of the self-dual background are
\begin{equation}\label{SigmaPleb1}
\begin{aligned}
    \Sigma^{11} &= 8 \, d y^1 \wedge d y^2 ,\\
    \Sigma^{22} &= 8\, d \bar{y}^{\bar{1}} \wedge d \bar{y}^{\bar{2}}, \\
    \Sigma^{12} &= 4\sum_{j = 1, 2} \sum_{\bj = \bar{1}, \bar{2}} \partial_j \partial_{\bj} \Omega_{j \bj} d y^j \wedge d\bar{y}^{\bj}.
\end{aligned}
\end{equation}
Note that $\Sigma^{11}$ and $\Sigma^{22}$ don't actually depend on $\Omega$, implying \eqref{Sigma0Sigma} and \eqref{Sigma0alphabeta}. Let us also take the opportunity to write down one extra equation,
\begin{equation}\label{oneextraeq}
    \Sigma^{12} \wedge \widetilde{\Sigma}^{\dot A \dot B}_0 =   (\alpha \partial)^{(\dot A} (\beta \partial)^{\dot B)} \Omega \, \times 32\,  d y^1 \wedge d y^2 \wedge d \bar{y}^{\bar{1}} \wedge d \bar{y}^{\bar{2}}
\end{equation}
which can be computed straightforwardly from the expressions \eqref{SigmaPleb1}.

Now that we have computed all of the required objects individually, let us plug them into the MHV generating function \eqref{MHVgenerating} and compute
\begin{equation}
\begin{aligned}
     \int \Sigma^{AB} \wedge \gamma^{(1)}_{AE} \wedge \gamma^{(2) E}_{B}  &\propto \epsilon_1 \epsilon_2 \int \Sigma^{AB} \wedge \frac{\alpha_A \alpha_E \alpha_C \txi_{\dot C} \theta_0^{C \dot C}}{[\txi \talpha]} \wedge \frac{\beta_B \beta^E \beta_D \txi_{\dot D} \theta_0^{D \dot D}}{[\txi \tbeta]} e^{i(p_1 + p_2) \cdot y} \\
     &\propto \epsilon_1 \epsilon_2 \lr{\alpha \beta} \int  \frac{ \alpha_C \beta_D \txi_{\dot C} \txi_{\dot D} }{[\txi \talpha] [\txi \tbeta]} \, \Sigma^{12} \wedge \theta_0^{C \dot C} \wedge \theta_0^{D \dot D} e^{i(p_1 + p_2) \cdot y} \\
     & \propto \epsilon_1 \epsilon_2  \lr{\alpha \beta} \int  \frac{ \alpha_C \beta_D \txi_{\dot C} \txi_{\dot D} }{[\txi \talpha] [\txi \tbeta]} \, \Sigma^{12} \wedge ( \vep^{C D}  \widetilde{\Sigma}_0^{\dot C \dot D} + \vep^{\dot C \dot D} \Sigma_0^{CD} ) e^{i(p_1 + p_2) \cdot y} \\
     & \propto \epsilon_1 \epsilon_2 \lr{\alpha \beta}^2 \int \mathrm{dvol}  \frac{ \txi_{\dot C} \txi_{\dot D} }{[\txi \talpha] [\txi \tbeta]} (\alpha \partial)^{(\dot C} (\beta \partial)^{\dot D)} \Omega \, \, e^{i(p_1 + p_2) \cdot y}.
\end{aligned}
\end{equation}
The above steps follow by simple spinor contractions. The only non-trivial step was our use of the equation \eqref{oneextraeq} in the last equality. Here $\mathrm{dvol}$ is the volume 4-form of the self-dual background, which is given by $\mathrm{dvol} = - 4 i \, d y^1 \wedge d y^2 \wedge d \bar{y}^{\bar{1}} \wedge d \bar{y}^{\bar{2}}$ we can also denote $\propto d^4 y$.

Now let us simplify $(\alpha \partial)^{(\dot C} (\beta \partial)^{\dot D)} \Omega$. The top piece of the perturbiner expansion for $\Omega$ is simply proportional to $e^{i(p_3 + \ldots + p_n) \cdot y}$ via \eqref{OmegaHalfSoft}. For brevity let us define $P_+ \equiv p_3 + \ldots + p_n$ to be the total momenta of the positive-helicity gravitons, so
\begin{equation}
\begin{aligned}
    (\alpha \partial)^{\dot C} (\beta \partial)^{\dot D} \frac{\partial^{n-2}}{\partial \epsilon_3 \ldots \partial \epsilon_n} \Omega \Big\rvert_{\epsilon_i = 0 \, \forall \, i = 3,\ldots,n} = - (\langle \alpha | P_+)^{\dot C}  (\langle \beta | P_+)^{\dot D}\frac{\partial^{n-2}}{\partial \epsilon_3 \ldots \partial \epsilon_n} \Omega \Big\rvert_{\epsilon_i = 0 \, \forall \, i = 3,\ldots,n}.
\end{aligned}
\end{equation}
Plugging this back into the MHV generating function, we get
\begin{equation}\label{MHVfinalOmega}
\begin{aligned}
    &\mathcal{M}^{\rm tree}_{\rm MHV}(1^-,2^-,3^+,\ldots,n^+) \propto \frac{\partial^{n}}{\partial \epsilon_1 \ldots \partial \epsilon_n} \int \Sigma^{AB} \wedge \gamma^{(1)}_{AC} \wedge \gamma^{(2) C}_{B} \Big\rvert_{\epsilon_i = 0 \, \forall\, i = 1,\ldots,n} \\
    &\propto  \lr{\alpha \beta}^2 \frac{\partial^{n-2}}{\partial \epsilon_3 \ldots \partial \epsilon_n}  \int d^4 y  \frac{ \langle \alpha| P_+ | \txi ] \langle \beta| P_+ | \txi ] }{[\txi \talpha] [\txi \tbeta]}   \Omega \, \, e^{i(p_1 + p_2) \cdot y} \Big\rvert_{\epsilon_i = 0 \, \forall \, i = 3,\ldots,n} .
\end{aligned}
\end{equation}
Now let us use the fact that, due to the integral over spacetime, there is an implicit $\delta^4(p_1 + p_2 + p_3 + \ldots + p_n)$ in the expression enforcing momentum conservation, which is equivalent to $|\alpha\rangle [\talpha| + |\beta \rangle [\tbeta| + P_+ = 0$. Inserting this into the above expression, we have
\begin{equation}\label{MHVbeautiful}
\begin{aligned}
    &\mathcal{M}^{\rm tree}_{\rm MHV}(1^-,2^-,3^+,\ldots,n^+) 
    \propto  \lr{\alpha \beta}^4 \frac{\partial^{n-2}}{\partial \epsilon_3 \ldots \partial \epsilon_n}  \int d^4 y \,  \Omega \, \, e^{i(p_1 + p_2) \cdot y} \Big\rvert_{\epsilon_i = 0 \, \forall \, i = 3,\ldots,n}.
\end{aligned}
\end{equation}
This equation, due to \cite{Adamo:2021bej}, is a beautiful result: the MHV amplitude itself is nothing but the Fourier transform of the Kähler scalar of the positive-helicity gravitons! Because we already know $\Omega$ of the self-dual background is equal to the half-soft function \eqref{OmegaHalfSoft} due to the fact that it must solve the Berends-Giele relation \eqref{psiheavenly}, this calculation explains why the half-soft function, which we originally introduced in section \ref{sec:BDPR} as a two-spinor generalization of the all-plus current $\mathcal{J}_{\rm G}$, is \textit{also} equal to the MHV amplitude, as in \eqref{halfsoftMHV}.

\section{Color factors of tree-level and one-loop amplitudes}\label{app:color}

In this appendix we will review how gauge theory color factors work at tree-level and one-loop-level, for the uninitiated. Say $N_c$ is the number of colors. We normalize our $\mathfrak{su}(N_c)$ generators such that
\begin{equation}\label{TaTb}
    \tr( T^a T^b) = \delta^{ab}.
\end{equation}
With this normalization, the matrices satisfy a completeness relation
\begin{equation}\label{suncompleteness}
    T^a_{ij} T^a_{k \ell} = \delta_{i \ell} \delta_{jk} - \frac{1}{N_c} \delta_{ij} \delta_{k \ell}.
\end{equation}
See for instance \cite{Haber:2019sgz} for a proof. The $\mathfrak{su}(N_c)$ generators also satisfy
\begin{equation}\label{trace0}
    \tr(T^a) = 0.
\end{equation}
We also, as always, have the cyclic property of the trace
\begin{equation}\label{cyclicityofTr}
    \tr(T^{a_1} T^{a_2} \ldots T^{a_n}) = \tr(T^{a_2} \ldots T^{a_n} T^{a_1} ).
\end{equation}
If our generators lived in $\mathfrak{u}(N_c)$ instead of $\mathfrak{su}(N_c)$, then the identity matrix would be included in the list of generators and the completeness relation would be
\begin{equation}\label{uncompleteness}
    T^a_{ij} T^a_{k \ell} = \delta_{i \ell} \delta_{jk}
\end{equation}
which is a simpler formula. (Note that the identity element does not satisfy \eqref{trace0}.)

Even though our actual gauge algebra is $\mathfrak{su}(N_c)$, there is no harm in proceeding as if it is $\mathfrak{u}(N_c)$ if we want to use the simpler completeness relation. This is because the $\mathfrak{u}(1)$ photon decouples from the rest of the particles in the Lagrangian. We will however only consider external gluons which live in the original $\mathfrak{su}(N_c)$.

Now, with the structure constants
\begin{equation}
    [T^a, T^b] = i f^{abc} T^c
\end{equation}
\eqref{TaTb} implies
\begin{equation}\label{TTTf}
    \tr( [T^a, T^b] T^c ) = i f^{abc}.
\end{equation}
From the cyclicity of the trace, one can show
\begin{equation}
    \tr( [T^a, T^b] T^c ) = \tr(  [ T^b , T^c ] T^a )
\end{equation}
which means that, whenever we have chosen our generators to satisfy \eqref{TaTb}, we also have the cyclic identity for the structure constants
\begin{equation}
    f^{abc} = f^{bca}
\end{equation}
which means $f^{abc}$ are totally antisymmetric in $abc$.

Let us now review the well-known story of color-ordering in tree-level and one-loop pure-gluon scattering amplitudes built out of cubic vertices. Our presentation will follow \cite{Dixon:2011xs, Bern:1994zx}. See also \cite{DelDuca:1999rs}.

\begin{figure}[h]
    \centering
    \includegraphics[width=0.85\linewidth]{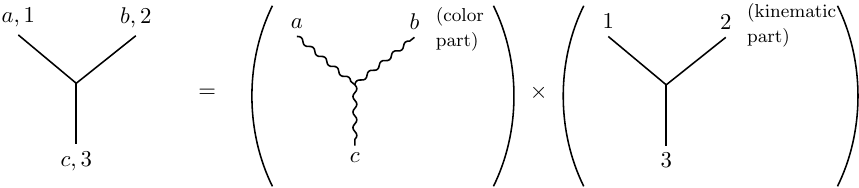}
    \caption{Splitting Feynman diagrams into a color and a kinematic part}
    \label{fig:feynsplit}
\end{figure}

A three-point gauge theory vertex is comprised of two parts: a structure constant $i f^{abc}$ times a kinematic part. See, for instance our \ppm vertex in equation \ref{ppmsdymvertex}. Both parts are individually anti-symmetric in the three particles, and multiplied together they are completely symmetric. We will visually distinguish between these two factors by writing ``color part'' or ``kinematic part'' in the corner of all the diagrams. We will also draw the color diagrams with squiggly lines. See for instance figure \ref{fig:feynsplit}. 

\begin{figure}[h]
    \centering
    \includegraphics{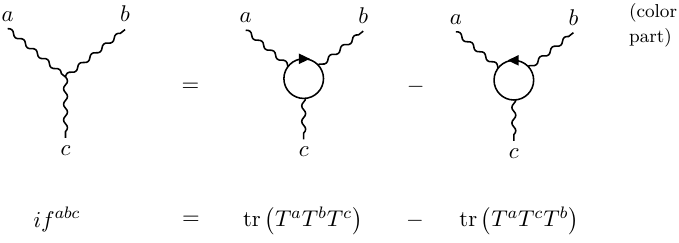}
    \caption{How to split up the structure constant into a sum of trace terms.}
    \label{fig:fabccircle}
\end{figure}

Let us now focus on the color part of the diagram. If we write $i f^{abc} = \tr(T^a T^b T^c) - \tr(T^b T^a T^c)$, we will split up each vertex into a difference of two vertices corresponding to the two color orderings, notated with an oriented circle. See figure \ref{fig:fabccircle}.

\begin{figure}[h]
    \centering
    \includegraphics{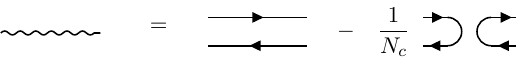}
    \caption{Graphical representation of $\mathfrak{su}(N_c)$ completeness relation \eqref{suncompleteness}.}
    \label{fig:squigglypropagator}
\end{figure}

We can then use the completeness relation \eqref{suncompleteness} to reduce products of traces like $\tr( ... T^a) \tr(T^a...)$ where an internal color $T^a$ is being summed over into a single trace like $\tr(... \, ...)$. We represent this diagrammatically in figure \ref{fig:squigglypropagator}. This is the t'Hooft double line notation \cite{tHooft:1973alw}.

Of course, we can drop the $\propto \frac{1}{N_c}$ piece from figure \ref{fig:squigglypropagator} due to the $\mathfrak{u}(1)$ decoupling and only keep the double line. (If we wanted to keep the second piece, we would find that all of the terms associated with it would cancel pairwise anyway.) Note that the $\mathfrak{u}(N_c)$ completeness relation \eqref{uncompleteness} can be expressed as
\begin{equation}
    \tr( X T^a) \tr(T^a Y) = \tr(XY).
\end{equation}

\begin{figure}[h]
    \centering
    \includegraphics[width=0.8\textwidth]{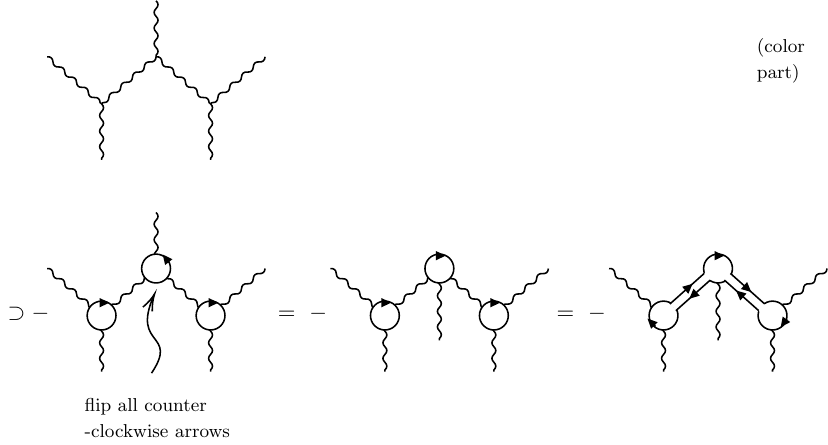}
    \caption{Procedure for handling color ordering: flip all counterclockwise circles to be clockwise, which puts all external colors in a unique a clockwise pattern. All signs from the diagrams will be cancelled once the corresponding sign from the kinematic diagram flip is accounted for as well.}
    \label{fig:flippy}
\end{figure}

\begin{figure}[h]
    \centering
    \includegraphics[width=0.35\linewidth]{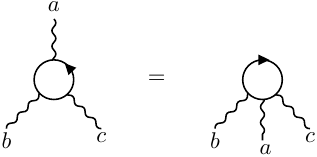}
    \caption{$\tr(T^a T^b T^c) = \tr(T^a T^b T^c)$.}
    \label{fig:fabc}
\end{figure}

Now, consider a sum of color factor diagrams where some of the vertices are a clockwise ``$+$'' circle and some of the vertices are a counterclockwise ``$-$'' circle. There is, of course, a unique way to flip all of the counterclockwise circles into clockwise circles. By ``flip'', we mean reverse the direction of the arrow and also switch the placement of one of the legs. See figure \ref{fig:flippy}. This ``flip'' operation does not actually induce any sign, due to $\tr(T^a T^b T^c) = \tr(T^a T^b T^c)$. See figure \ref{fig:fabc}.

After we ``flip'' all of the counterclockwise circles to be clockwise, the external particles will all have a well-defined cyclic color ordering. When we perform this same flip on the kinematic part of the diagram, we introduce a sign which happily cancels out the original sign of the counterclockwise circle.

This implies that when we sum over all tree-level Feynman diagrams, the expression will take the form of a sum over color orderings of the external particles, with an overall color factor $\tr(T^{a_1} \ldots T^{a_n})$. We say that the full amplitude $\mathcal{A}$ can be written as a sum of ``partial amplitudes'' $A$ via
\begin{equation}
    \mathcal{A}^{\rm tree}(1, \ldots, n) = \sum_{\sigma \in S_n / \mathbb{Z}_n} \tr(T^{\sigma(1)} \ldots T^{\sigma(n)} ) A^{\rm tree}(\sigma(1), \ldots, \sigma(n) ) .
\end{equation}
The rules to compute the partial amplitudes require one to fix the the positions of the external particles in a fixed clockwise color-ordering and remove the factors of $i f^{abc}$ from the Feynman rules. Additionally, due to the antisymmetry of the kinematic vertices, if we mirror the diagram we have the reflection identity
\begin{equation}
    A^{\rm tree}(1,2, \ldots, n\!-\!1, n) = (-1)^n A^{\rm tree}(n, n\!-\!1, \ldots, 2,1).
\end{equation}

\begin{figure}[h]
    \centering
    \includegraphics{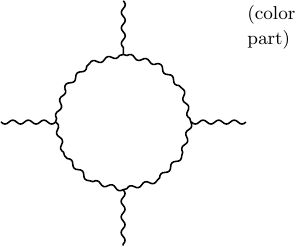}
    \caption{Color part of a 4-pt box.}
    \label{fig:squiggly4pt}
\end{figure}

\begin{figure}[h]
    \centering
    \includegraphics{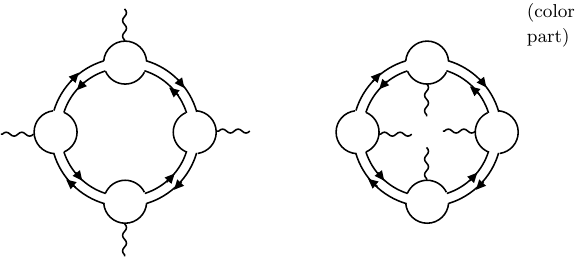}
    \caption{One type of diagram that descends from figure \ref{fig:squiggly4pt}, where the legs all point in or out. }
    \label{fig:circle4ptinout}
\end{figure}

\begin{figure}[h]
    \centering
    \includegraphics{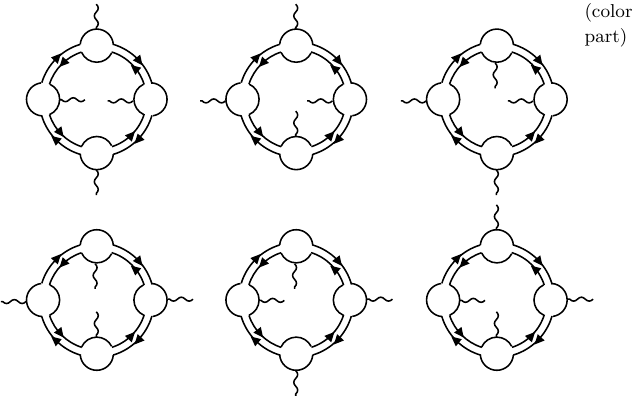}
    \caption{Another type of diagram that descends from figure \ref{fig:squiggly4pt}, where some legs point in and some legs point out.}
    \label{fig:circle4ptsomeinout}
\end{figure}

Now let us consider what happens at one-loop level. Consider, for instance, figure \ref{fig:squiggly4pt}.
Here there are four external legs connected to a central loop. We now turn this diagram into a sum over $2^4 =16$ diagrams with color ordered vertices. We then take the counterclockwise vertices and forcibly ``flip'' them, which makes the counterclockwise-outwards-pointing external particles now point inwards and be oriented clockwise. There will be two kinds of diagrams that arise from this process. The first kind will have all particles pointing completely outwards or completely inwards, such as in figure \ref{fig:circle4ptinout}. 
The second kind of diagram will have some of the legs pointing outward and some of the legs pointing inwards, such as in figure \ref{fig:circle4ptsomeinout}. (If only one leg is pointing inwards or outwards then the contribution will be zero because $\tr(T^a) = 0$.)

Taken together, one-loop amplitudes can be written as
\begin{equation}
\begin{aligned}
    & \mathcal{A}^\text{1-loop}(1, \ldots, n) = \sum_{\sigma \in S_n / \mathbb{Z}_n} N_c \tr(T^{\sigma(1)} \ldots T^{\sigma(n)} ) A^{\text{1-loop}}_{n;1}(\sigma(1), \ldots, \sigma(n)) \\
    & + \sum_{j=2}^{\lfloor\frac{n}{2} \rfloor + 1} \sum_{\sigma \in S_n /\mathbb{Z}_{j-1} \times \mathbb{Z}_{n-j+1} }  \tr(T^{\sigma(1) } \ldots T^{\sigma(j-1)}) \tr(T^{\sigma(j)} \ldots T^{\sigma(n)}) A^{\text{1-loop}}_{n;j} (\sigma(1) \ldots \sigma(n)).
\end{aligned}
\end{equation}
All of the single-trace terms come from the diagrams where all legs are pointing outward or inward. (The inward diagrams just amount to color-order-reversed versions of the outward diagrams. In order to compare them directly, one can turn an inward-diagram into an outward-diagram by moving one of the edges of the loop through `infinity' on the plane, which flips the circle inside-out without changing any orientations or incurring extra signs.) The double trace terms meanwhile come from diagrams where some legs point out and some point in.

The factor of $N_c$ in front of the single-trace terms comes from the fact that the inner or outer ring with no external particles contributes $\Tr(I) = N_c$. This term dominates in the large $N_c$ limit. 

The single-trace partial amplitudes $A_{n;1}$ are once again computed from planar, color ordered, Feynman diagrams where in the Feynman rules one drops the $if^{abc}$ vertex factors. The double-trace partial amplitudes $A_{n;j}$ for $j \geq 2$ are determined from $A_{n;1}$ by the formula
\begin{equation}\label{Anjeq}
    A_{n;j}^{\text{1-loop}}(1,2,\ldots,j\!-\!1;j, \ldots, n) = (-1)^{j-1} \sum_{\sigma \in \text{COP}\{\alpha \} \{\beta\} } A_{n;1}^{\text{1-loop}}(\sigma(1), \ldots, \sigma(n)).
\end{equation}
Here, $\alpha = \{j-1, \ldots, 2, 1\}$ and $\beta = \{j, \ldots, n\}$. Note the ordering of the set $\alpha$ is reversed. The color ordered permutations $\text{COP}\{\alpha\} \{\beta\}$ is the set of all permutations that respects the cyclic color ordering of both $\alpha$ and $\beta$, with $n$ held fixed in the last position. If $\alpha = \{3,2,1\}$ and $\beta = \{4,5\}$, then all the permutations in $\text{COP}\{\alpha\} \{\beta\}$ are
\begin{equation}
\begin{array}{cccc}
    (3,2,1,4,5),\; & (3,2,4,1,5),\; & (3,4,2,1,5),\; & (4,3,2,1,5),  \\
    (2,1,3,4,5),\; & (2,1,4,3,5),\; & (2,4,1,3,5),\; & (4,2,1,3,5),  \\
    (1,3,2,4,5),\; & (1,3,4,2,5),\; & (1,4,3,2,5),\; & (4,1,3,2,5).
\end{array}
\end{equation}

Let's give the proof of \eqref{Anjeq}. All one-loop diagrams look like trees coming off of a central loop, which we were instructed to flip to be inwards-pointing if the vertex circle at the base of the tree was counterclockwise.

It is clear by inspection that the form of \eqref{Anjeq} is basically correct, because all the diagrams that get added to the double-trace partial amplitudes have trees on an inside track (the $\alpha$ trees) and trees on an outside track (the $\beta$ trees). The factor of $(-1)^{j-1}$ comes from the signs involved in flipping the inside trees to the outside where they can be compared with the single-trace diagrams which are being summed over.

However there is one thing we have to check in order to prove \eqref{Anjeq}. We have to check that if there are Feynman diagrams in $A_{n;1}$ where some of the elements of $\alpha$ and $\beta$ belong to the same tree, then they are ``projected out,'' summing to zero, by the sum over $\text{COP}\{\alpha\} \{\beta\}$. This is because the only kinds of diagrams that contribute to the double-trace terms are diagrams where the trees with all-$\alpha$ legs point inwards and the trees with all-$\beta$ legs point outwards.

\begin{figure}[h]
    \centering
    \includegraphics[width=0.9\linewidth]{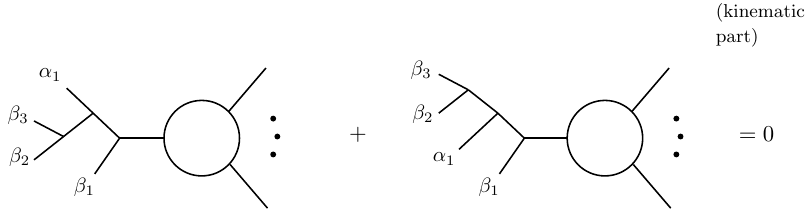}
    \caption{Partial amplitudes that have trees with leaves in both $\alpha$ and $\beta$ will cancel when summed over $\text{COP}\{\alpha\}\{\beta\}$ due to the antisymmetry of the stripped vertex. Here is one such cancelling pair.}
    \label{fig:cancellingpair}
\end{figure}

This is thankfully not hard to see. Consider a tree where some of the nodes belong to $\alpha$ and some to $\beta$. Notice that the trees can all be paired up in such a way that they sum to zero, due to the antisymmetry of the kinematic Feynman rules. An example is drawn in figure \ref{fig:cancellingpair}.

When we group the trees into pairs that cancel, those trees can be thought of as differing by elements of $\text{COP}\{\alpha\}\{\beta\}$ where one of the, say, elements of $\alpha$ has moved past a block of elements in $\beta$ at the highest level in the tree. The double-trace structure is invisible to such moves, but the color-stripped kinematic diagrams have a relative minus sign between the the two. Because any Feynman diagram with both $\alpha_i$'s and $\beta_i$'s sharing a tree is projected out by this sum, we have now proven \eqref{Anjeq}.

We conclude the section by noting that the suggestive-looking double-lined diagrams with some external lines pointed inwards and some pointed outwards have an interpretation in string theory as vertex operators being placed on the inner-boundary or outer-boundary of an annulus-shaped worldsheet \cite{Bern:1990ux}.

\section{Derivation of Berends-Giele SDG current}\label{sec:berendssdg}

In this appendix we show that the tree formula for the all-plus SDG Berends-Giele current \eqref{JGtree}, copied below
\begin{equation}\label{JGtreecopy}
\begin{aligned}
    \mJ_{\rm G}(1,\ldots,n) = \left(-\frac{\kappa}{2}\right)^{n-1} \left( \prod_{a = 1}^n \frac{1}{\lr{\alpha a}^4} \right) \sum_{\text{trees}} \sum_{\text{edges } (i j)} \frac{[ij]}{\langle i j \rangle} \lr{\alpha i}^2 \lr{\alpha j}^2
\end{aligned}
\end{equation}
solves the recursion relation \eqref{gravrecursion}, also copied below
\begin{equation}\label{gravrecursioncopy}
    p_N^2 \mJ_{\rm G}(N) = -\frac{\kappa}{2}\sum_{A, B | A \cup B = N }   \langle \alpha | p_A p_B | \alpha \rangle^2 \mJ_{\rm G}(A) \mJ_{\rm G}(B).
\end{equation}
We will show this by plugging \eqref{JGtreecopy} into both sides of the equation. We'll use a proof from \cite{mypaper}. Another diagrammatic proof can be found in \cite{Hasuwannakit:2025agr,Krasnov:2013wsa}.

On the RHS of \eqref{gravrecursioncopy}, we have two factors of disjoint trees $\mJ_{\rm G}(A)$ and $\mJ_{\rm G}(B)$ we are summing over, where one of the trees has nodes in the set $A$ and the other tree has nodes in the set $B$.

\begin{figure}[h]
    \centering
    \includegraphics{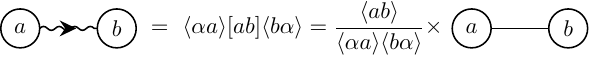}
    \caption{Definition of squiggly edge.}
    \label{fig:newrules1}
\end{figure}

However, there is also a factor of $\lr{\alpha|p_A p_B|\alpha}^2$ in the sum. Let's try to understand this factor graphically. Note that
\begin{equation}
    \lr{\alpha | p_A p_B |\alpha} = \sum_{a \in A} \sum_{b \in B}\lr{\alpha a}[ab]\lr{b \alpha}.
\end{equation}
Pictorially, we can think of the factor $\lr{\alpha | p_A p_B |\alpha}^2$ as drawing two ``squiggly'' edges from one tree with nodes in $A$ to another tree with nodes in $B$, where each squiggly edge from $a \in A$ to $b \in B$ corresponds to a factor of $\lr{\alpha a}[ab]\lr{b \alpha}$. See figure \ref{fig:newrules1}.

A squiggly edge is the same as a normal edge, but multiplied by a factor of $\frac{\lr{ab} }{\lr{\alpha a} \lr{b \alpha}}$. Note that the squiggly edge is oriented, as it is antisymmetric under the interchange $a \leftrightarrow b$. 

\begin{figure}[h]
    \centering
    \includegraphics{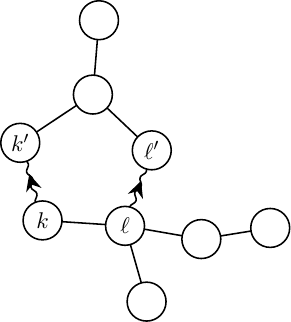}
    \caption{Pictorial representation of the RHS of \eqref{gravrecursioncopy}.}
    \label{fig:cycleproof1}
\end{figure}

Our sum of disjoint trees can now be thought of as a sum of graphs with a single cycle in them, where there are two squiggly edges on the cycle. If $(kk')$ and $(\ell\ell')$ are two distinct squiggly edges in the cycle, then we write the RHS of \eqref{gravrecursioncopy} as
\begin{equation}
    \text{RHS} = \left(-\frac{\kappa}{2} \right)^{n-1} \left( \prod_{a=1}^n \frac{1}{\lr{\alpha a}^4 }\right) \sum_{\substack{\text{one loop}\\\text{graphs}}} \left( \sum_{\text{edges } (i,j)} \frac{[ij]}{\lr{ij}} \lr{\alpha i}^2 \lr{\alpha j}^2 \right)  2 \sum_{\substack{(kk') \in \text{cycle} \\ (\ell\ell') \in \text{cycle} \\ (kk') \neq (\ell\ell')} }\frac{\lr{kk'}}{\lr{\alpha k} \lr{ \alpha k'}} \frac{\lr{\ell \ell'}}{\lr{\alpha \ell} \lr{ \alpha \ell'}}.
\end{equation}
The factor of 2 comes from the indistinguishability of the edges $(kk')$ and $(\ell \ell')$. (Note that if we include $(kk')$, $(\ell \ell')$ as a term in the sum, then we don't also include $(\ell \ell')$, $(kk')$ in the sum.) Note that both squiggly edges must point from the same component to the other, i.e. from an $A$ tree to a $B$ tree. We draw one such graph from ``RHS'' in figure \ref{fig:cycleproof1}.

\begin{figure}[h]
    \centering
    \includegraphics{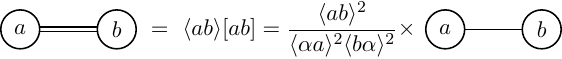}
    \caption{Definition of double-lined edge}
    \label{fig:newrules2}
\end{figure}

Let us now look at the LHS of \eqref{gravrecursioncopy} with the tree formula plugged in for $\mJ_{\rm G}(N)$. Note that
\begin{equation}
    p_N^2 = \sum_{\substack{a \in N, b \in N \\a \neq b}} [ab]\lr{ab}
\end{equation}
so we can think of the quantity $p_N^2 \mJ_{\rm G}(N)$ as a sum over tree graphs where we have added in one extra edge $(ab)$ which has its own special factor of $[ab]\lr{ab}$. We will draw this as a double-lined edge, shown in figure \ref{fig:newrules2}.
Note that the double-lined edge is a standard edge multiplied by $\frac{\lr{ab}^2}{\lr{\alpha a}^2 \lr{\alpha b}^2 }$. 

\begin{figure}[h]
    \centering
    \includegraphics{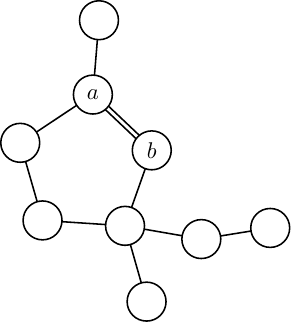}
    \caption{Pictorial representation of the LHS of \eqref{gravrecursioncopy}.}
    \label{fig:cycleproof2}
\end{figure}

When we add the double-lined edge into the pre-existing tree graph from $\mJ_{\rm G}(N)$, the edge closes the tree into a one loop graph where $(ab)$ is an edge in the cycle. The LHS of \eqref{gravrecursioncopy} is then
\begin{equation}
    \text{LHS} = \left(-\frac{\kappa}{2} \right)^{n-1} \left( \prod_{a=1}^n \frac{1}{\lr{\alpha a}^4 }\right) \sum_{\substack{\text{one loop}\\\text{graphs}}} \left( \sum_{\text{edges } (ij) } \frac{[ij]}{\lr{ij}} \lr{\alpha i}^2 \lr{\alpha j}^2 \right) \sum_{(ab) \in \text{cycle}} \left( \frac{\lr{ab}}{\lr{\alpha a} \lr{ \alpha b}} \right)^2.
\end{equation}
We draw one such graph from ``LHS'' in figure \ref{fig:cycleproof2}.

We can therefore show that the LHS equals the RHS if
\begin{equation}\label{cycle2}
    \sum_{(ab)\in \text{cycle}} \left( \frac{\lr{ab}}{\lr{\alpha a} \lr{ \alpha b}} \right)^2 \overset{?}{=} 2 \sum_{\substack{(kk') \in \text{cycle} \\ (\ell\ell')\in \text{cycle} \\ (kk') \neq (\ell\ell')}} \frac{\lr{kk'}}{\lr{\alpha k} \lr{ \alpha k'}} \frac{\lr{\ell\ell'}}{\lr{\alpha \ell} \lr{ \alpha \ell'}}.
\end{equation}
However, from the Schouten identity \eqref{schouten3}, we know that
\begin{equation}\label{cycleab}
    \sum_{(ab)\in \text{cycle}} \frac{\lr{a b}}{\lr{\alpha a} \lr{ \alpha b}} = 0
\end{equation}
and squaring the above equation gives \eqref{cycle2}, completing the proof of \eqref{JGtreecopy}. (Note that $(kk')$ and $(\ell \ell')$ both point from one connected component to the other. We need to reverse the direction of one arrow in order to make contact with the cyclic cancellation in \eqref{cycleab}.  This introduces a minus sign necessary to make the equality \eqref{cycle2} work.)

\section{Derivation of collinear-limits recursion relation for SDG current}\label{sec:collinearsdg}

In this appendix we write down the proof of the collinear-limit based recursive formula for the all-plus gravity current, \eqref{JGpole}. Proofs can also be found in \cite{Krasnov:2013wsa,Delfino:2014xea,mypaper,Guevara:2025tsm}.

We write the formula again below.
\begin{equation}\label{JGpolecopy}
\begin{aligned}
    &\mJ_{\rm G}(|1\rangle, |1], \ldots, |n\rangle, |n] ) \\
    &= - \frac{\kappa}{2} \, \sum_{i = 1}^{n-1} \frac{\lr{\alpha i}^2}{\lr{\alpha n}^2} \frac{[ni]}{\lr{ni}} \mJ_{\rm G}(|1\rangle, |1], \ldots, |i \rangle, |i] + \frac{\lr{\alpha n}}{\lr{\alpha i}} |n], \ldots, |n\!-\!1\rangle, |n\!-\!1]).
\end{aligned}
\end{equation}
We will prove this using the tree-formula \eqref{JGtreecopy}.

Let us think, diagrammatically, how to represent the RHS of \eqref{JGpolecopy}. How can we represent a special node with angle bracket $|i\rangle$ and square bracket $|i] + \frac{\lr{\alpha n}}{\lr{\alpha i}} |n]$? Looking at the tree formula \eqref{JGtreecopy}, we see that the square brackets only appear in the numerator. So, if some other node $j$ connects to this special node, then the square bracket term in the numerator is $[ji] + \frac{\lr{\alpha n}}{\lr{\alpha i}} [jn]$. This term can be thought of as the sum of \textit{two} different trees, one where $j$ connects to a node $i$ and one where it more or less connects to a node $n$.

However, because the whole term from this edge, accounting for the denominator, is $\frac{1}{\lr{ji}} ([ji] + \frac{\lr{\alpha n}}{\lr{\alpha i}} [jn])$, we see that the tree where $j$ connects to $n$ differs from a normal edge in two ways: (1) the denominator is $\lr{ji}$ instead of $\lr{jn}$, and (2) there is also a factor of $\frac{\lr{\alpha n}}{\lr{\alpha i}}$  included.

\begin{figure}[h]
    \centering
    \includegraphics{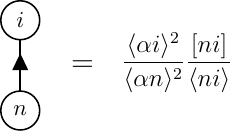}
    \caption{Arrowed edge with special rule.}
    \label{fig:arrow1}
\end{figure}

Let us, therefore, define a new type of edge. We draw it in figure \ref{fig:arrow1}. It will have an arrow that points from $n$ to an adjacent node $i$. We say the edge itself corresponds to a factor of $\frac{\lr{\alpha i}^2}{\lr{\alpha n}^2}\frac{[ni]}{\lr{ni}}$, which comes from the prefactor in \eqref{JGpolecopy}.

\begin{figure}[h]
    \centering
    \includegraphics{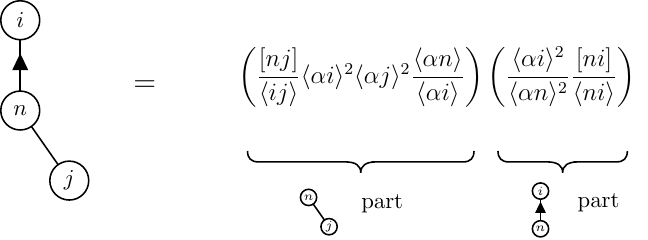}
    \caption{Rule for nodes which connect to $n$, when $n$ has an arrowed edge.}
    \label{fig:arrow2}
\end{figure}

The special rule for this edge is that, if some other node $j$ connects to $n$, then that edge corresponds to a factor of $\frac{[nj]}{\lr{ij}} \lr{\alpha i}^2 \lr{\alpha j}^2 \times \frac{\lr{\alpha n}}{\lr{\alpha i}}$. This differs from a normal edge because $|n\rangle$ is replaced with $|i\rangle$ and we additionally multiply by a factor of $\frac{\lr{\alpha n}}{\lr{\alpha i}}$. The arrow should be thought of as redirecting the angle bracket in the denominator from $|n\rangle$ to $|i\rangle$. See figure \ref{fig:arrow2}.

So, we can rewrite the RHS of \eqref{JGpole} as a sum over trees with \textit{one arrowed edge} that points from $n$ to some adjacent node $i$. ($i$ can be any adjacent node because we are summing over $i$ in \eqref{JGpole}.)

\begin{figure}[h]
    \centering
    \includegraphics{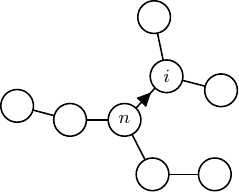}
    \caption{An example of an ``arrowed tree,'' where an arrow points from $n$ to one neighbor.}
    \label{fig:arrowintree}
\end{figure}

We will call all of these trees, with one arrow pointing from $n$ to an adjacent node, ``arrowed trees.'' An example is drawn in figure \ref{fig:arrowintree}.

We can now write
\begin{equation}\label{eqD3}
\begin{aligned}
    -&\frac{\kappa}{2} \, \sum_{i = 1}^{n-1} \frac{\lr{\alpha i}^2}{\lr{\alpha n}^2} \frac{[ni]}{\lr{ni}} \mJ_{\rm G}(|1\rangle, |1], \ldots, |i \rangle, |i] + \frac{\lr{\alpha n}}{\lr{\alpha i}} |n], \ldots, |n\!-\!1\rangle, |n\!-\!1]) \\
    &= \left( - \frac{\kappa}{2} \right)^{n-1} \left( \prod_{a =1}^{n-1} \frac{1}{\lr{\alpha a}^4} \right)\sum_{\substack{\text{arrowed} \\ \text{trees}}} \sum_{\text{edges } (ij)} W_{(ij)}
\end{aligned}
\end{equation}
where we define the ``weights'' $W_{(ij)}$ of the edges by
\begin{equation}
\begin{aligned}
    W_{(ij)} &= \dfrac{[ij]}{\lr{ij}} \lr{\alpha i}^2 \lr{\alpha j}^2  \hspace{0.5 cm} &&\text{if } i\neq n, j \neq n \\
    W_{(ni)} &= \dfrac{\lr{\alpha i}^2}{\lr{\alpha n}^2} \dfrac{[ni]}{\lr{ni}} && \text{if } (ni) \text{ is the arrowed edge } n \to i \\
    W_{(nj)} &= \dfrac{[nj]}{\lr{ij}} \lr{\alpha i}^2 \lr{\alpha j}^2 \dfrac{\lr{\alpha n}}{\lr{\alpha i}}  && \text{ if } n \to i \text{ is the arrowed edge and } j \neq i.
\end{aligned}
\end{equation}

Now, when we sum over arrowed trees, we can also imagine ourselves taking a pre-existing tree, without an arrow, and then summing over all neighbors of $n$, turning the edge from $n$ to that neighbor into an arrowed edge.

\begin{figure}[h]
    \centering
    \includegraphics{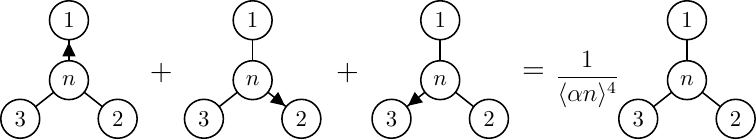}
    \caption{The partial fractions formula implies that if we sum over all possible positions of the arrowed edge, the result is the same as the diagram with no arrow multiplied by $\frac{1}{\lr{\alpha n}^4}$.}
    \label{fig:arrowsum}
\end{figure}

However, it turns out that if we sum over all the placements of the arrow, it just equals the contribution of the tree without the arrow! This is depicted in \ref{fig:arrowsum}. (Note that the RHS contains an extra factor of $\frac{1}{\lr{\alpha n}^4}$ necessary to make \eqref{eqD3} match the desired expression.)

Factoring out all the common prefactors, the validity of this sum boils down to the ``partial fractions'' identity
\begin{equation} \label{eq:usefulID}
    \sum_{i = 1}^m \left( \frac{\langle \alpha i \rangle}{\langle \alpha n \rangle} \right)^{m-1} \frac{1}{\langle  n i \rangle} \prod_{\substack{j = 1\\ j \neq i}}^m \frac{1}{\langle  i j \rangle} = \prod_{i = 1}^m \frac{1}{\langle n i \rangle}.
\end{equation}
This is a well-known identity, although we wrote down an inductive proof of it using the Schouten identity in Appendix A of \cite{Guevara:2025tsm}.

\section{Derivation of SDYM double-off-shell-current}\label{sec:derivationdoubleoffshell}

\subsection{4-dimensional current}\label{sec:derivationdoubleoffshell1}

In this appendix we prove the closed-form expression \eqref{JYMdoublesoln} for the SDYM double-off-shell-current in exactly 4 dimensions. While we struggled to locate where exactly Mahlon wrote down the original proof, we did manage to come up with one nonetheless.

We copy the final result below as
\begin{equation}\label{JYMdoublesoln2}
     J_{\rm YM}(\ell \, ; \, 1, \ldots, n) =(-i \sqrt{2} g)^{n} \frac{-i}{\lr{\alpha 1} \lr{12} \ldots \lr{n-1,n} \lr{\alpha n}} \left( \sum_{k=1}^{n} \frac{ \lr{\alpha |(\ell+p_{1,k} ) k|\alpha } }{(\ell + p_{1,k-1})^2(\ell + p_{1,k})^2 } \right) .
\end{equation}
Here $p_{i,j} \equiv p_i + p_{i+1} + \ldots + p_{j-1} + p_j$ and $p_{i,i-1} = 0$. Specifically, we will show that the above solution for the double-off-shell current satisfies the Berends-Giele recursion relation
\begin{equation}\label{JQNYMrecursion2}
    (\ell+p_N)^2 J_{\rm YM}(\ell\, ;\, N) = i \sqrt{2} g \sum_{\substack{j=0\\ A = \{1, \ldots, j\} \\ B=\{j+1, \ldots ,n\}  }}^{n-1} \lr{ \alpha | (\ell+p_A) p_B | \alpha } J_{\rm YM}(\ell\,;\,A) J_{\rm YM}(B)
\end{equation}
where $N = \{1, \ldots, n\}$ and of course the single-off-shell-current is
\begin{equation}
    J_{\rm YM}(N ) = (-i \sqrt{2} g)^{n-1} \frac{1}{\lr{\alpha 1}} \frac{1}{\lr{12} \ldots \lr{n-1,n}} \frac{1}{\lr{\alpha n}}.
\end{equation}
The base case $(j=0)$ for the double-off-shell current is
\begin{equation}
    J_{\rm YM}(\ell) = \frac{i}{\ell^2}.
\end{equation}
Let us plug the above expressions into \eqref{JQNYMrecursion2} and remove an overall factor of $\frac{(-i \sqrt{2} g)^n (-i)}{\lr{\alpha 1} \lr{12} \ldots \lr{n-1,n} \lr{\alpha n} }$. On the left hand side, we get
\begin{equation}
    (**) = (\ell + p_{1,n})^2  \left( \sum_{k=1}^{n} \frac{ \br{(\ell+p_{1,k} ) k} }{(\ell + p_{1,k-1})^2(\ell + p_{1,k})^2 } \right).
\end{equation}
On the right hand side, we have
\begin{equation}
    (*) = \frac{1}{\ell^2} \br{\ell p_{1,n}}  -  \sum_{\substack{k=1 }}^{n-1} \sum_{k=1}^j \br{(\ell + p_{1,j})p_{j+1,n} } \frac{\lr{j,j+1}}{\lr{\alpha j} \lr{\alpha ,j+1}}   \frac{ \br{(\ell + p_{1,k})k}}{(\ell + p_{1,k-1})^2(\ell + p_{1,k})^2}
\end{equation}
where the first term is the $j = 0$ term.

We shall now manipulate $(*)$ and show that it is equal to $(**)$. We use the color \textcolor{blue}{blue} to highlight what changes from line to line.

First, let us replace $p_{j+1,n}$ with a sum over $b$ where $j < b \leq n$.
\begin{equation}
\begin{aligned}
    (*) &= \frac{1}{\ell^2} \br{\ell p_{1,n}}  -  \sum_{ \textcolor{blue}{ 1 \leq k\leq j<b\leq n} }  \br{(\ell + p_{1,j})\textcolor{blue}{b} } \frac{\lr{j,j+1}}{\lr{\alpha j} \lr{\alpha ,j+1}}   \frac{ \br{(\ell + p_{1,k})k}}{(\ell + p_{1,k-1})^2(\ell + p_{1,k})^2}
\end{aligned}
\end{equation}
Now we break up $\br{(\ell + p_{1,j})b }$ into two parts, with $p_{1,j} = p_{1,k} + p_{k+1,j}$.
\begin{equation}
\begin{aligned}
    (*) &= \frac{1}{\ell^2} \br{\ell p_{1,n}}  -  \sum_{ 1 \leq k\leq j<b\leq n}  \br{(\ell \textcolor{blue}{+ p_{1,k} } ) b } \frac{\lr{j,j+1}}{\lr{\alpha j} \lr{\alpha ,j+1}}   \frac{ \br{(\ell + p_{1,k})k}}{(\ell + p_{1,k-1})^2(\ell + p_{1,k})^2} \\
    & -  \sum_{ 1 \leq k\leq j<b\leq n}  \br{ \textcolor{blue}{p_{k+1,j}} b } \frac{\lr{j,j+1}}{\lr{\alpha j} \lr{\alpha ,j+1}}   \frac{ \br{(\ell + p_{1,k})k}}{(\ell + p_{1,k-1})^2(\ell + p_{1,k})^2} 
\end{aligned}
\end{equation}
Now we do the $j$ sum in the first line, which we can do with the telescoping Schouten (eikonal) identity, and replace $p_{k+1,j}$ with a sum over $a$ where $k < a \leq j$ in the second line.
\begin{equation}
\begin{aligned}
    (*) &= \frac{1}{\ell^2} \br{\ell p_{1,n}}  -  \sum_{\textcolor{blue}{ 1 \leq k<b\leq n}}  \br{(\ell + p_{1,k}) b } \textcolor{blue}{\frac{\lr{k b}}{\lr{\alpha k} \lr{\alpha b}} }   \frac{ \br{(\ell + p_{1,k})k}}{(\ell + p_{1,k-1})^2(\ell + p_{1,k})^2} \\
    & -  \sum_{ \textcolor{blue}{1 \leq k<  a \leq j<b\leq n} } \br{ \, \textcolor{blue}{a}\,  b } \frac{\lr{j,j+1}}{\lr{\alpha j} \lr{\alpha ,j+1}}   \frac{ \br{(\ell + p_{1,k})k}}{(\ell + p_{1,k-1})^2(\ell + p_{1,k})^2} 
\end{aligned}
\end{equation}
Now do the $j$ sum in the second line, again with the telescoping Schouten (eikonal) identity, to get
\begin{equation}
\begin{aligned}
    (*) &= \frac{1}{\ell^2} \br{\ell p_{1,n}}  -  \sum_{ 1 \leq k<b\leq n}  \br{(\ell + p_{1,k}) b } \frac{\lr{k b}}{\lr{\alpha k} \lr{\alpha b}}   \frac{ \br{(\ell + p_{1,k})k}}{(\ell + p_{1,k-1})^2(\ell + p_{1,k})^2} \\
    & +  \sum_{ \textcolor{blue}{1 \leq k<  a < b\leq n}} \textcolor{blue}{[a b ] \lr{ab} }  \frac{ \br{(\ell + p_{1,k})k}}{(\ell + p_{1,k-1})^2(\ell + p_{1,k})^2} .
\end{aligned}
\end{equation}
Then the we do the $a$ and $b$ sum in the second line.
\begin{equation}\label{D11}
\begin{aligned}
    (*) &= \frac{1}{\ell^2} \br{\ell p_{1,n}}  -  \sum_{ 1 \leq k<b\leq n}  \br{(\ell + p_{1,k}) b } \frac{\lr{k b}}{\lr{\alpha k} \lr{\alpha b}}   \frac{ \br{(\ell + p_{1,k})k}}{(\ell + p_{1,k-1})^2(\ell + p_{1,k})^2} \\
    & +  \sum_{k=1}^{n-2}  \textcolor{blue}{p_{k+1,n}^2 } \frac{ \br{(\ell + p_{1,k})k}}{(\ell + p_{1,k-1})^2(\ell + p_{1,k})^2}
\end{aligned}
\end{equation}
Now, note the following Clifford identity (proved in appendix \ref{sec:lemma}).
\begin{equation}
    \br{\ell k}  \br{(\ell+k)b}  = \frac{\lr{\alpha k} \lr{\alpha b}}{\lr{kb}} \Big(  (\ell+k)^2 \br{\ell(k+b)} - (\ell+k+b)^2 \br{\ell k} - \ell^2 \br{(\ell+k)b} \Big) 
\end{equation}
Take the identity and shift $\ell \to \ell + p_{1,k-1}$. It then becomes
\begin{equation}
\begin{aligned}
    \br{(\ell+p_{1,k-1})k}  \br{(\ell+p_{1,k-1}+k)b} =&\; \frac{\lr{\alpha k}\lr{\alpha b}}{\lr{kb}} \Big( (\ell + p_{1,k})^2 \br{(\ell + p_{1,k-1})(k+b) } \\
    &\;\;\;\;\;\;\;\;\;\;\;\;\;\;\;\;- (\ell + p_{1,k} + b)^2 \br{(\ell + p_{1,k})k} \\
    &\;\;\;\;\;\;\;\;\;\;\;\;\;\;\;\;- (\ell + p_{1,k-1})^2 \br{(\ell + p_{1,k}) b} \Big).
\end{aligned}
\end{equation}
We now plug this into the first line of \eqref{D11} to get
\begin{equation}\label{F14}
\begin{aligned}
    (*) &= \frac{1}{\ell^2} \br{\ell p_{1,n}}   \\ & \textcolor{blue}{ -  \sum_{ 1 \leq k<b\leq n}   \frac{ (\ell + p_{1,k})^2  \br{(\ell + p_{1,k-1})(\mathbf{k} + b) }  }{(\ell + p_{1,k-1})^2 (\ell + p_{1,k})^2} }
     \\ & \textcolor{blue}{+  \sum_{ 1 \leq k<b\leq n}   \frac{   (\ell + p_{1,k} + b)^2 \br{(\ell + p_{1,k})k} }{(\ell + p_{1,k-1})^2(\ell + p_{1,k})^2} } \\
    & \textcolor{blue}{+  \sum_{ 1 \leq k<b\leq n}   \frac{  (\ell + p_{1,k-1})^2  \br{(\ell + p_{1,k}) b}  }{(\ell + p_{1,k-1})^2(\ell + p_{1,k})^2} } \\
    & +  \sum_{k=1}^{n-2}  p_{k+1,n}^2  \frac{ \br{(\ell + p_{1,k})k}}{(\ell + p_{1,k-1})^2(\ell + p_{1,k})^2}.
\end{aligned}
\end{equation}
Now, take the $k$ in the $k+b$ part of the second term above (which we've bolded to draw attention to it) and put it in the third term. Let's also change the upper limit of the sum in the fifth term from $n-2$ to $n-1$, because the $k=n-1$ summand is zero anyway.
\begin{equation}
\begin{aligned}
    (*) &= \frac{1}{\ell^2} \br{\ell p_{1,n}}   \\ &  -  \sum_{ 1 \leq k<b\leq n}   \frac{ (\ell + p_{1,k})^2  \br{(\ell + p_{1,k-1}) b }  }{(\ell + p_{1,k-1})^2 (\ell + p_{1,k})^2} 
     \\ & +  \sum_{ 1 \leq k<b\leq n}   \frac{   ( (\ell + p_{1,k} + b)^2 \textcolor{blue}{- (\ell + p_{1,k})^2 } ) \br{(\ell + p_{1,k})k} }{(\ell + p_{1,k-1})^2(\ell + p_{1,k})^2}  \\
    & +  \sum_{ 1 \leq k<b\leq n}   \frac{  (\ell + p_{1,k-1})^2  \br{(\ell + p_{1,k}) b}  }{(\ell + p_{1,k-1})^2(\ell + p_{1,k})^2}  \\
    & +  \sum_{k=1}^{ \textcolor{blue}{n-1}}  p_{k+1,n}^2  \frac{ \br{(\ell + p_{1,k})k}}{(\ell + p_{1,k-1})^2(\ell + p_{1,k})^2}
\end{aligned}
\end{equation}
Next we replace in $(\ell + p_{1,k} + b)^2 - (\ell + p_{1,k})^2 = 2 b \cdot (\ell + p_{1,k})$ for the third term. Then, we perform the sums over $b$ in the second, third, and fourth term. We get
\begin{equation}
\begin{aligned}
    (*) &= \frac{1}{\ell^2} \br{\ell p_{1,n}}   \\ &  -  \sum_{\textcolor{blue}{ 1 \leq k<  n} }  \frac{ (\ell + p_{1,k})^2  \br{(\ell + p_{1,k-1}) \textcolor{blue}{\mathbf{p_{k+1,n}}}  }  }{(\ell + p_{1,k-1})^2 (\ell + p_{1,k})^2} 
     \\ & +  \sum_{\textcolor{blue}{ 1 \leq k < n} }  \frac{   \textcolor{blue}{ 2 p_{k+1,n} \cdot (\ell + p_{1,k}) }  \br{(\ell + p_{1,k})k} }{(\ell + p_{1,k-1})^2(\ell + p_{1,k})^2}  \\
    & +  \sum_{\textcolor{blue}{ 1 \leq k < n} }  \frac{  (\ell + p_{1,k-1})^2  \br{(\ell + p_{1,k}) \textcolor{blue}{p_{k+1,n}} }  }{(\ell + p_{1,k-1})^2(\ell + p_{1,k})^2}  \\
    & +  \sum_{1 \leq k < n}  p_{k+1,n}^2  \frac{ \br{(\ell + p_{1,k})k}}{(\ell + p_{1,k-1})^2(\ell + p_{1,k})^2}.
\end{aligned}
\end{equation}
Now take notice of the $p_{k+1,n}$ term that we have bolded. We want this to be $p_{k,n}$. We therefore add and subtract the $k$ term to the whole expression to accomplish this.
\begin{equation}
\begin{aligned}
    (*) &= \frac{1}{\ell^2} \br{\ell p_{1,n}}   \\ &  -  \sum_{ 1 \leq k<  n}   \frac{ (\ell + p_{1,k})^2  \br{(\ell + p_{1,k-1}) \textcolor{blue}{p_{k,n}}  }  }{(\ell + p_{1,k-1})^2 (\ell + p_{1,k})^2} 
     \\ & +  \sum_{ 1 \leq k < n}   \frac{    2 p_{k+1,n} \cdot (\ell + p_{1,k})   \br{(\ell + p_{1,k})k} }{(\ell + p_{1,k-1})^2(\ell + p_{1,k})^2}  \\
    & +  \sum_{ 1 \leq k < n}   \frac{  (\ell + p_{1,k-1})^2  \br{(\ell + p_{1,k}) p_{k+1,n} }  }{(\ell + p_{1,k-1})^2(\ell + p_{1,k})^2}  \\
    & +  \sum_{1 \leq k < n}  p_{k+1,n}^2  \frac{ \br{(\ell + p_{1,k})k}}{(\ell + p_{1,k-1})^2(\ell + p_{1,k})^2} \\
    & + \textcolor{blue}{\sum_{ 1 \leq k<  n}   \frac{ (\ell + p_{1,k})^2  \br{(\ell + p_{1,k-1}) k  }  }{(\ell + p_{1,k-1})^2 (\ell + p_{1,k})^2} }.
\end{aligned}
\end{equation}
Now it's time for some simplifications. One can unite the third, fifth, and sixth term and factor out an overall $(\ell + p_{1,n})^2$ using the relation $(\ell + p_{1,n})^2 = (\ell + p_{1,k-1})^2 + 2 p_{k+1,n} \cdot (\ell + p_{1,k}) + p_{k+1,n}^2$. Let us also cancel out the propagators in the remaining two sums.
\begin{equation}\label{F18}
\begin{aligned}
    (*) &= \frac{1}{\ell^2} \br{\ell p_{1,n}}   \\ &  -  \sum_{ 1 \leq k<  n}   \frac{ \cancel{ (\ell + p_{1,k})^2 } \br{(\ell + p_{1,k-1}) p_{k,n}  }  }{(\ell + p_{1,k-1})^2 \cancel{(\ell + p_{1,k})^2} } \\
    & +  \sum_{ 1 \leq k < n}   \frac{  \cancel{(\ell + p_{1,k-1})^2 } \br{(\ell + p_{1,k}) p_{k+1,n} }  }{ \cancel{(\ell + p_{1,k-1})^2} (\ell + p_{1,k})^2}  \\
    & \textcolor{blue}{+ (\ell + p_{1,n})^2 \sum_{ 1 \leq k<  n}   \frac{   \br{(\ell + p_{1,k-1}) k  }  }{(\ell + p_{1,k-1})^2 (\ell + p_{1,k})^2} }.
\end{aligned}
\end{equation}
Let's now bring together the first two sums. This is trivially a telescoping sum and the lower boundary cancels the first term in our expression.
\begin{equation}
\begin{aligned}
    (*) &= \cancel{\frac{1}{\ell^2} \br{\ell p_{1,n}} }  \\ & \underbrace{ \textcolor{blue}{ +  \sum_{ 1 \leq k<  n} \left(  - \frac{   \br{(\ell + p_{1,k-1}) p_{k,n}  }  }{(\ell + p_{1,k-1})^2 }  +  \frac{   \br{(\ell + p_{1,k}) p_{k+1,n} }  }{ (\ell + p_{1,k})^2}\right) }  }_{  \cancel{-\frac{\br{\ell p_{1,n} }}{\ell^2} } + \frac{\br{ (\ell + p_{1,n-1})n  }}{(\ell +p_{1,n-1})^2} } \\
    & + (\ell + p_{1,n})^2 \sum_{ 1 \leq k<  n}   \frac{   \br{(\ell + p_{1,k-1}) k  }  }{(\ell + p_{1,k-1})^2 (\ell + p_{1,k})^2} 
\end{aligned}
\end{equation}
Let us now clean up the expression and multiply the first term by $1 = \frac{(\ell + p_{1,n})^2}{(\ell + p_{1,n})^2}$ to bring it into a standardized form.
\begin{equation}\label{F20}
\begin{aligned}
    (*) = \frac{(\ell + p_{1,n})^2}{(\ell + p_{1,n})^2} \times \frac{\br{ (\ell + p_{1,n-1})n  }}{(\ell +p_{1,n-1})^2}
    + (\ell + p_{1,n})^2 \sum_{ 1 \leq k<  n}   \frac{   \br{(\ell + p_{1,k-1}) k  }  }{(\ell + p_{1,k-1})^2 (\ell + p_{1,k})^2}
\end{aligned}
\end{equation}
Combining the two terms, we have
\begin{equation}
\begin{aligned}
    (*) &= (\ell + p_{1,n})^2 \sum_{ 1 \leq k \leq   n}   \frac{   \br{(\ell + p_{1,k-1}) k  }  }{(\ell + p_{1,k-1})^2 (\ell + p_{1,k})^2} = (**)
\end{aligned}
\end{equation}
and this completes the proof.

\subsection{$4-2\epsilon$-dimensional current}\label{sec:derivationdoubleoffshell2}

In this section we compute the $4-2\epsilon$ dimensional SDYM double-off-shell current $J_{\rm YM}(L;N)$ \eqref{JYMdoublesolnmu}. This current is defined in by going up ``one step'' in the recursion relation for the double-off-shell current and replacing all the 4-dimensional propagators with the $4-2\epsilon$-dimensional ones. We write
\begin{equation}
    J_{\rm YM}(L \, ; \, N) \equiv \frac{i \sqrt{2} g}{(L + p_N)^2}  \sum_{\substack{j=0\\ A = \{1, \ldots, j\} \\ B=\{j+1, \ldots ,n\}  }}^{n-1} \lr{ \alpha | (\ell+p_A) p_B | \alpha } J_{\rm YM}(\ell\,;\,A) \Big\rvert_{\ell \, \mapsto L} J_{\rm YM}(B).
\end{equation}
Note that the substitution $\ell \mapsto L$ does not affect the numerators of $J_{\rm YM}(\ell ; A)$, only the propagators in the denominator via the relation $(L+p)^2 = (\ell + p)^2 - \mu^2$. Our goal then is to separate out all the extra terms generated by the identity
\begin{equation}
    \frac{(\ell + p)^2}{(L + p)^2} = 1 + \frac{\mu^2}{(L + p)^2}.
\end{equation}

Anticipating this moment, we have conveniently isolated the only steps in the previous subsection where such a division occurs. The first place occurs at equation \eqref{F18}. The second place occurs at the end, after our penultimate equation \eqref{F20}, when we are to divide by $(L + p_N)^2$ which cancels out an overall factor of $(\ell + p_N)^2$. Note also that we, in the final step of our proof in the last section, also multiplied a term by a factor of $1 = \frac{(\ell + p_{1,n})^2}{(\ell + p_{1,n})^2}$, which we must now replace with $1 = \frac{(L + p_{1,n})^2}{(L + p_{1,n})^2}$.

Let us bring all these terms together. We port the terms over from \eqref{F20} in \textcolor{blue}{blue} and write the additional terms generated by the cancellations of \eqref{F18} in \textcolor{purple}{purple}. The result is
\begin{equation}
\begin{aligned}
    J_{\rm YM}(L \; ;  N) &= \frac{(-i \sqrt{2} g)^n (-i) }{\lr{\alpha 1} \lr{12} \ldots \lr{n-1,n} \lr{\alpha n}} \Bigg(  \\
    & \textcolor{blue}{\frac{1}{(L + p_{1,n})^2} \times \frac{\br{ (\ell + p_{1,n-1})n  }}{(L +p_{1,n-1})^2}
    + \frac{(\ell + p_{1,n})^2 }{(L + p_{1,n})^2 }\sum_{ 1 \leq k<  n}   \frac{ \br{(\ell + p_{1,k-1}) k } }{(L + p_{1,k-1})^2 (L + p_{1,k})^2} } \\
    & \textcolor{purple}{ + \frac{\mu^2}{(L+p_{1,n})^2} \sum_{ 1 \leq k<  n}  \left( - \frac{ \br{(\ell + p_{1,k-1} ) p_{k,n}  }  }{(L + p_{1,k-1})^2 (L + p_{1,k})^2  }  +   \frac{   \br{(\ell + p_{1,k} ) p_{k+1,n} }  }{ (L + p_{1,k-1})^2 (L + p_{1,k})^2 }  \right) } \Bigg).
\end{aligned}
\end{equation}
Using the elementary simplification
\begin{equation}
\begin{aligned}
    \br{(\ell + p_{1,k-1}) k} - \br{(\ell + p_{1,k-1}) p_{k,n}} + \br{(\ell + p_{1,k}) p_{k+1,n}} = -\br{ p_{k, n} k}
\end{aligned}
\end{equation}
the expression is equal to
\begin{equation}
\begin{aligned}
    J_{\rm YM}(L \; ;  N) &= \frac{(-i \sqrt{2} g)^n (-i) }{\lr{\alpha 1} \lr{12} \ldots \lr{n-1,n} \lr{\alpha n}} \Bigg( \sum_{ 1 \leq k \leq n}   \frac{ \br{(\ell + p_{1,k-1}) k } }{(L + p_{1,k-1})^2 (L + p_{1,k})^2}  \\
    &  - \frac{\mu^2}{(L+p_{1,n})^2} \sum_{ 1 \leq k \leq  n}   \frac{ \br{ p_{k,n} k } }{(L + p_{1,k-1})^2 (L + p_{1,k})^2}  \Bigg)
\end{aligned}
\end{equation}
which is what we wanted to show.

\subsection{A useful Clifford algebra identity}\label{sec:lemma}

In this appendix we are going to prove a very useful identity. First, it will be useful for us to define a certain spinor object constructed out of four momenta. The object is defined as
\begin{equation}
    \lr{\alpha | p_1 p_2 p_3 p_4 |\alpha} \equiv - \, \alpha_A \, (p_1^\mu \, \sigma_\mu^{A \dot A}) ( p_{2,\nu} \, \sigma^\nu_{B \dot A} ) (p_3^\rho \, \sigma_\rho^{B \dot B}) ( p_{4,\gamma} \, \sigma^\gamma_{C \dot B} )\, \alpha^C.
\end{equation}
Note the minus sign out front. Recalling that $\lr{ij} = \lambda_i^A \lambda_{j A}$, $[ij] = \tlambda_i^{\dot A} \tlambda_{j,\dot A}$, the minus sign is there so that, if all the momenta are on-shell, we get the natural equation
\begin{equation}
\begin{aligned}
    \lr{\alpha | p_1 p_2 p_3 p_4 |\alpha} &\equiv - \alpha_A \, (p_1^\mu \, \sigma_\mu^{A \dot A}) ( p_{2,\nu} \, \sigma^\nu_{B \dot A} ) (p_3^\rho \, \sigma_\rho^{B \dot B}) ( p_{4,\gamma} \, \sigma^\gamma_{C \dot B} )\, \alpha^C \\
    &= - \underbrace{(\alpha_A \lambda_1^A)}_{-1} (\tlambda_1^{\dot A} \tlambda_{2 \dot A}) \underbrace{(\lambda_{2B} \lambda_3^B)}_{-1}(\tlambda_3^{\dot B} \tlambda_{4 \dot B}) \underbrace{(\lambda_{4C} \alpha^C)}_{-1} \\
    &= \lr{\alpha 1} [12]\lr{23} [34] \lr{4\alpha}.
\end{aligned}
\end{equation}
This sign is not present in the two-momentum object because
\begin{equation}
\begin{aligned}
    \lr{\alpha | p_1 p_2 |\alpha} &= \alpha_A \, (p_1^\mu \, \sigma_\mu^{A \dot A}) ( p_{2,\nu} \, \sigma^\nu_{B \dot A} )\, \alpha^B \\
    &= \underbrace{(\alpha_A \lambda_1^A)}_{-1} (\tlambda_1^{\dot A} \tlambda_{2 \dot A}) \underbrace{(\lambda_{2 B}  \alpha^B)}_{-1} \\
    &= \lr{\alpha 1} [12] \lr{2 \alpha}.
\end{aligned}
\end{equation}
Using the Clifford algebra relation $\{ \sigma^\mu, \sigma^\nu\} = 2 \eta^{\mu \nu}$, and accounting for the relative sign in between the four-momentum and two-momentum object, for any momenta $P_1$, $P_2$, $P_3$ we have
\begin{equation}\label{fourmomentumclifford}
\begin{aligned}
    \br{ P_3 P_1 P_2 P_3 } &= -P_3^2 \br{P_1 P_2} + (2 P_3 \cdot P_2) \br{P_1 P_3} - (2 P_3 \cdot P_1) \br{P_2 P_3}.
\end{aligned}
\end{equation}

Now say $p_1^2 = p_2^2 = 0$ and $\ell^2 \neq 0$. We are interested in expanding the quantity $\br{\ell p_1 } \br{(\ell+ p_1) p_2}$, which is a product of two-momentum objects, as a sum of such objects. We first massage the expression as
\begin{equation}
\begin{aligned}
    \br{\ell p_1 } \br{(\ell+ p_1) p_2} 
    &=
    \frac{\lr{1 \alpha} \lr{2 \alpha} }{\lr{12}} \lr{\alpha | (\ell+p_1) p_1 p_2 (\ell+p_1) |\alpha} .
\end{aligned}
\end{equation}
Now we plug in to \eqref{fourmomentumclifford} to get the following equation.
\begin{equation}
\begin{aligned}
    &= \frac{\lr{\alpha 1} \lr{\alpha 2}}{\lr{12}} \left( -(\ell+p_1)^2 \br{p_1 p_2} - 2 (\ell + p_1) \cdot p_2 \br{\ell p_1} + 2 \ell \cdot p_1 \br{(\ell+p_1) p_2 } \right) \\
    &= \frac{\lr{\alpha 1} \lr{\alpha 2}}{\lr{12}} \left( -(\ell\!+\!p_1)^2 \br{p_1 p_2} - ((\ell\!+\!p_1\!+\!p_2)^2 - (\ell\!+\!p_1)^2) \br{\ell p_1} + ( (\ell\!+\!p_1)^2 - \ell^2) \br{(\ell\!+\!p_1) p_2 } \right) \\
    &= \frac{\lr{\alpha 1} \lr{\alpha 2}}{\lr{12}} \big( \!\!-\!(\ell\!+\!p_1)^2 \big(\br{p_1 p_2} \!-\! \br{\ell p_1} \!-\! \br{(\ell+p_1)p_2}\big) \!-\! (\ell\!+\!p_1\!+\!p_2)^2  \br{\ell p_1} \!-\! \ell^2 \br{(\ell\!+\!p_1)p_2 } \big) 
\end{aligned}
\end{equation}
After one last simplification, we have now derived the very handy identity
\begin{empheq}[box=\fbox]{align} \nonumber
  \;\;\; \vphantom{\Big\rvert}&\br{\ell p_1}  \br{(\ell+p_1)p_2} \\ \nonumber
    &= \frac{\lr{\alpha 1} \lr{\alpha 2}}{\lr{12}} \Big(  (\ell+p_1)^2 \br{\ell(p_1+p_2)} - (\ell+p_1+p_2)^2 \br{\ell p_1} - \ell^2 \br{(\ell+p_1) p_2 } \Big) . \;\;\; \vphantom{\Bigg\rvert}
\end{empheq}
\vspace{-0.5 cm}
\begin{equation}\label{lemmatoprove}
\textcolor{white}{.}
\end{equation}

\section{More properties of the SDYM one-loop-all-plus formula}\label{sec:OLAPprop}

In this appendix we review some basic properties of the SDYM one-loop-all-plus single-trace partial amplitude formula 
\begin{equation}\label{YM1LAPformula2}
    A^{\text{1-loop}}_{n;1} =  -\frac{i}{3} \frac{ (-i \sqrt{2} g)^n}{(4 \pi)^2} \frac{1}{\lr{12} \ldots \lr{n1}} \sum_{1 \leq i_1 < i_2 < i_3 < i_4 \leq n} \lr{i_1 i_2}[i_2 i_3] \lr{i_3 i_4} [i_4 i_1].
\end{equation}

\subsection{Cyclic invariance}

The numerator of \eqref{YM1LAPformula2} does not display manifest cyclic invariance amongst the particles due to the sum of the form $1 \leq \ldots \leq n$, so in this subsection we show that cyclic invariance 
\begin{equation}\label{eqG2}
    A^{\text{1-loop}}_{n;1}(1^+,2^+, \ldots, (n-1)^+, n^+) = A^{\text{1-loop}}_{n;1}(n^+,1^+,2^+\ldots,(n-1)^+)
\end{equation}
does hold on the support of momentum conservation. This elaborates on the discussion in \cite{Bern:1993qk}.

First we decompose
\begin{equation}
    \sum_{1 \leq i_1 < i_2 < i_3 < i_4 \leq n} \lr{i_1 i_2}[i_2 i_3] \lr{i_3 i_4} [i_4 i_1] = \frac{1}{2}(E_n + O_n)
\end{equation}
into the parity even and odd pieces
\begin{equation}
    \begin{aligned}
        E_n &= \sum_{1 \leq i_1 < i_2 < i_3 < i_4 \leq n} \tr(\slashed{p_{i_1}} \slashed{p_{i_2}} \slashed{p_{i_3}} \slashed{p_{i_4}} ),\\
        O_n &= \sum_{1 \leq i_1 < i_2 < i_3 < i_4 \leq n} \tr(\slashed{p_{i_1}} \slashed{p_{i_2}} \slashed{p_{i_3}} \slashed{p_{i_4}} \gamma^5 ).
    \end{aligned}
\end{equation}
Here we use
\begin{equation}\label{trtr5}
\begin{aligned}
    \tr(\slashed{p_{1}} \slashed{p_{2}} \slashed{p_{3}} \slashed{p_{4}} ) &= \lr{12}[23]\lr{34} [41] + [12]\lr{23}[34]\lr{41},\\
    \tr(\slashed{p_{1}} \slashed{p_{2}} \slashed{p_{3}} \slashed{p_{4}} \gamma^5 ) &= \lr{12}[23]\lr{34} [41] - [12]\lr{23}[34]\lr{41}.
\end{aligned}
\end{equation}
Note also the relations
\begin{equation}
\begin{aligned}
    \tr(\gamma^\mu \gamma^\nu \gamma^\rho \gamma^\sigma) &= 4 (\eta^{\mu \nu} \eta^{\rho \sigma} - \eta^{\mu \rho} \eta^{\nu \sigma} + \eta^{\mu \sigma} \eta^{\nu \rho} ), \\
    \tr(\gamma^\mu \gamma^\nu \gamma^\rho \gamma^\sigma \gamma^5) &= - 4 i \vep^{\mu \nu \rho \sigma},
\end{aligned}
\end{equation}
which can be shown to be equivalent to the previous identities using Schouten and \eqref{epsilonidentity}.

\begin{figure}[h]
    \centering
    \includegraphics{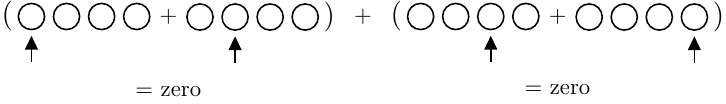}
    \caption{Step in argument that $O_n$ is cyclic invariant}
    \label{fig:fourdots}
\end{figure}

It is easy to see that $E_n$ is cyclically symmetric using the cyclicity of the trace. Therefore we only need to focus on $O_n$. Only the terms where $i_4 = n$ will not obviously be equal in \eqref{eqG2}. So we need to show
\begin{equation}
    \sum_{1 \leq i_1 < i_2 < i_3 \leq n-1} \tr( \slashed{p_{i_1}} \slashed{p_{i_2}} \slashed{p_{i_3}} \slashed{p_n}\gamma^5) \overset{?}{=} \sum_{1 \leq i_1 < i_2 < i_3 \leq n-1} \tr( \slashed{p_n} \slashed{p_{i_1}} \slashed{p_{i_2}} \slashed{p_{i_3}} \gamma^5)
\end{equation}
which using $\slashed{p_n} \gamma^5 = -\gamma^5 \slashed{p_n}$ becomes
\begin{equation}
    \sum_{1 \leq i_1 < i_2 < i_3 \leq n-1} \tr( \slashed{p_{i_1}} \slashed{p_{i_2}} \slashed{p_{i_3}} \slashed{p_n}\gamma^5) \overset{?}{=} 0.
\end{equation}
Using momentum conservation, we can replace $p_n$ with a (negative) sum of all the other momenta. Then using the antisymmetry of the epsilon pseudotensor, we can remove all the momenta which coincide with one of $i_1$, $i_2$, $i_3$. The result is
\begin{equation}
    \sum_{1 \leq i_1 < i_2 < i_3 \leq n-1} \sum_{i_4 \neq i_1, i_2, i_3, n} \vep_{\mu \nu \rho \sigma} p_{i_1}^\mu p_{i_2}^\nu p_{i_3}^\rho p_{i_4}^\sigma \overset{?}{=} 0.
\end{equation}
However, this sum is zero as we'll now explain. Imagine you pick some distinct $i_1$, $i_2$, $i_3$ from $1$ to $n-1$ that are in order. Then you pick a distinct $i_4$ from $1$ to $n-1$. Let's say you want to re-order the indices so that $i_4$ is placed in ascending order with the rest of the indices. So for instance if $i_2 < i_4 < i_3$, you would reorder $\vep (i_1,i_2,i_3,i_4)$ to $- \vep (i_1,i_2,i_4,i_3)$ at the cost of a minus sign. This organizes all the terms in a unique way. After doing this, they all cancel out for the following reason. Picking $i_1,i_2,i_3$ in order and then $i_4$ is the same thing as just picking four distinct integers from $1$ to $n-1$, and \textit{then} singling out one of the integers to be designated as the ``special'' $i_4$. However, the choice of $i_4$ will cancel out with its neighbor as shown in figure \ref{fig:fourdots}, which proves the whole sum is zero.

\subsection{Holomorphic collinear limit of planar partial amplitude}\label{holcolplanarsdym}

In this section we check that for one-loop-all-plus,
\begin{equation}
    \lim_{\lr{i, i+1} \to 0} A^{\text{1-loop}}_{n;1}(1, \ldots, n) = \frac{- i \sqrt{2} g }{\lr{i, i+1}} A^{\text{1-loop}}_{n-1;1}(1, \ldots, |i\rangle, |i] + |i{+}1], \ldots, n).
\end{equation}
Note that analogous relation also holds for the full amplitude. In fact, the appropriate formula holds for the single-trace and double-trace parts of the amplitude independently. (Note \eqref{YM1LAPformula2} was shown to hold for the full collinear limit $\lr{i,i+1} \to 0$, $[i,i+1] \to 0$ in \cite{Bern:1993qk}, which is actually how the all-$n$ formula for $A^{\text{1-loop}}_{n;1}$ was originally conjectured.)

Because \eqref{YM1LAPformula2} is cyclically symmetric, we assume for simplicity that $i \neq n$. In this case, it is easy to check that
\begin{equation}
\begin{aligned}
    \lim_{\lr{i, i+1} \to 0}  A^{\text{1-loop}}_{n;1}(1, \ldots, n) &= \frac{- i \sqrt{2} g }{\lr{i, i+1}} A^{\text{1-loop}}_{n-1;1}(1, \ldots, |i\rangle, |i] + |i{+}1], \ldots, n) \\
    &+ \lim_{\lr{i, i+1} \to 0}  -\frac{i}{3} \frac{ (-i \sqrt{2} g)^n}{(4 \pi)^2} \frac{1}{\lr{12} \ldots \lr{n1}} \sum_{\substack{1 \leq i_1 < i \\  i+1 < i_4 \leq n} } t(i_1, i, i+1, i_4).
\end{aligned}
\end{equation}
The first term is what we desire, and the second term is defined by
\begin{equation}
\begin{aligned}
    \lim_{\lr{i, i+1} \to 0 }t(i_1, i, i+1, i_4) &= \lim_{\lr{i, i+1} \to 0}\sum_{\substack{1 \leq i_1 < i \\  i+1 < i_4 \leq n} } \lr{i_1 i} [i, i{+}1] \lr{i{+}1, i_4}[i_4 i_1] \\
    &= \lim_{\lr{i, i+1} \to 0}[i, i+1] \sum_{\substack{1 \leq i_1 < i \\  i+1 < i_4 \leq n} } \lr{i+1 | i_4 i_1 | i}.
\end{aligned}
\end{equation}
In the full collinear limit $[i,i+1] \to 0$ and this is trivially zero. In the holomorphic collinear limit, however, to show it is zero one must use momentum conservation
\begin{equation}
    \sum_{i+1 < i_4 \leq n} p_{i_4} = - p_i - p_{i+1} - \sum_{1 \leq i_1 < i} p_{i_1}
\end{equation}
and then plug it in the previous formula. The $p_i+p_{i+1}$ will go away when hit by $\langle i+1|$, so
\begin{equation}
    \lim_{\lr{i, i+1} \to 0} t(i_1, i, i+1, i_4) = -\lim_{\lr{i, i+1} \to 0}[i, i+1] \sum_{1 \leq i_1 < i} \sum_{1 \leq i_1' < i} \lr{i+1 | i_1' i_1 | i} = 0.
\end{equation}
This disappears by the antisymmetric identity $\lr{\alpha | PP  |\alpha} = 0$, completing the proof.

\section{Relationship between classical solutions, Berends-Giele currents, and amplitudes}\label{sec:classicalsolns}

In this appendix we will review the connection between tree-level off-shell Berends-Giele currents and classical solutions \cite{LipinskiJusinskas:2026ctz}. For simplicity, consider the scalar $\varphi^3$ theory with action
\begin{equation}
    S[\varphi] = \int d^4 x \left(\frac{1}{2} (\partial \varphi)^2 - \frac{g}{3!} \varphi^3 \right)
\end{equation}
and classical equation of motion
\begin{equation}
    \Box \varphi + \frac{g}{2} \varphi^2 = 0.
\end{equation}
We start by following the usual discussion, found e.g. in \cite{Monteiro:2011pc}. Denoting the sourced action as
\begin{equation}
    S_J[\varphi] = S[\varphi] + \int d^4 x J(x) \, \varphi(x),
\end{equation}
we define the sourced path integral $Z[J]$ and its generating function $W[J]$ via
\begin{equation}
    Z[J] = \int \mD \varphi \, e^{i S_J[\varphi]} = e^{i W[J]}.
\end{equation}
Denoting correlators within the sourced path integral as
\begin{equation}
    \lr{\ldots}_J \equiv \int \mD \varphi \, e^{i S_J[\varphi]} (\ldots)
\end{equation}
we have the standard expression
\begin{equation}
    \lr{\varphi(x_1) \ldots \varphi(x_n)}_0 = (-i)^n \frac{\delta}{ \delta J(x_1)} \ldots \frac{\delta}{ \delta J(x_n)} Z[J] \bvert_{J = 0}.
\end{equation}
The correlators can be expressed as a sum over all Feynman diagrams where the sources are external legs. If we only sum over connected Feynman diagrams instead of all diagrams, then on the RHS $Z$ is replaced with $i W$:
\begin{equation}\label{connectedW}
    \lr{\varphi(x_1) \ldots \varphi(x_n)}_{0, \rm{connected}} = (-i)^n \frac{\delta}{ \delta J(x_1)} \ldots \frac{\delta}{ \delta J(x_n)} i W[J] \bvert_{J = 0}.
\end{equation}
At tree-level (i.e. $\hbar \to 0$) the path integral is equal to its on-shell value. If we define $\phiclJ$ as classical solution which extremizes the sourced action given vanishing boundary conditions (due to the implicit $i \epsilon$) then we have
\begin{equation}
    W[J] = S_J[\phiclJ]
\end{equation}
which implies
\begin{equation}
    \frac{\delta}{\delta J(x)} W[J] = \phiclJ(x) + \int d^4 y \left( \frac{\delta S_J}{\delta \varphi(y)}[\phiclJ ]\right) \frac{\delta \phiclJ(y)}{\delta J(x)}.
\end{equation}
Because the action is stationary when varied around the classical solution,
\begin{equation}
    \frac{\delta}{\delta J(x)} W[J] = \phiclJ(x).
\end{equation}
Plugging this into \eqref{connectedW}, we have
\begin{equation}\label{connectedeq}
    \lr{\varphi(x) \varphi(x_1) \ldots \varphi(x_n)}_{0, \rm{connected}} = (-i)^n \frac{\delta}{\delta J(x_1)} \ldots \frac{\delta}{\delta J(x_n)} \phiclJ(x) \bvert_{J = 0}.
\end{equation}
Fourier transforming both $\varphi(x_i)$ and $J(x_i)$ via
\begin{equation}
    \varphi(x) = \int \frac{d^4 k}{(2 \pi)^4} \;  e^{i k \cdot x} \tvarphi(k), \hspace{1 cm} J(x) = \int \frac{d^4 k}{(2 \pi)^4} \;  e^{i k \cdot x} \tJ(k),
\end{equation}
we have from the chain rule
\begin{equation}
    \int d^4 x \, e^{-i k \cdot x} \frac{\delta}{\delta J(x)} = (2 \pi)^4 \frac{\delta}{\delta \tJ(-k)}.
\end{equation}
If we Fourier transform \eqref{connectedeq} and amputate the external legs,
\begin{equation}\label{connectedJ}
\begin{aligned}
    &(-i)^n (k_1^2 \ldots k_n^2) \lr{\varphi(x)  \tvarphi(k_1) \ldots \tvarphi(k_n) }_{0, \rm{connected}} \\
    &= (2 \pi)^{4 n} \; (-k_1^2) \frac{\delta}{\delta \tJ(-k_1)} \ldots (-k_n^2) \frac{\delta}{\delta \tJ(-k_n)} \phiclJ(x) \bvert_{\tJ = 0}.
\end{aligned}
\end{equation}

We now wish to remove the currents from the RHS of the equation. Because we are differentiating with respect to $\tilde{J}$ and then setting them equal to zero, let us define a current $j(x)$ to be a sum of plane waves multiplied by infinitesimal parameters $\epsilon_i$ along with the prefactors $(-k_i^2)$:
\begin{equation}
    j(x) = \sum_{i=1}^n \epsilon_i (-k_i^2)  e^{-i k \cdot x} , \hspace{1 cm} \tilde{j}(k) = \sum_{i=1}^n \epsilon_i (- k_i^2 )\,  (2 \pi)^4 \delta^{(4)}(k + k_i).
\end{equation}
With this current, \eqref{connectedJ} becomes
\begin{equation}
    (-i)^n (k_1^2 \ldots k_n^2) \lr{\varphi(x)  \tvarphi(k_1) \ldots  \tvarphi(k_n) }_{0, \rm{connected}} = \frac{\partial}{\partial \epsilon_1} \ldots \frac{\partial}{\partial \epsilon_n} \phiclj(x) \Big\vert_{\epsilon_i = 0\; \forall i}.
\end{equation}
If we want to solve
\begin{equation}
    \Box \varphi_{{\rm{cl}},j} + \frac{g}{2} \varphi_{{\rm{cl}},j}^2 = j
\end{equation}
we can Taylor expand the solution in the coupling in $g$ via
\begin{equation}
    \varphi_{{\rm{cl}},j} = \sum_{m=0}^\infty g^m \varphi_{{\rm{cl}},j}^{(m)}
\end{equation}
and recursively solve for the coefficients through
\begin{equation}
    \varphi^{(0)}_{{\rm{cl}},j} = \sum_{i=1}^n \epsilon_i e^{-i k_i \cdot x}, \hspace{1 cm} \Box \varphi_{{\rm{cl}},j}^{(m+1)} = - \frac{1}{2} \sum_{l=0}^m \varphi_{{\rm{cl}},j}^{(l)}\varphi_{{\rm{cl}},j}^{(m-l)}.
\end{equation}
The above equation is nothing more than the Berends-Giele recursion relation. Note that $\varphi^{(m)}_{{\rm{cl}},j}$ is of order $\mathcal{O}(\epsilon^m)$.

Let us now take the on-shell limits $k_i^2 \to 0$ for $i = 1 , \ldots, n$. In this limit, $\varphi_{{\rm{cl}},j}$ clearly becomes a solution to the unsourced equations of motion.

We define the ``perturbiner expansion'' $\varphi^{\rm(ptb)}$ to be $\varphi_{{\rm{cl}},j}$ where we take the $\epsilon_i$'s to square to zero.
\begin{equation}
    \varphi^{\rm(ptb)} \equiv \lim_{k_i^2 \to 0} \varphi_{{\rm{cl}},j} \Big\vert_{\epsilon_i^2 = 0 \; \forall i}
\end{equation}
In other words, the perturbiner expansion is a solution to the equation of motion where the leading (free) term is a sum of on-shell plane waves multiplied by infinitesimal parameters
\begin{equation}
    \varphi^{\rm(ptb)} = \sum_{i=1}^n \epsilon_i \; e^{-i k_i \cdot x} + \ldots
\end{equation}
and the higher-order terms are found by recursively plugging the solution into the interacting equations of motion. While the infinitesimal parameters are defined such that each individual parameter squares to zero, the product of distinct parameters is not zero. (If you wish, this definition is adopted for economy because we will be setting all $\epsilon_i$ to zero at the end of the calculation anyway after we differentiate by each of them once.)
\begin{equation}
    (\epsilon_i)^2 = 0 \hspace{1 cm} \epsilon_{i} \epsilon_{i'} \neq 0 \text{ for } i\neq i'
\end{equation}

Having defined the perturbiner classical solutions, we may now write
\begin{equation}
    \lim_{k_i^2 \to 0} (-i)^n( k_1^2 \ldots k_n^2 )\lr{\varphi(x) \tvarphi(k_1) \ldots \tvarphi(k_n) }_{0, \rm{connected}} = \frac{\partial}{\partial \epsilon_1} \ldots \frac{\partial}{\partial \epsilon_n} \varphi^{\rm (ptb)}(x) \Big\vert_{\epsilon_i = 0\; \forall i}.
\end{equation}
Fourier transforming $x$ to the off-shell momentum $k$, we get
\begin{equation}
    \boxed{ \lim_{k_i^2 \to 0} (-i)^n (k_1^2 \ldots k_n^2) \lr{\tilde{\varphi}(k) \tvarphi(k_1) \ldots  \tvarphi(k_n) }_{0, \rm{connected}} =  \frac{\partial}{\partial \epsilon_1} \ldots \frac{\partial}{\partial \epsilon_n} \tilde{\varphi}^{\rm (ptb)}(k) \Big\vert_{\epsilon_i = 0\; \forall i}.} 
\end{equation}
The LHS of the above equation is the definition of a Berends-Giele current with one off-shell leg, and the RHS is the top piece of a classical perturbiner solution. Furthermore, if we amputate the final leg and take $k^2 \to 0$, we get
\begin{equation}\label{Aptb}
    \boxed{ \mathcal{A}(k, k_1, \ldots, k_n) = - i \lim_{k^2 \to 0} k^2 \frac{\partial}{\partial \epsilon_1} \ldots \frac{\partial}{\partial \epsilon_n} \tilde{\varphi}^{\rm (ptb)}(k) \Big\vert_{\epsilon_i = 0\; \forall i} }
\end{equation}
in the all-outgoing convention.

Let's look at the example of the 3-pt amplitude in $\varphi^3$ theory. The full perturbiner expansion with two seed particles is easily solved to be
\begin{equation}
    \tilde{\varphi}^{\rm (ptb)}(x) = \epsilon_1 e^{-i k_1 \cdot x} +  \epsilon_2 e^{-i k_2 \cdot x} + \epsilon_1 \epsilon_2 \frac{g }{(k_1 + k_2)^2} e^{-i (k_1 + k_2) \cdot x}
\end{equation}
and plugging this in to \eqref{Aptb} we get
\begin{equation}
    \mathcal{A}(k, k_1, k_2) = - i g \, (2 \pi)^4 \delta^{(4)}(k_1 + k_2 + k)
\end{equation}
where in the above expression we've restored the often-suppressed delta function which is produced naturally from the formula.

This concludes our discussion of perturbiner expansions and amplitudes, with \eqref{Aptb} being the equation which relates the two. Having said this, there is also another way that tree-level amplitudes can be calculated using perturbiners. It involves simply plugging the perturbiner expansion into the action itself. The position $x$ of the solution is integrated over spacetime when computing the action, producing the delta function, so only the on-shell legs comprise the particles in the amplitude. (See for instance \eqref{MHVgenerating} and \eqref{MHVbeautiful} for the gravity MHV amplitude.) It is interesting how there are two quite different formulae which both relate perturbiners to amplitudes.

\bibliography{oneloopbib.bib}
\bibliographystyle{jhep}

\end{document}